\documentclass[a4paper]{aa}
\usepackage{txfonts,graphicx,natbib,aalongtable}
\usepackage{color}
\bibpunct{(}{)}{;}{a}{}{,}
\newcommand{\pq}{\ensuremath{P_Q}}
\newcommand{\pu}{\ensuremath{P_U}}
\newcommand{\pqe}{\ensuremath{P_Q^{\rm \,E}}}
\newcommand{\pue}{\ensuremath{P_U^{\rm \,E}}}
\newcommand{\pl}{\ensuremath{P}}
\newcommand{\px}{\ensuremath{P_{X}}}
\newcommand{\thetae}{\ensuremath{\theta^{\rm \,E}}}
\newcommand{\thetaG}{\ensuremath{\theta^{\rm \,G}}}

\newcommand{\fo}{\ensuremath{f^\parallel}}
\newcommand{\fe}{\ensuremath{f^\perp}}
\newcommand{\lmax}{\ensuremath{\lambda_\mathrm{\,max}}}

\newcommand{\pmax}{\ensuremath{P_\mathrm{\,max}}}

\newcommand{\snr}{\ensuremath{S/N}}

\usepackage{amstext}

\begin{document}

\title{Large Interstellar Polarisation Survey}

\subtitle{LIPS I: FORS2 spectropolarimetry in the Southern Hemisphere\thanks{Table 2 and reduced data (intensity and polarisation spectra)
    are available in electronic form at the CDS via anonymous ftp to cdsarc.u-strasbg.fr (130.79.128.5)
    or via http://cdsweb.u-strasbg.fr/cgi-bin/qcat?J/A+A/}}
       \author{
        Stefano~Bagnulo        \inst{1}
       \and
        Nick~L.J.~Cox          \inst{2}
        \and
        Aleksandar~Cikota      \inst{3}
        \and
        Ralf Siebenmorgen      \inst{3} 
        \and
        Nikolai~V. Voshchinnikov\inst{4}
        \and
        Ferdinando Patat     \inst{3}
        \and
        Keith T. Smith    \inst{5}
        \and
        Jonathan V.\ Smoker \inst{6}
        \and
        Stefan Taubenberger \inst{3,7}
        \and
        Lex Kaper \inst{2}
        \and
        Jan Cami \inst{8}
        \and
        the LIPS collaboration}

\institute{Armagh Observatory and Planetarium,
           College Hill,
           Armagh BT61 9DG,
           Northern Ireland, U.K.
           \email{sba@arm.ac.uk}
           \and
           Anton Pannekoek Institute for Astronomy,
           University of Amsterdam, NL-1090 GE Amsterdam, The Netherlands
           \and
           European Southern Observatory,
           Karl-Schwarzschild-Str. 2,
           D-85748 Garching b. M\"{u}nchen,
           Germany
           \and
           Sobolev Astronomical Institute,
           St. Petersburg University,
           Universitetskii prosp. 28,
           St. Petersburg, 198504 Russia
           \and
           AAAS Science International, 
           Clarendon House, Clarendon Road, 
           Cambridge CB2 8FH, UK
           \and
           European Southern Observatory, Alonso de Cordova 3107,
           Casilla 19001, Vitacura, Santiago 19, Chile
           \and
           Max-Planck-Institut f\"{u}r Astrophysik, Karl-Schwarzschild-Str. 1,
           D-85741 Garching b. M\"{u}nchen, Germany
           \and
           Department of Physics and Astronomy and Centre for Planetary
           Science and Exploration (CPSX), The University of Western
           Ontario, London, ON N6A 3K7, Canada
           }

\authorrunning{S. Bagnulo et al.}
\titlerunning{Large Interstellar Polarisation survey (LIPS). I.}

\date{Received: 2017-06-28 / Accepted: 2017-10-02}
\abstract{
  Polarimetric studies of light transmitted through
  interstellar clouds may give constraints on the properties of the
  interstellar dust grains. Traditionally, broadband linear
  polarisation (BBLP) measurements have been considered an important
  diagnostic tool for the study of the interstellar dust, while
  comparatively less attention has been paid to spectropolarimetric
  measurements. However, spectropolarimetry may offer stronger
  constraints than BBLP, for example by revealing narrowband
  features, and by allowing us to distinguish the contribution of
  dust from the contribution of interstellar gas. Therefore, we have
  decided to carry out a Large Interstellar Polarisation Survey (LIPS)
  using spectropolarimetric facilities in both hemispheres. Here we
  present the results obtained in the Southern Hemisphere with the
  FORS2 instrument of the ESO Very Large Telescope.  Our spectra cover the
  wavelength range 380--950\,nm at a spectral resolving power of about 880.
  We have produced a publicly available catalogue of 127 linear
  polarisation spectra of 101 targets. We also provide the
  Serkowski-curve parameters, as well as the wavelength gradient of the
  polarisation position angle for the interstellar polarisation along
  76 different lines of sight. In agreement with previous literature,
    we found that the best-fit parameters
    of the Serkowski-curve are not independent of each other. However, the
    relationships that we obtained are not always consistent with what
has been    found in previous studies.
  }

\maketitle
\section{Introduction}
A full characterisation of the interstellar medium (ISM) is of crucial
importance to understand the chemical history of the Universe and
various fundamental astrophysical processes, such as the formation and
evolution of the stars and their planetary systems. Most of the heavy
elements are present in the ISM in the form of dust
grains. Non-spherical dust grains, aligned along a preferred
direction, linearly polarise the stellar background radiation at
wavelengths close to the grain size. Therefore, linear polarimetric
measurements of stars that are not intrinsically polarised may set
constraints on the chemical composition, shape, and size distribution
of the interstellar dust, as well as on the mechanisms responsible
for dust grain alignment.

Polarimetry of the interstellar medium started with the discovery work
by \citet{Hiltner49}, \citet{Hall49}, and \citet{Dom49}, culminating
later into major surveys such as those by \citet{Seretal75}, who
performed broadband linear polarimetry (BBLP) in the optical filters
$UBVR$, and by \citet{Whietal92}, who extended BBLP measurements to
the full optical and near-IR $UBVRIJHK$ filter set \citep[the
    results of many BBLP surveys have been collected in a catalogue
    by][]{Heiles00}. BBLP surveys have also  been carried out in just
  one filter \citep[e.g.][]{Leroy93,Sanetal11}, and have been used to
  explore the boundaries of the Local Cavity
  \citep{Leroy99,Sanetal11} and its possible interaction with a
  nearby superbubble seen in the direction of the Galactic Centre
  \citep{Sanetal11}. However, since the fraction of interstellar
polarisation depends on wavelength, multi-filter measurements are
crucial for the physical interpretation of the observations. The
amount of linear polarisation is linked to the column density (in very
simple terms: the higher the amount of aligned dust along the line of sight, the higher the
polarisation) and to the optical properties of the dust, while the
wavelength at which the polarisation reaches its maximum depends on
the grain size (or size distribution) \citep{Caretal73}. The
polarisation position angle, if due to a single interstellar cloud
with a homogeneous magnetic field, does not change with wavelength,
and is determined by the direction of the interstellar magnetic field
in the plane of the sky.

While BBLP surveys -- historically more common than
spectropolarimetric surveys -- give substantial insight into the
behaviour of interstellar polarisation, better constraints may be
obtained with a more refined sampling of the polarisation as a function
of wavelength. Compared to multicolour BBLP measurements,
spectropolarimetry may allow us to better test if the well-known
Serkowski curve provides good fits to the interstellar polarisation,
or if polarisation spectra present narrowband features that deviate
from it, maybe systematically. Another advantage of spectropolarimetry
versus multicolour BBLP is that the former technique allows us to
distinguish the contribution of  dust from that of gas.
Finally, we note that although much effort has been dedicated to
the observations of the ISM, many lines of sight have never been
observed in polarimetric mode.

Spectropolarimetry is therefore the obvious choice for modern
polarimetric surveys, when this mode is available at the telescope.
Some initial low-resolution polarisation spectra were presented first
by \citet{WolNan71} and later by \citet{WolSmi84}.  \citet{Andetal96}
presented spectropolarimetric data in the UV and optical range for 38
objects at a spectral resolution of $\sim 20$\,nm. In the past 25
years, optical spectropolarimetry of the ISM was obtained with the
observations of more than 200 stars at the Pine Bluff Observatory near
Madison, Wisconsin, USA, using the HPOL spectropolarimeter. The
observed spectral range was 320--775\,nm, then extended to 1050\,nm,
with a spectral resolution of 2.5\,nm (resolving power $R~\sim$~300),
then refined to 1\,nm ($R \sim$~775). The database of processed data
is publicly available\footnote{See {http://www.sal.wisc.edu/HPOL/} and 
CDS catalogue\\ \ {\tt J/AcA/49/59}}, and
details of the project are given in \citet{Weitenbeck99} and
\citet{Meaetal12}.  More in-depth insight into the shape, size
distribution, and composition of interstellar grains can come from
detailed physical models that treat the interaction of dust with
photons. For example, \citet{Sieetal14} presented state-of-the-art models
that simultaneously reproduce the observed linear-polarisation curve
and the UV/visual extinction curve. These recent studies show that
models are more advanced than most ISM polarisation data and make
clear that medium-resolution linear
spectropolarimetry with a high signal-to-noise ratio (S/N) brings a new perspective to the issue of the dust
size distribution and composition in the Galactic ISM. Hence a
spectropolarimetry survey is required to develop a comprehensive
picture of the exact details of the cosmic dust -- initially limited
to Galactic environments -- and to exhaustively test the new models in
their applicability to a statistically representative sample (rather
than concentrating on a handful of targets).

These considerations have prompted us to initiate a new
spectropolarimetric survey, the Large Interstellar Polarisation Survey
(LIPS) in both hemispheres with the FOcal Reducer/low-dispersion
Spectrograph (FORS2) instrument of the Very Large Telescope (VLT) of
the European Southern Observatory (ESO) and the Intermediate-dispersion Spectrograph and Imaging System (ISIS) instrument of the
William Herschel Telescope (WHT). LIPS fills a gap in our knowledge of
dust properties in diffuse lines of sight, the gaseous content of which
is being studied with the ESO Diffuse Interstellar Band Large
Exploration Survey \citep[EDIBLES,][]{Coxetal17}.  The main
goal of EDIBLES is to create a detailed inventory of interstellar gas
in diffuse clouds and to derive accurate physical conditions in order
to determine molecular properties of the unidentified diffuse
interstellar bands \citep{Sarre06,CamCox14}.  Together with existing
UV spectroscopy, LIPS and EDIBLES constitute an extensive dataset for
interstellar extinction and polarisation, which provides a
comprehensive simultaneous view of the properties of the interstellar
gas and dust in diffuse interstellar lines of sight.

This paper presents the FORS2 survey in the Southern Hemisphere.

\section{Observations}\label{sec:obs}

\subsection{Target list}

\begin{figure}
  \includegraphics*[scale=0.27,trim={2.5cm 0.5cm 0.3cm 0.8cm},clip]{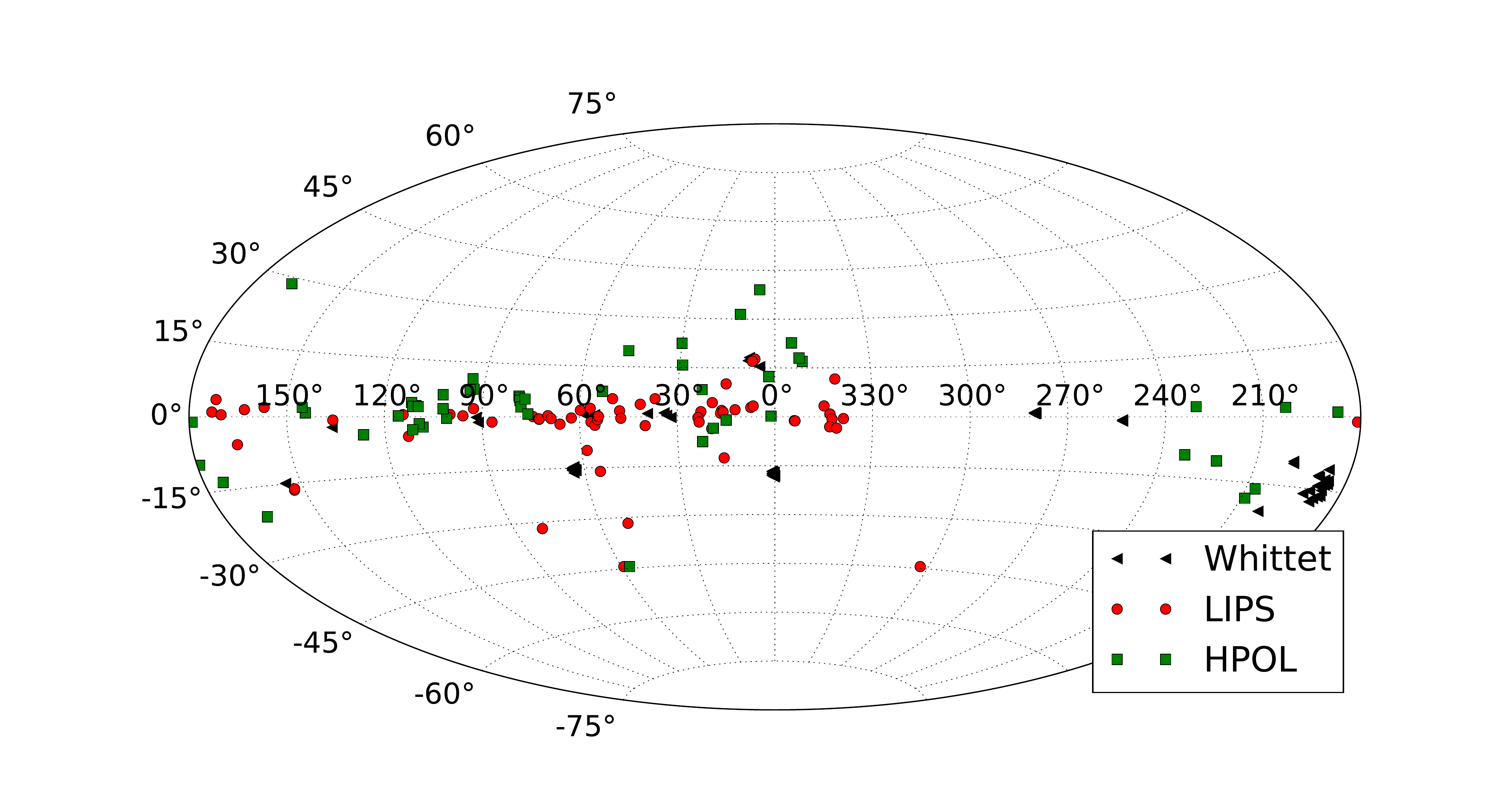}
  \caption{\label{Fig_Coordinates} Location of our targets in the Galaxy (red circles)
    compared to those of other major surveys: HPOL (blue circles) and \citet{Whietal92}
  (black triangles).}
\end{figure}

The targets of the FORS2 survey are a set of O- and B-type stars selected from the EDIBLES target
list, excluding the brightest stars ($V \leq 5$ mag) and those
accessible with ISIS at the WHT (La Palma), although a few northern
targets ($V \sim 8$~mag) were included for cross-calibration and
characterising instrumental effects.  The
initial set of selected targets covers a wide range of interstellar
properties in terms of dust extinction (E($B-V$) = 0 to 2~mag, $R_V$ =
1.5 to 6), molecular abundances ( f($H_2$)= 0.0 to 0.8), radiation
fields, gas, and dust temperatures \citep{Coxetal17}.  To
expand the parameter space of dust and environments, additional
sightlines in the Magellanic Clouds were added, as well as a few faint
($V \geq 9$) highly reddened Galactic targets that are too faint for
the EDIBLES survey.  The full target list/observing log is given in
Table~\ref{Tab_Log}, and their position in Galactic coordinates is
shown with red symbols in Fig.~\ref{Fig_Coordinates}, which also
shows the position of the targets of two other major surveys.

\subsection{Instrument and instrument settings}

FORS2 is a multi-purpose instrument capable of imaging and
low-resolution spectroscopy in the optical; it is equipped with polarimetric
optics. It is attached at the Cassegrain focus of Unit\,1, Antu, of
the ESO VLT of the Paranal Observatory. The instrument is described in
\citet{AppRup92} and \citet{Appetal98}.

The polarimetric optics are arranged according to the optical design
described by \citet{App67}. The polarimetric module consists of a
retarder waveplate ($\lambda/2$ for observations of linear
polarisation, and $\lambda/4$ for observations of circular
polarisation) followed by a Wollaston prism.  The Wollaston prism
splits the beam into two components, one linearly polarised along the
principal plane of the Wollaston (the parallel beam \fo), and one
linearly polarised perpendicularly to that plane (the perpendicular
beam \fe). Following the scheme by \citet{Scaretal83}, a
6.8\arcmin\ $\times$ 22\arcsec\ Wollaston mask prevents the
superposition of the two beams.

\subsubsection{Grism}

Grism 300\,V was employed for our observations. With a
0.5\arcsec\ slit width, it provides a spectral resolving power of about
880. The order-separating filter GG435 may be used to cut off the
incoming radiation at $\lambda \la 435$\,nm to avoid second-order contamination
from the blue part of the spectrum into the red part (at $\lambda \ga
650$\,nm). However, since our targets are mostly highly reddened, when
planning the observations, we assumed that it would probably
be safe to observe without order-separating filter, and cover the
wavelength range down to $\sim 360$\,nm, which is
the blue limit of the spectral range covered by the instrument
\citep[for further considerations on the effects of second-order
  contamination on spectropolarimetric measurements, see][]{Patetal10}.
For each star, we therefore decided to obtain an observing
series with order-separating filter GG435, and one observing series
without it. Observing both with and without order-separating filter
allowed us to verify this hypothesis (see Sect.~\ref{Sect_Results}),
and to have some redundancy in the range $435-650$\,nm that
would serve as a quality check. Our expectation was that
the polarised spectra obtained with and without order-separating
filter in the wavelength range $435-650$\,nm would be identical within
photon-noise uncertainties, while at $\lambda \ga 650$\,nm, a possible
difference between the two would be ascribed to contamination from the
second order in the spectra obtained without filter. In fact, since CCD
sensitivity drops dramatically at $\lambda \la 380$\,nm, the useful
spectral range covered by our observations was finally $380-930$\,nm
with no filter, and $435-930$\,nm with the filter GG435 inserted.

\subsubsection{CCD and CCD readout}
In service mode, the detector installed in FORS2 is a mosaic composed
of two 2k $\times$ 4k E2V CCDs (pixel size of $15 \times 15\,\mu$m$^2$,
pixel scale of 0.125\arcsec), optimised for the red. Most of our
observations were obtained without rebinning, in order to maximise the
\snr\ and avoid saturation towards the bright targets, while for certain
fainter targets, we used the $2\times2$ rebinning mode. 

\subsection{Observing strategy and data reduction}
Our observations were obtained setting the retarder waveplate at four
position angles: $0\degr$, $22.5\degr$, $45\degr$, and $67.5\degr$. This
implementation of the so-called ``beam-swapping technique'' allowed
us to minimise instrumental effects, as explicitly suggested in the
FORS1/2 manual and thoroughly discussed by \citet{Bagetal09},
for instance.

Stokes~$Q$ and $U$ parameters are defined as in \citet{Shurcliff62}.
In the following, we consider the ratios $Q/I$ and $U/I$,
adopting the notation
\begin{equation}
\pq' = \frac{Q}{I}\ \ {\rm and}\ \ \pu' = \frac{U}{I} \;.
\label{Eq_Pq_Pu_Def}
\end{equation}
The reduced Stokes parameters $\pq'$\ and $\pu'$\ were measured by combining
the photon counts of the parallel and perpendicular beams
(\fo\ and \fe, respectively) observed at retarder
waveplate positions $\alpha = 0^\circ$, $22.5^\circ$, $45^\circ$, and
$67.5^\circ$, as given by the following formula:
\begin{equation}
\px' = 
\frac{1}{2}
\Bigg\{ \left(\frac{\fo - \fe}{\fo + \fe}\right)_{\alpha= \phi_0} -
        \left(\frac{\fo - \fe}{\fo + \fe}\right)_{\alpha=\phi_0+45^\circ}
\Bigg\} \ ,
\label{Eq_Stokes}
\end{equation}
where $\phi_0 = 0$ if $X = Q$ and $\phi_0=22.5^\circ$
if $X = U$.
The uncertainty due to photon-noise on $\px'$ is 
\begin{equation}
\begin{array}{rcl}
\sigma^2_{P_X} & = &
  \left(\left(\frac{\fe}{(\fo + \fe)^2}\right)^2 \sigma^2_{\fo} +
  \left(\frac{\fo}{({\fo + \fe})^2}\right)^2
    \sigma^2_{\fe}\right)_{\alpha=\phi_0 } + \\
             &   &
  \left(\left(\frac{\fe}{({\fo + \fe})^2}\right)^2 \sigma^2_{\fo} +
  \left(\frac{\fo}{({\fo + \fe})^2}\right)^2
  \sigma^2_{\fe}\right)_{\alpha=\phi_0+45^\circ} \; . \\
\end{array}
\label{Eq_Sigma_QU}
\end{equation}
Assuming that the fluxes in both beams and
for all retarder waveplate positions are similar, we obtain
\begin{equation}
\sigma_{\px} = \frac{1}{S/N}\ ,
\end{equation}
where {\it S/N} is the signal-to-noise ratio accumulated
in all beams used to calculate \px.

\subsection{Correction for the chromatism (wavelength dependence) 
of the retarder waveplate and for instrumental polarisation}
The position angle of the retarder waveplate identifies the direction
along which the electric field propagates with no phase shift (this is
the fast axis of the retarder waveplate).  This direction is slightly
wavelength dependent, and this deviation introduces a rotation of the
measured position angle of the polarisation that may be corrected as
explained, for example in Sect.~4.2 of
\citeauthor{Bagetal09} (\citeyear{Bagetal09}; see also
\citealt{Patetal11}), that is,
\begin{equation}
  \begin{array}{rcl}
  \pq &=& \phantom{-}\pq' \cos2\epsilon + \pu' \sin2\epsilon \\
  \pu &=& -\pq' \sin2\epsilon + \pu' \cos2\epsilon \\
\end{array}
\label{Eq_Rotation}
,\end{equation}
where $\epsilon$ is the deviation angle tabulated in the instrument
webpages. The fraction of linear polarisation
\begin{equation}
\pl\ = \sqrt{\pq^2+\pu^2} 
\end{equation}
is not affected by the problem of the
chromatism of the retarder waveplate (i.e. $\pl=\pl'$), while the
measured position angle $\theta'$ and the ``true'' position angle $\theta$
are simply related by
\begin{equation}
  \theta = \theta' - \epsilon \; .
  \label{Eq_theta}
\end{equation}

\citet{Fosetal07} and \citet{Sieetal14} noted that FORS2
spectropolarimetric observations are affected by a small amount of
wavelength-dependent instrumental polarisation (mainly in $Q$), which
has been further analysed and analytically quantified by \citet{Ciketal17}
for grism 300V as
\begin{equation}
  \begin{array}{rcl}
    \pq^{\rm \ instr.} = 9.66\,\times\,10^{-8}\,\lambda + 3.29\,\times\,10^{-5}\\
    \pu^{\rm \ instr.} = 7.28\,\times\,10^{-8}\,\lambda - 4.54\,\times\,10^{-4}\\
  \end{array}
\label{Eq_Instr}
,\end{equation}
where $\lambda$ must be expressed in \AA. \citet{Ciketal17} derived Eqs.~(\ref{Eq_Instr})
from observations taken with the instrument position angle $\chi$ set to zero
(i.e. with the parallel beam of the Wollaston prism parallel to the
north celestial meridian). Various experiments presented
by \citet{Sieetal14} show that FORS2 instrumental polarisation tends to be constant
in the instrument reference system; when we wish to express the Stokes
parameters in a reference system other than the instrumental one, instrumental
polarisation therefore has to be corrected before rotating the Stokes
parameters. We assume that Stokes parameters $\pq$ and $\pu$
were measured
with the instrument position angle on sky set to a value {\it PA}, counted counterclockwise
from the north celestial meridian (for instance, {\it PA} could correspond to
the parallactic angle), and already corrected for the
chromatism of the retarder waveplate. To transform the Stokes parameters $\pq',\pu'$
into a new reference system with its reference direction parallel to the north celestial
meridian, we have to apply
\begin{equation}
  \begin{array}{rcl}
    \pqe &=& \phantom{-}(\pq - \pq^{\rm \ instr.})\cos 2\chi + (\pu - \pu^{\rm \ instr.})\sin 2\chi \\
    \pue &=&          - (\pq - \pq^{\rm \ instr.})\sin 2\chi + (\pu - \pu^{\rm \ instr.})\cos 2\chi \; ,\\
  \end{array}
  \label{Eq_Corr}
\end{equation}
where \px\ are given by Eqs.~(\ref{Eq_Rotation}), $\px^{\rm \ instr.}$ are
given by Eqs.~(\ref{Eq_Instr}), and $\chi$ is the angle, counted
counterclockwise and looking at the source by which the old
(instrument) reference system has to be rotated to coincide with the
new (equatorial) reference system \citep[e.g.][]{Lanetal07}, that is,
$\chi = -PA$,\footnote{The instrument position angle counted
  counterclockwise on sky from the north celestial meridian,
  $\chi_{\rm instr}$, is given by the FORS2 fits-header keyword
  ADA.POSANG taken with the opposite sign ($\chi_{\rm instr} =
  -$\,ADA.POSANG), hence in Eqs.~(\ref{Eq_Corr}),  $\chi
  = $ADA.POSANG should be used.} while the fraction of linear polarisation and its
position angle are given by
\begin{equation}
  \begin{array}{rcl}
  \pl &=& \sqrt{(\pq - \pq^{\rm \ instr.})^2 + (\pu - \pu^{\rm \ instr.})^2} \\
  \thetae &=& \theta + \chi \\
\end{array}
  ,\end{equation}
where $\theta$ is given by Eq.~(\ref{Eq_theta}). 

Our final results can be given either using the reduced Stokes
parameters \pq\ and \pu\ or the fraction of linear polarisation
\pl\ and its position angle $\theta$. The latter representation is more
convenient when the interest is in a map of the polarisation and its
direction, for example, as function of the Galactic coordinates, as is our interest here.
In our case, it is also
convenient to refer the position angle of the polarisation to the
direction of the galactic pole. This angle \thetaG\ may be obtained
for the equatorial position angle \thetae\ as
\begin{equation}
  \thetaG = \thetae + \eta
,\end{equation}
where
\begin{equation}
  \cos \eta = 
  \frac{\cos(\delta) \sin(\delta_{\rm G}) -
                 \sin(\delta) \cos(\delta_{\rm G}) \cos(\alpha-\alpha_{\rm G})}{\cos(b)} \ ,
\end{equation}
where $b$, $\alpha$, and $\delta$ are the galactic latitude, right
ascension, and declination of the star, respectively, and $(\alpha_{\rm
  G},\delta_{\rm G})$ are the right ascension
and declination of the North Galactic Pole.

\section{Analysis}\label{Sect_Analysis}
\begin{figure}
\begin{center}
\scalebox{0.47}{
\includegraphics*[trim={1.7cm 6.2cm 1.0cm 2.6cm},clip]{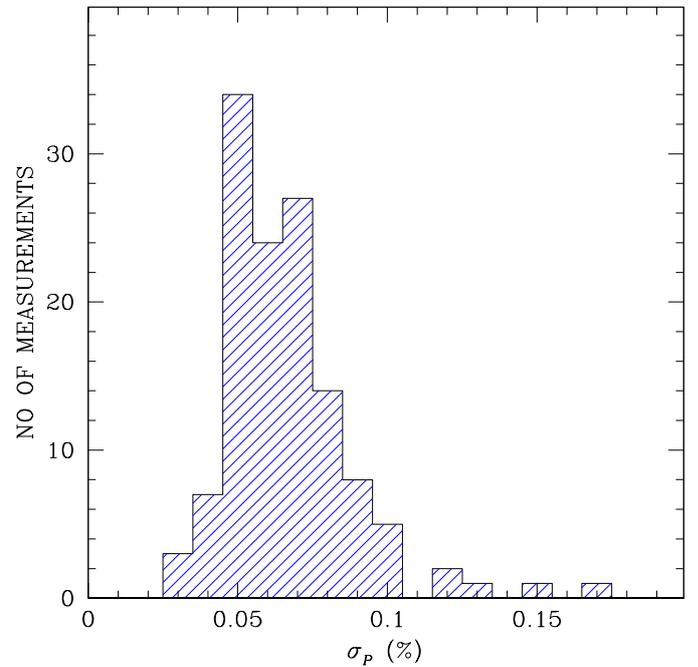}}
\end{center}
\caption{\label{Fig_SNR} Histogram of the photon-noise error bars of the linear polarisation
  \pl\ per nm, in the wavelength range where the \snr\ is highest (typically
  around 500\,nm).}
\end{figure}
Our dataset has low
photon-noise. Figure~\ref{Fig_SNR} shows the histogram of the error
bars of the fraction of polarisation \pl\ per nm, obtained after
averaging out the spectra obtained with and without order-separating
filter. If we were to integrate a broadband filter, these error bars would
decrease to the order of a few units in $10^{-4}$. As discussed below,
these uncertainties are hardly representative of the actual accuracy
of our measurements. In this section we describe how we discovered
various non-photon-noise errors, and the way we have tried to minimise
them.

\subsection{Uncertainties due to data reduction}
\begin{figure}
\begin{center}
\includegraphics*[trim={0.2cm 1.7cm 0.3cm 2.8cm},width=\columnwidth,clip]{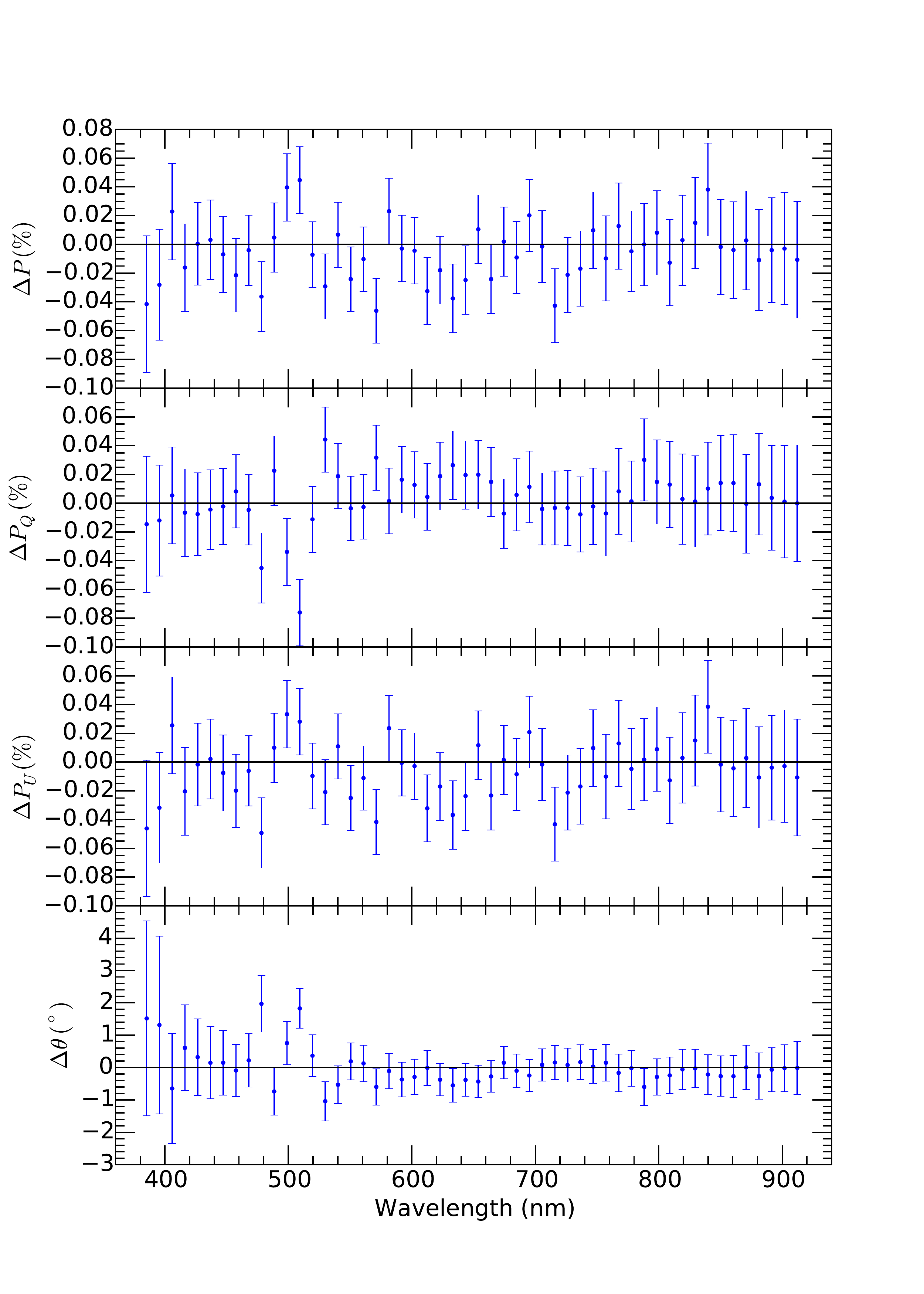}
\end{center}
\caption{\label{Fig_DeltaP} Residuals of independently reduced data, plotted with photon-noise error bars.}
\end{figure}
\citet{Bagetal12} have discussed the noise due to data reduction,
meaning that different but still reasonable choices in the way data
are treated lead to (slightly) different results. In case of measurements with a very high
\snr, these differences may sometimes be comparable to
photon noise, especially in the presence of sharp spectral features (such as
emission lines of the telluric O$_2$ bands). As a first check, we
decided to verify whether the differences between the products of independent
data reduction were random and well within the photon noise. A dozen
observations were reduced independently by two of us (SB and AC), and
results were compared.  The reduction methods differed mainly in the
way background and flux extraction were treated: in one case, no
background was subtracted (since the exposure time was very short) and
wavelength calibration was carried out on the extracted spectra; in
another case, the background was removed and wavelength calibration
performed on the 2D spectra. The final products were found perfectly
consistent in the continuum; an example is shown in
Fig.~\ref{Fig_DeltaP}. The two different reductions led to
discrepant results around the sharp features discussed in
Sect.~\ref{Sect_Spikes}, and they are due to a different spectral
sampling/rebinning.

\subsection{Inconsistencies between consecutive observations}\label{Sect_Incons}
\begin{figure*}
\begin{center}
\scalebox{0.65}{
\includegraphics*[angle=270,trim={0.8cm 0.7cm 0.3cm 0.8cm},clip]{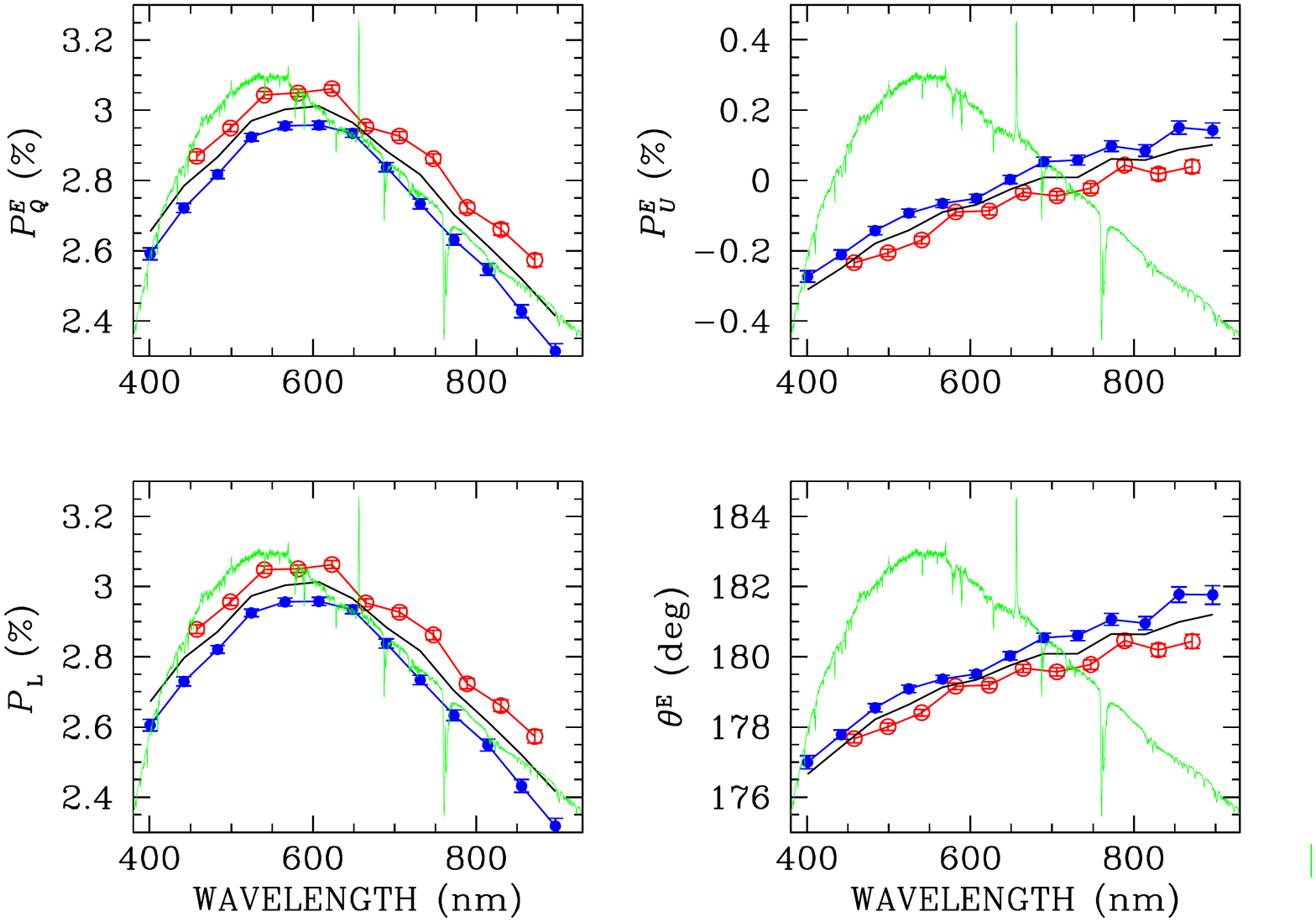}}
\end{center}
\caption{\label{Fig_HD149404} Observations of star HD\,149404 obtained
  on 2015-05-12. The top
  left panel shows \pqe\, , the top right panel shows \pue, the bottom
  left panel shows \pl,\ and the bottom right panel shows $\theta_E$.
  In each panel, red empty circles refer to the observations obtained
  with order-separating filter, and blue solid circles refer to the
  observations obtained without order separating filter, with a
  41.4\,nm wavelength bin size.  The black solid line is their average
  obtained from Eqs.~(\ref{Eq_Average}). The green solid line (in arbitrary
  units and identical in
  all panels) shows the stellar flux not
  corrected for instrument+telescope transmission function, and with a
  0.7\,nm bin size.  Error bars due to photon noise are smaller than
  or comparable to the symbol size, but the shape of the SED may help to
  visualise how the photon-noise error bars, which are proportional to the
  inverse of the S/N, change with wavelength.}
\end{figure*}
\begin{figure}
\begin{center}
\scalebox{0.45}{
\includegraphics*[trim={0.8cm 5.6cm 0.3cm 3.4cm},clip]{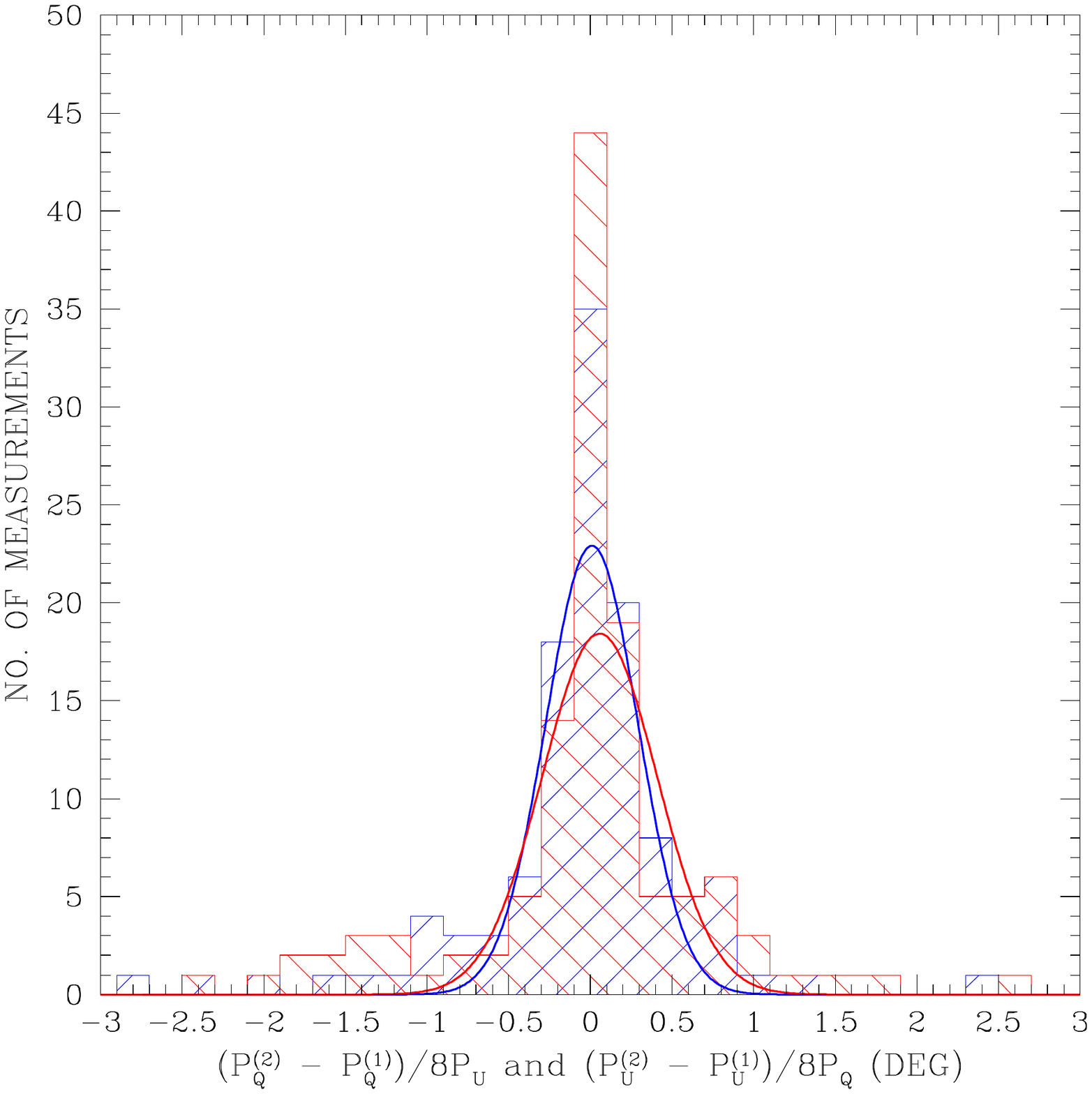}}
\end{center}
\caption{\label{Fig_Histos} Histograms of two distributions used to
  estimate the uncertainty due to an imperfect setting of the
  retarder waveplate (see text) and their best fits with Gaussian
  distributions. Blue lines correspond to the data obtained without
  order-separating filter, and red lines to the data obtained with
  GG435 filter. Best fits were obtained using data with $\vert \pq
  \vert \ge 0.5$\,\% and $\vert \pu \vert \ge 0.5$\,\%, respectively,
  as explained in the text.}
\end{figure}
All our stars were observed consecutively with and without order-separating filter.  In some cases, the
polarisation spectra measured in the wavelength range common to the
two instrument setups were found inconsistent among themselves, in that
the polarisation spectra obtained with and without filter would appear
systematically shifted by a quantity much larger than the photon-noise error
bars. The most dramatic example in our database is represented by the
observations of HD\,149404 obtained on 2015 May 5, shown in
Fig.~\ref{Fig_HD149404}, where the red empty circles refer to the
polarisation measured with the order-separating filter, the blue solid
circles to the polarisation measured without filter, and the solid black
line to their average calculated as explained later in this section. In
this example, the polarisation measured with the order-separating
filter appears systematically higher ($\Delta P \simeq 0.1$\,\%) than
that observed without filter.

HD\,149404 is a detached massive O-star binary (O7.5 If + ON9.7 I),
with a period of 9.81\,d \citep{MasCon79}. Previous literature data
show that HD\,149404 exhibits polarimetric variability, but on a much
longer timescale than the interval of time between our two
observations: \cite{Luna88} obtained BBLP in the B filter, finding that
the fraction of linear polarisation is nearly constant with time, but
the position angle changes periodically, so that the polarisation
variation in the \pq-\pu\ plane shows a nearly circular
pattern. However, it is unlikely that the differences observed in two
observing series obtained within a few minutes from each other have an
origin intrinsic to the source. We also note that the pairs of
observations of the same star obtained on 2016 January 30 are fully
consistent among themselves, therefore we conclude that the
variability observed on 2015 May 5 has an instrumental origin.

Numerical simulations performed by Cikota et al. (2017, presentation
at the ESO Calibration Workshop) suggested that this short-timescale
variability could be linked to an inaccurate setting of the
retarder waveplate.

The idea of Cikota et al. (2017) may be tested analytically starting
from Eqs.~(34) and (36) of \citet{Bagetal09}, assuming that the
chromatism of the retarder waveplate and of the Wollaston prism are
perfectly corrected, that is, by setting $\delta \alpha = \delta \beta=0$ in
the expressions of $\widehat{\mathcal{G}}_X$ of Eqs.~(34), assuming a
perfect flat-field ($\delta h = 0$) and assuming
that the phase shift introduced by the retarder waveplate is exactly
180\degr\ ($\delta \gamma = 0$), but associating with each position
$\alpha$ of the retarder waveplate an indetermination $\delta_\alpha$.
Using Eqs.~(34) and (36) of \citet{Bagetal09}, setting $N=1,$ and considering
  the retarder waveplate at the angles
$0\degr + \delta_{0}$,
$22.5\degr + \delta_{22.5}$,
$45\degr + \delta_{45}$,
$67.5\degr + \delta_{67.5}$,
we find that the measured parameters
$\widehat{\pq}, \widehat{\pu}$ are connected to the ``true''
parameters \pq\ and \pu\ via
\begin{equation}
  \begin{array}{rcl}
    \widehat{\pq} &=& \pq + 4\pu\ \frac{1}{2}\left(\delta_{0}  +\delta_{45}\right)\\
    \widehat{\pu} &=& \pu - 4\pq\ \frac{1}{2}\left(\delta_{22.5}+\delta_{67.5}\right) \ . \\ 
\end{array}
\label{Eq_QUsigma}
\end{equation}
Equations~(\ref{Eq_QUsigma}) show that the inaccuracy of the
setting of the retarder waveplate is responsible for a cross-talk
from \pq\ to \pu\ and vice versa. According to the signs of the actual reduced
Stokes parameters and of the difference between actual and nominal
position of the retarder waveplate, the measurement may be either
an over-estimate or an under-estimate of the ``true'' polarisation. In particular, we can write
\begin{equation}
  \begin{array}{rcl}
{\rm rms}^2\left(\widehat{\pq}-\pq\right) &=& 4 \pu^2 \left({\rm rms}^2(\delta_{0})  +{\rm rms}^2(\delta_{45})\right)\\
{\rm rms}^2\left(\widehat{\pu}-\pu\right) &=& 4 \pq^2 \left({\rm rms}^2(\delta_{22.5})+{\rm rms}^2(\delta_{67.5})\right)\\
  \end{array}
,\end{equation}
and assuming a constant rms $\delta$ for all waveplate positions, we obtain
\begin{equation}
  \begin{array}{rcl}
{\rm rms}(\widehat{\pq}) &=& 2 \sqrt{2}\, \vert \pu \vert\,  {\rm rms}(\delta)\\
{\rm rms}(\widehat{\pu}) &=& 2 \sqrt{2}\, \vert \pq \vert\,  {\rm rms}(\delta) \; . \\
  \end{array}
\end{equation}
The FORS2 User Manual states that the accuracy of the setting of
the retarder waveplate is $\sim 0.1\degr$. This means, for instance,
that the expected absolute uncertainty in the \pq\ measurement of a
source with $\pu = 1$\,\% is 0.005\,\%.  We note that by using
Eqs.~(34) and (36) of \citet{Bagetal09}, we have considered the case in
which the reduced Stokes parameters are calculated with the difference
method. Similar results would apply by adopting the ratio method, as
can be verified by using Eqs.~(40) and (41) of \citet{Bagetal09}.  In
the following, we test whether these uncertainties may be responsible for
the observed discrepancies.

In case of two consecutive measurements (1) and (2), we have
\begin{equation}
  \begin{array}{rcl}
{\rm rms}\left(\frac{\widehat{\pq}^{(2)}-\widehat{\pq}^{(1)}}{8 \pu}\right) &=& {\rm rms}(\delta) \\
{\rm rms}\left(\frac{\widehat{\pu}^{(2)}-\widehat{\pu}^{(1)}}{8 \pq}\right) &=& {\rm rms}(\delta)\; . \\
  \end{array}
  \end{equation}

From our dataset we calculated the BBLP value in the $V$ filter (see Sect.~\ref{Sect_Convolution})
and measured the distributions of the quantities
\[
\frac{180}{\pi}\frac{\widehat{\pq}^{\rm (2)} - \widehat{\pq}^{\rm (1)}}{8\pu}\ \ {\rm and} \ \
\frac{180}{\pi}\frac{\widehat{\pu}^{\rm (2)} - \widehat{\pu}^{\rm (1)}}{8\pq}\; ,
\]
where the ``true'' parameters \px\ ($X=Q,U$) were approximated by
$\left(\widehat{\px}^{\rm (1)} + \widehat{\px}^{\rm (2)}\right)/2$.
We report these distributions in form of histograms in
Fig.~\ref{Fig_Histos}. These distributions were found approximately
centred about zero (as expected), but could not be represented by
Gaussian distributions with $\sigma = 0.1^\circ$. We then decided to
remove the points for which the denominator was $\le 0.5$\,\%, finding
the Gaussian best fits with $\sigma = 0.28\degr$ and 0.35\degr,
respectively. Although these values are still $\text{about
three} $ times higher
than the uncertainty declared in the FORS User Manual, our considerations analytically support the conclusion by Cikota et
al. (2017) that an inaccuracy of the setting of the retarder
waveplate is responsible for the discrepancies that we found in our
dataset. 

\subsection{Spikes in the polarisation spectra}\label{Sect_Spikes}
Our polarisation spectra exhibit some departures from a smooth
behaviour, some of which are due to cosmic rays. Cosmic rays are generally
not well removed during flux extraction because models of the line
spread function (LSF) may not be accurate enough compared to the small
polarimetric signals that we intend to measure, leaving some spurious
spikes randomly distributed in wavelength.

Spikes appear more frequently in proximity of sharp spectral lines
(e.g.\ strong H emission lines), or telluric bands (e.g.\ O$_2$-A band
at 760\,nm), in the form of a sharp increase or decrease of the
polarisation, and/or a sudden change in position angle. In several
cases, polarimetric spikes appear in the observations obtained with
the order-separating filter, but not in those obtained without filter,
or vice versa. Only rarely are they present in both datasets, and their
shape and amplitude largely depends on spectral sampling and
rebinning. Most of these features are probably spurious and are caused by a tiny
offset that is introduced either by imperfect wavelength calibration or by
instrument flexures, which creates a difference in the fluxes measured in
the spectral bins of different beams when the flux changes rapidly with wavelength. Similar problems were found in
FORS circular polarisation measurements and were discussed by
\citet{Bagetal13}. However, there might be situations in which the spikes
that we have detected are real. For instance, in stars with emission lines (e.g. Be
stars), the observed linear polarisation is probably due to a combination
of the contribution from interstellar and disc scattering. Strong
emission lines may dilute the radiation polarised in the circumstellar
disc, causing a change in polarisation.  To determine which of these spikes
(if any) are real or if real features are hidden behind
spurious spikes requires observations with much higher spectral
resolution obtained with a very stable instrument, such as ESPaDOnS
at the CHFT or HARPSpol of the La Silla Observatory.

\subsection{Presenting the results}
\subsubsection{Averaging the observations}\label{Sect_Aver}
The uncertainties introduced by random misalignments of the retarder
waveplate discussed in Sect.~\ref{Sect_Incons} may be reduced by
increasing the number of waveplate positions at which the retarder
  waveplate is set during the observing sequence, for example, by considering a
  sequence 0\degr, 22.5\degr, \ldots, 157.5\degr\ instead of
  0\degr, 22.5\degr, 45\degr, and 67.5\degr. Unfortunately, 
our observing strategy was decided before we discovered the accuracy
problem of the retarder waveplate setting, therefore our observations
were carried out using only four positions of the retarder waveplate.
However, the accuracy of our polarisation spectra may still be improved by 
adopting the average of the measurements obtained with and without order-separating filter. The spectral range 380--435\,nm is covered only by
the observations without order-separating filter. In this interval
range, we therefore adopted the sum of the reduced Stokes parameters measured
without filter and the average of the half-difference between the
\pq\ and \pu\ values measured with and without filter in the common
wavelength range ($\lambda \ge 435$\,nm).  In other words, we have adopted
\begin{equation}
\begin{array}{lcl}
    \px = \frac{1}{2}\left(\px^{\rm (2)} + \px^{\rm (1)}\right)    &{\rm if}& \lambda \ge 435\,{\rm nm}\\
    \px = \px^{\rm (2)} +
    \frac{1}{2N}\ \sum_{i=1}^{N}
    \left(\px^{\rm (1)}(\lambda_i)-\px^{\rm (2)}(\lambda_i)\right) &{\rm if}& \lambda < 435\,{\rm nm} \; , \\
\end{array}
\label{Eq_Average}
\end{equation}
where the index (1) refers to the observations obtained with order-separating
filter, and the index (2) refers to those obtained without filter.

\subsubsection{Rebinning}
Most of our targets are very bright, and it is in principle possible to reach
extremely high \snr\ with a very short exposure time. The ADU converter
sets an upper limit to the \snr\ that may be reached per pixel bin, which in general
depends on CCD gain, on the full-well capacity, on the seeing conditions
(when seeing is worse, the flux may be spread over more
pixels, and this increases the flux that may be measured before
saturation is reached, simply by increasing the exposure time), and on the number of
frames that are obtained
per target.  For each frame with 1\arcsec\ seeing conditions and with a converting
factor 1.25\,e$^{-}$\,ADU$^{-1}$, the maximum {\it S/N} that can be reached
in the wavelength bin that corresponds to the dispersion bin is $\sim 800$.
Adding the flux accumulated in two beams and
in two frames, this corresponds to an error bar $\sigma_X \sim
0.06$\,\%. Typically, our frames were exposed at 20-30\,\% of the
maximum, and each observing series consisted of two frames per Stokes parameter.
Therefore, our typical error bar per spectral bin in each Stokes
parameter was $0.12-0.15$\,\%. Our spectral resolution of 880 is preserved up to
a rebinning of two 0.161\,nm pixels, but when no narrrow spectral features are
present in our spectra, it makes sense to consider heavier rebinning.
We experimented by combining
the signals of 2, 4, 8, 16, 32, 64, 128, and 256 spectral bins, and for our
figures we decided to adopt a 20.6\,nm bin for a formal
uncertainty due to photon noise of 0.01-0.02\,\%.

We should note that rebinning must be performed on the fluxes \fo\ and
\fe\ before combining the flux to obtain the \pq\ and \pu\ spectra since
\[
\frac{\sum_{i=1}^{N} \left(\fo(\lambda_i)-\fe(\lambda_i)\right)}{\sum_{i=1}^{N} \left(\fo(\lambda_i)+\fe(\lambda_i)\right)}
\neq
\sum_{i=1}^{N}\frac{\fo(\lambda_i)-\fe(\lambda_i)}{\fo(\lambda_i)+\fe(\lambda_i)} \ .
\]

\begin{figure}
\begin{center}
\includegraphics*[trim={0.6cm 5.7cm 1.2cm 3.4cm},width=\columnwidth,clip]{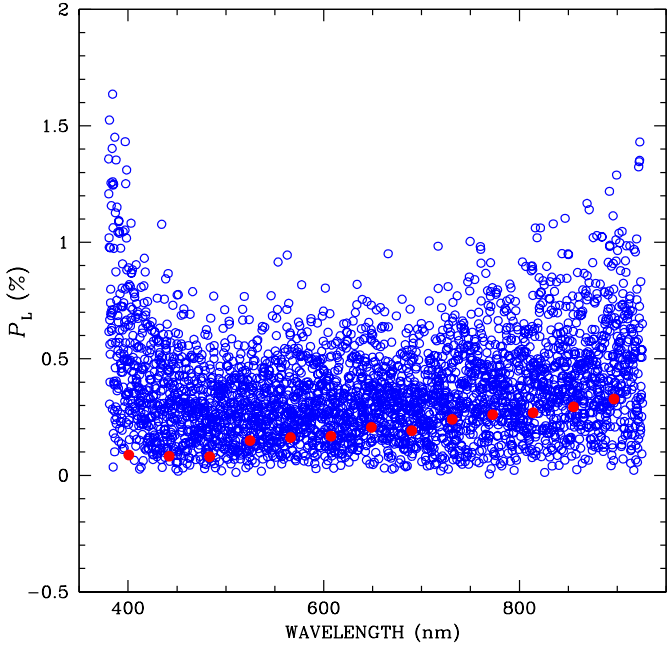}
\end{center}
\caption{\label{Fig_LowPol} Observations of the low-polarisation star HD\,101065.
  The empty blue circle shows the linear polarisation with the original bin size of 0.16\,nm.
  Red solid circles show the polarisation values obtained after rebinning \fo\ and \fe\ into
  $\sim 20$\,nm size bins.}
\end{figure}

We note also that in case of low polarisation
signals, the fraction of linear polarisation pertaining to a larger
bin size may appear to be lower than the average of the polarisation signal
plotted with a smaller bin size. An example of this situation is given in
Fig.~\ref{Fig_LowPol}, which shows a polarisation spectrum with a
wavelength bin size of 0.16\,nm (corresponding to one CCD pixel) and
the polarisation spectrum of the same star rebinned to a 20.6\,nm bin
size. The reason for this apparent inconsistency is that the \pl\ distribution
may be approximated with a Gaussian distribution only for $\pl \gg
\sigma_P$, a condition that may not be satisfied by the signal
accumulated in a small wavelength bin. For the same reason, the
best fit to \pl\ with a Serkowski curve with a least-squares technique
must be performed on spectra that have been rebinned so that for each
spectral bin $\pl \gg \sigma_P$. In previous literature, for
example,
\citet{Whietal92}, this problem was mitigated by adopting the
correction formula derived by
\citet{WarKro74} and by \citet{ClaStu86}, namely
\begin{equation}
  \pl^{\rm \ (c)} = \pl\,\left[1-\left(\frac{\sigma_p}{\pl}\right)^2\right]^\frac{1}{2} \;,
  \label{Eq_Correction}
\end{equation}
but in our case, after rebinning at 20.6\,nm, we found
$\sigma_P/P \ll 1$ was always valid, in particular, that
$\sigma_P/P \le 10$ in all spectral bins of all our targets,
even in the bluest spectral regions where the {\it S/N} is lowest.

\subsubsection{Synthetic broadband linear polarisation}\label{Sect_Convolution}
For comparison with previous and future BBLP measurements, we
calculated for each filter $F$ of $BVRI$
\begin{equation}
  \begin{array}{rcl}
\pq(F)&=&\frac{\int_{0}^{\infty}\mathrm{d}{\lambda}\
\pq({\lambda})\, I_Q({\lambda})\, T_{\rm F}({\lambda})}{\int_{0}^{\infty}\mathrm{d}{\lambda}\ I_Q({\lambda})\,T_{\rm F}({\lambda})}\\
\pu(F)&=&\frac{\int_{0}^{\infty}\mathrm{d}{\lambda}\
\pu({\lambda})\, I_U(\lambda)\, T_{\rm F}({\lambda})}{\int_{0}^{\infty}\mathrm{d}{\lambda}\ I_U({\lambda})\,T_{\rm F}({\lambda})}\\
\end{array}
  ,\end{equation}
where $T_{F}$ is the transmission function of the $F$ filter
\footnote{\tt http://www.eso.org/sci/facilities/paranal/\\instruments/fors/inst/Filters/curves.html},
and
\begin{equation}
\begin{array}{rcl}
I_Q &=& \left(\fo + \fe\right) \vert_{\alpha = 0^\circ} + 
        \left(\fo + \fe\right) \vert_{\alpha = 45^\circ} \\[2mm]
I_U &=& \left(\fo + \fe\right) \vert_{\alpha = 22.5^\circ} + 
        \left(\fo + \fe\right) \vert_{\alpha = 67.5^\circ} \; .\\
\end{array}
\end{equation}

\subsection{Fitting the polarisation spectra with the Serkowski curve}\label{Sect_Serkowski}
The wavelength dependence of linear polarisation due to interstellar dust is usually well
fit by the Serkowski curve introduced by
\citet{Serkowski73}
\begin{equation}
\frac{\pl(\lambda)}{\pmax} = \exp \left[ -K \ \ln^2
    \left( \frac{\lmax}{\lambda} \right) \right]\,,
\label{Eq_Serkowski}
\end{equation}
where \lmax\ is the wavelength where the polarisation reaches its
maximum \pmax, and $K$ is a constant that controls the half-width of
the curve. The values of \pmax\ (generally of a few units per cent at
most) and \lmax\ (typically between 400 and 800\,nm) depend on the
position of the star (distance and galactic coordinates).

Although alternative functional forms have been
proposed to describe the interstellar polarisation curve (see for example
\citeauthor{WolSmi84} \citeyear{WolSmi84} and
\citeauthor{Efimov09} \citeyear{Efimov09}),
only the Serkowski-curve parameters
are presented here for comparison to the literature.  Linear
polarisation spectra are publicly available~\footnote{See footnote to the title.} to
facilitate future comparisons, dust modelling, and analysis.

We have applied a least-squares technique by
minimising the expression
\begin{equation}
  \chi^2 = \sum_i^{N}
  \frac{\left(y_i - A - B\,x_i - C\,x_i^2\right)^2}{\sigma^2_i}
\label{Eq_Chi}
,\end{equation}
where
\[
\begin{array}{rcl}
x_i &=& \ln(\lambda_i)\\
y_i &=& \ln\big(\pl(\lambda_i)\big) \\
\sigma_i &=& \sigma_{\pl}/\pl(\lambda_i)\\
A &=& \ln \pmax - K \ln^2 \lmax \\
B &=& 2 K \ln \lmax\\
C &=& -K,\\
\end{array}
\]
hence
\[
\begin{array}{rcl}
K &=& -C\; ,\\
\lmax &=& \exp\left(-\frac{B}{2C}\right)\; ,\\
\pmax &=& \exp\left(A - \frac{B^2}{4C}\right)\; ,\\
\end{array}
\]
and the parameter errors are
\[
\begin{array}{rcl}
    \sigma^2_K &=& \sigma^2_{CC} \\[2mm]
\sigma^2_\lambda &=& \frac{1}{4C^2}\exp\left(-\frac{B}{C}\right)\
                   \big(C^2\,\sigma^2_{BB} + B^2\,\sigma^2_{CC} -
                   2BC\sigma_{BC} \big) \\[2mm]
   \sigma^2_{p} &=& \frac{1}{16\,C^4}\exp\left(2A-\frac{B^2}{2C}\right)\
                   \Big(16 C^4 \sigma^2_{AA} + 4B^2C^2 \sigma^2_{BB} +B^4\, \sigma^2_{CC} + \\
               & & - 16\,BC^3 \sigma^2_{AB} + 8\,B^2C^2 \sigma^2_{AC} -
                   4\,B^3C\, \sigma^2_{BC} \Big), \\
\end{array}
\]
where $\sigma_{XY}$ ($X,Y=A,B,C$) are the elements of the inverse of
the $\chi^2$ matrix. Since photon noise alone underestimates the
actual error bars, the best-fit parameter uncertainties
were finally multiplied by $(\chi^2/\nu)^{1/2}$, where $\chi^2$ is
given by Eq.~(\ref{Eq_Chi}) and $\nu$ is the number of degrees of
freedom, as discussed by \citep{Bagetal12}.

\subsection{Fitting the polarisation position angle}
In many cases, the polarisation position angle was found to be slightly wavelength
dependent. In order to quantify the deviations from a constant value, we
decided to fit the position angle $\thetae(\lambda)$ with a straight line
and to consider the slope of this line in addition to the Serkowski parameters.

\section{Results}\label{Sect_Results}
Table~\ref{Tab_Results} shows the values obtained by calculating from
our polarisation spectra synthetic BBLP values in the $BVRI$ filters,
as well as the best-fit parameters $K$, \lmax\, , \pmax\ , and
d$\thetae$/d$\lambda$.  Photon-noise error bars in the broadband
values are virtually negligible ($\sim 10^{-4}$), and the accuracy of
our measurements is limited by non-photon noise, as discussed in
Sect.~\ref{Sect_Incons}.  All our polarisation spectra are shown in
the figures of the Appendix in the form of \pl\ and \thetae\ plots. We set the same
\pl\ -axis range (1.5\,\%) and \thetae\ range (40\degr) to facilitate
visual comparison. In some cases,
the \pl\ range was expanded to 3.0\,\% or 3.5\,\%, and the
\thetae\ range to 80\degr\ or 120\degr. To facilitate the comparison
between plots with different $y$-scales, we adopted identical
tick sizes throughout (0.1/0.5\,\%). We note also that the figures
are sorted
by RA, but with some exceptions, in order to display in the same row
the observations of the same star obtained in different
epochs. Stellar fluxes (not corrected for atmosphere and instrument+telescope
transmission function) are also plotted in arbitrary units both in
the \pl\ and in the \thetae\ panels.

Around the
telluric O$_2$~A-band (759-770\,nm) in some polarisation spectra, we note a depolarisation of the
continuum. In other cases, the polarisation across a sharp spectral emission line
increases and changes direction. In these cases, the best-fit
parameters of the Serkowski curve were calculated after
interpolating the continuum. These spectral regions were not used
to calculate the BBLP measurements given in
Table~\ref{Tab_Results} or for the best fit with the Serkowski curve.

We did not fit any of the stars with
$\pmax \le 0.70$\,\% with the Serkowski curve. We finally note that the following stars
have very low (but non-zero) polarisation values:
HD\,50820 (0.1--0.2\,\%),
HD\,101065 (0.1--0.2\,\%),
HD\,105416 (0.2\,\%),
HD\,115842 (0.4\,\%), 
HD\,134591 (0.3--0.4\,\%),
and HD\,152424 (0.0--0.4\,\%).


\section{Quality checks}
\subsection{Consistency of the observations of stars observed at more than
one epoch}\label{Sect_Cons_One}
\begin{figure}
\begin{center}
\scalebox{0.47}{
\includegraphics*[trim={0.7cm 5.6cm 1.7cm 3.2cm},clip]{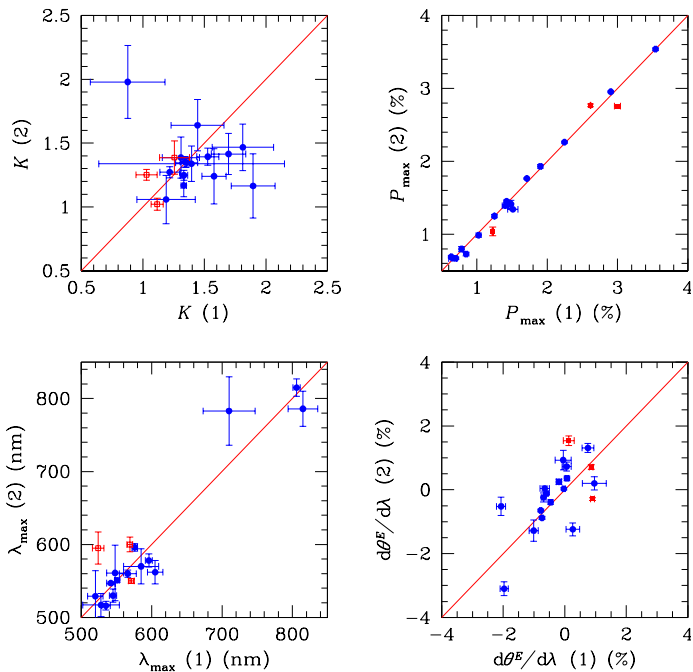}}
\end{center}
\caption{\label{Fig_SerComp} Comparison of the best-fit parameters of the Serkowski
  curve and the position-angle gradient from pairs of observations of the same stars
  obtained with FORS2 at different epochs. The red empty squares correspond to three
  stars that are likely polarimetric variables, namely
  HD\,149404, HD\,153919, and HD\,116852 (see Sect.~\ref{Sect_TimeVar}).}
\end{figure}

Some of our targets have been observed at more than one epoch:
25 stars were observed twice and one target was observed three
times. The consistency between these observations can be checked in
Table~\ref{Tab_Results} and by visual inspection of the figures in
the Appendix. A direct comparison of the Serkowski curve parameters and
of the position-angle gradient is shown in Fig.~\ref{Fig_SerComp}.
Three of the targets are probably polarimetrically variable
(see Sect.~\ref{Sect_TimeVar}) and are highlighted with special
symbols in the figure. Except for these three cases, the
consistency of the $\lmax$ and \pmax\ parameters recovered from different datasets is
generally very good, while the consistency of the $K$ parameter and of
the gradient of the polarisation angle appears less satisfactory.

\subsection{Consistency with previous work}\label{Sect_Cons_Two}
\begin{figure*}[th!]
\begin{center}
\includegraphics[trim={0.2cm 1.7cm 0.3cm 2.8cm},width=\columnwidth,clip]{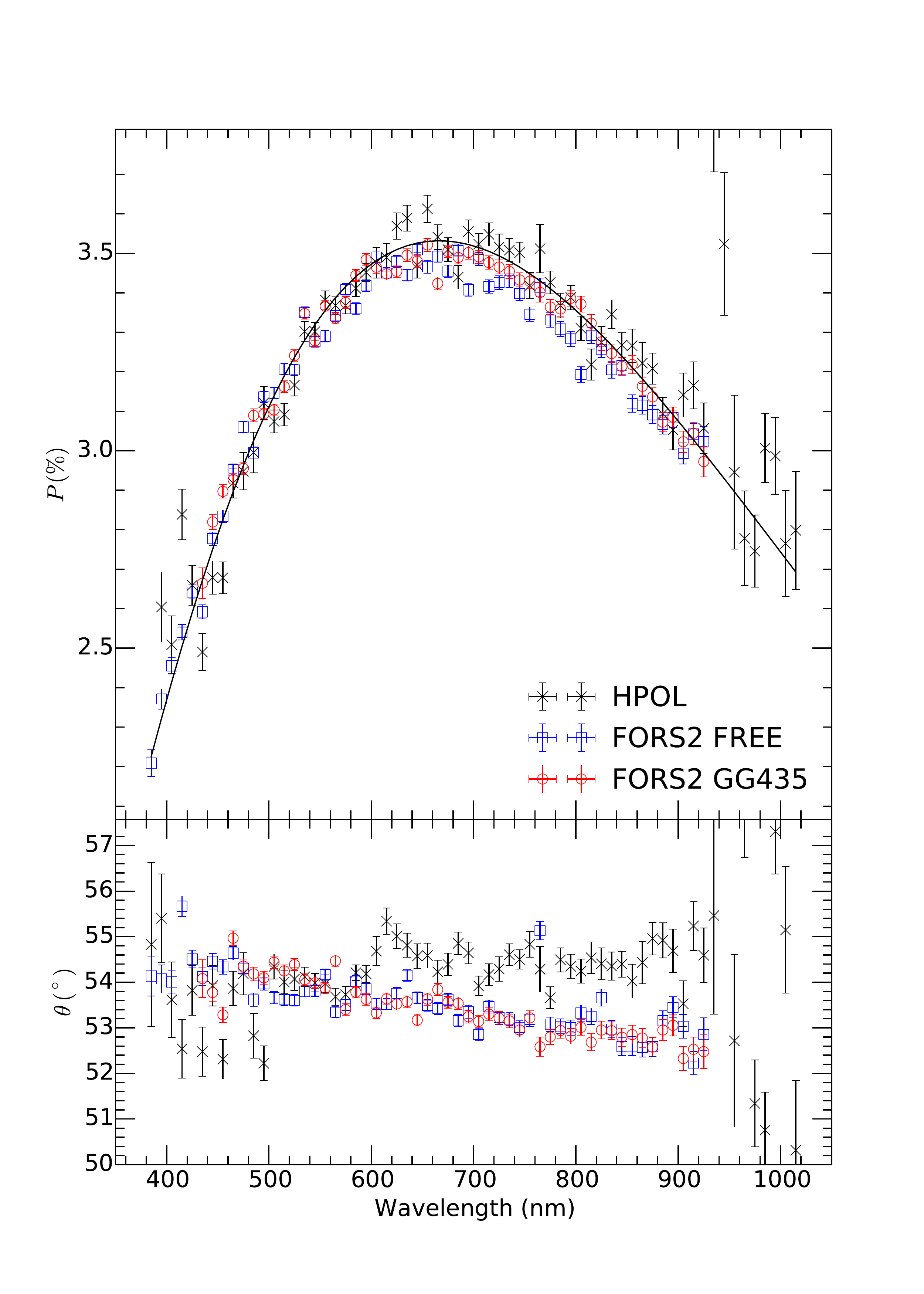}
\includegraphics[trim={0.2cm 1.7cm 0.3cm 2.8cm},width=\columnwidth,clip]{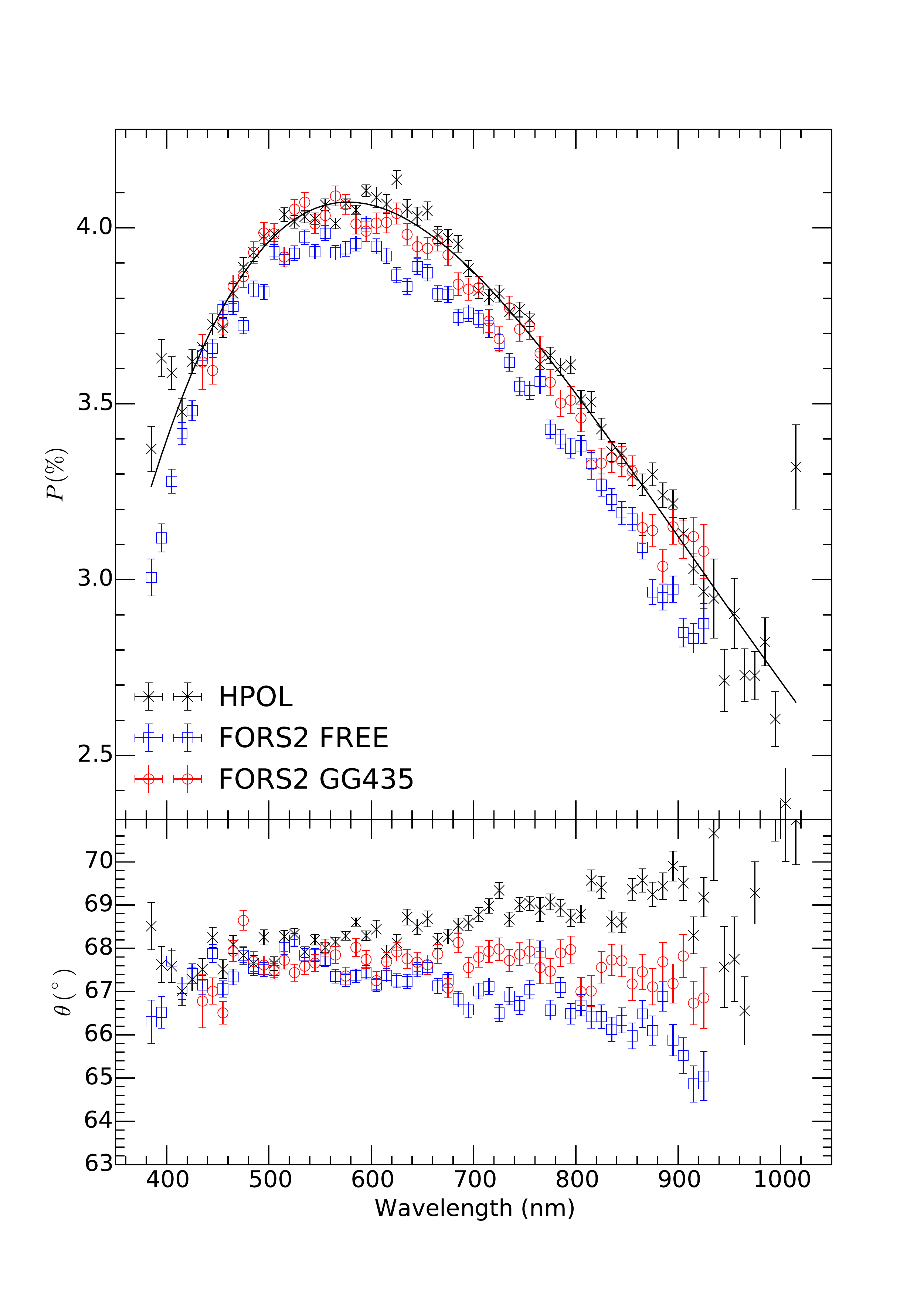}
\end{center}
\caption{\label{Fig_Comparison} Comparison between FORS2 (blue and red symbols) and HPOL (black symbols) data.
  Black solid lines are best fits with the Serkowski curve to the HPOL data.
  {\it Left panel:} HD\,147888. {\it Right panel:}  HD\,161056.}
\end{figure*}

Only a few stars in our sample have previously been observed in
polarimetric mode. Six stars have been observed with HPOL, and
Fig.~\ref{Fig_Comparison} shows the comparison between our
observations and HPOL observations for two of them, HD\,147888 and
HD\,161056.  HD\,37903 was also observed in spectropolarimetric mode
with FORS2 by \citet{Sieetal14}. For another five stars, BBLP
measurements were available in the literature. A good but not perfect
agreement was found, as demonstrated in Table~\ref{Tab_Previous}, which
shows a comparison between the Serkowski best-fit parameters obtained
from this work and those from previous works.

\section{Discussion}
In this section we present an initial qualitative discussion of
our results, in particular in comparison with previous works.

\subsection{Galactic polarisation map}
\begin{figure}
  \includegraphics*[scale=0.34,trim={3cm 0.5cm 0.3cm 0.8cm},clip]{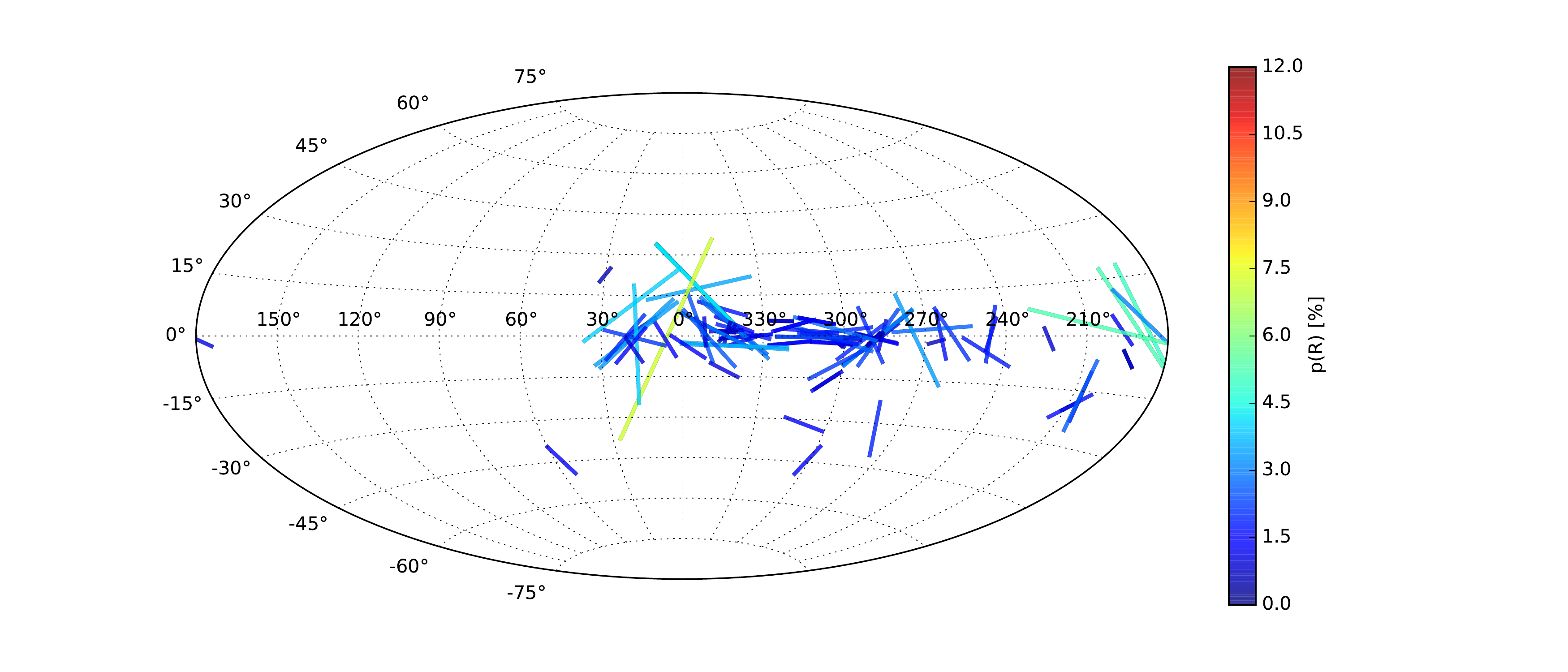}
  \includegraphics*[scale=0.34,trim={3cm 0.5cm 0.3cm 0.8cm},clip]{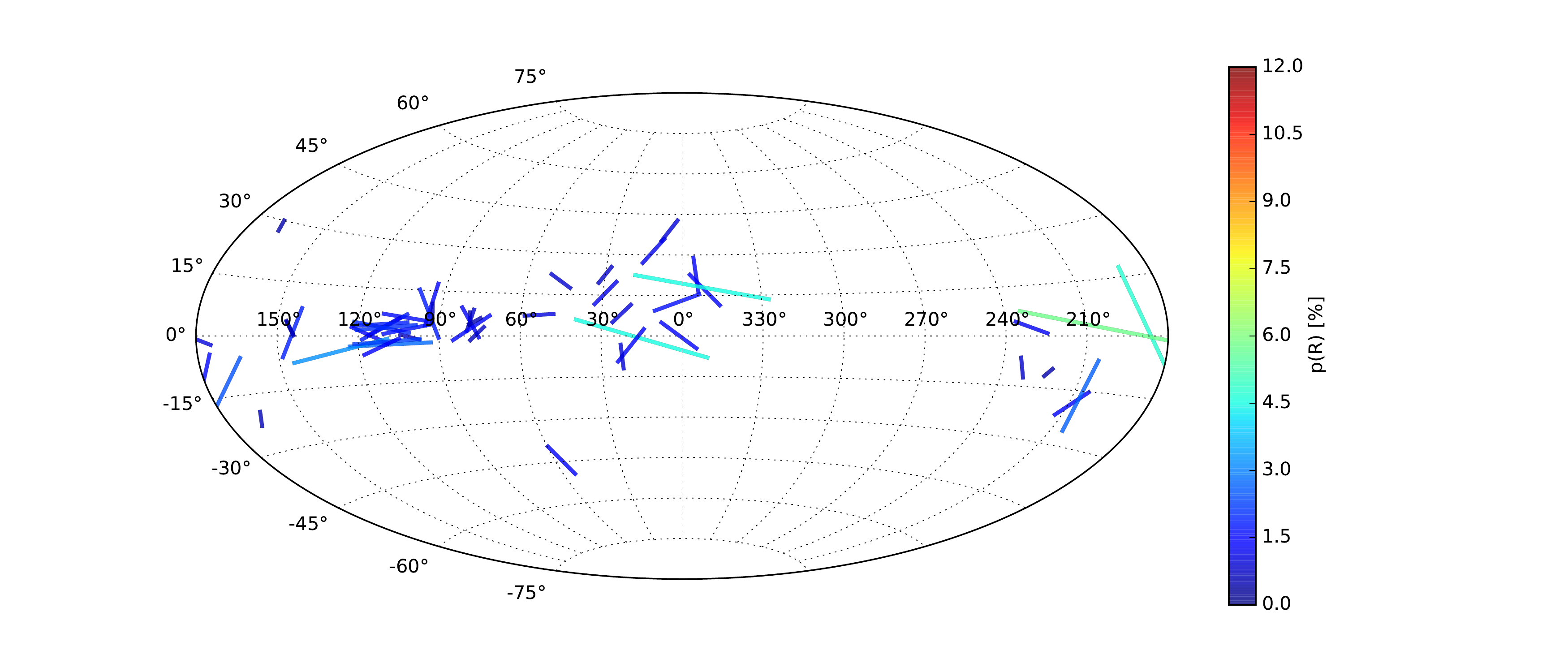}
  \includegraphics*[scale=0.34,trim={3cm 0.5cm 0.3cm 0.8cm},clip]{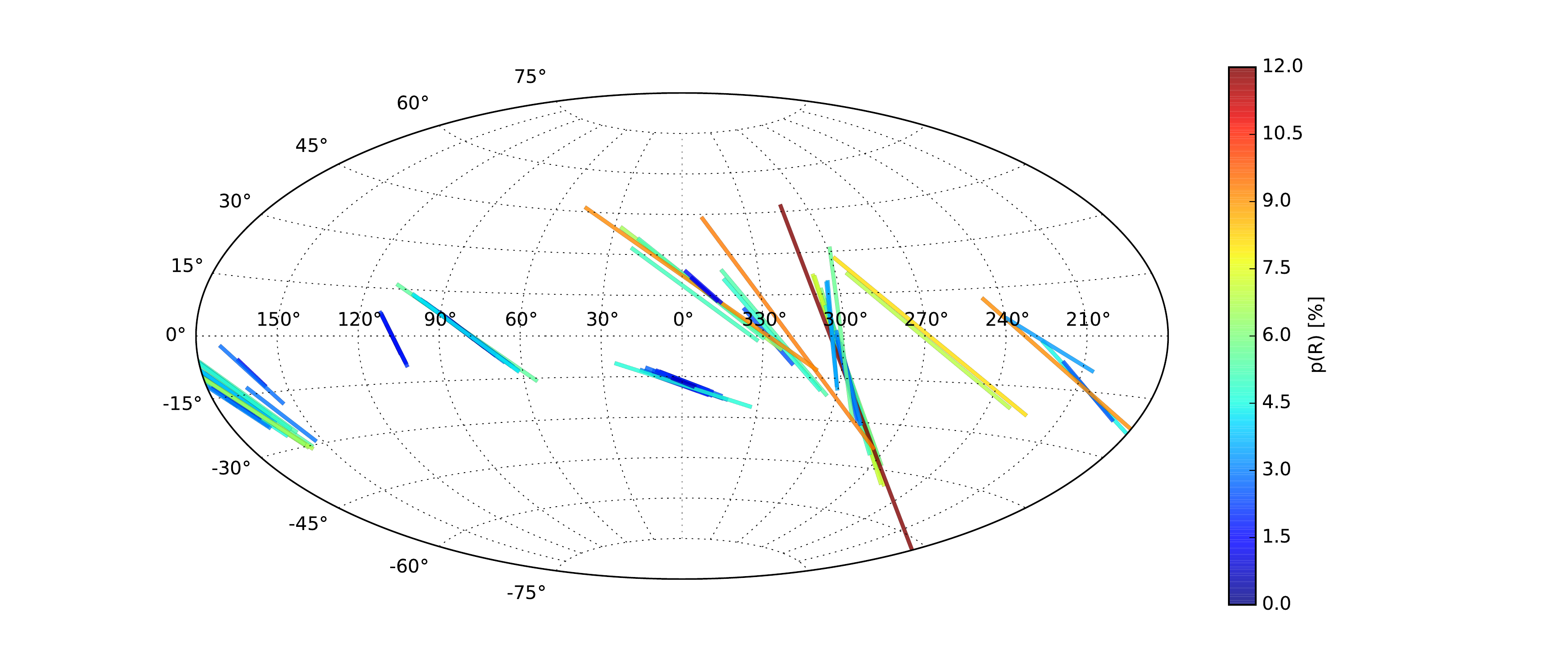}
  \caption{\label{Fig_Polmap} Polarisation map of our Galaxy from our data
    (top panel), from HPOL data (middle panel), and from \citet{Whietal92} (bottom panel). The
  length of the segments is directly proportional to the fraction of linear polarisation.}
\end{figure}

The most immediate way to globally visualise our data is to plot a
polarisation map as a function of galactic coordinates. This is done in
Fig.~\ref{Fig_Polmap} for our new FORS2 data (top panel), the data
from HPOL (middle panel), and those from \citet{Whietal92} (bottom
panel). Pending accurate Gaia distances for the targets in this survey
\citep{GaiaPrusti}, we do not show distance-longitude distribution plots.

\subsection{Relationships between the best-fit parameters $K$ and \lmax\ of the Serkowski curve}
\begin{figure}
  \includegraphics*[scale=0.32,trim={0.1cm 0.1cm 0.5cm 2.1cm},clip]{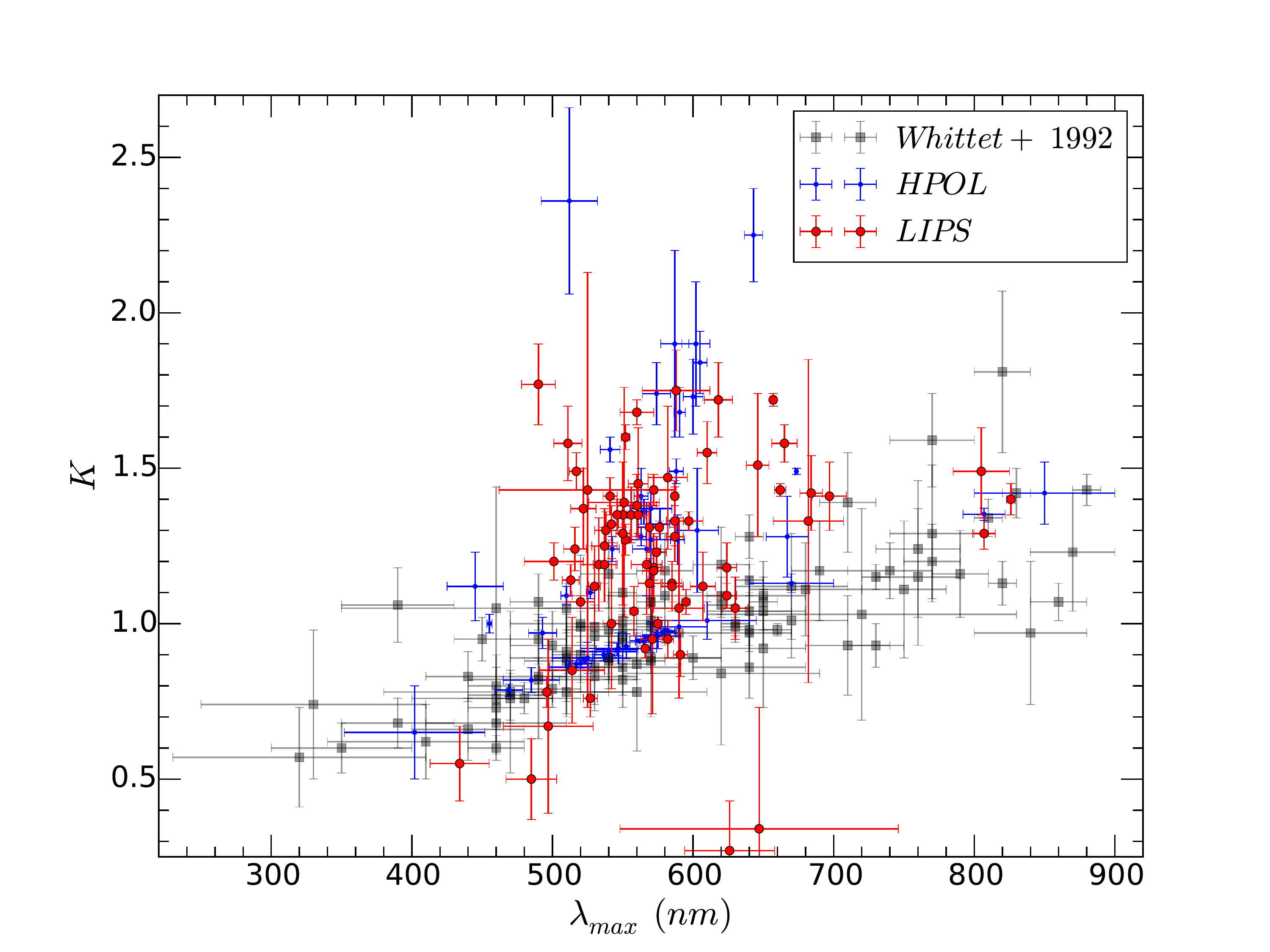}
  \caption{\label{Fig_KL} $K$ vs.\ \lmax\ from the best fit obtained with the Serkowski curve using the targets
  of this survey (red symbols), HPOL data (blue symbols), and data from \citet{Whietal92} (black symbols).}
\end{figure}
\begin{figure}
  \includegraphics*[scale=0.32,trim={0.1cm 0.1cm 0.5cm 2.1cm},clip]{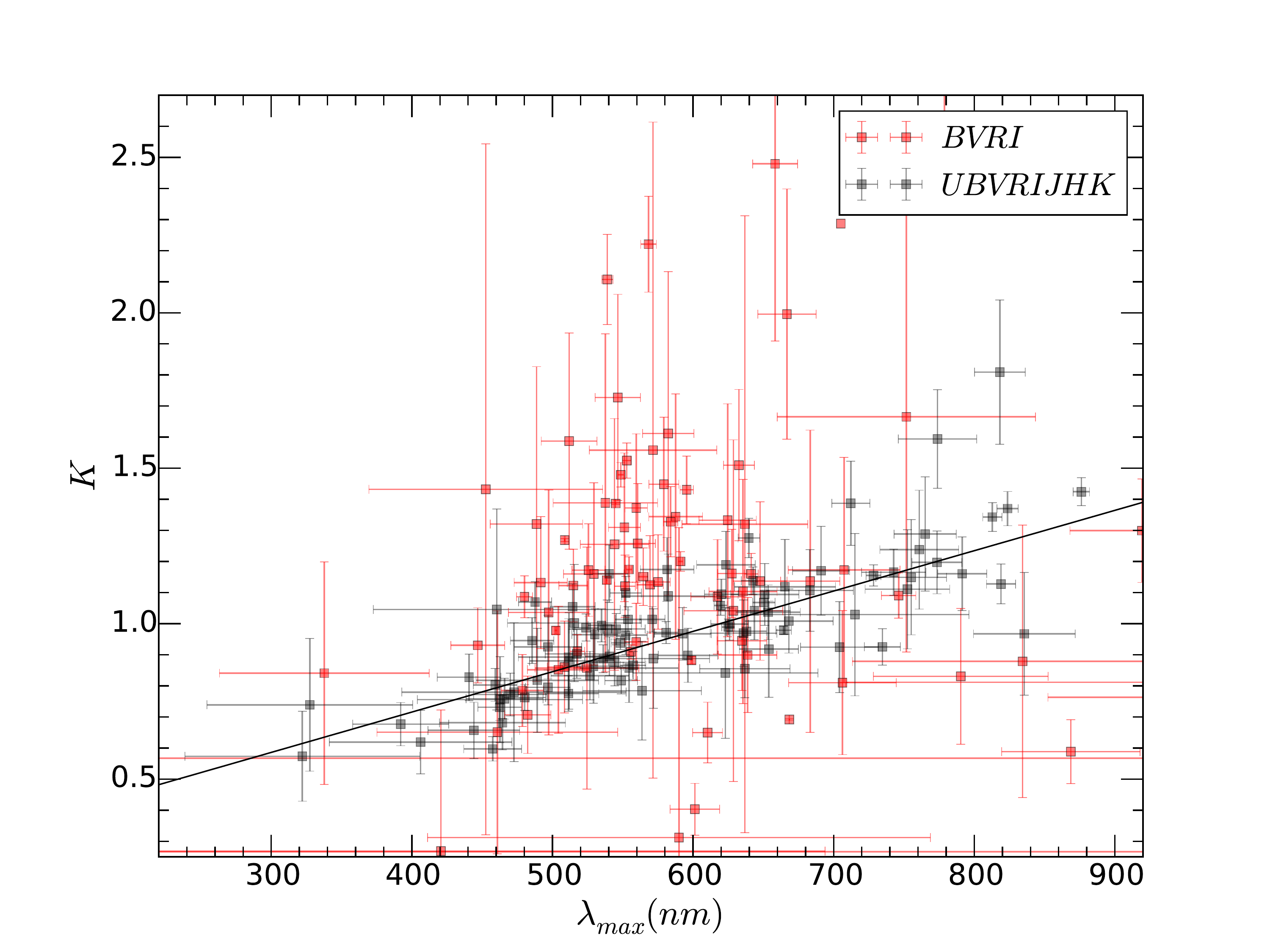}
  \caption{\label{Fig_BVRI} Relationship between $K$ and \lmax\ using
    all {\it UBVRIJHK} filters of \citet{Whietal92} (black symbols,
    as in Fig.~\ref{Fig_KL}),
    and using only their $BVRI$ photometry (red symbols).}
\end{figure}
\begin{figure*}
  \includegraphics*[scale=0.32]{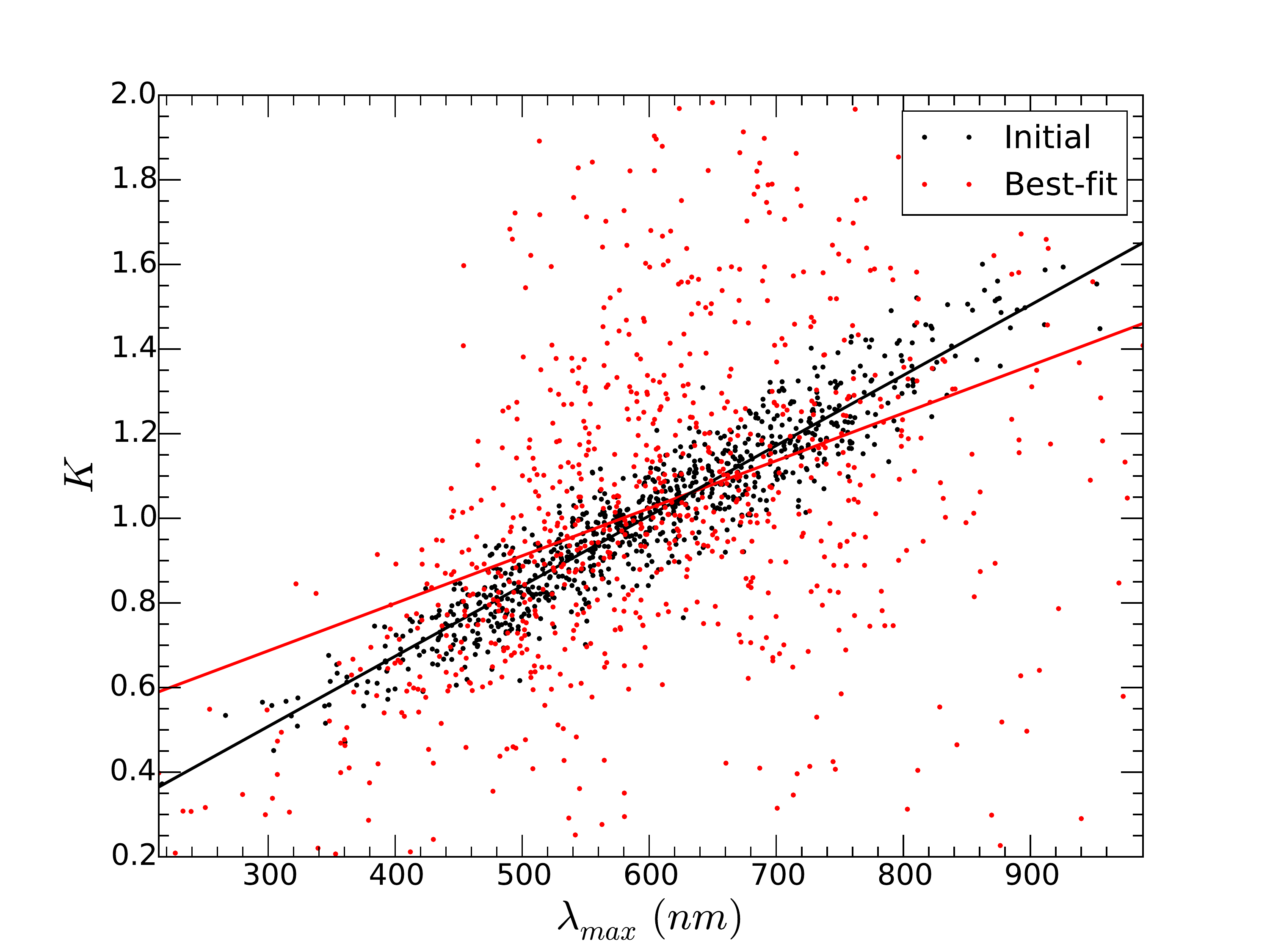}
  \includegraphics*[scale=0.32]{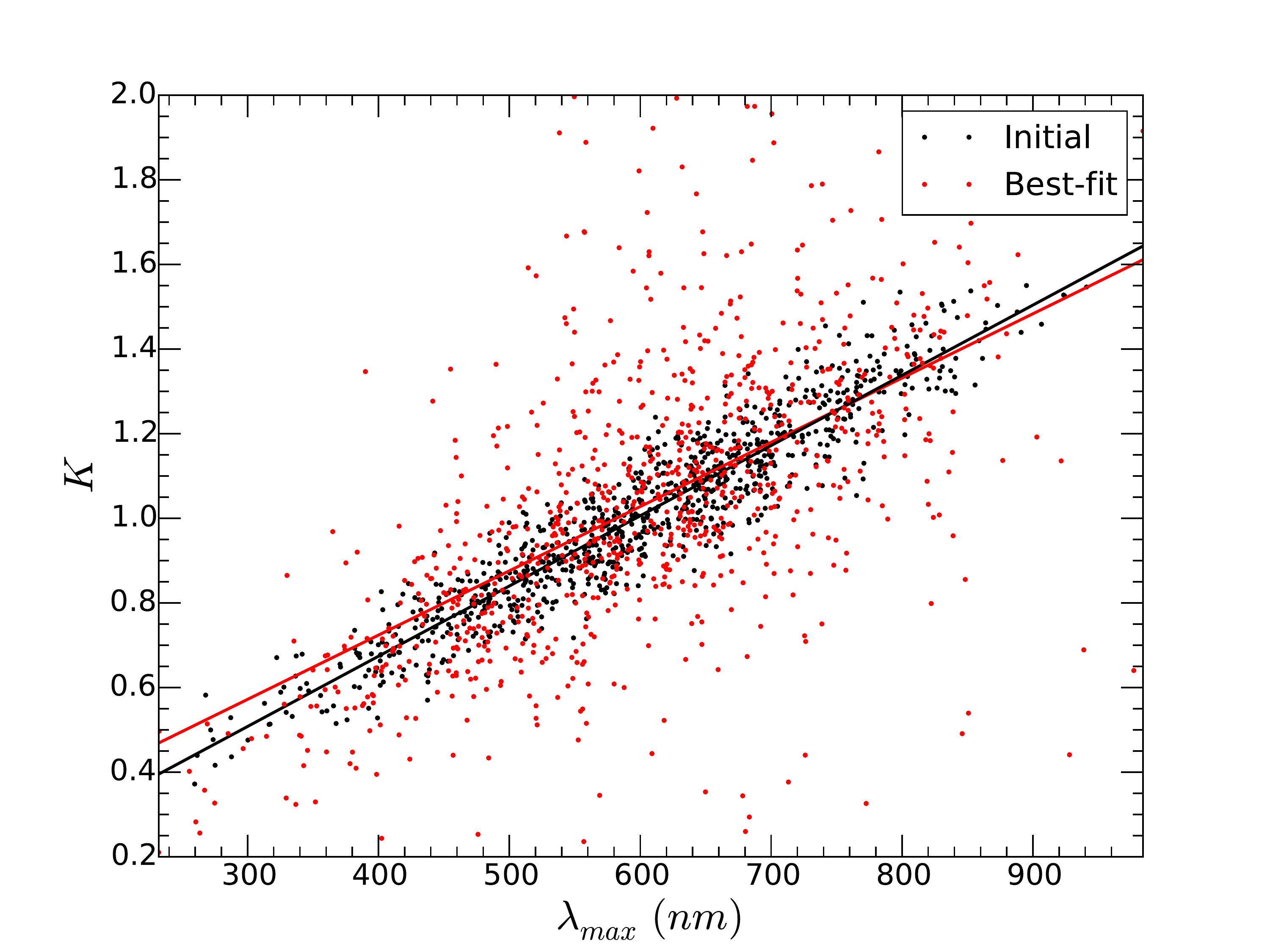}
  \caption{\label{Fig_Test} Simulation of the recovery of the free parameters of the Serkowski curve
    from a numerical dataset simulating noisy data. {\it Left panel:}
    using a 100\,nm step size. {\it Right panel:} using a 20\,nm step size.
    The figure is explained in the text.}
\end{figure*}
In this section we discuss the relationship between the
free-parameters $K$ and \lmax\ of Eq.~(\ref{Eq_Serkowski}) obtained as
explained in Sect.~\ref{Sect_Serkowski} from our dataset, and we
compare our results with those previously obtained in the literature.

In some earlier works, the value of $K = 1.15$ originally introduced
by \citet{Serkowski73} was assumed as a standard. Eventually, it was
found that infrared polarimetric data could not be fitted by
adopting
this fixed value for $K$ \citep{DykJon78,Wiletal80}, and $K$ was
considered as a free parameter. \citet{Wiletal80} found a correlation
between \lmax\ and $K$ : the normalised polarisation curve
would be narrowing as \lmax\ increases. \citet{Wiletal80} suggested to
adopt
\[
K=1.7\,\lmax\ ,
\]
and later, \citet{Whietal92} suggested the relationship
\begin{equation}
K = (0.01 \pm 0.05) + (1.66 \pm 0.09)\, \lmax \; ,
\label{Eq_Whittet}
\end{equation}
where in both Eqs.\ \lmax\ is expressed in $\mu$m.
Figure~\ref{Fig_KL} shows the relationship between best-fit values of
$K$ and \lmax\ from our survey and those obtained from the
dataset by \citet{Whietal92} and HPOL. It clearly appears that
our data (red circles) and those from HPOL (blue circles) are located
in the same region of the $\lmax-K$ plane, but none of these two datasets
follows the linear trend seen in \citet{Whietal92} data (black symbols).

The question now is whether this discrepancy is due to different
features of the ISM in the regions probed by the different surveys, or
whether it is an artefact due to either photon noise or systematic
error. We should keep in mind that the results of Sects.~\ref{Sect_Cons_One}
and \ref{Sect_Cons_Two} suggest that while the \pmax\ and
\lmax\ values retrieved from different datasets are generally
consistent among themselves, the recovery of the $K$ parameter,
representing the width of the polarisation spectrum, seems
less robust. We note that these inconsistencies cannot be due to the
issues with the positionning of the retarder waveplate discussed in
Sect.~\ref{Sect_Incons}, as they would not
affect the ratio $\pl(\lambda)/\pmax$ of Eq.~(\ref{Eq_Serkowski}).

To answer to this question, we first studied whether the observed
discrepancy may be an artefact caused by the shorter wavelength range of our observations compared to those of \citet{Whietal92}, which
extends into the $U$ filter and into the near-IR. 
\citet{ClaAlr83} indeed suggested that the estimated $\lmax-K$ correlation
depends on the wavelength range in which the observations are carried
out. For instance, by adopting the $UBVR$ polarimetric data from
\citet{Seretal75}, \citet{ClaAlr83} found that $K$ and \lmax\ are
anti-correlated. By considering another $UBVR$ literature dataset,
they again obtained a linear relationship with a negative slope, which is
opposite to what has been found by \citet{Wiletal80}. However, using all
available data within the wavelength range from 330 to 950\,nm,
\citet{ClaAlr83} found a positive correlation between $K$ and \lmax.
Inspired by these earlier tests, we performed a similar experiment. We fitted the \citet{Whietal92} dataset discarding data
obtained with the $U$ and $JHK$ filters, and considering only those
obtained with the {\it BVRI} filters. With this subset, we were still able to
recover a loosely linear relationship between $K$ and \lmax\, but
opposite to that given by Eq.~(\ref{Eq_Whittet}) and closer
to the relationship found with our survey (see Fig.~\ref{Fig_BVRI} and
compare it with Fig.~\ref{Fig_KL}).  This result is in agreement with
the previous studies of \citet{ClaAlr83}, and suggests that the lack
of consistency between our results and those from \citet{Whietal92}
may indeed be an artefact that is due to the different wavelength coverage of
the datasets.

To further investigate this issue, we performed Monte Carlo
simulations in order to test which types of observational constraints
are needed to retrieve a linear relationship between $K$ and \lmax.
We generated a number of initial Serkowski-curve
parameter values \lmax, \pmax\ , and $K$ randomly selected within these
limits: $470\,{\rm nm} \le \lmax \le 730\,{\rm nm}$, and $1.4 \le \pmax
\le\ 5.8$.  The parameter $K$ was then calculated via
Eq.~(\ref{Eq_Whittet}).  We then generated artificial polarisation
spectra with 100\,nm wavelength bins in the spectral range 400 to
900\,nm using the Serkowski curve, and we scattered all data points
according to a Gaussian distribution with $\sigma=0.1$\,\%.
 Then we fitted our artificially generated
polarisation spectra with a Serkowski curve, adopting \lmax, \pmax\ and
$K$ as free parameters, and we used our best-fit parameters to create
a new $\lmax-K$ plot. The results are shown in the left panel of Fig.~\ref{Fig_Test},
 where the black solid line represents the
relationship given by \citet{Whietal92}, the black dots correspond to
the $K$ and \lmax\ parameters scattered around this relationship, the
red dots show their values as recovered after our numerical simulation
of photon noise, and the red solid line represents the best fit to the
recovered $K$ and \lmax\ parameters. These results are consistent with
our previous finding that measuring BBLP in
only four filters does not allow us to recover the linear relationship
between $K$ and \lmax, while when adding three additional data points
in the near-IR, we were able to recover it.

Next we performed a numerical simulation to investigate the LIPS
database. We repeated the same experiment as before, but using a
20\,nm step size instead of 100\,nm, still associating a 0.1\,\% error
bar per spectral bin. For comparison with our real dataset, we note
that the typical residual of the best-fit Serkowski curve to the LIPS
data is $\lesssim 0.05$\,\%. The result of this test is shown in the
right panel of Fig.~\ref{Fig_Test}.  Although the red straight line in
the right panel of Fig.~\ref{Fig_Test} slightly deviates from the
original distribution of black dots, the results of our second test
suggest that if the polarisation that we have measured with FORS2
followed the linear relationship by \citet{Whietal92}, then we would still have been able to recover a similar (although not identical)
linear relationship.  Therefore the results of this final experiment
seem to contradict the idea that the discrepancies between our results
and those found by \citet{Whietal92} are due to the limited wavelength
range of the spectropolarimetric observations compared to the sample
by \citet{Whietal92}.

\begin{figure}
  \includegraphics*[scale=0.62,trim={0.1cm 0.2cm 0.5cm 1.2cm},clip]{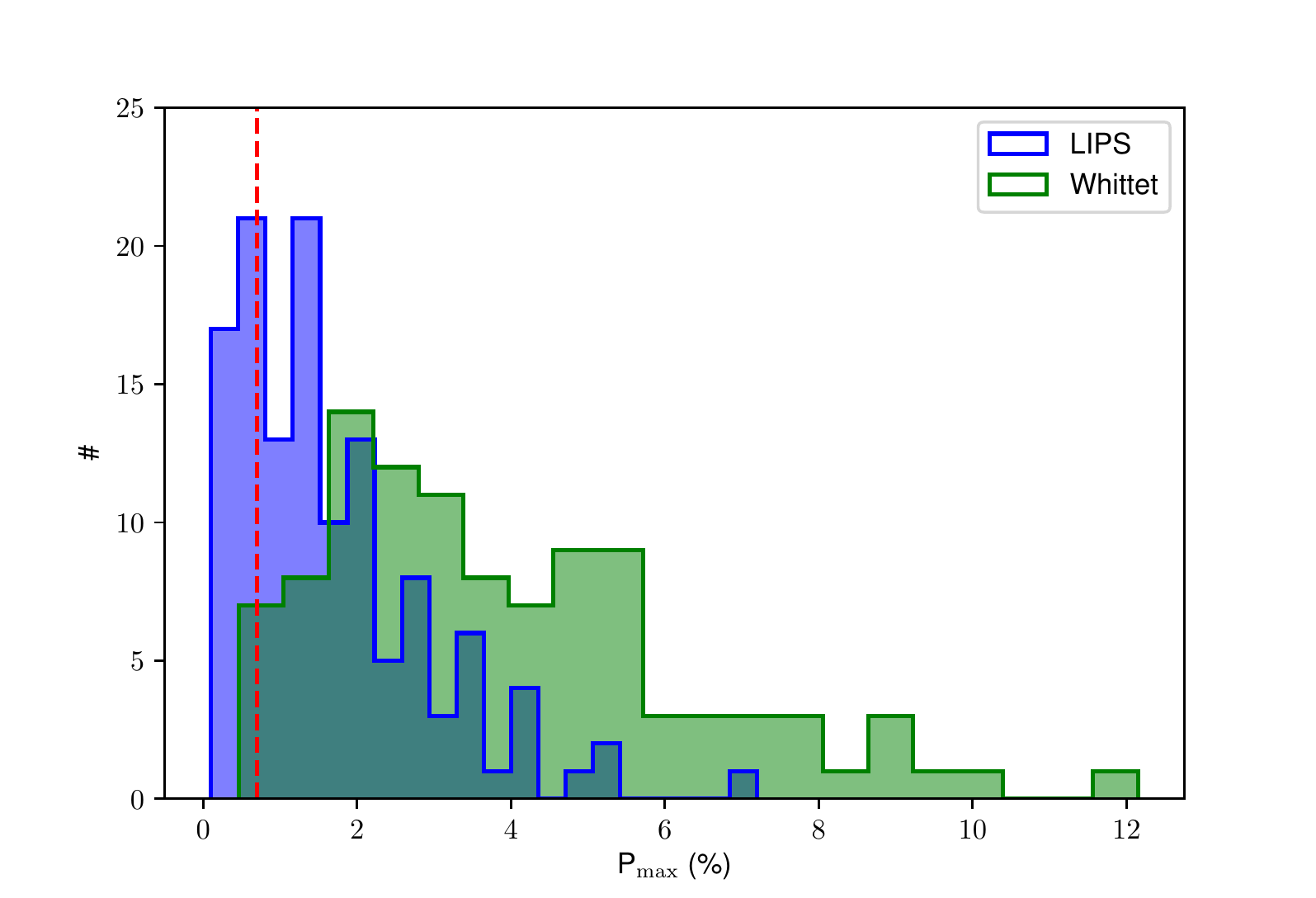}\\
  \caption{\label{Fig_Histo_Serk} Distribution
    of \pmax\ for our sample and for that of \citet{Whietal92}. The red dashed
     line at 0.70\,\% shows the \pmax\ limit below which we did not calculate the best-fit
    parameters of the Serkowski curve.
  }
\end{figure}

In conclusion, if the results by \citet{Whietal92} are correct, it
means that the shape of our polarisation spectra is affected by errors
that are substantially larger than photon noise, probably $\la
0.1$\,\% in the BBLP filters, and we suggest that in order to retrieve
meaningful $K$ values of the Serkowski-curve parameters, it is
desirable to obtain a stronger constraint in the blue and red spectral
regions than is possible with FORS optical
spectropolarimetry. However, the observed discrepancies between our
sample and that of \citet{Whietal92} may also be due to the fact that
different surveys probe regions of the ISM with different dust grain
features \citep[e.g.][]{Vos12}. The differences between the surveys
are also evident in the \pmax\ histogram of Fig.~\ref{Fig_Histo_Serk}.

\subsection{Relationships between \lmax, \pmax\ and the extinction}
\begin{figure}
  \includegraphics*[scale=0.5,trim={0.4cm 0.0cm 0.5cm 0.9cm},clip]{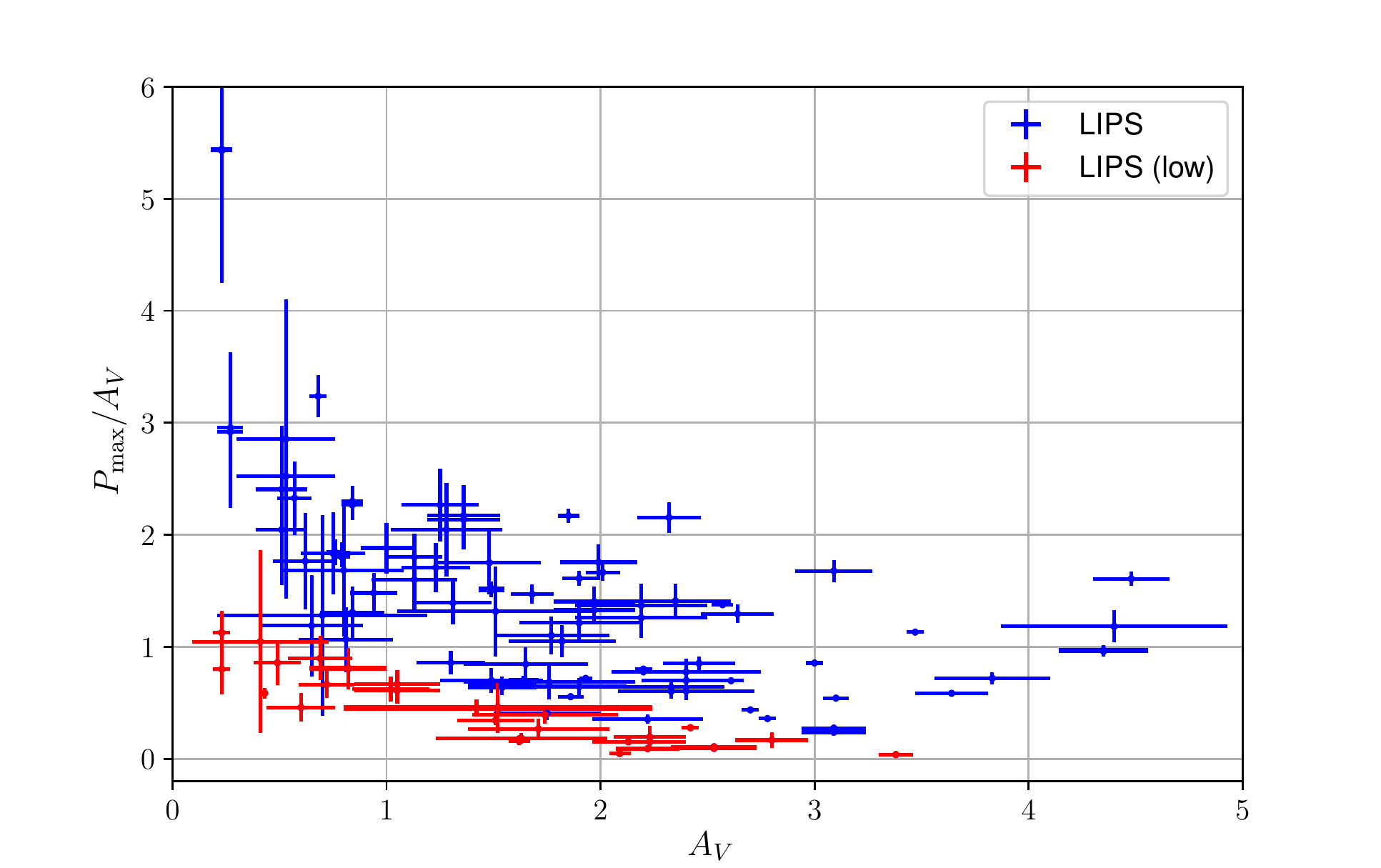}\\
  \caption{\label{Fig_PAV} Relationship between polarisation efficiency and
    visual extinction. The red symbols (``LIPS low'') refer to data with $\pl \le 0.7$\,\%.
  }
\end{figure}
\begin{figure}
  \includegraphics*[scale=0.5,trim={0.4cm 0.0cm 0.5cm 0.9cm},clip]{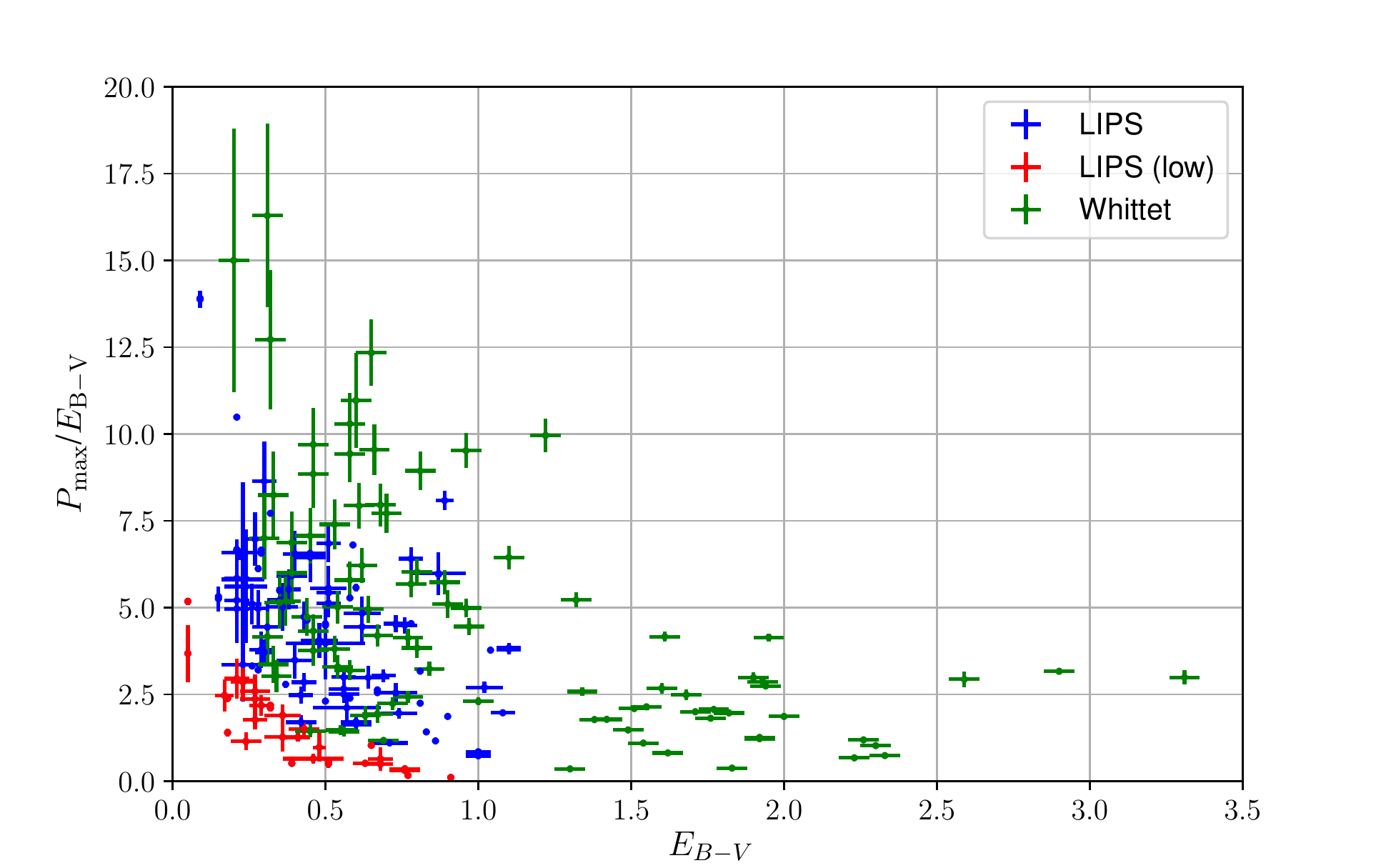}\\
  \caption{\label{Fig_PEBV} Relationship between maximum polarisation and
    $E(B-V)$. ``LIPS (low)'' refer to stars for which $\pl \le 0.7$\,\%.
  }
\end{figure}
\begin{figure}
  \scalebox{0.47}{
\includegraphics*[trim={1.7cm 6.2cm 1.0cm 2.6cm},clip]{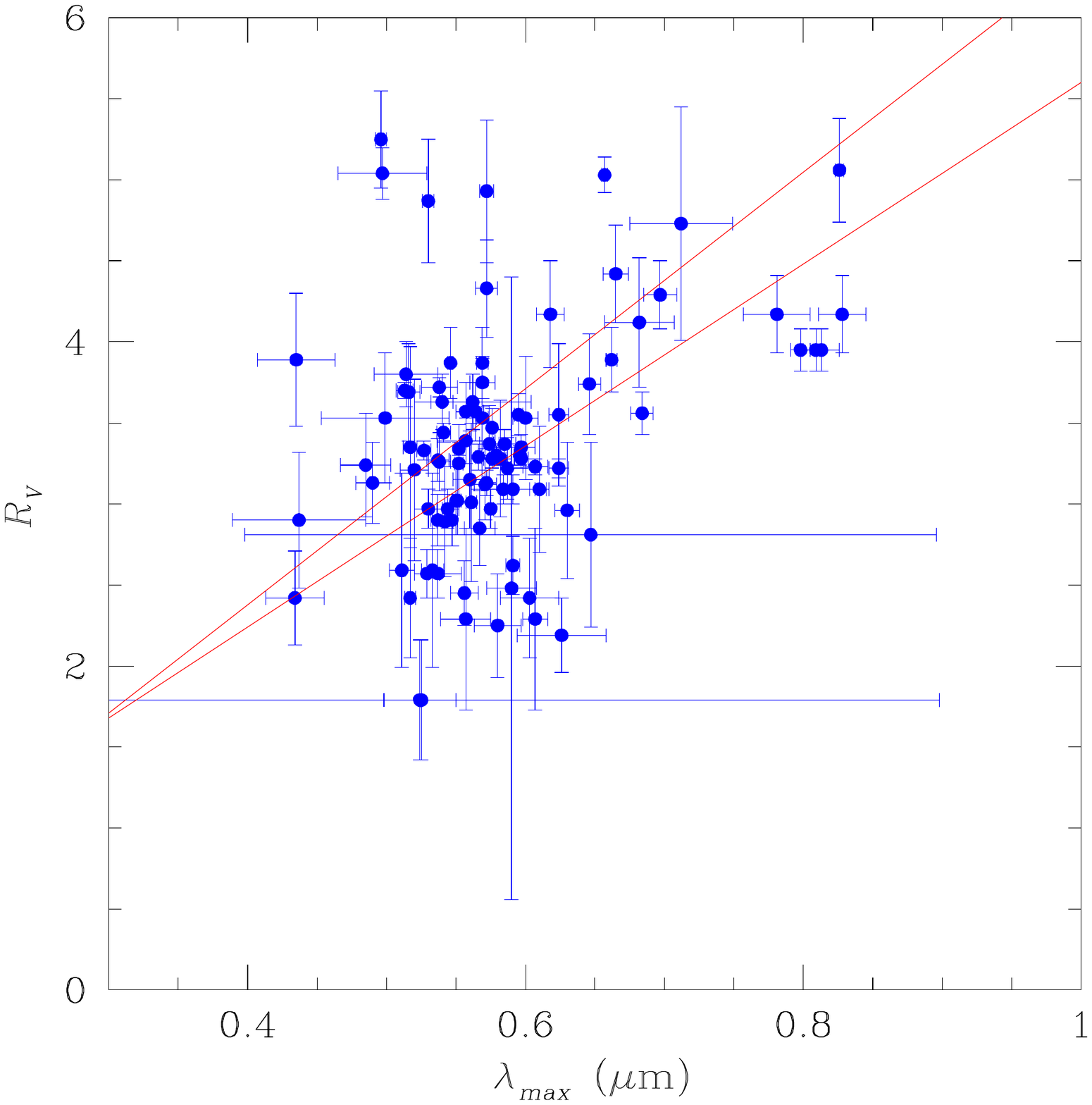}}
  \caption{\label{Fig_Rlmax} The relationship between $R_V$ and \lmax.
    Symbols refer to the data from our survey, and the solid lines
    are the relationships of Eqs.~(\ref{Eq_WhiVan}) and (\ref{Eq_ClaMat}).
  }
\end{figure}
The \lmax\ and \pmax\ parameters retrieved from the best fit seem
accurate enough to allow us a meaningful discussion of their
relationships with other features of the ISM, like for instance the
visual extinction.

\subsubsection{Polarising efficiency $\pmax/A_V$ vs.\ visual extinction $A_V$}
Figure~\ref{Fig_PAV} shows the parameter $\pmax/A_V$ 
, the historically defined {\it \textup{polarising efficiency}}
\citep{Tametal87,Whietal94,Greetal95}, plotted versus the total visual
extinction $A_V$. The polarisation efficiency decreases as visual
extinction increases, in agreement with what was found by
\citet{Whietal94} (see their Fig.~7), by \citet{Geretal95}, who
proposed the relationship $p_\mathrm{max}/A_V = 1.38 \,A_V^{-0.56}$
(see their Fig.~1), and \citet{Whietal01}, who proposed the
relationship $p_\mathrm{max}/A_V = 2.5\,A_V^{-0.71}$ (see their
Fig.~10).

\subsubsection{Polarising efficiency $\pmax/E(B-V)$\ vs.\ selective extinction $E(B-V)$}

Another quantity representing the polarising efficiency is the ratio
$\pmax/E(B-V)$, where $E(B-V) = A_B - A_V$ is the selective
extinction. \citet{Seretal75} found that this parameter is always \la
9\,\%\,mag$^{-1}$. The relationship between $\pmax/E(B-V)$\ and
$E(B-V)$ depends on the dust grain size and has been discussed by
\citet{Vosetal16}.  Figure~\ref{Fig_PEBV} shows this relationship for
our dataset.

\subsubsection{Total-to-selective extinction ratio $R_V$ vs. \lmax}
\citet{WhiVan78} derived the linear relationship
\begin{equation}
  R_V = (5.6 \pm 0.3)\,\lmax
  \label{Eq_WhiVan}
,\end{equation}
where $R_V = A_V/E(B-V)$ is the ratio between total and selective
extinction, and \lmax\ is expressed in $\mu$m, while \citet{ClaMat88} suggested 
\begin{equation}
   R_V = -0.29 \pm 0.74 + (6.67 \pm 1.17)\,\lmax
\label{Eq_ClaMat}
.\end{equation}
The validity of this relationship was not confirmed in later studies
\citep[e.g.][]{Whietal01}, and our data do not follow any of these
relationships either (see Fig.~\ref{Fig_Rlmax}).

\subsection{Comments on special cases}

\subsubsection{Position angle varying with wavelength}
If polarisation is due to dichroic extinction from
interstellar dust, we would expect a fairly smooth function
$\pl(\lambda)$ that is reasonably well reproduced by the Serkowski curve, with
the position angle independent of $\lambda$. 

Of special interest are stars for which the position angle of
the polarisation is not constant with wavelength. These cases may be
identified from inspection of the last column of
Table~\ref{Tab_Results}, which provides the wavelength gradients of
the polarisation position angle. This may be due to the presence of
more than one dust cloud along the line of sight, with different
characteristics, but we should note that the steepest gradients are
usually associated with stars with low polarisation (e.g. the blue
supergiant HD\,154368 and the main-sequence O star HD\,93222).
Particularly puzzling is the case of the Be stars HD\,155851 and
HD\,172694, which show a sharp change in slope in both the
polarisation and the position angle at wavelengths longer than
820\,nm. The observed polarisation is probably given by the
superposition of the contribution from the circumstellar disc and the
ISM dust, and the change in polarisation, which corresponds to the
Paschen edge, almost certainly originates in the disc. The
polarimetric properties (including variability) of Be stars have been
studied in the past \citep[e.g.][]{Bjorkman94}, but to our best
knowledge, the observed sharp change in polarisation at longer
wavelength has not previously been reported.

\subsubsection{Time variability}\label{Sect_TimeVar}
Twenty-two stars of our sample were observed twice and one was
observed three times. These stars can be checked for variability.

From our observations, it appears that the high-mass X-binary
HD\,153919 shows variability in the fraction of linear polarisation,
consistent with the observations by \citet{van80}, \citet{Oesetal82},
and \citet{DolTap84}, who found the star to be periodically variable with
an amplitude of $\sim 0.3$\,\% .

The blue supergiant HD\,149404 (observed twice by us) shows
variability both in \pl\ and \thetae\ (the \thetae\ variability was
already detected in a previous work, see the discussion in
Sect.~\ref{Sect_Incons}).

Star HD\,116852 shows a discrepancy of 0.2\,\% between the
polarisation measured in two different observing epochs, which may
also be due to intrinsic variability.

\section{Conclusions}
Using the FORS2 instrument of the ESO VLT, we have obtained moderate-resolution ($R \sim 880$) polarisation spectra with
high \snr\ for 101
diffuse interstellar lines of sight, and made them publicly available
-- at different rebinning steps -- at CDS. We used the polarisation
spectra to estimate broadband linear polarisation values in the {\it
  BVRI} filters and for an analysis with the Serkowski curve. In
particular, we fitted the polarisation
spectra of all stars with $\pmax \ge 0.7$\,\% (76 targets) with
the Serkowski curve,
and, in addition, we fitted the wavelength-dependent position angle
with a first-order polynomial, estimating for each star the newly
introduced parameter d$\theta$/d$\lambda$ (measured in
\degr/100\,nm). As expected, we often found this parameter to be
consistent with zero, but with exceptions, especially for stars with
low polarisation. Remarkable exceptions are represented by the
Be stars
HD\,155851 and HD\,172694, which exhibit a sharp change in slope in
the polarisation and its position angle at wavelengths $\ga 820$\,nm,
in proximity of the Paschen edge.

Uncertainties due to photon noise are very small (typically $\le 0.1$\,\%
per nm, reduced to a few units in $10^{-4}$ when integrating in a
broadband filter). However, while processing the data, we discovered
some small discrepancies larger than photon noise between datasets
obtained within a very short interval of time. We have ascribed these
discrepancies to an imperfectly accurate
setting of the retarder waveplate. This issue was mitigated by
combining data obtained with and without order-separating filter.
Spectra are also affected by cosmic rays, and, corresponding to sharp
features in intensity (such as emission lines or the O$_2$-A band),
exhibit polarisation spikes that we believe to be spurious (at least
most of them). In spite of these caveats, our Large Interstellar
Polarisation Survey (LIPS) of the Southern Hemisphere significantly
extends the current dataset of moderate-resolution spectropolarimetry
of the diffuse ISM, particularly for sightlines with relatively low
interstellar reddening. Owing to systematic errors, we estimate that the
accuracy of our BBLP measurements is of the order of 0.1\,\%.

In previous works, it was found that the parameters of the Serkowski
curve are not independent of each other; for example, \citet{Whietal92} found
a linear relationship between the wavelength \lmax\ at which the
polarisation reaches maximum and the width $K$ of the polarisation
spectrum. Our results deviate significantly from the trends found by
\citet{Whietal92}. We have investigated this issue by means of
numerical simulations, without reaching a clear conclusion. On the
one hand, the $K$ parameter of the Serkowski curve (which is related
to the width of the polarisation spectrum) may not be sufficiently
well constrained by our observations. However, the observed deviations
may also be attributed to real physical differences -- either in dust
grains or interstellar conditions -- between the different regions
probed with LIPS and the sample by \citet{Whietal92}.

We have also performed a preliminary investigation of the relationship
between polarisation and interstellar extinction, improving a
previously found relationship between polarisation efficiency and
visual (or selective) extinction. In agreement with previous works, we
 were unable to confirm the existence of a linear relationship between the
ratio of total-to-selective extinction and the wavelength at which the
polarisation reached maximum.

Moderate-resolution spectropolarimetry adds new information that can
help us to better understand the properties of dust and/or ISM
conditions. Our Southern Hemisphere FORS2 data will be combined with
the Northern Hemisphere ISIS data, which will be described in a
separate publication. This combined LIPS dataset will then enable a
number of detailed studies of polarisation properties and modelling of
dust grains.

\begin{acknowledgements}
This work is based on observations collected at the European
Organisation for Astronomical Research in the Southern Hemisphere
under ESO programmes 095.C-0855 and 096.C-0159, PI=N.L.J.\ Cox.
Some of the data presented in this paper were obtained from the
Mikulski Archive for Space Telescopes (MAST). STScI is operated by the
Association of Universities for Research in Astronomy, Inc., under
NASA contract NAS5-26555. Support for MAST for non-HST data is
provided by the NASA Office of Space Science via grant NNX09AF08G and
by other grants and contracts. NVV acknowledges the support from RFBR
grant 16-02-00194 and RFBR–DST grant 16-52-45005.

\end{acknowledgements}

\bibliography{sbabib}


\begin{tiny}
\begin{longtable}{lrrrll@{$\,\pm$\,}rl@{$\,\pm$\,}rl@{$\,\pm$\,}rcrrr}
  \caption{\label{Tab_Log}
    Target list with essential parameters and observing log.
    Coefficients in Cols.~6 to 8 are taken from the references of Col.~9.
    The exposure time corresponds to the sum of all
    exposures obtained with and without order-separating filter (typically evenly
  split between the two settings).}\\
\hline\hline
\multicolumn{1}{l}{  Star     } & 
\multicolumn{1}{c}{  RA       } & 
\multicolumn{1}{c}{  DEC      } & 
\multicolumn{1}{l}{  $V$      } & 
\multicolumn{1}{l}{Spectral type} & 
\multicolumn{2}{c}{ E(B-V)    } & 
\multicolumn{2}{c}{ $R_V$     } & 
\multicolumn{2}{c}{ $A_V$     } & 
\multicolumn{1}{c}{ REF       } & 
\multicolumn{1}{c}{  DATE     } & 
\multicolumn{1}{c}{  UT       } & 
\multicolumn{1}{c}{  Exp      } \\ 
            & 
            & 
            & 
            & 
            & 
\multicolumn{2}{c}{}& 
\multicolumn{2}{c}{}& 
\multicolumn{2}{c}{}& 
            & 
\multicolumn{1}{c}{           } & 
\multicolumn{1}{c}{           } & 
\multicolumn{1}{c}{  (sec)    } \\
\hline
\endfirsthead
\caption{continued.} \\
\hline\hline
\multicolumn{1}{l}{  Star     } & 
\multicolumn{1}{c}{  RA       } & 
\multicolumn{1}{c}{  DEC      } & 
\multicolumn{1}{l}{  $V$      } & 
\multicolumn{1}{l}{Spectral type} & 
\multicolumn{2}{c}{ E(B-V)   } & 
\multicolumn{2}{c}{ $R_V$     } & 
\multicolumn{2}{c}{ $A_V$     } & 
\multicolumn{1}{c}{ REF       } & 
\multicolumn{1}{c}{  DATE     } & 
\multicolumn{1}{c}{  UT       } & 
\multicolumn{1}{c}{  Exp      } \\ 
            & 
            & 
            & 
            & 
            & 
\multicolumn{2}{c}{}& 
\multicolumn{2}{c}{}& 
\multicolumn{2}{c}{}& 
            & 
\multicolumn{1}{c}{           } & 
\multicolumn{1}{c}{           } & 
\multicolumn{1}{c}{  (sec)    } \\
\hline
\endhead
\hline
\endfoot
AzV18         &00:47:12.0&  $-$73:06:33.2&       12.48&  B2.5Ia         & 0.17 & 0.03   &    2.90 & 0.42 &      0.49 & 0.11 &    C  &2015-10-14 &04:19&  720\\  
AzV456        &01:10:55.0&  $-$72:42:56.3&       12.89&  O9.7Ib         & 0.26 & 0.03   &    2.19 & 0.23 &      0.57 & 0.08 &    C  &2015-10-14 &03:33&  720\\  
HD 36982      &05:35:09.8&  $-$05:27:53.3&        8.46&  B1.5Vp         & 0.36 & 0.06   &    4.84 & 0.52 &      1.74 & 0.34 &    V  &2015-10-17 &05:14&   32\\  
HD 37021      &05:35:16.1&  $-$05:23:06.9&        7.96&  B0 V           & 0.48 & 0.02   &    5.84 & 0.26 &      2.80 & 0.17 &    V  &2015-10-17 &05:27&   24\\  
HD 37061      &05:35:31.4&   $-5$:16:02.6&        6.83&  B0.5 V         & 0.56 & 0.04   &    4.29 & 0.21 &      2.40 & 0.21 &    V  &2015-10-14 &05:27&   24\\  
HD 38029      &05:36:55.2&  $-$69:11:37.7&       11.56&  WC+O6/6.5III   & 0.27 & 0.03   &    3.80 & 0.20 &      1.00 & 0.12 &  C06  &2015-10-14 &04:45&  480\\  
HD 37367      &05:39:18.3&     29:12:54.8&        5.99&  B2IV$-$V       & 0.42 & 0.04   &    3.55 & 0.44 &      1.49 & 0.24 &    V  &2015-10-16 &08:54&    8\\  
HD 37903      &05:41:38.4&  $-$02:15:32.5&        7.83&  B1.5 V         & 0.35 & 0.04   &    3.74 & 0.31 &      1.31 & 0.18 &    V  &2015-10-17 &05:40&   16\\  
HD 38087      &05:43:00.6&  $-$02:18:45.4&        8.29&  B5 V           & 0.30 & 0.04   &    4.93 & 0.44 &      1.48 & 0.24 &    V  &2015-10-17 &07:21&   24\\  
HD 39680      &05:54:44.7&     13:51:17.1&        7.99&  O6V:[n]pevar &\multicolumn{2}{c}{0.32}&4.73& 0.72&     1.52 & 0.72 &    P  &2015-10-14 &08:50&   24\\  
              &          &               &            &                 & \multicolumn{6}{c}{}                              &       &2015-12-01 &04:18&   48\\  %
HD 251204     &06:05:05.7&     23:23:38.5&       10.20&  B0             & 0.78 & 0.04   &    2.97 & 0.12 &      2.32 & 0.15 &    V  &2015-10-14 &08:17&  160\\  
HD 252325     &06:09:00.3&     20:38:25.9&       10.80&  B1:V: &\multicolumn{2}{c}{0.87}&    3.55 & 0.13 &      3.09 & 0.18 &    V  &2015-10-14 &08:33&  240\\  
HD 43384      &06:16:58.7&     23:44:27.3&        6.25&  B3Iab &\multicolumn{2}{c}{0.58}&    3.29 & 0.08 &      1.90 & 0.08 &    W  &2015-10-14 &09:05&   12\\  
HD 45314      &06:27:15.8&     14:53:21.2&        6.64&  O9:npe         & 0.43 & 0.04   &    4.42 & 0.30 &      1.90 & 0.22 &    V  &2015-10-17 &07:36&   16\\  
Walker 67     &06:40:37.2&     09:47:29.7&       10.79&  B1V            & 0.87 & 0.09   &    5.06 & 0.32 &      4.40 & 0.53 &    V  &2015-10-17 &07:52&  240\\  
HD 49787      &06:49:55.5&  $-$05:30:47.5&        7.54&  B1 V:pe&\multicolumn{2}{c}{0.18}&   2.39 & 0.20 &\multicolumn{2}{c}{0.43}&W&2015-10-17 &06:11&   12\\  
HD 50820      &06:54:42.0&  $-$01:45:23.4&        6.21&  B3IVe+K2II&\multicolumn{2}{c}{0.77}&4.39 & 0.08 &      3.38 & 0.08 &    W  &2015-10-17 &06:24&   12\\  
HD 54662      &07:09:20.2&  $-$10:20:47.6&        6.21&  O7Vzvar?&\multicolumn{2}{c}{0.26}&  3.12 & 0.22 &      0.81 & 0.22 &    W  &2015-10-17 &06:41&   12\\  
HD 61827      &07:39:49.3&  $-$32:34:42.1&        8.20&  B3 Iab&\multicolumn{2}{c}{0.81}&    3.22 & 0.06 &      2.61 & 0.06 &    W  &2015-05-22 &23:00&   32\\  
CD$-$28 5205  &07:58:42.9&  $-$28:26:19.8&       11.16&  B3/4           & 0.57 & 0.11   &    3.09 & 0.39 &      1.76 & 0.40 &    V  &2015-05-12 &00:43&  400\\  
HD 66194      &07:58:50.6&  $-$60:49:28.1&        5.81&  B2 Vne&\multicolumn{2}{c}{0.18}&    2.25 & 0.32 &      0.41 & 0.32 &    V  &2015-05-22 &23:17&    4\\  
LS 908        &07:59:12.0&  $-$28:34:04.3&       11.59&  O9             & 0.64 & 0.07   &    2.85 & 0.23 &      1.82 & 0.25 &    V  &2015-05-12 &01:07&  480\\  
HD 73882      &08:39:09.5&  $-$40:25:09.3&        7.19&  O8.5IV         & 0.69 & 0.04   &    3.56 & 0.13 &      2.46 & 0.17 &    V  &2015-05-22 &23:35&   12\\  
HD 75309      &08:47:28.0&  $-$46:27:04.0&        7.84&  B2Ib/II        & 0.29 & 0.04   &    3.53 & 0.40 &      1.02 & 0.18 &    V  &2015-05-23 &00:44&   20\\  
HD 75860      &08:50:53.2&  $-$43:45:05.4&        7.58&  B2C Iab&\multicolumn{2}{c}{0.90}&   3.44 & 0.06 &      3.10 & 0.06 &    W  &2015-05-22 &23:54&   20\\  
HD 76868      &08:58:59.3&     03:39:22.0&        8.04&  B5/7e &\multicolumn{2}{c}{0.39}&    5.70 & 0.15 &      2.22 & 0.15 &    W  &2015-05-29 &23:36&   32\\  
HD 79186      &09:11:04.4&  $-$44:52:04.4&        5.00&  B5Ia           & 0.40 & 0.04   &    3.21 & 0.56 &      1.28 & 0.26 &    V  &2015-05-23 &00:11&    2\\  
HD 80558      &09:18:42.4&  $-$51:33:38.3&        5.93&  B6Ia  &\multicolumn{2}{c}{0.60}&    3.35 & 0.08 &      2.01 & 0.08 &    W  &2015-05-23 &00:27&    4\\  
HD 89137      &10:15:40.1&  $-$51:15:24.1&        7.97&  ON9.7II(n)     & 0.27 & 0.04   &    2.68 & 0.28 &      0.72 & 0.13 &    V  &2015-05-23 &01:02&   32\\  
HD 91824      &10:34:46.6&  $-$58:09:22.0&        8.15&  O7V((f))z      & 0.24 & 0.07   &    3.35 & 0.62 &      0.80 & 0.28 &    V  &2015-05-24 &01:55&   32\\  
HD 91983      &10:35:54.2&  $-$58:15:27.4&        8.58&  O9.5/B0I/II    & 0.29 & 0.04   &    2.89 & 0.34 &      0.84 & 0.15 &    V  &2016-01-30 &05:37&   32\\  
HD 93160      &10:44:07.3&  $-$59:34:30.6&        7.60&  O7III((f))     & 0.39 & 0.05   &    4.87 & 0.38 &      1.90 & 0.28 &    V  &2015-05-24 &02:12&   20\\  
HD 93205      &10:44:33.7&  $-$59:44:15.5&        7.75&  O3.5V((f))+O8V & 0.38 & 0.04   &    3.25 & 0.24 &      1.23 & 0.16 &    V  &2015-05-24 &01:06&   24\\  
HD 93222      &10:44:36.2&  $-$60:05:29.0&        8.10&  O7V((f))z      & 0.36 & 0.06   &    4.76 & 0.48 &      1.71 & 0.33 &    V  &2015-05-24 &01:22&   32\\  
HD 303308     &10:45:05.8&  $-$59:40:06.4&        8.17&  O4.5V((fc))    & 0.45 & 0.05   &    3.02 & 0.21 &      1.36 & 0.17 &    V  &2015-05-24 &01:39&   32\\  
              &          &               &            &                 & \multicolumn{6}{c}{}                              &       &2016-01-30 &04:47&   24\\  %
HD 93632      &10:47:12.5&  $-$60:05:49.8&        9.10&  O5Ifvar        & 0.56 & 0.05   &    4.17 & 0.24 &      2.33 & 0.25 &    V  &2015-05-12 &01:29&   80\\  
              &          &               &            &                 & \multicolumn{6}{c}{}                              &       &2015-12-29 &08:01&   48\\  %
HD 93843      &10:48:37.8&  $-$60:13:25.5&        7.33&  O5III(fc)      & 0.27 & 0.05   &    3.89 & 0.41 &      1.05 & 0.20 &    V  &2015-05-12 &01:45&   12\\  
              &          &               &            &                 & \multicolumn{6}{c}{}                              &       &2015-12-29 &07:03&    8\\  %
HD 94493      &10:53:15.1&  $-$60:48:53.2&        7.27&  B0.5 Ib        & 0.23 & 0.04   &    3.57 & 0.45 &      0.82 & 0.18 &    V  &2015-05-12 &02:02&   12\\  
              &          &               &            &                 & \multicolumn{6}{c}{}                              &       &2016-01-26 &05:19&   12\\  %
HD 96715      &11:07:32.8&  $-$59:57:48.7&        8.30&  O4V((f))z      & 0.43 & 0.04   &    2.62 & 0.18 &      1.13 & 0.13 &    V  &2015-05-20 &04:42&   32\\  
HD 97484      &11:12:04.5&  $-$61:05:42.9&        8.52&  O8V            & 0.60 & 0.05   &    2.57 & 0.15 &      1.54 & 0.16 &    V  &2015-05-17 &02:13&   40\\  
              &          &               &            &                 & \multicolumn{6}{c}{}                              &       &2016-02-14 &03:26&   32\\  %
HD 99953      &11:29:15.1&  $-$63:33:14.2&        6.57&  B1 Ia          & 0.48 & 0.06   &    3.69 & 0.30 &      1.77 & 0.27 &    V  &2015-05-25 &01:33&    8\\  
HD 101065     &11:37:37.0&  $-$46:42:34.9&        8.03&  F8/G0p&\multicolumn{2}{c}{0.91}&    2.30 & 0.05 &      2.09 & 0.05 &    W  &2015-05-25 &01:49&   32\\  
HD 103779     &11:56:57.5&  $-$63:14:56.7&        7.22&  B0.5 Iab       & 0.21 & 0.04   &    3.29 & 0.43 &      0.69 & 0.15 &    V  &2015-05-23 &01:21&   12\\  
HD 104705     &12:03:23.9&  $-$62:41:45. &        7.83&  B0.5 III       & 0.23 & 0.07   &    2.81 & 0.57 &      0.65 & 0.24 &    V  &2015-05-23 &01:39&   20\\  
HD 105416     &12:08:14.7&  $-$48:41:32.9&        5.34&  B9.5/A0V&\multicolumn{2}{c}{0.05}&  4.62 & 0.75 &      0.23 & 0.04 &    W  &2015-05-12 &03:58&    2\\  
              &          &               &            &                 & \multicolumn{6}{c}{}                              &       &2016-02-07 &04:24&    4\\ %
HD 108639     &12:29:09.5&  $-$60:48:17.5&        7.81&  B1III&\multicolumn{2}{c}{0.35}&\multicolumn{2}{c}{}&\multicolumn{2}{c}{}&X&2016-01-23 &04:51&   16\\  
HD 109399     &12:35:16.5&  $-$72:43:00.8&        7.67&  B0.5 Ib&\multicolumn{2}{c}{0.21}&   3.26 & 0.18 &      0.68 & 0.04 &    W  &2015-05-23 &01:56&   20\\  
HD 111934     &12:53:37.6&  $-$60:21:25.4&        6.92&  B1.5 Ib        & 0.51 & 0.06   &    2.45 & 0.20 &      1.25 & 0.18 &    V  &2015-04-07 &08:54&   12\\  
HD 112272     &12:56:33.7&  $-$64:21:39.2&        7.39&  B0.5 Ia        & 1.00 & 0.04   &    3.09 & 0.09 &      3.09 & 0.15 &    V  &2015-04-07 &09:11&   16\\  
              &          &               &            &                 & \multicolumn{6}{c}{}                              &       &2015-12-29 &07:37&   12\\  %
CPD$-$63 2495 &13:02:47.7&  $-$63:50:08.7&       10.08&  O9.5Ve         & 1.02 & 0.06   &    3.75 & 0.15 &      3.83 & 0.27 &    V  &2015-04-07 &07:50&  160\\  
HD 113904     &13:08:07.2&  $-$65:18:21.7&        5.53&  O9 III&\multicolumn{2}{c}{0.21}&    3.63 & 0.17 &      0.76 & 0.04 &    W  &2015-04-07 &09:27&    3\\  
              &          &               &            &                 & \multicolumn{6}{c}{}                              &       &2015-12-29 &07:20&    6\\  %
HD 114886     &13:14:44.4&  $-$63:34:51.8&        6.89&O9\,III+O9.5\,III&\multicolumn{2}{c}{0.29}&2.90&0.16&    0.84 & 0.05 &    W  &2015-05-12 &02:23&   12\\  
              &          &               &            &                 & \multicolumn{6}{c}{}                              &       &2016-01-30 &05:06&   24\\  %
HD 115071     &13:16:04.8&  $-$62:35:01.5&        7.97&  O9.5III+B0Ib&\multicolumn{2}{c}{0.44}&\multicolumn{2}{c}{}&\multicolumn{2}{c}{}&X&2016-01-26&05:42&24\\  
HD 115842     &13:20:48.3&  $-$55:48:02.5&        6.09&  B0.5Ia/ab&\multicolumn{2}{c}{0.51}& 3.18 & 0.10 &      1.62 & 0.05 &    W  &2015-05-12 &03:39&    4\\  
              &          &               &            &                 & \multicolumn{6}{c}{}                              &       &2016-01-30 &05:22&    8\\  %
HD 116852     &13:30:23.5&  $-$78:51:20.5&        8.47&  O8.5II-III((f))& 0.21 & 0.04   &    2.42 & 0.37 &      0.51 & 0.12 &    V  &2015-05-12 &05:03&   40\\  
              &          &               &            &                 & \multicolumn{6}{c}{}                              &       &2016-02-22 &05:02&   32\\  %
HD 119159     &13:42:56.1&  $-$56:46:04.7&        6.00&  B1/2II&\multicolumn{2}{c}{0.09}&    2.59 & 0.60 &      0.23 & 0.05 &    W  &2015-05-12 &03:23&    4\\  
              &          &               &            &                 & \multicolumn{6}{c}{}                              &       &2016-01-28 &07:56&    8\\  %
HD 122879     &14:06:25.2&  $-$59:42:57.2&        6.50&  B0Ia           & 0.36 & 0.05   &    3.15 & 0.30 &      1.13 & 0.20 &    V  &2015-05-12 &04:13&    8\\  
HD 124314     &14:15:01.6&  $-$61:42:24.4&        6.64&  O6 IV(n)((f))&\multicolumn{2}{c}{0.50}&2.97&0.12&      1.49 & 0.06 &    W  &2015-05-12 &04:28&    8\\  
              &          &               &            &                 & \multicolumn{6}{c}{}                              &       &2016-02-22 &05:33&   16\\  %
HD 129557     &14:45:11.0&  $-$55:36:05.9&        6.09&  B2III          & 0.23 & 0.07   &    2.29 & 0.56 &      0.53 & 0.23 &    V  &2015-05-11 &08:48&    4\\  
              &          &               &            &                 & \multicolumn{6}{c}{}                              &       &2016-02-22 &04:46&    8\\  %
HD 133518     &15:06:56.0&  $-$52:01:47.2&        6.39&  B2 Vp &\multicolumn{2}{c}{0.15}&    1.79 & 0.37 &      0.27 & 0.06 &    W  &2015-05-11 &09:06&    4\\  
              &          &               &            &                 & \multicolumn{6}{c}{}                              &       &2016-02-12 &07:10&    8\\  %
HD 134591     &15:11:51.0&  $-$34:45:47.4&        8.37&  B5III          & 0.24 & 0.05   &    2.51 & 0.40 &      0.60 & 0.16 &    V  &2015-05-12 &02:47&   40\\  
              &          &               &            &                 & \multicolumn{6}{c}{}                              &       &2016-02-22 &05:18&   32\\  %
HD 135591     &15:18:49.1&  $-$60:29:46.8&        5.46&  O8IV((f))&\multicolumn{2}{c}{0.22}& 3.57 & 0.18 &      0.79 & 0.04 &    W  &2015-05-12 &04:46&    3\\  
              &          &               &            &                 & \multicolumn{6}{c}{}                              &       &2016-01-29 &08:42&    6\\  %
HD 147683     &16:24:43.7&  $-$34:53:37.5&        7.05&  B3: Vn&\multicolumn{2}{c}{0.28}&\multicolumn{2}{c}{}&\multicolumn{2}{c}{}&X&2016-02-12 &08:08&    8\\  
HD 147888     &16:25:24.3&  $-$23:27:36.8&        6.74&  B3 V:          & 0.51 & 0.04   &    3.89 & 0.20 &      1.99 & 0.18 &    V  &2016-01-30 &08:40&   24\\  
HD 147889     &16:25:24.3&  $-$24:27:56.6&        7.90&  B2 V           & 1.10 & 0.04   &    3.95 & 0.13 &      4.35 & 0.21 &    V  &2015-05-13 &09:53&   32\\  
              &          &               &            &                 & \multicolumn{6}{c}{}                              &       &2015-05-24 &03:21&   32\\  %
              &          &               &            &                 & \multicolumn{6}{c}{}                              &       &2016-01-28 &08:58&   24\\  %
HD 148379     &16:29:42.3&  $-$46:14:35.6&       5.37&   B2Iab          & 0.73 & 0.07   &    3.28 & 0.36 &      2.40 & 0.35 &    V  &2015-05-11 &09:26&    2\\  
HD 148688     &16:31:41.8&  $-$41:49:01.7&       5.39&   B1Iaeqp&\multicolumn{2}{c}{0.58}&   3.33 & 0.06 &      1.93 & 0.03 &    W  &2015-05-11 &09:42&    2\\  
HD 148937     &16:33:52.4&  $-$48:06:40.5&        6.71&  O6f?p &\multicolumn{2}{c}{0.67}&    3.28 & 0.06 &      2.20 & 0.04 &    W  &2015-05-12 &05:20&    8\\  
              &          &               &            &                 & \multicolumn{6}{c}{}                              &       &2016-02-12 &07:52&   16\\  %
HD 149404     &16:36:22.6&  $-$42:51:31.9&        5.47&  O8.5Iab(f)p    & 0.62 & 0.06   &    3.53 & 0.38 &      2.19 & 0.31 &    V  &2015-05-12 &05:38&    3\\  
              &          &               &            &                 & \multicolumn{6}{c}{}                              &       &2016-01-30 &08:57&    6\\  %
HD 150136     &16:41:20.4&  $-$48:45:46.7&        5.65&O3.5$-$4 III(f*)+O&\multicolumn{2}{c}{0.50}&3.27&0.13&   1.64 & 0.07 &    W  &2015-05-20 &05:00&    3\\  
HD 151804     &16:51:33.7&  $-$41:13:49.9&        5.22&  O8Iaf          & 0.30 & 0.03   &    4.33 & 0.30 &      1.30 & 0.16 &    V  &2015-05-24 &03:39&    3\\  
HD 151805     &16:51:35.7&  $-$41:46:35.5&        9.01&  B1Ib           & 0.43 & 0.05   &    3.29 & 0.30 &      1.42 & 0.21 &    V  &2016-02-22 &08:24&   48\\  
HD 152235     &16:53:58.9&  $-$41:59:39.6&        6.38&  B0.5Ia         & 0.71 & 0.06   &    3.13 & 0.25 &      2.22 & 0.26 &    V  &2015-05-24 &03:58&    4\\  
HD 152248     &16:54:10.1&  $-$41:49:30.1&        6.05&  O7Iabf+O7Ib(f) & 0.41 & 0.04   &    3.68 & 0.26 &      1.51 & 0.18 &    V  &2015-05-24 &04:15&    4\\  
HD 152249     &16:54:11.6&  $-$41:50:57.2&        6.45&  OC9Iab         & 0.46 & 0.10   &    3.54 & 0.45 &      1.63 & 0.40 &    V  &2015-05-24 &04:31&    8\\  
HD 152408     &16:54:58.5&  $-$41:09:03.1&        5.77&  O8Iape         & 0.42 & 0.05   &    4.17 & 0.33 &      1.75 & 0.25 &    V  &2015-05-24 &04:47&    3\\  
HD 152424     &16:55:03.3&  $-$42:05:27.0&        6.27&  OC9.2Ia        & 0.68 & 0.04   &    3.28 & 0.15 &      2.23 & 0.17 &    V  &2015-05-24 &05:04&    4\\  
              &          &               &            &                 & \multicolumn{6}{c}{}                              &       &2016-02-12 &07:34&   12\\  %
HD 153919     &17:03:56.8&  $-$37:50:38.9&        6.51&  O6Iafcp        & 0.51 & 0.04   &    3.87 & 0.22 &      1.97 & 0.19 &    V  &2015-04-12 &08:58&    8\\  
              &          &               &            &                 & \multicolumn{6}{c}{}                              &       &2016-02-22 &07:26&   16\\  %
HD 154043     &17:05:18.9&  $-$47:04:08.5&        7.10&  B2Iab          &\multicolumn{2}{c}{0.78}&3.30 & 0.06 & 2.57 & 0.05 &    W  &2015-04-12 &08:39&   12\\  
              &          &               &            &                 & \multicolumn{6}{c}{}                              &       &2016-02-02 &09:08&    8\\%
HD 154368     &17:06:28.4&  $-$35:27:03.8&        6.13&  O9.2Iab        & 0.76 & 0.05   &    3.33 & 0.15 &      2.53 & 0.20 &    V  &2015-04-12 &09:19&    4\\
              &          &               &            &                 & \multicolumn{6}{c}{}                              &       &2016-02-22 &08:44&    8\\%
HD 155806     &17:15:19.2&  $-$33:32:54.3&        5.53&  O7.5V((f))z(e) &\multicolumn{2}{c}{0.28}&2.48 & 1.92 & 0.70 & 0.49 &    P  &2015-05-24 &05:36&    3\\
HD 155851     &17:15:33.8&  $-$32:41:23.1&        8.13&  B0Vne &\multicolumn{2}{c}{0.32}&    5.25 & 0.30 & 1.68 & 0.10 &    W  &2015-05-24 &05:19&   32\\  
HD 156201     &17:17:45.5&  $-$35:13:27.0&        8.01&  B0.5 Ia&\multicolumn{2}{c}{0.86}&   3.23 & 0.05 & 2.78 & 0.04 &    W  &2015-05-25 &02:06&   32\\  
HD 157038     &17:22:39.2&  $-$37:48:16.7&        6.44&  B3Iap  &\multicolumn{2}{c}{0.81}&   3.70 & 0.05 & 3.00 & 0.04 &    W  &2015-05-25 &02:32&    8\\  
HD 157978     &17:26:19.0&     07:35:44.3&        6.04&  G0/K0 II:+A1   &\multicolumn{2}{c}{0.65}&3.72 & 0.06 & 2.42 & 0.04 &    W  &2016-03-14 &09:27&    8\\  
HD 161056     &17:43:47.0&  $-$07:04:46.6&        6.32&  B1.5 V  &\multicolumn{2}{c}{0.59}&  3.13 & 0.08 & 1.85 & 0.05 &    W  &2016-02-17 &08:57&   12\\  
HD 163181     &17:56:16.1&  $-$32:28:30.0&        6.61&  O9.5Ia/ab      & 0.74 & 0.06   &    3.24 & 0.32 &      2.40 & 0.32 &    V  &2015-05-25 &02:57&    8\\  
HD 164073     &18:02:00.6&  $-$48:48:37.7&        8.03&  B3III/IV       & 0.21 & 0.04   &    2.96 & 0.42 &      0.62 & 0.15 &    V  &2015-05-25 &03:19&   32\\  
HD 164740     &18:03:38.3&  $-$24:22:35.0&        9.10&  O7: V + sec    & 0.89 & 0.03   &    5.03 & 0.11 &      4.48 & 0.18 &    V  &2015-04-07 &07:29&  160\\  
HD 315023     &18:04:20.6&  $-$24:13:54.9&       10.03&  OB             & 0.40 & 0.06   &    4.12 & 0.40 &      1.65 & 0.29 &    V  &2015-04-07 &08:11&  160\\  
HD 165319     &18:05:58.8&  $-$14:11:53.0&        8.04&  O9.7Ib C &\multicolumn{2}{c}{0.83} &3.25 & 0.05 &      2.70 & 0.04 &    W  &2015-10-16 &01:15&   24\\  
HD 167838     &18:17:37.7&  $-$15:25:50.6&        6.73&  B5 Ia          & 0.63 & 0.04   &    3.39 & 0.14 &      2.13 & 0.17 &    V  &2016-03-13 &09:06&   16\\  
BD$-$13 4920  &18:18:26.2&  $-$13:50:05.2&       10.06&  B1V            & 0.73 & 0.04   &    3.22 & 0.19 &      2.35 & 0.26 &    V  &2015-04-07 &08:30&  160\\  
HD 168076     &18:18:36.4&  $-$13:48:02.0&        8.25&  O4III(f)       & 0.76 & 0.04   &    3.47 & 0.12 &      2.64 & 0.17 &    V  &2015-05-30 &09:49&   32\\  
HD 169454     &18:25:15.2&  $-$13:58:42.3&        6.71&  B1Ia           & 1.08 & 0.04   &    3.37 & 0.09 &      3.64 & 0.17 &    V  &2016-03-13 &09:22&   16\\  
HD 170740     &18:31:25.7&  $-$10:47:45.0&        5.72&  B2 V           & 0.50 & 0.13   &    3.01 & 0.49 &      1.51 & 0.46 &    V  &2016-03-15 &09:06&   12\\  
HD 170938     &18:32:37.8&  $-$15:42:05.9&        7.99&  B1 Ia &\multicolumn{2}{c}{1.04}&    3.34 & 0.04 &      3.47 & 0.04 &    W  &2015-11-08 &00:08&   16\\  
HD 171957     &18:38:04.5&  $-$14:00:17.2&        6.47&  B9 IV &\multicolumn{2}{c}{0.27} &\multicolumn{2}{c}{}&\multicolumn{2}{c}{}& X  &2015-11-08 &00:41&   12\\  
HD 172694     &18:42:16.6&  $-$15:51:20.8&        8.12&  B1: Vne&\multicolumn{2}{c}{0.37}&   5.04 & 0.16 &      1.86 & 0.06 &    W  &2015-06-02 &04:27&  160\\  
HD 203532     &21:33:54.6&  $-$82:40:59.1&        6.38&  B3 IV          & 0.28 & 0.03   &    3.37 & 0.24 &      0.94 & 0.11 &    V  &2015-06-02 &06:22&    8\\  
HD 210121     &22:08:11.9&  $-$03:31:52.8&        7.68&  B7II           & 0.31 & 0.05   &    2.42 & 0.29 &      0.75 & 0.15 &    V  &2015-10-02 &03:02&   24\\  
\hline
\end{longtable}
\end{tiny}

\noindent
Key to references: 
V=\citet{Valetal04}; W=\citet{Wegner03}; C=\citet{Caretal05};
C06=\citet{Coxetal06}; P=\citet{Patetal03}; G=\citet{Goretal03}; 
X=No R(V) data.

\newpage
 \begin{tiny}
\begin{longtable}{lcrrrrrrrrr@{\,$\pm$\,}lr@{\,$\pm$\,}lr@{\,$\pm$\,}lr@{\,$\pm$\,}ll}
  \caption{\label{Tab_Results} BBLP values in the BVRI filters, parameters of the Serkowski curves, and
    wavelength gradients for the stars observed in this LIPS survey. 
    Best-fit parameters are omitted for stars with $\pl \le 0.7$\,\%.
    The last column refers to some comments listed at the end of the Table.}\\
\hline\hline
                  &                          
                  &                          
                  \multicolumn{2}{c}{$B$}&   
                  \multicolumn{2}{c}{$V$}&   
                  \multicolumn{2}{c}{$R$}&   
                  \multicolumn{2}{c}{$I$}&   
                  \multicolumn{9}{c}{}       \\ 
                  &                          
                  &                          
                  \multicolumn{1}{c}{\pl}&   
                  \multicolumn{1}{c}{$\thetae$}& 
                  \multicolumn{1}{c}{\pl}&   
                  \multicolumn{1}{c}{$\thetae$ }& 
                  \multicolumn{1}{c}{\pl}&    
                  \multicolumn{1}{c}{$\thetae$}& 
                  \multicolumn{1}{c}{\pl}& 
                  \multicolumn{1}{c}{$\thetae$}& 
                  \multicolumn{2}{c}{$K$}      & 
                  \multicolumn{2}{c}{\lmax}   & 
                  \multicolumn{2}{c}{\pmax}   & 
                  \multicolumn{2}{c}{d$\theta$/d$\lambda$}& 
                  \multicolumn{1}{l}{C}   \\ 
                  Star &                     
                  Date &                     
                  \multicolumn{1}{c}{(\%)}   & 
                  \multicolumn{1}{c}{(\degr)}& 
                  \multicolumn{1}{c}{(\%)}   &  
                  \multicolumn{1}{c}{(\degr)}& 
                  \multicolumn{1}{c}{(\%)}   &  
                  \multicolumn{1}{c}{(\degr)}& 
                  \multicolumn{1}{c}{(\%)}   & 
                  \multicolumn{1}{c}{(\degr)}& 
                  \multicolumn{2}{c}{}       & 
                  \multicolumn{2}{c}{(nm)}  & 
                  \multicolumn{2}{c}{(\%)}  & 
                  \multicolumn{2}{l}{($^\circ$/100\,nm)} & 
                                            \\    
\hline
\endfirsthead
\caption{continued.} \\
\hline\hline
                  &                          
                  &                          
                  \multicolumn{2}{c}{$B$}&   
                  \multicolumn{2}{c}{$V$}&   
                  \multicolumn{2}{c}{$R$}&   
                  \multicolumn{2}{c}{$I$}&   
                  \multicolumn{9}{c}{}       \\ 
                  &                          
                  &                          
                  \multicolumn{1}{c}{\pl}&   
                  \multicolumn{1}{c}{$\thetae$}& 
                  \multicolumn{1}{c}{\pl}&   
                  \multicolumn{1}{c}{$\thetae$ }& 
                  \multicolumn{1}{c}{\pl}&    
                  \multicolumn{1}{c}{$\thetae$}& 
                  \multicolumn{1}{c}{\pl}& 
                  \multicolumn{1}{c}{$\thetae$}& 
                  \multicolumn{2}{c}{$K$}      & 
                  \multicolumn{2}{c}{\lmax}   & 
                  \multicolumn{2}{c}{\pmax}   & 
                  \multicolumn{2}{c}{d$\theta$/d$\lambda$}& 
                  \multicolumn{1}{l}{C}   \\ 
                  Star &                     
                  Date &                     
                  \multicolumn{1}{c}{(\%)}   & 
                  \multicolumn{1}{c}{(\degr)}& 
                  \multicolumn{1}{c}{(\%)}   &  
                  \multicolumn{1}{c}{(\degr)}& 
                  \multicolumn{1}{c}{(\%)}   &  
                  \multicolumn{1}{c}{(\degr)}& 
                  \multicolumn{1}{c}{(\%)}   & 
                  \multicolumn{1}{c}{(\degr)}& 
                  \multicolumn{2}{c}{}       & 
                  \multicolumn{2}{c}{(nm)}  & 
                  \multicolumn{2}{c}{(\%)}  & 
                  \multicolumn{2}{c}{($^\circ$/100\,nm)} & 
                                            \\    
\hline
\endhead
\hline
\endfoot
AzV\,18        &2015-10-14& 0.41&105.7&  0.40&105.4&  0.36&105.1&  0.30&105.9& \multicolumn{6}{c}{}         &$-$0.37& 0.52 &   \\        
AzV\,456       &2015-10-14& 1.24&162.3&  1.32&161.1&  1.33&161.6&  1.30&161.1&  0.27&0.16& 626&32& 1.33&0.02&$-$0.30& 0.23 &   \\        
HD\,36982      &2015-10-17& 0.57& 62.2&  0.68& 59.9&  0.65& 57.5&  0.58& 52.8& \multicolumn{6}{c}{}         &$-$2.56& 0.26 &   \\        
HD\,37021      &2015-10-17& 0.26& 95.2&  0.34&101.7&  0.38&105.7&  0.45&110.2& \multicolumn{6}{c}{}         &$ $3.82& 0.34 &   \\        
HD\,37061      &2015-10-14& 1.25& 64.6&  1.54& 61.6&  1.64& 59.5&  1.64& 57.5&  1.41&0.11& 697&12& 1.68&0.00&$-$1.85& 0.07 &   \\        
HD\,38029      &2015-10-14& 1.82&105.2&  1.87&104.9&  1.77&105.1&  1.62&105.3&  0.85&0.17& 514&23& 1.88&0.01&$ $0.23& 0.07 &   \\        
HD\,37367      &2015-10-16& 0.88& 14.2&  1.02& 15.2&  1.03& 15.5&  0.97& 15.6&  1.18&0.08& 624& 7& 1.04&0.02&$ $0.27& 0.10 &   \\        
HD\,37903      &2015-10-17& 1.45&119.9&  1.74&121.0&  1.80&121.6&  1.72&122.4&  1.51&0.23& 646& 8& 1.82&0.01&$ $0.69& 0.06 &   \\        
HD\,38087      &2015-10-17& 2.37&118.4&  2.57&118.1&  2.51&118.6&  2.30&119.3&  1.18&0.04& 572& 5& 2.59&0.01&$ $0.30& 0.07 &   \\        
HD\,39680      &2015-10-14& 0.44&163.6&  0.60&163.0&  0.68&162.6&  0.70&162.9& \multicolumn{6}{c}{}         &$ $0.27& 0.21 & 1 \\        
               &2015-12-01& 0.39&166.1&  0.56&163.9&  0.63&163.0&  0.65&163.4& \multicolumn{6}{c}{}         &$-$1.09& 0.17 & 1 \\        
HD\,251204     &2015-10-14& 4.64&153.8&  4.97&154.3&  4.88&154.6&  4.51&155.0&  1.00&0.02& 575& 2& 5.00&0.01&$ $0.33& 0.03 &   \\        
HD\,252325     &2015-10-14& 4.69&162.5&  5.13&162.6&  5.08&162.8&  4.73&163.1&  1.07&0.04& 595& 3& 5.18&0.01&$ $0.19& 0.02 &   \\        
HD\,43384      &2015-10-14& 2.88&170.5&  3.05&171.1&  2.98&171.4&  2.75&171.7&  0.92&0.03& 566& 2& 3.06&0.01&$ $0.27& 0.02 &   \\        
HD\,45314      &2015-10-17& 0.91&162.8&  1.15&160.1&  1.21&159.2&  1.17&158.8&  1.58&0.06& 665& 9& 1.23&0.00&$-$0.79& 0.13 &   \\        
Walker 67      &2015-10-17& 3.01& 18.7&  4.16& 17.2&  4.81& 16.4&  5.16& 16.0&  1.40&0.05& 826& 3& 5.20&0.04&$-$0.58& 0.06 &   \\        
HD\,49787      &2015-10-17& 0.23& 10.7&  0.26&  4.2&  0.26&  4.5&  0.28&  8.6& \multicolumn{6}{c}{}         &$ $0.07& 0.82 &   \\        
HD\,50820      &2015-10-17& 0.15&112.7&  0.15&117.9&  0.13&126.3&  0.13&138.2& \multicolumn{6}{c}{}         &$ $7.05& 0.62 &   \\        
HD\,54662      &2015-10-17& 0.79&145.0&  0.86&145.1&  0.84&146.3&  0.79&148.6&  0.95&0.24& 571& 8& 0.86&0.01&$ $1.10& 0.13 &   \\        
HD\,61827      &2015-05-22& 1.59&  4.0&  1.78&  3.5&  1.80&  3.5&  1.71&  3.6&  1.09&0.04& 624& 7& 1.82&0.00&$-$0.11& 0.06 &   \\        
CD$-$28 5205   &2015-05-12& 1.01&100.7&  1.18&100.6&  1.17&100.5&  1.09&100.7&  1.55&0.10& 610& 7& 1.21&0.01&$-$0.06& 0.10 &   \\        
HD\,66194      &2015-05-22& 0.42&144.0&  0.43&144.0&  0.42&149.4&  0.39&156.5& \multicolumn{6}{c}{}         &$ $4.05& 0.34 &   \\        
LS\,908        &2015-05-12& 1.74&110.0&  1.90&110.0&  1.85&109.7&  1.67&109.7&  1.19&0.16& 567&11& 1.91&0.01&$ $0.02& 0.12 &   \\        
HD\,73882      &2015-05-22& 1.58&164.6&  1.95&164.1&  2.07&164.0&  2.02&164.1&  1.42&0.12& 684& 8& 2.10&0.02&$-$0.09& 0.08 &   \\        
HD\,75309      &2015-05-23& 0.61& 52.1&  0.62& 54.3&  0.58& 53.0&  0.51& 50.1& \multicolumn{6}{c}{}         &$-$0.43& 0.33 &   \\        
HD\,75860      &2015-05-22& 1.57&142.0&  1.66&142.3&  1.57&142.3&  1.38&142.3&  1.41&0.06& 541& 5& 1.68&0.01&$ $0.06& 0.04 &   \\        
HD\,76868      &2015-05-29& 0.20& 27.6&  0.19& 23.0&  0.21& 18.2&  0.22& 14.2& \multicolumn{6}{c}{}         &$-$3.51& 0.37 &   \\        
HD\,79186      &2015-05-23& 2.52& 46.8&  2.60& 46.9&  2.46& 46.5&  2.17& 46.2&  1.07&0.02& 520&10& 2.62&0.00&$-$0.22& 0.05 &   \\        
HD\,80558      &2015-05-23& 2.88&164.0&  3.30&163.3&  3.27&163.3&  3.00&163.7&  1.33&0.03& 597&10& 3.34&0.06&$ $0.14& 0.04 & 2 \\        
HD\,89137      &2015-05-23& 0.37& 32.6&  0.44& 31.0&  0.47& 28.7&  0.46& 27.6& \multicolumn{6}{c}{}         &$-$1.52& 0.22 &   \\        
HD\,91824      &2015-05-24& 1.27& 96.9&  1.33& 96.2&  1.23& 96.0&  1.04& 95.7&  1.49&0.06& 517& 5& 1.35&0.01&$-$0.23& 0.06 &   \\        
HD\,91983      &2016-01-30& 1.05&130.0&  1.09&132.2&  1.05&134.8&  0.97&137.2&  1.00&0.21& 542& 6& 1.10&0.01&$ $2.16& 0.10 &   \\        
HD\,93160      &2015-05-24& 2.22&114.5&  2.29&114.7&  2.18&115.0&  1.92&115.9&  1.12&0.13& 530& 4& 2.31&0.01&$ $0.42& 0.03 &   \\        
HD\,93205      &2015-05-24& 1.96&100.5&  2.09& 99.5&  2.01& 99.2&  1.79& 98.4&  1.27&0.05& 552& 3& 2.10&0.01&$-$0.47& 0.05 &   \\        
HD\,93222      &2015-05-24& 0.79&127.1&  0.68&134.7&  0.57&146.5&  0.53&162.2& \multicolumn{6}{c}{}         &$ $9.34& 0.38 & 3 \\        
HD\,303308     &2015-05-24& 2.69& 99.0&  2.88& 99.0&  2.76& 98.9&  2.42& 98.9&  1.35&0.04& 550& 3& 2.90&0.01&$-$0.01& 0.05 &   \\        
               &2016-01-30& 2.73&100.1&  2.93& 99.9&  2.81&100.0&  2.49&100.0&  1.35&0.04& 551& 3& 2.95&0.01&$ $0.04& 0.04 &   \\        
HD\,93632      &2015-05-12& 0.75& 54.9&  1.15& 52.0&  1.35& 49.3&  1.47& 47.6&  1.57&0.10& 828&17& 1.49&0.01&$-$1.96& 0.12 &   \\        
               &2015-12-29& 0.93& 60.5&  1.17& 54.6&  1.33& 51.5&  1.40& 48.2&  1.41&0.19& 781&24& 1.40&0.05&$-$2.76& 0.12 &   \\        
HD\,93843      &2015-05-12& 0.64&101.5&  0.61&100.0&  0.53& 98.2&  0.44& 94.7& \multicolumn{6}{c}{}         &$-$2.15& 0.15 &   \\        
               &2015-12-29& 0.66&101.5&  0.64&100.8&  0.55&100.1&  0.43& 99.7& \multicolumn{6}{c}{}         &$-$0.44& 0.29 &   \\        
HD\,94493      &2015-05-12& 0.61&107.9&  0.62&106.0&  0.55&106.7&  0.44&108.4& \multicolumn{6}{c}{}         &$-$0.04& 0.29 &   \\        
               &2016-01-26& 0.66&107.8&  0.63&107.2&  0.55&109.2&  0.42&111.9& \multicolumn{6}{c}{}         &$ $1.13& 0.30 &   \\        
HD\,96715      &2015-05-20& 1.85&  3.4&  2.03&  2.5&  2.00&  2.2&  1.88&  2.4&  0.90&0.07& 591& 5& 2.03&0.01&$-$0.19& 0.08 &   \\        
HD\,97484      &2015-05-17& 0.97& 55.0&  1.02& 57.1&  0.95& 57.9&  0.86& 57.8&  1.23&0.19& 537&17& 1.03&0.01&$ $0.51& 0.15 &   \\        
               &2016-02-14& 0.93& 58.7&  0.98& 59.8&  0.92& 61.3&  0.82& 62.8&  1.15&0.11& 529& 9& 0.98&0.03&$ $1.31& 0.17 &   \\        
HD\,99953      &2015-05-25& 1.88&107.6&  1.93&107.2&  1.80&107.8&  1.55&108.4&  1.24&0.07& 516& 8& 1.95&0.00&$ $0.47& 0.05 &   \\        
HD\,101065     &2015-05-25& 0.02&  6.7&  0.10& 13.5&  0.12&  8.7&  0.16&  4.2& \multicolumn{6}{c}{}         &$-$2.87& 0.40 &   \\        
HD\,103779     &2015-05-23& 0.59& 77.2&  0.61& 75.1&  0.54& 72.0&  0.45& 69.9& \multicolumn{6}{c}{}         &$-$2.89& 0.17 &   \\        
HD\,104705     &2015-05-23& 0.69& 89.4&  0.76& 88.3&  0.79& 88.6&  0.71& 86.8&  0.34&0.39& 647&99& 0.77&0.07&$-$1.66& 0.45 &   \\        
HD\,105416     &2015-05-12& 0.18& 91.2&  0.21& 79.6&  0.16& 76.9&  0.12& 66.2& \multicolumn{6}{c}{}         &$-$4.78& 0.83 &   \\        
               &2016-02-07& 0.31& 86.1&  0.25& 89.1&  0.23& 88.5&  0.18& 88.2& \multicolumn{6}{c}{}         &$ $0.85& 0.41 &   \\        
HD\,108639     &2016-01-23& 1.72& 90.7&  1.92& 90.7&  1.83& 90.9&  1.64& 91.3&  1.29&0.23& 550& 5& 1.92&0.04&$ $0.21& 0.08 &   \\        
HD\,109399     &2015-05-23& 2.08&114.5&  2.19&114.5&  2.09&114.7&  1.81&115.5&  1.30&0.06& 538& 8& 2.20&0.01&$ $0.29& 0.06 &   \\        
HD\,111934     &2015-04-07& 2.61& 74.9&  2.82& 74.2&  2.71& 73.8&  2.40& 73.7&  1.35&0.09& 556&10& 2.83&0.02&$-$0.22& 0.05 &   \\        
HD\,112272     &2015-04-07& 0.73& 56.7&  0.84& 56.3&  0.81& 55.0&  0.71& 53.8&  1.93&0.13& 584&23& 0.85&0.02&$-$1.12& 0.13 &   \\        
               &2015-12-29& 0.68& 48.2&  0.72& 51.8&  0.71& 50.3&  0.64& 46.6&  1.56&0.19& 591&24& 0.73&0.02&$-$2.01& 0.19 &   \\        
CPD$-$63 2495  &2015-04-07& 2.49& 76.2&  2.75& 72.3&  2.63& 70.6&  2.36& 69.4&  1.31&0.08& 569& 9& 2.75&0.04&$-$1.56& 0.12 &   \\        
HD\,113904     &2015-04-07& 1.33& 83.1&  1.38& 82.4&  1.35& 81.8&  1.27& 81.2&  0.79&0.39& 540& 8& 1.40&0.05&$-$0.01& 0.15 & 4 \\        
               &2015-12-29& 1.19& 81.9&  1.38& 83.8&  1.30& 84.0&  1.12& 84.3&  1.99&0.34& 562&42& 1.39&0.02&$ $0.72& 0.13 & 4 \\        
HD\,114886     &2015-05-12& 1.76& 74.1&  1.89& 73.9&  1.80& 73.3&  1.58& 72.4&  1.31&0.07& 537& 5& 1.90&0.02&$-$0.54& 0.06 &   \\        
               &2016-01-30& 1.78& 74.9&  1.93& 74.8&  1.83& 74.5&  1.62& 74.3&  1.34&0.04& 547& 4& 1.93&0.02&$-$0.14& 0.04 &   \\        
HD\,115071     &2016-01-26& 1.87& 78.4&  2.03& 79.3&  1.94& 79.1&  1.69& 79.3&  1.35&0.05& 546& 4& 2.04&0.02&$-$0.01& 0.04 &   \\        
HD\,115842     &2015-05-12& 0.31&137.4&  0.28&142.6&  0.27&153.0&  0.34&158.0& \multicolumn{6}{c}{}         &$ $6.73& 0.86 &   \\        
               &2016-01-30& 0.29&117.9&  0.29&125.9&  0.29&135.5&  0.29&141.6& \multicolumn{6}{c}{}         &$ $7.18& 0.41 &   \\        
HD\,116852     &2015-05-12& 1.14&115.4&  1.23&115.4&  1.15&116.3&  0.99&116.9&  1.20&0.09& 517& 4& 1.23&0.01&$ $0.23& 0.13 &   \\        
               &2016-02-22& 0.84&111.9&  1.03&114.8&  1.02&116.6&  0.92&117.7&  1.56&0.11& 603&21& 1.04&0.06&$ $1.57& 0.17 &   \\        
HD\,119159     &2015-05-12& 1.16& 71.0&  1.24& 68.8&  1.17& 67.8&  0.99& 67.2&  1.56&0.07& 533& 8& 1.25&0.01&$-$0.65& 0.09 &   \\        
               &2016-01-28& 1.16& 69.8&  1.24& 69.0&  1.17& 68.2&  1.00& 67.1&  1.19&0.20& 511& 9& 1.25&0.02&$-$0.30& 0.13 &   \\        
HD\,122879     &2015-05-12& 1.66& 70.0&  1.79& 69.6&  1.71& 69.0&  1.48& 68.7&  1.68&0.04& 560&12& 1.81&0.01&$-$0.57& 0.07 &   \\        
HD\,124314     &2015-05-12& 2.12& 70.0&  2.22& 70.0&  2.14& 70.2&  1.92& 70.1&  1.19&0.18& 544& 8& 2.24&0.01&$ $0.10& 0.07 &   \\        
               &2016-02-22& 2.15& 69.7&  2.25& 70.5&  2.11& 70.6&  1.85& 71.0&  1.30&0.06& 530&10& 2.27&0.01&$ $0.40& 0.06 &   \\        
HD\,129557     &2015-05-11& 1.21& 80.4&  1.49& 80.6&  1.47& 79.7&  1.38& 79.1&  1.44&0.26& 607& 9& 1.51&0.07&$-$0.77& 0.13 &   \\        
               &2016-02-22& 1.20& 80.1&  1.34& 80.3&  1.27& 80.3&  1.11& 80.8&  1.51&0.20& 557&18& 1.34&0.01&$ $0.08& 0.07 &   \\        
HD\,133518     &2015-05-11& 0.69& 58.4&  0.78& 58.2&  0.73& 58.3&  0.60& 58.9&  1.61&1.27& 525&99& 0.79&0.05&$ $0.81& 0.26 & 5 \\        
               &2016-02-12& 0.80& 57.0&  0.80& 58.9&  0.74& 58.6&  0.66& 57.3&  1.25&0.14& 524&26& 0.80&0.04&$ $0.28& 0.18 & 6 \\        
HD\,134591     &2015-05-12&     &     &  0.28&105.0&  0.25&102.6&  0.26& 97.2& \multicolumn{6}{c}{}         &$-$1.94& 1.15 & 7 \\        
               &2016-02-22& 0.30&111.7&  0.29&112.6&  0.25&116.5&  0.22&122.4& \multicolumn{6}{c}{}         &$ $2.54& 1.02 &   \\        
HD\,135591     &2015-05-12& 1.28& 62.7&  1.42& 63.0&  1.36& 62.5&  1.20& 62.6&  1.53&0.08& 564& 6& 1.42&0.01&$-$0.21& 0.06 &   \\        
               &2016-01-29& 1.32& 64.4&  1.45& 64.3&  1.39& 64.6&  1.24& 65.3&  1.37&0.27& 557& 7& 1.45&0.02&$ $0.16& 0.08 &   \\        
HD\,147683     &2016-02-12& 1.58& 28.1&  1.71& 27.6&  1.66& 27.1&  1.50& 27.4&  1.13&0.08& 569& 8& 1.71&0.03&$-$0.16& 0.12 &   \\        
HD\,147888     &2016-01-30& 2.73& 54.4&  3.31& 53.9&  3.45& 53.5&  3.33& 53.0&  1.43&0.02& 662& 4& 3.49&0.01&$-$0.38& 0.02 &   \\        
HD\,147889     &2015-05-13& 2.60&177.5&  3.48&176.4&  3.97&175.5&  4.21&174.6&  1.32&0.02& 809& 4& 4.22&0.00&$-$0.72& 0.03 &   \\        
               &2015-05-24& 2.59&178.7&  3.47&176.8&  3.96&175.9&  4.16&175.0&  1.37&0.04& 798& 7& 4.18&0.01&$-$0.78& 0.04 &   \\        
               &2016-01-28& 2.70&178.9&  3.49&177.2&  3.95&176.1&  4.16&175.1&  1.18&0.09& 813&13& 4.18&0.01&$-$0.83& 0.04 &   \\        
HD\,148379     &2015-05-11& 1.82& 29.5&  1.85& 28.4&  1.82& 27.8&  1.69& 27.2&  0.95&0.06& 582& 4& 1.86&0.09&$-$0.37& 0.07 & 8 \\        
HD\,148688     &2015-05-11& 1.32& 17.7&  1.38& 20.1&  1.34& 20.3&  1.22& 20.5&  0.76&0.06& 527& 5& 1.39&0.01&$ $0.86& 0.17 &   \\        
HD\,148937     &2015-05-12& 1.54& 48.5&  1.70& 47.9&  1.66& 47.2&  1.50& 46.8&  1.31&0.11& 576& 6& 1.71&0.01&$-$0.47& 0.06 &   \\        
               &2016-02-12& 1.56& 48.0&  1.74& 48.1&  1.72& 47.6&  1.58& 47.4&  1.36&0.15& 597& 5& 1.76&0.01&$-$0.30& 0.05 &   \\        
HD\,149404     &2015-05-12& 2.70&177.7&  2.98&178.9&  2.93&179.7&  2.67&  0.5&  1.02&0.08& 569& 4& 3.00&0.04&$ $0.86& 0.05 & 9 \\        
               &2016-01-30& 2.44&  3.9&  2.71&  3.4&  2.71&  3.1&  2.50&  2.5&  1.24&0.03& 600& 9& 2.76&0.01&$-$0.32& 0.03 & 9 \\        
HD\,150136     &2015-05-20& 1.04& 56.6&  1.15& 58.6&  1.09& 59.1&  0.98& 58.5&  1.19&0.12& 537& 8& 1.15&0.04&$ $0.44& 0.16 &   \\        
HD\,151804     &2015-05-24& 1.04& 43.2&  1.10& 43.2&  1.09& 42.1&  0.98& 41.3&  1.17&0.12& 572& 8& 1.11&0.01&$-$0.65& 0.08 &   \\        
HD\,151805     &2016-02-22& 0.63& 29.4&  0.64& 34.3&  0.60& 36.5&  0.52& 37.8& \multicolumn{6}{c}{}         &$ $2.39& 0.39 &   \\        
HD\,152235     &2015-05-24& 0.77&117.3&  0.75&114.7&  0.66&113.5&  0.53&112.9&  1.77&0.13& 490&12& 0.78&0.01&$-$0.80& 0.21 &   \\        
HD\,152248     &2015-05-24& 0.50&118.6&  0.50&116.2&  0.45&115.4&  0.35&115.2& \multicolumn{6}{c}{}         &$-$1.38& 0.27 &   \\        
HD\,152249     &2015-05-24& 0.20& 66.8&  0.28& 64.2&  0.30& 59.7&  0.29& 57.7& \multicolumn{6}{c}{}         &$-$3.09& 0.38 &   \\        
HD\,152408     &2015-05-24& 0.55& 35.3&  0.69& 35.7&  0.71& 35.2&  0.63& 33.7&  1.72&0.12& 618&10& 0.71&0.03&$-$0.81& 0.15 &   \\        
HD\,152424     &2015-05-24& 0.31&109.9&  0.24&107.3&  0.14&108.9&  0.05&141.8& \multicolumn{6}{c}{}         &$ $3.61& 1.95 & 10\\        
               &2016-02-12& 0.38& 89.6&  0.31& 87.6&  0.27& 82.9&  0.21& 76.3& \multicolumn{6}{c}{}         &$-$4.73& 0.29 & 10\\        
HD\,153919     &2015-04-12& 2.40&  8.1&  2.60&  9.9&  2.55& 10.7&  2.32& 11.4&  1.09&0.12& 569& 4& 2.62&0.01&$ $0.78& 0.06 & 11\\        
               &2016-02-22& 2.65&  8.5&  2.75& 10.2&  2.67& 10.7&  2.41& 11.5&  0.99&0.04& 546& 2& 2.77&0.02&$ $0.60& 0.05 & 11\\        
HD\,154043     &2015-04-12& 3.12& 34.8&  3.51& 34.1&  3.45& 33.4&  3.17& 32.1&  1.30&0.03& 596& 4& 3.54&0.01&$-$0.79& 0.03 &   \\        
               &2016-02-02& 3.21& 36.8&  3.51& 36.2&  3.43& 35.4&  3.13& 34.7&  1.26&0.03& 579& 8& 3.54&0.02&$-$0.65& 0.04 &   \\        
HD\,154368     &2015-04-12& 0.39&166.1&  0.30&158.8&  0.27&148.8&  0.25&138.0& \multicolumn{6}{c}{}         &$-$7.96& 0.32 & 12\\        
               &2016-02-22& 0.37&166.0&  0.28&152.0&  0.27&142.9&  0.28&133.5& \multicolumn{6}{c}{}         &$-$8.29& 0.41 & 12\\        
HD\,155806     &2015-05-24& 0.77&153.3&  0.90&154.6&  0.90&156.0&  0.82&156.8&  1.05&0.29& 590&18& 0.90&0.02&$ $1.35& 0.14 &   \\        
HD\,155851     &2015-05-24& 2.48&147.5&  2.44&146.5&  2.31&145.9&  2.12&145.8&  0.78&0.05& 496& 4& 2.47&0.01&$-$0.78& 0.08 & 13\\        
HD\,156201     &2015-05-25& 0.87&133.8&  0.98&129.0&  0.98&127.9&  0.91&127.8&  1.12&0.11& 607& 9& 1.00&0.03&$-$0.54& 0.22 &   \\        
HD\,157038     &2015-05-25& 2.50&170.5&  2.54&168.4&  2.36&166.7&  2.08&165.4&  1.14&0.05& 513& 6& 2.57&0.02&$-$1.47& 0.10 &   \\        
HD\,157978     &2016-03-14& 0.62& 83.1&  0.67& 81.5&  0.61& 81.0&  0.53& 80.0& \multicolumn{6}{c}{}         &$-$0.61& 0.09 &   \\        
HD\,161056     &2016-02-17& 3.59& 67.4&  3.98& 67.6&  3.87& 67.4&  3.47& 67.1&  1.43&0.05& 572& 2& 4.02&0.04&$-$0.15& 0.03 &   \\        
HD\,163181     &2015-05-25& 1.43&173.0&  1.43&175.1&  1.38&176.8&  1.29&178.6&  0.50&0.13& 485&18& 1.45&0.01&$ $1.69& 0.09 &   \\        
HD\,164073     &2015-05-25& 0.91&  1.4&  1.08&  1.3&  1.08&  1.0&  1.02&  0.4&  1.05&0.10& 630& 9& 1.09&0.02&$-$0.26& 0.10 &   \\        
HD\,164740     &2015-04-07& 5.48& 93.8&  6.78& 94.6&  7.09& 94.9&  6.74& 95.2&  1.72&0.02& 657& 2& 7.20&0.02&$ $0.33& 0.02 &   \\        
HD\,315023     &2015-04-07& 1.07&151.9&  1.29&151.0&  1.38&151.1&  1.35&151.7&  1.33&0.52& 682&25& 1.39&0.02&$ $0.21& 0.06 &   \\        
HD\,165319     &2015-10-16& 1.14& 58.4&  1.16& 60.8&  1.07& 61.7&  0.87& 63.2&  1.58&0.12& 511&10& 1.18&0.01&$ $1.25& 0.13 &   \\        
HD\,167838     &2016-03-13& 0.29& 91.6&  0.32& 97.2&  0.30&101.7&  0.27&108.5& \multicolumn{6}{c}{}         &$ $4.76& 0.21 &   \\        
BD$-$13 4920   &2015-04-07& 2.92& 71.2&  3.28& 71.4&  3.22& 71.5&  2.91& 71.5&  1.41&0.03& 587& 2& 3.31&0.00&$ $0.07& 0.03 &   \\        
HD\,168076     &2015-05-30& 3.07& 67.1&  3.38& 66.4&  3.30& 65.8&  2.99& 65.3&  1.31&0.02& 576& 1& 3.41&0.01&$-$0.52& 0.03 &   \\        
HD\,169454     &2016-03-13& 1.91& 16.7&  2.12& 14.9&  2.07& 13.6&  1.92& 12.5&  1.12&0.08& 585& 8& 2.13&0.01&$-$1.03& 0.05 &   \\        
HD\,170740     &2016-03-15& 1.84& 78.2&  1.97& 77.4&  1.90& 76.8&  1.71& 76.2&  1.35&0.06& 561& 4& 1.99&0.01&$-$0.58& 0.04 &   \\        
HD\,170938     &2015-11-08& 3.62&117.6&  3.90&118.4&  3.68&119.2&  3.17&120.0&  1.60&0.04& 552& 3& 3.93&0.02&$ $0.63& 0.04 &   \\        
HD\,171957     &2015-11-08& 1.49& 76.6&  1.55& 77.1&  1.45& 76.8&  1.23& 75.9&  1.20&0.06& 501&21& 1.58&0.01&$-$0.36& 0.08 &   \\        
HD\,172694     &2015-06-02& 1.01&141.8&  1.02&146.3&  0.98&147.9&  0.89&148.0&  0.67&0.28& 497&32& 1.03&0.04&$ $2.58& 0.23 & 14\\        
HD\,203532     &2015-06-02& 1.29&127.9&  1.38&127.8&  1.35&128.1&  1.23&128.8&  1.23&0.05& 574& 7& 1.39&0.03&$ $0.24& 0.09 &   \\        
HD\,210121     &2015-10-02& 1.33&154.9&  1.34&155.9&  1.26&157.0&  1.12&157.7&  0.55&0.12& 434&21& 1.38&0.02&$ $0.86& 0.07 &   \\        

\hline

\end{longtable}
\end{tiny}

\begin{small}

\begin{enumerate}
\item[1.] HD\,39680 on 2015-10-04 and 2015-12-01: depolarisation and change of position angle in the Paschen line spectral region. 
\item[2.] HD 80558 on 2015-05-23: polarisation spectra obtained with and without order-separating filter show a remarkable difference ($\Delta P \sim 0.1$\,\%).
\item[3.] HD\,93222 on 2015-05-24 shows an increass of polarisation at $\lambda \ga 820$\,nm, and the position angle has a strong
 wavelength gradient. 
\item[4.] HD\,113904 is a triple system and on 2015-04-07 and on 2015-12-29 shows differences in the intensity and polarisation spectra.  
\item[5.] HD\,133518 on 2015-05-11: inconsistencies in the polarisation measured with and without order separating filter at
  longer wavelengths.
\item[6.] HD133518 on 2016-02-12: observations without filter show an artifact in one of the beams of the frame obtained at PA=67.5\degr, in
the wavelength range 796.7 to 820.6\,nm. We have arbitrarily fixed this problem by replacing the fluxes of both beams in this
wavelength range with those obtained at the same retarder PA measured in the series obtained with the order-separating filter in.
\item[7.] HD\,134591 on 2015-05-12 was observed only with order-separating filter.
\item[8.] HD\,148379 on 2015-05-11: polarisation spectra measured with and without order-separating filter show
large inconsistencies ($\Delta P \ga 0.1$\,\%).
\item[9.] HD\,149404 shows variability in polarisation (both in value and in position angle).
\item[10.] HD\,152424 variability in the intensity spectrum. Position angles measured on 2015-05-24 has larger errors.
\item[11.] HD\,153919 shows some variability in the fraction of linear polarisation (but not in the position angle).
\item[12.] HD\,154368 shows a strong variation of the position angle with wavelength in both observing dates.
\item[13.] HD\,155851 on 2015-05-24:  at $\lambda \ge 820$\,nm the polarisation decreases, and the gradient of its position angle changes its sign.
\item[14.] HD\,172694 on 2015-06-02 shows a sudden increase of polarisation at $\lambda \ge 820$\,nm. In the same wavelength region, the gradient of the polarisation position angle changes its sign, however the intensity spectrum does not show anything
remarkable in that interval.
\end{enumerate}

\end{small}

\newpage
\begin{table*}[h]
\centering
\caption{\label{Tab_Previous} Previously published Serkowski parameters for sightlines in common with LIPS.}
\begin{tabular}{llllll}
\hline\hline
          &                  &              & \multicolumn{3}{c}{Best-fit parameters} \\ 
STAR     &  Reference       & Instrument    & \multicolumn{1}{c}{$K$}                  &
                                              \multicolumn{1}{c}{\lmax\ (nm)}          &
                                              \multicolumn{1}{c}{\pmax\ (\%)}          \\
\hline
HD 37903  &  (This work)     &HPOL          & 0.73 $\pm$ 0.11 &  657 $\pm$ 10  & 1.95 $\pm$ 0.01   \\ 
          &\citet{Efimov09}  &              & 1.35            &  680           & 2.00              \\
          &  This work       &FORS2         & 1.51 $\pm$ 0.23 &  646 $\pm$  8  & 1.82 $\pm$ 0.01   \\ [2mm]
HD 38087  &  (This work)     & HPOL         & 1.25 $\pm$ 0.21 &  585 $\pm$  12 & 2.79 $\pm$ 0.02   \\
          &  This work       & FORS2        & 1.18 $\pm$ 0.04 &  572 $\pm$  5  & 2.59 $\pm$ 0.01   \\ [2mm]
HD 43384  &\citet{Wiletal80} & FORS2        & 0.97 $\pm$ 0.04 &  560 $\pm$ 10  & 3.00 $\pm$ 0.04   \\ 
          &  This work       & FORS2        & 0.92 $\pm$ 0.03 &  566 $\pm$  2  & 3.06 $\pm$ 0.01   \\ [2mm]
Walker 67 &\citet{Wiletal80} &              & 1.39 $\pm$ 0.09 &  810 $\pm$  2  & 5.31 $\pm$ 0.06   \\	
          & This work        &FORS2         & 1.40 $\pm$ 0.05 &  826 $\pm$  3  & 5.20 $\pm$ 0.04   \\ [2mm]
HD 108639 &\citet{Heiles00}  &              & 0.87 $\pm$ 0.13 &  520 $\pm$ 10  & 1.91 $\pm$ 0.02   \\
          & This work        &FORS2         & 1.29 $\pm$ 0.23 &  550 $\pm$  5  & 1.92 $\pm$ 0.04   \\ [2mm]
HD 147888 &\citet{Wiletal82} &              & 1.29 $\pm$ 0.05 &  720 $\pm$ 10  & 3.75 $\pm$ 0.07   \\	    
          & \citet{Maretal92}&              & 1.28 $\pm$ 0.04 &  700 $\pm$ 10  & 3.63 $\pm$ 0.04   \\
          &  (This work)      &HPOL         & 1.51 $\pm$ 0.04 &  672 $\pm$ 20  & 3.53 $\pm$ 0.01   \\
          & This work        &FORS2         & 1.43 $\pm$ 0.02 &  662 $\pm$  4  & 3.49 $\pm$ 0.01   \\ [2mm]
HD 147889 & \citet{Wiletal80}&              & 1.32 $\pm$ 0.05 &  800 $\pm$ 10  & 4.09 $\pm$ 0.07   \\  
          & \citet{Maretal92}&              & 1.25 $\pm$ 0.03 &  780 $\pm$ 10  & 4.02 $\pm$ 0.04   \\
          & This work   &FORS2 (2015-05-13) & 1.32 $\pm$ 0.02 &  809 $\pm$  4  & 4.22 $\pm$ 0.01   \\
          & This work   &FORS2 (2015-05-25) & 1.37 $\pm$ 0.04 &  798 $\pm$  7  & 4.18 $\pm$ 0.01   \\
          & This work   &FORS2 (2016-01-28) & 1.18 $\pm$ 0.09 &  815 $\pm$ 13  & 4.18 $\pm$ 0.01   \\ [2mm]
HD 161056 & \citet{Wiletal80}&              & 0.96 $\pm$ 0.09 &  560 $\pm$  2  & 3.8  $\pm$ 0.1    \\
          &  (This work)     &HPOL          & 1.3  $\pm$ 0.02 &  578 $\pm$  1  & 4.08 $\pm$ 0.01   \\
          & This work        &FORS2         & 1.43 $\pm$ 0.05 &  572 $\pm$  2  & 4.02 $\pm$ 0.04   \\ [2mm]
HD 164740 &\citet{Wiletal82} &              & 1.24 $\pm$ 0.12 &  670 $\pm$ 20  & 7.75 $\pm$ 0.15   \\ 
          & \citet{Maretal92}&              & 1.24 $\pm$ 0.07 &  670 $\pm$ 10  & 7.45 $\pm$ 0.14   \\
          & This work        &FORS2         & 1.72 $\pm$ 0.02 &  657 $\pm$  2  & 7.20 $\pm$ 0.02   \\ [2mm]
HD 210121 & \citet{Larson99} &              & 0.66 $\pm$ 0.09 &  380 $\pm$ 30  & 1.32 $\pm$ 0.04   \\ 
          &  (This work)     &HPOL          & 0.65 $\pm$ 0.15 &  402 $\pm$ 50  & 1.37 $\pm$ 0.06   \\
          & This work        &FORS2         & 0.55 $\pm$ 0.12 &  434 $\pm$ 21  & 1.38 $\pm$ 0.02   \\ [2mm]
HD 251204 &  (This work)     &HPOL          & 1.25 $\pm$ 0.11 &  584 $\pm$  7  & 4.95 $\pm$ 0.03   \\
          & This work        &FORS2         & 1.00 $\pm$ 0.02 &  575 $\pm$  2  & 5.00 $\pm$ 0.01   \\ 

\hline
\end{tabular}
\tablefoot{
  \citet{Wiletal80} and \citet{Maretal92} report updated Serkowski parameters from compilations
  of broad-band polarimetric data. The best-fit parameters from HPOL were obtained in this work
  using data made available at the Mikulski Archive for Space Telescopes (MAST).}
\end{table*}

\newpage

\noindent
  \includegraphics*[scale=0.42,trim={1.1cm 6.0cm 0.1cm 2.8cm},clip]{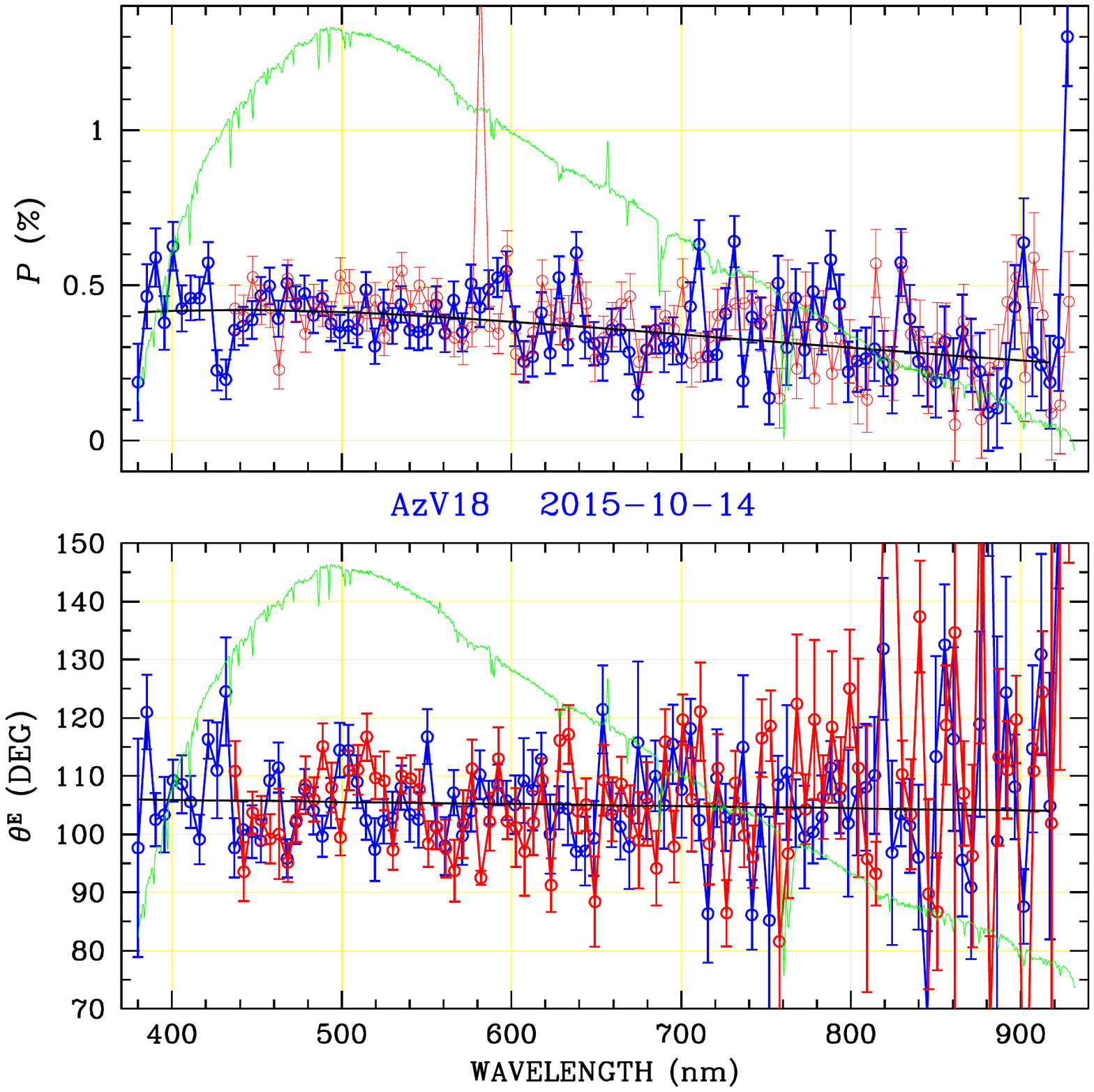}
  \includegraphics*[scale=0.42,trim={1.1cm 6.0cm 0.3cm 2.8cm},clip]{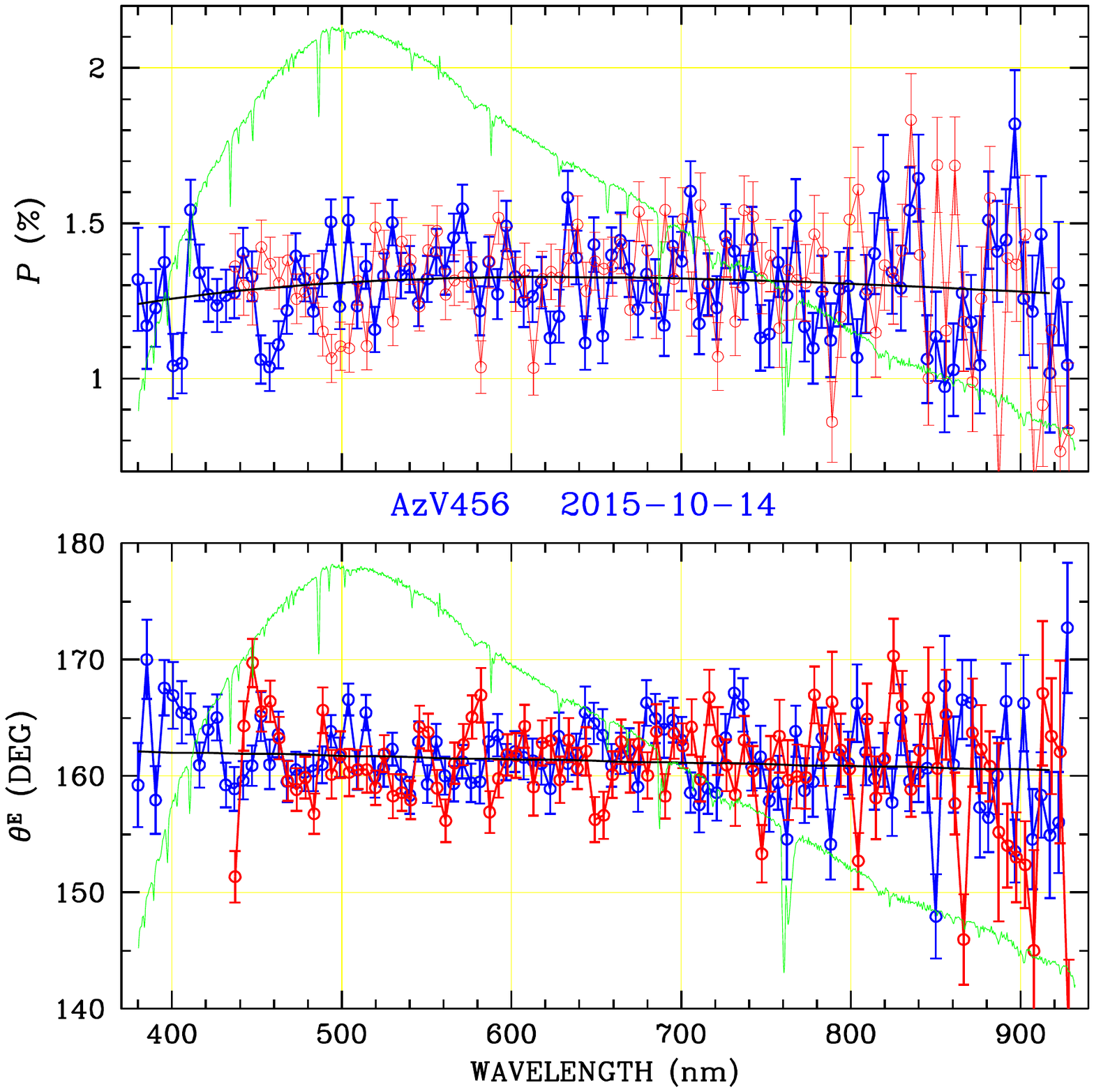}\\
  \includegraphics*[scale=0.42,trim={1.1cm 6.0cm 0.1cm 2.8cm},clip]{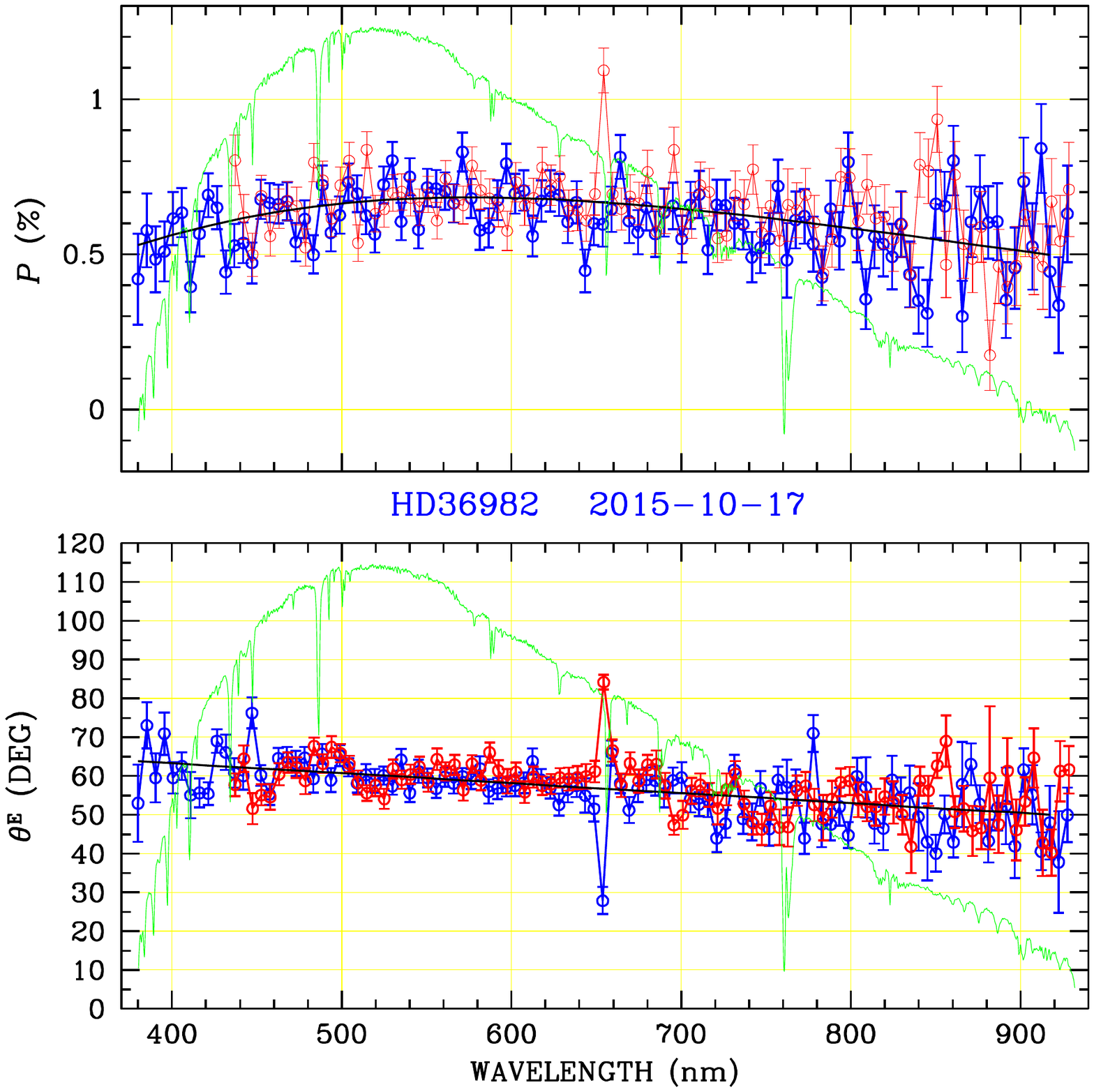}
  \includegraphics*[scale=0.42,trim={1.1cm 6.0cm 0.3cm 2.8cm},clip]{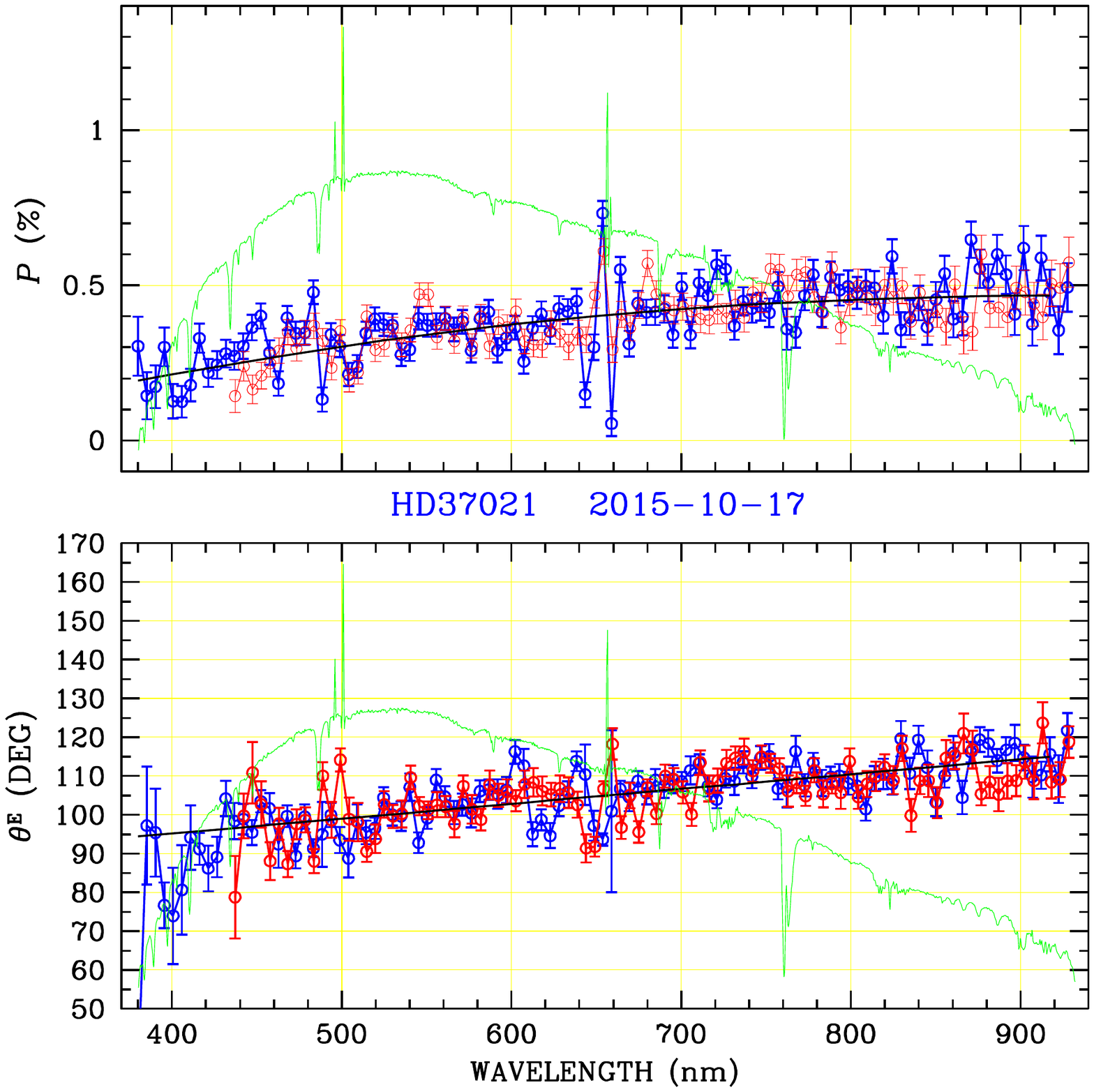}\\
  \includegraphics*[scale=0.42,trim={1.1cm 6.0cm 0.3cm 2.8cm},clip]{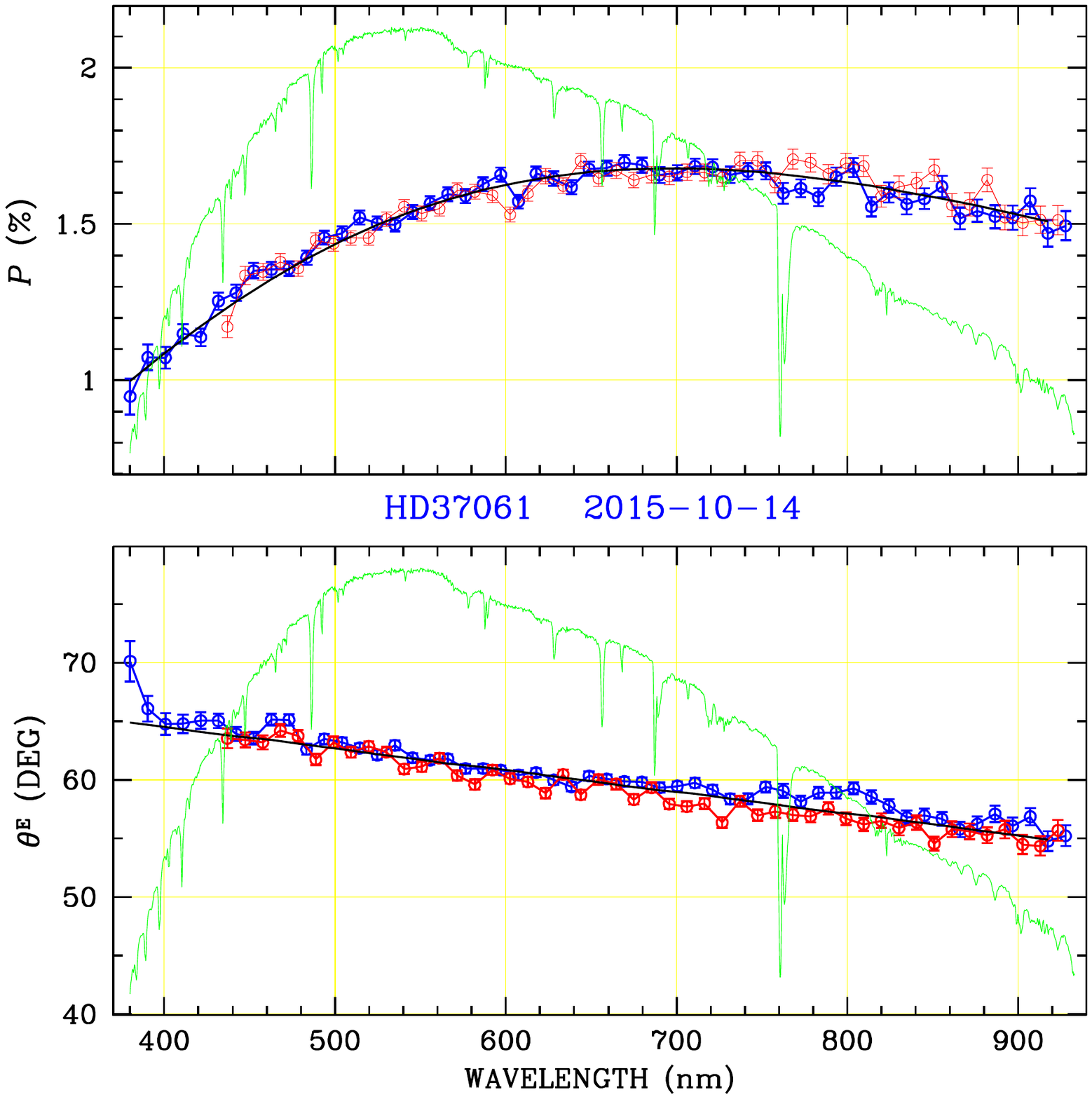}
  \includegraphics*[scale=0.42,trim={1.1cm 6.0cm 0.1cm 2.8cm},clip]{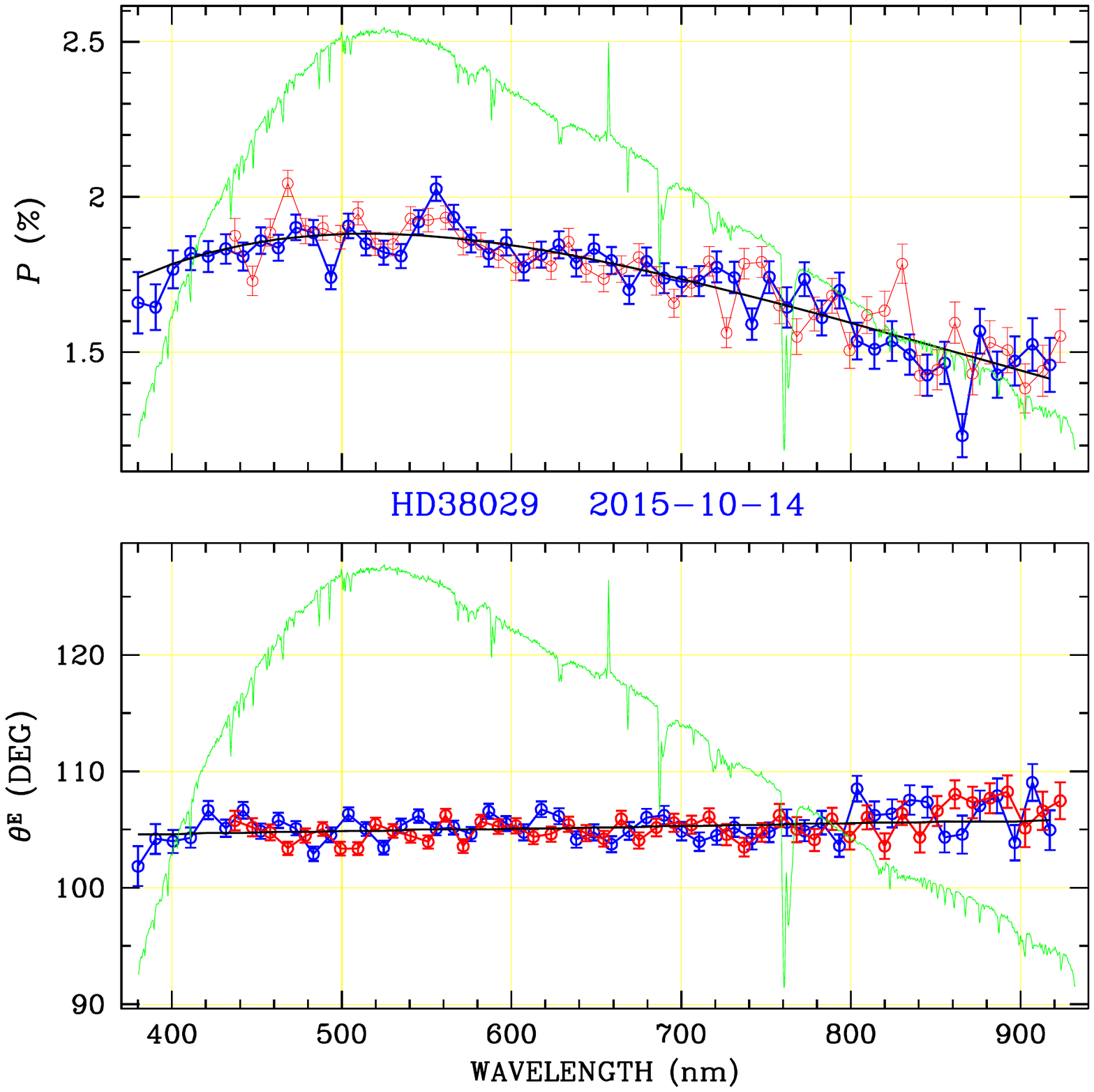}
\newpage

\noindent
  \includegraphics*[scale=0.42,trim={1.1cm 6.0cm 0.3cm 2.8cm},clip]{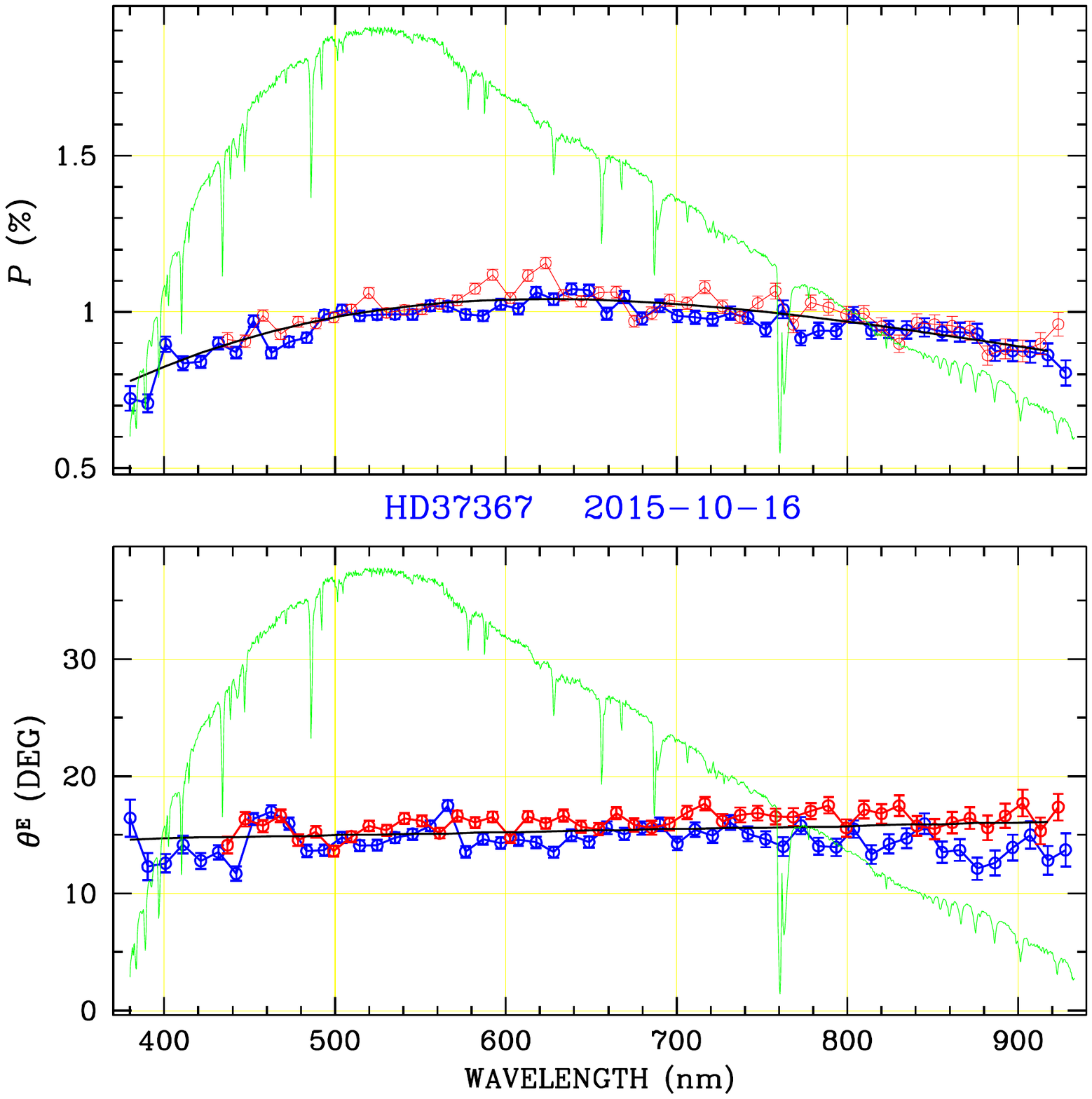}
  \includegraphics*[scale=0.42,trim={1.1cm 6.0cm 0.1cm 2.8cm},clip]{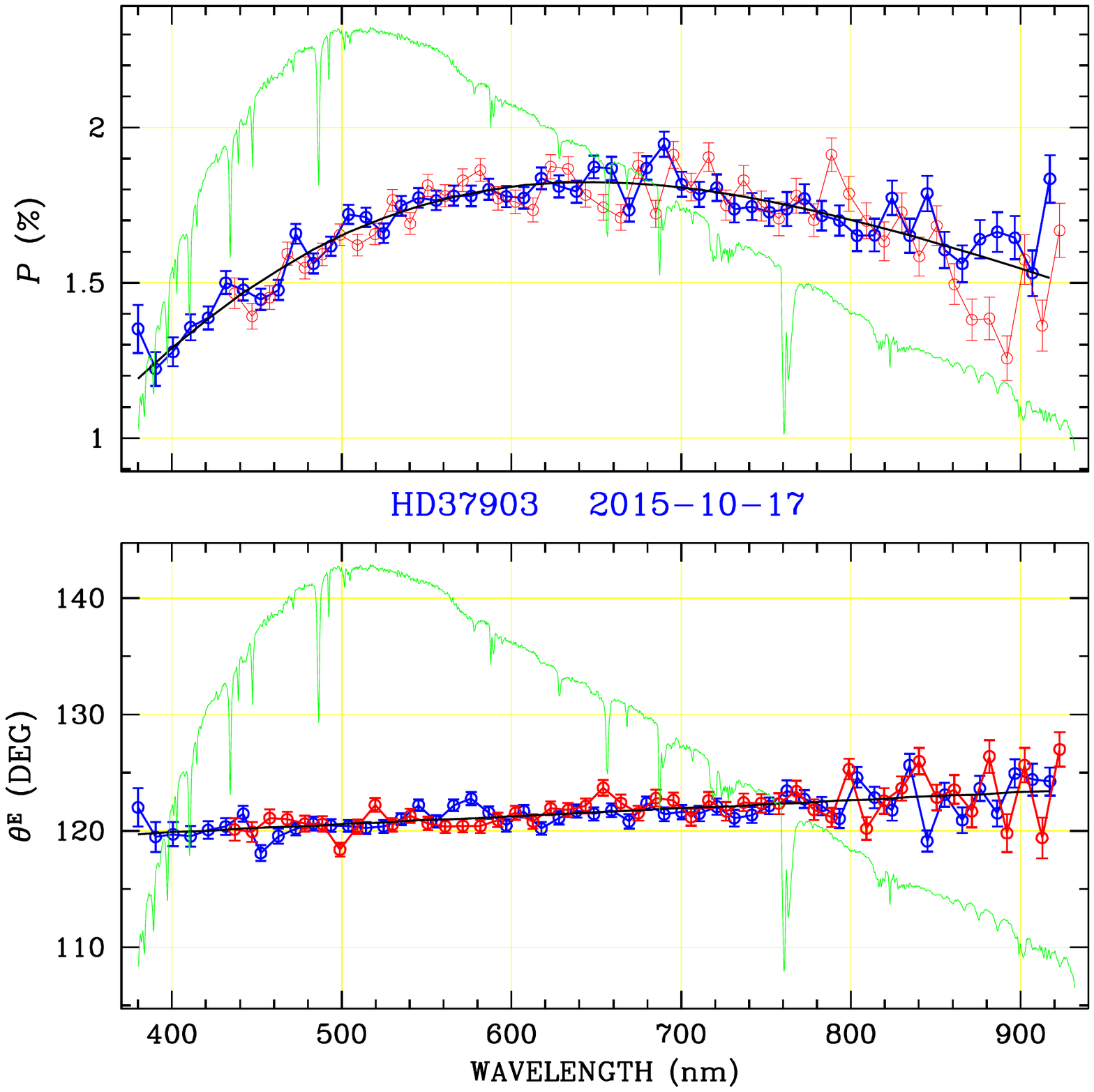}\\
  \includegraphics*[scale=0.42,trim={1.1cm 6.0cm 0.1cm 2.8cm},clip]{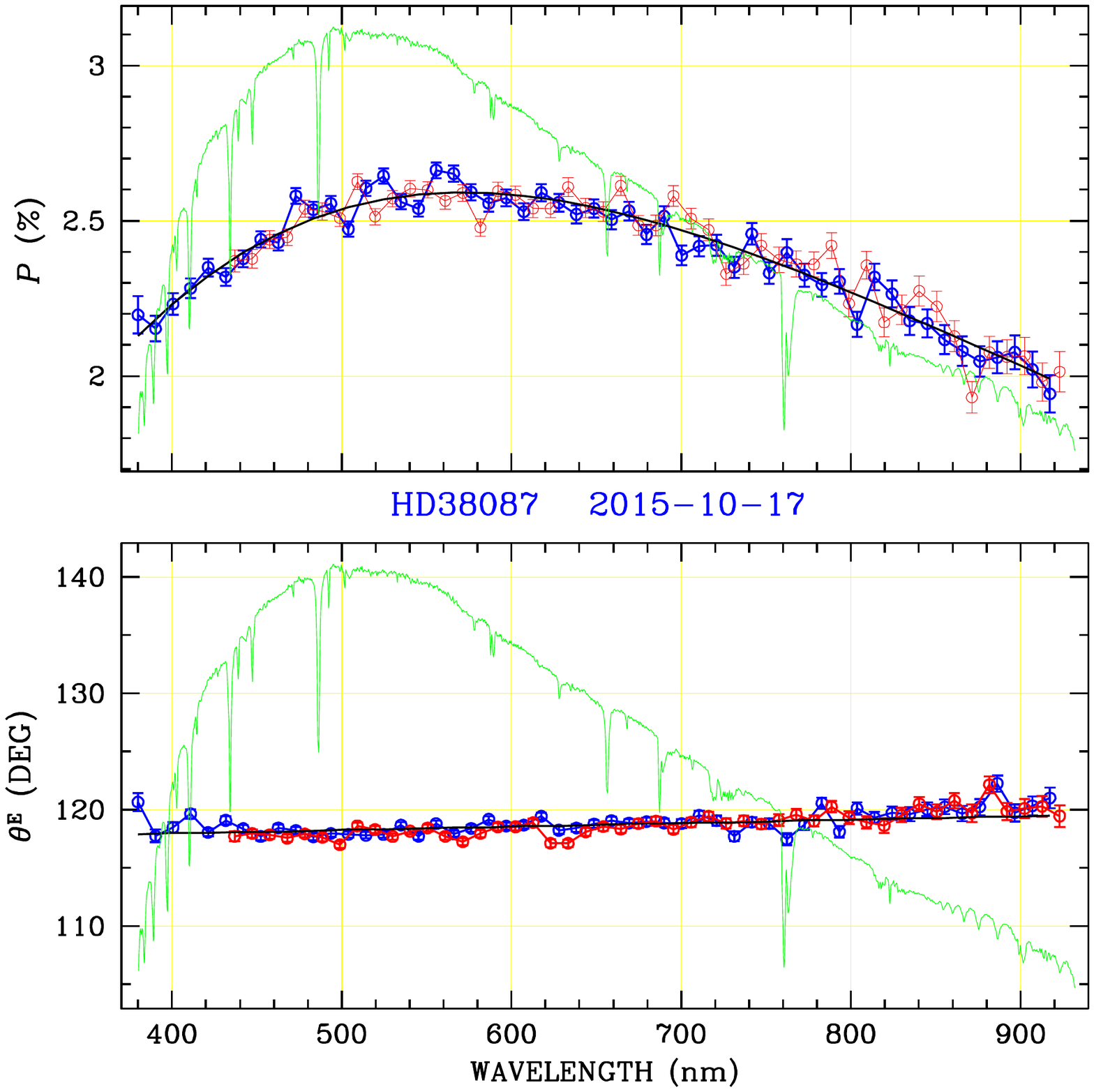}
  \includegraphics*[scale=0.42,trim={1.1cm 6.0cm 0.1cm 2.8cm},clip]{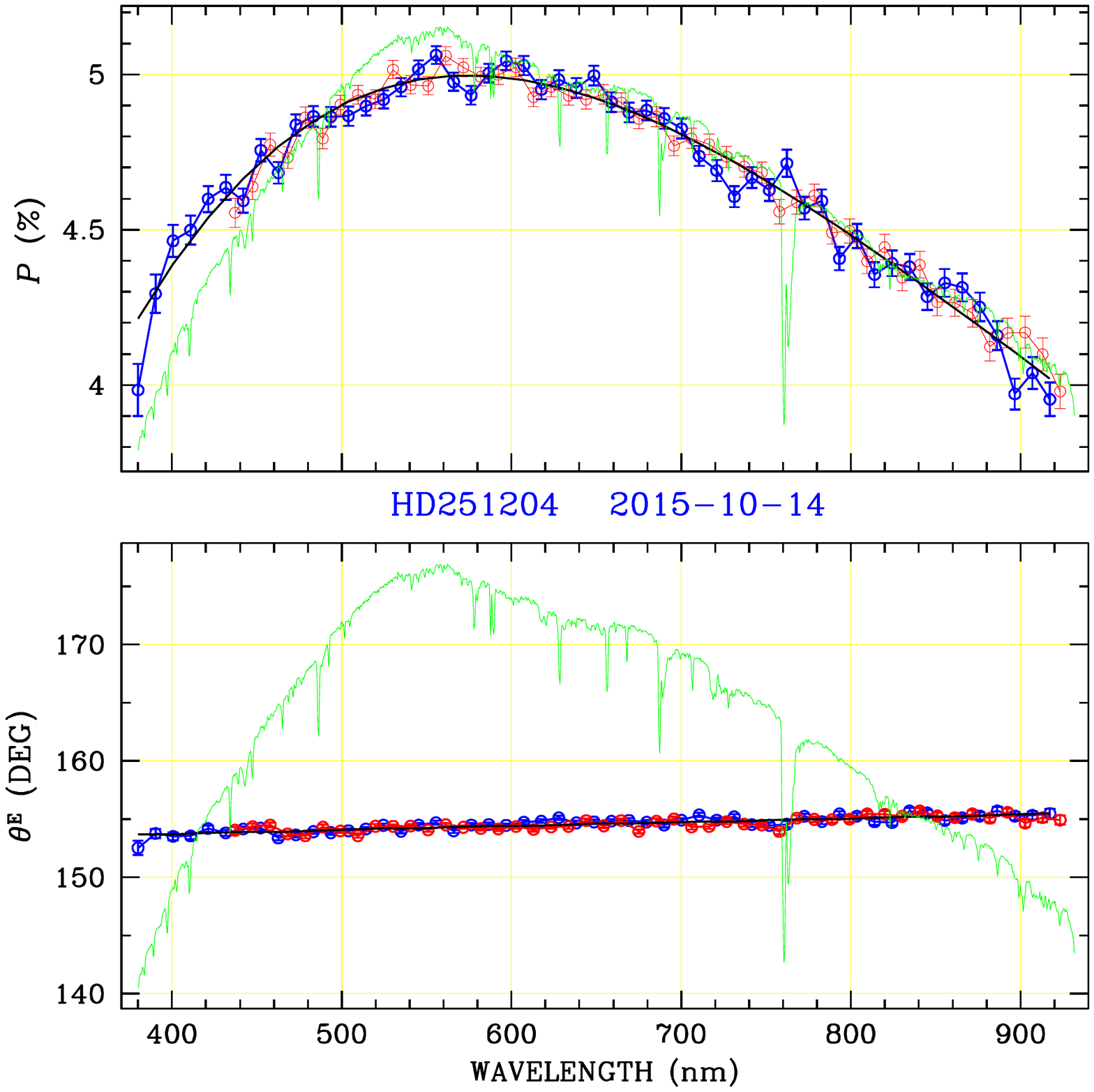}\\
  \includegraphics*[scale=0.42,trim={1.1cm 6.0cm 0.1cm 2.8cm},clip]{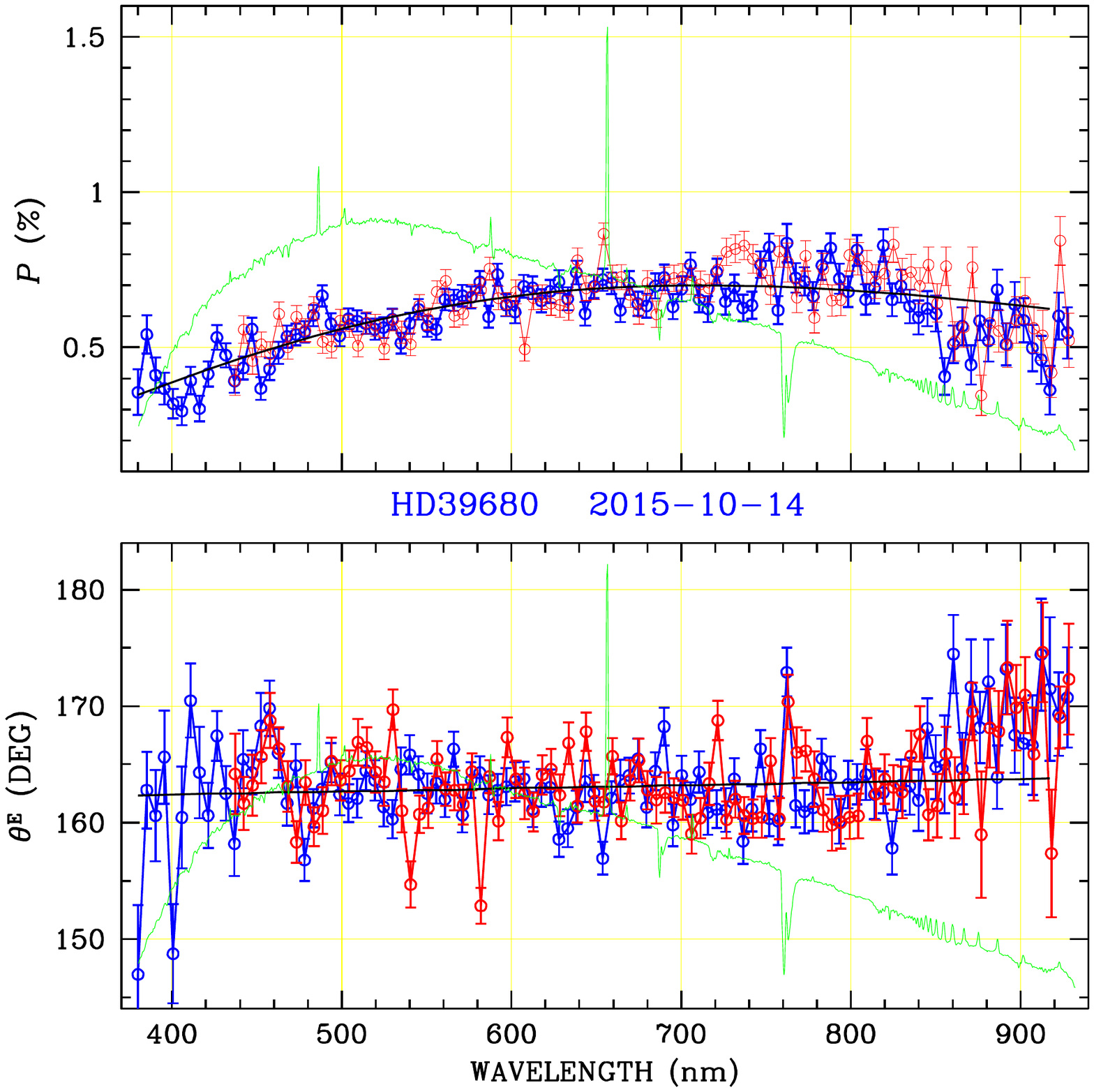}
  \includegraphics*[scale=0.42,trim={1.1cm 6.0cm 0.1cm 2.8cm},clip]{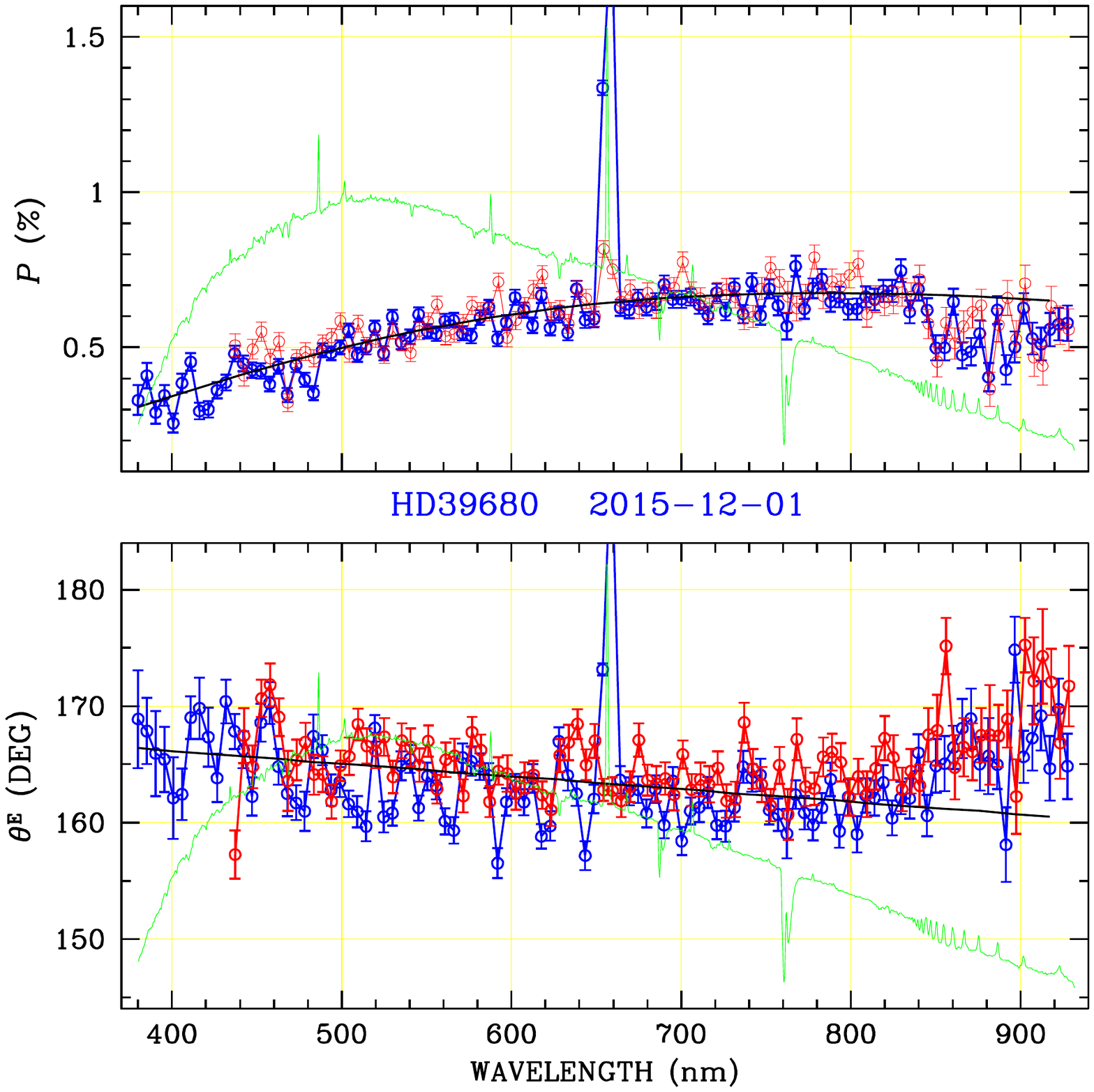}
\newpage

\noindent
  \includegraphics*[scale=0.42,trim={1.1cm 6.0cm 0.1cm 2.8cm},clip]{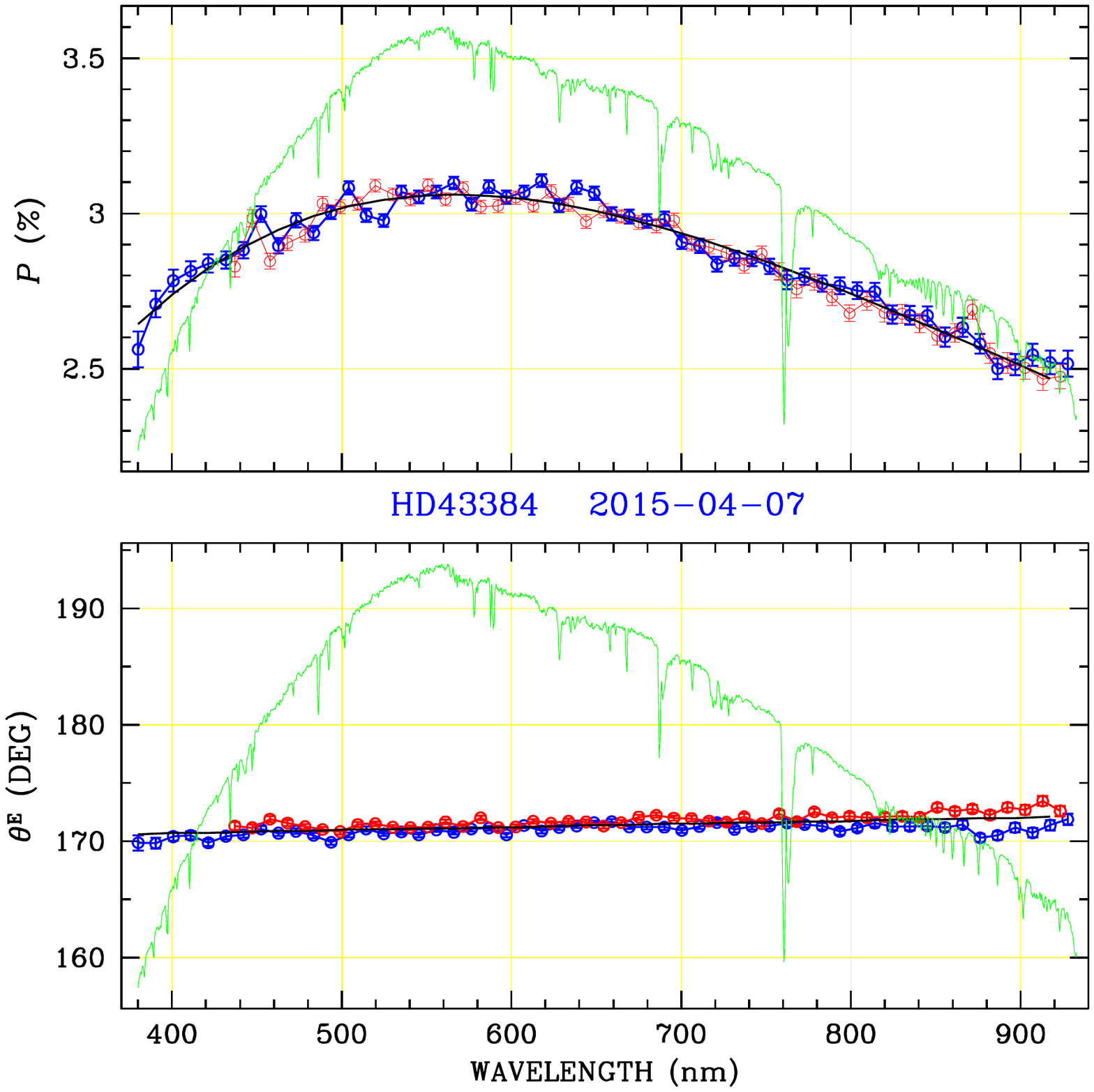}
  \includegraphics*[scale=0.42,trim={1.1cm 6.0cm 0.1cm 2.8cm},clip]{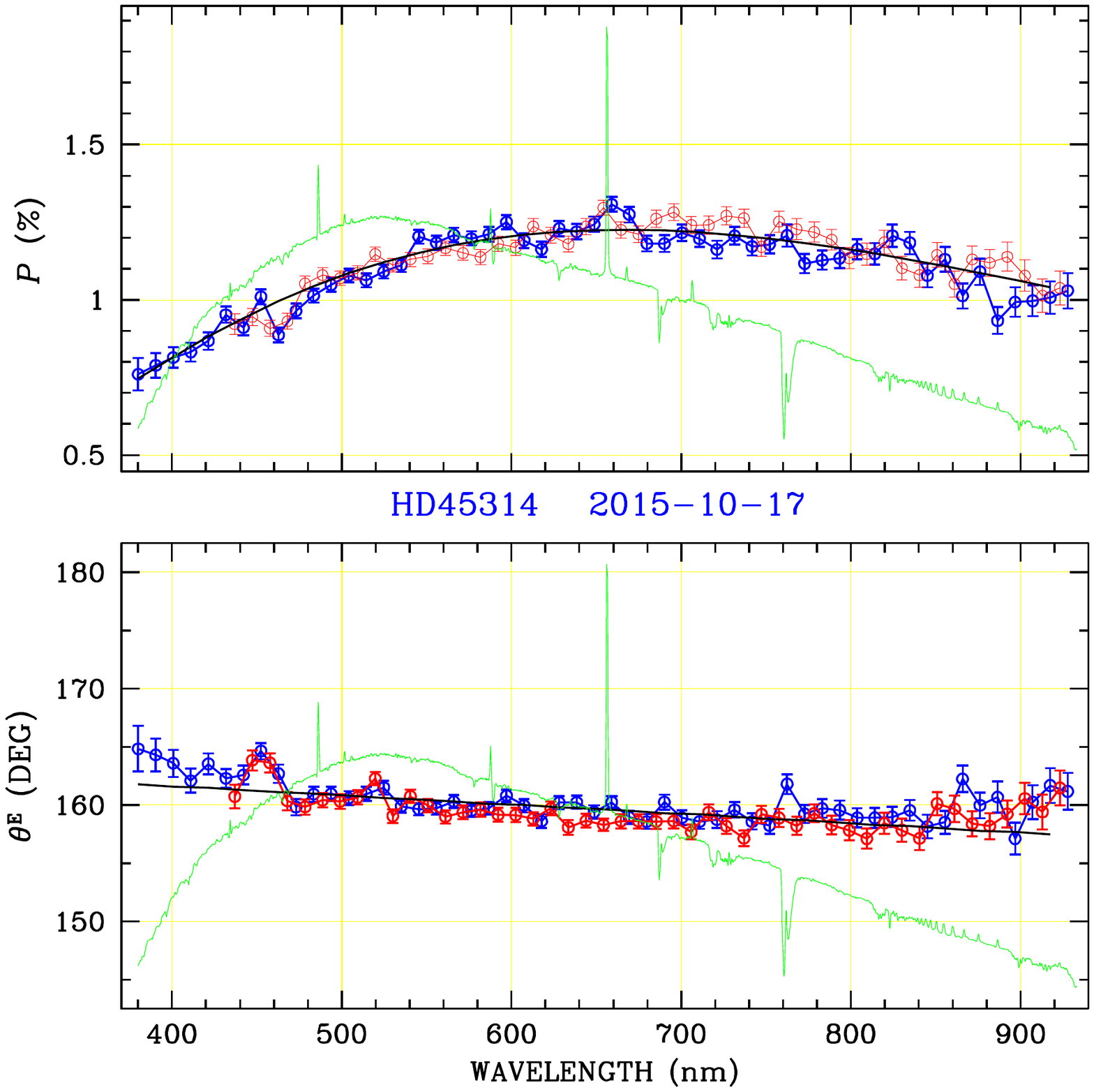}\\
  \includegraphics*[scale=0.42,trim={1.1cm 6.0cm 0.1cm 2.8cm},clip]{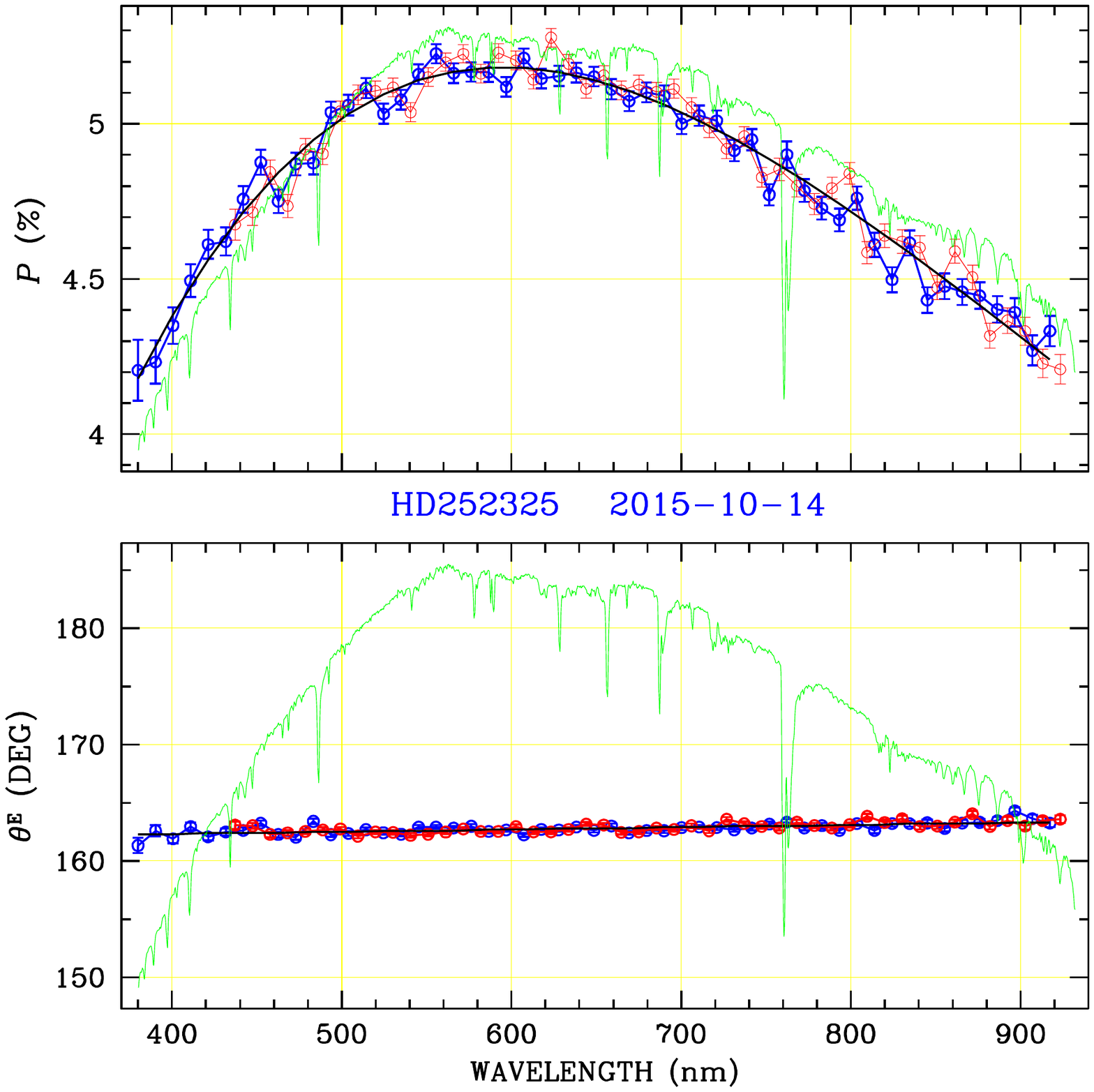}
  \includegraphics*[scale=0.42,trim={1.1cm 6.0cm 0.1cm 2.8cm},clip]{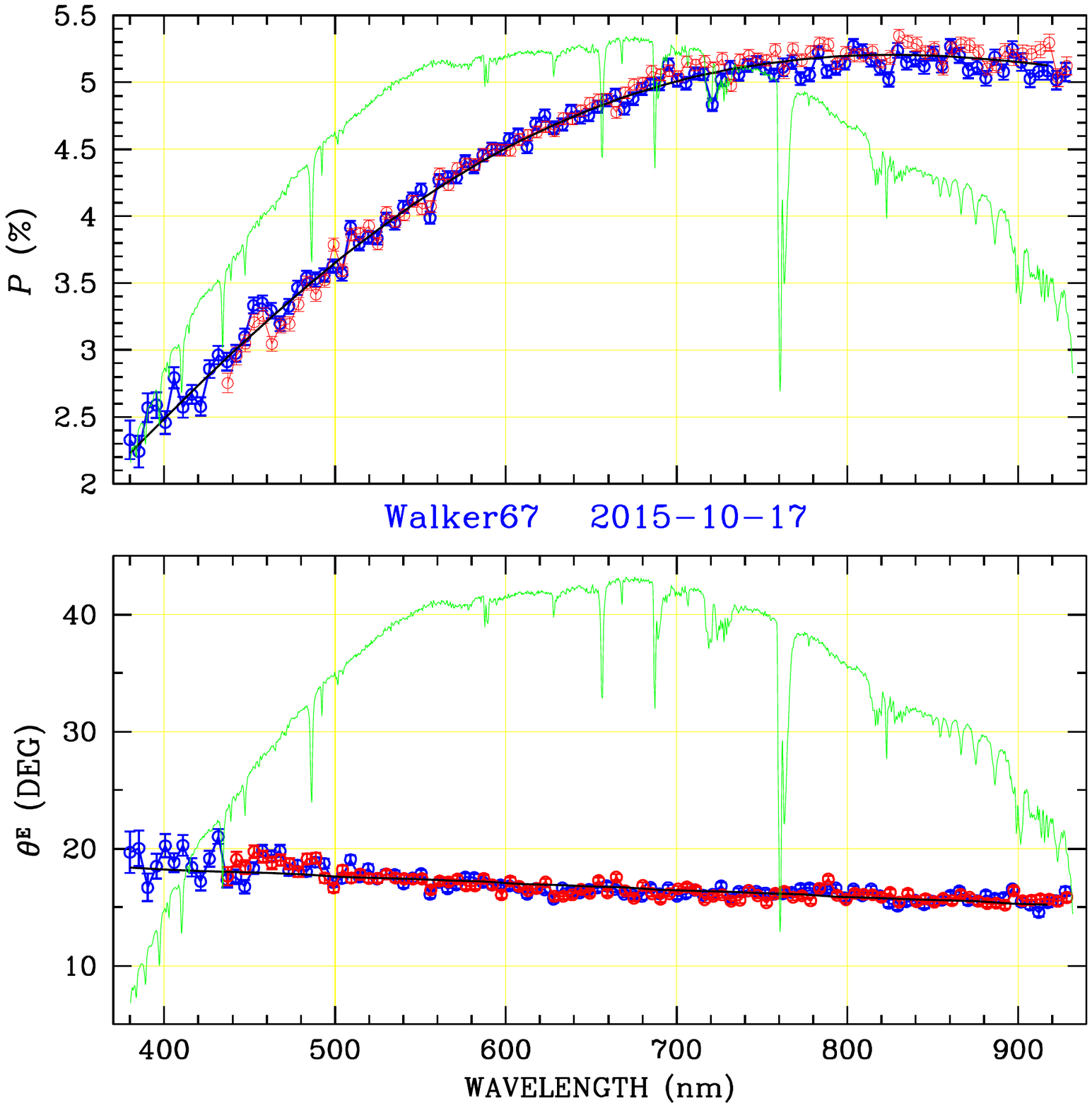}\\
  \includegraphics*[scale=0.42,trim={1.1cm 6.0cm 0.1cm 2.8cm},clip]{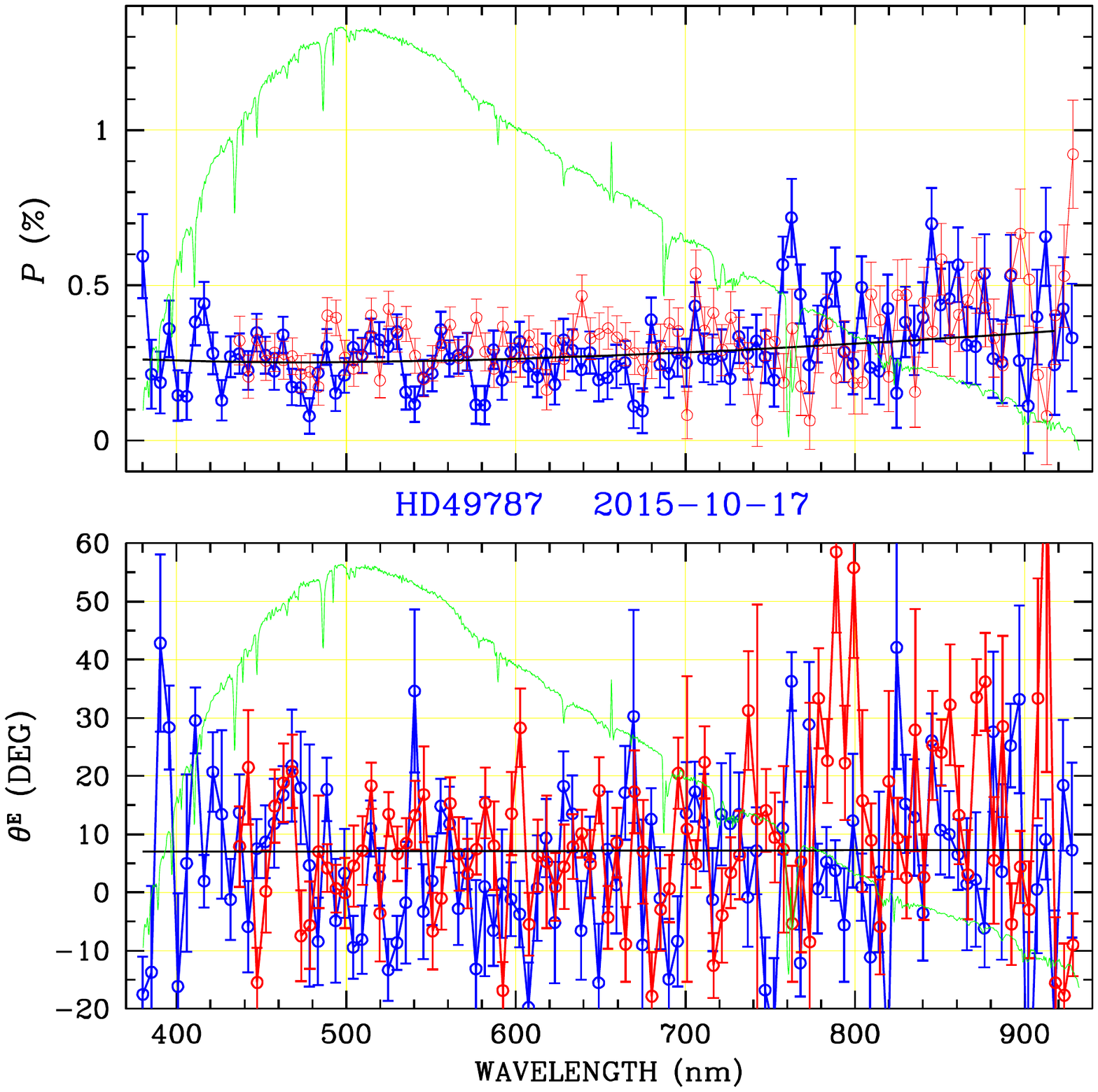}
  \includegraphics*[scale=0.42,trim={1.1cm 6.0cm 0.1cm 2.8cm},clip]{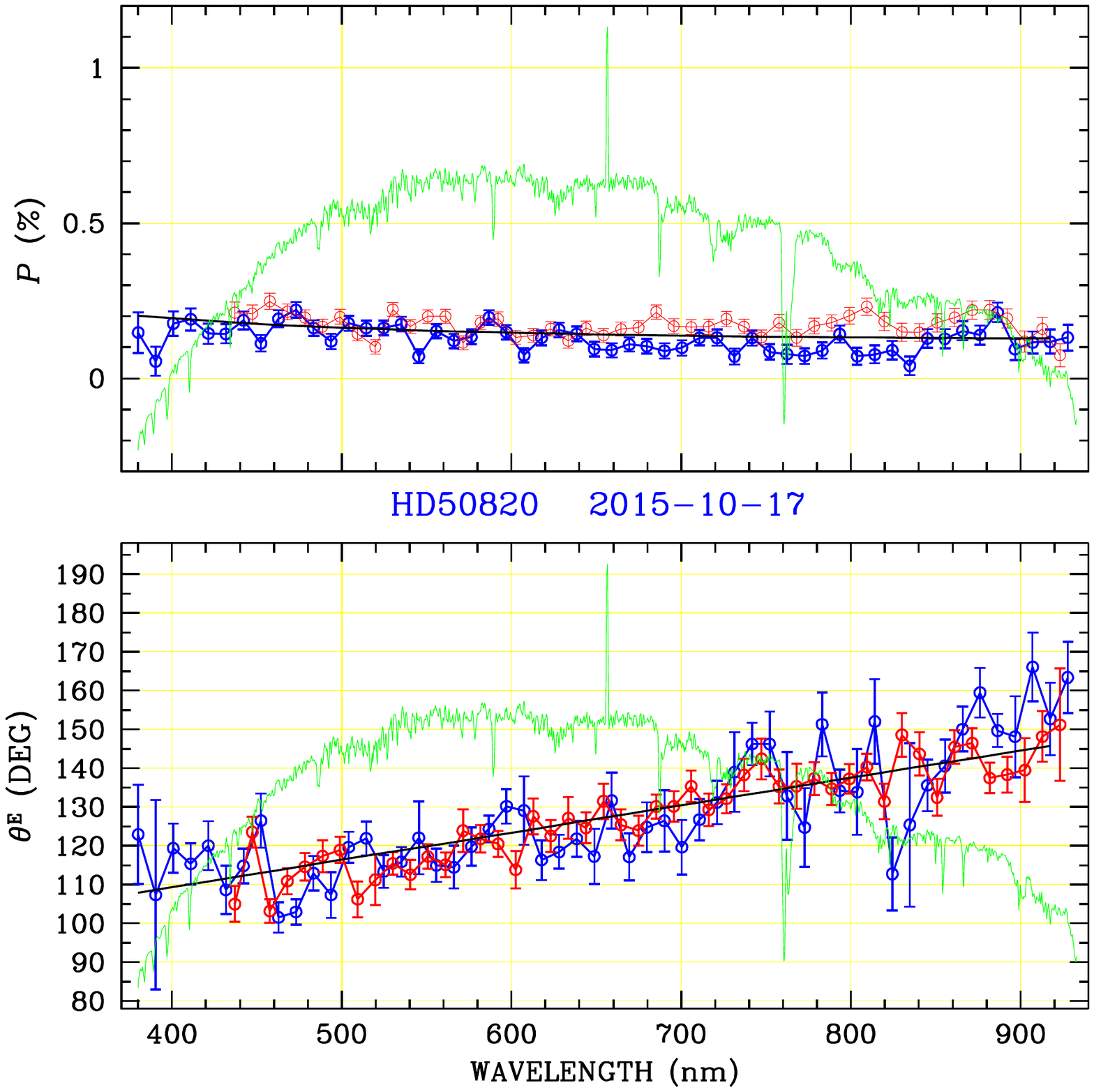}
\newpage

\noindent
  \includegraphics*[scale=0.42,trim={1.1cm 6.0cm 0.1cm 2.8cm},clip]{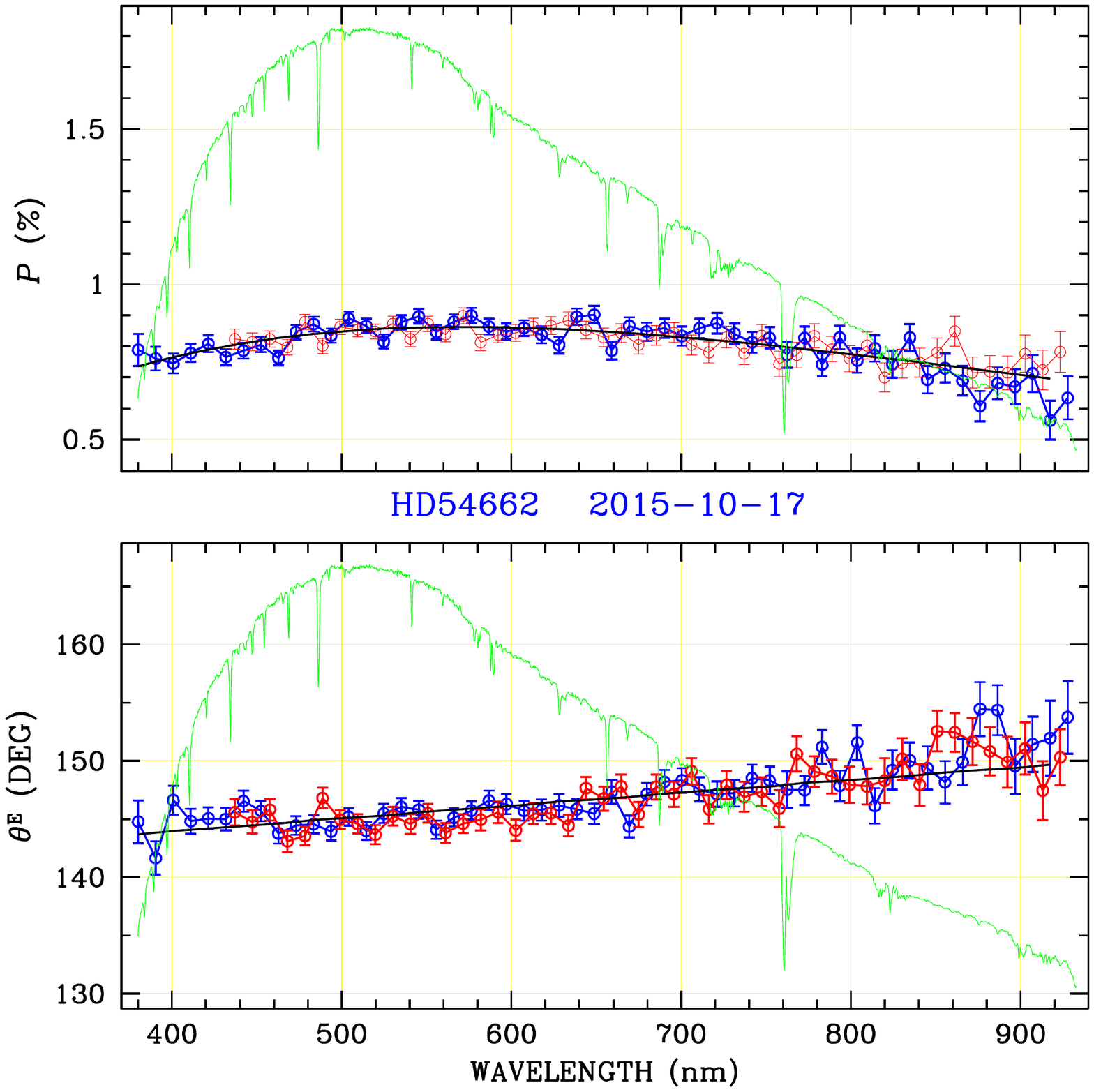}
  \includegraphics*[scale=0.42,trim={1.1cm 6.0cm 0.1cm 2.8cm},clip]{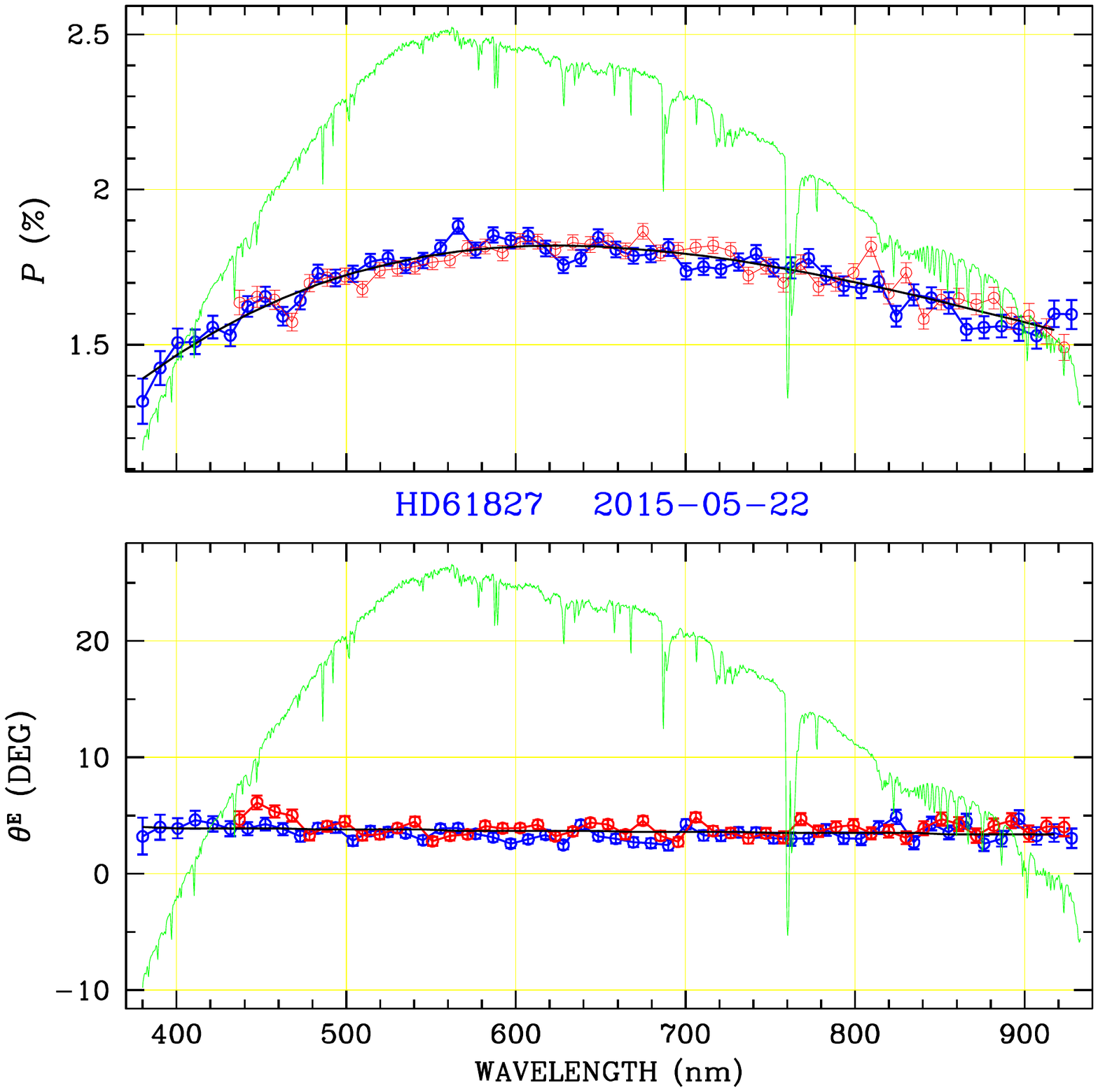}\\
  \includegraphics*[scale=0.42,trim={1.1cm 6.0cm 0.1cm 2.8cm},clip]{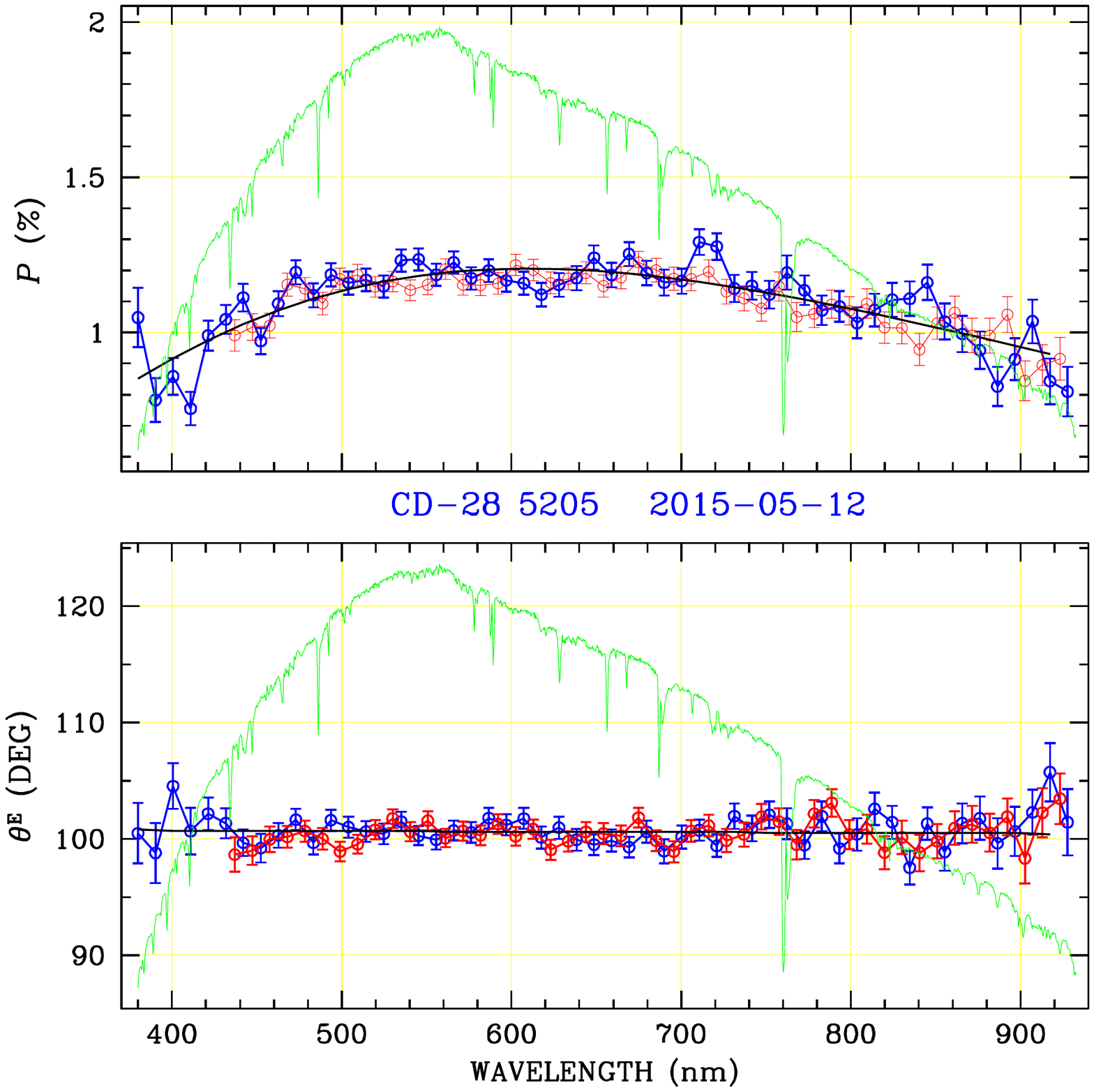}
  \includegraphics*[scale=0.42,trim={1.1cm 6.0cm 0.1cm 2.8cm},clip]{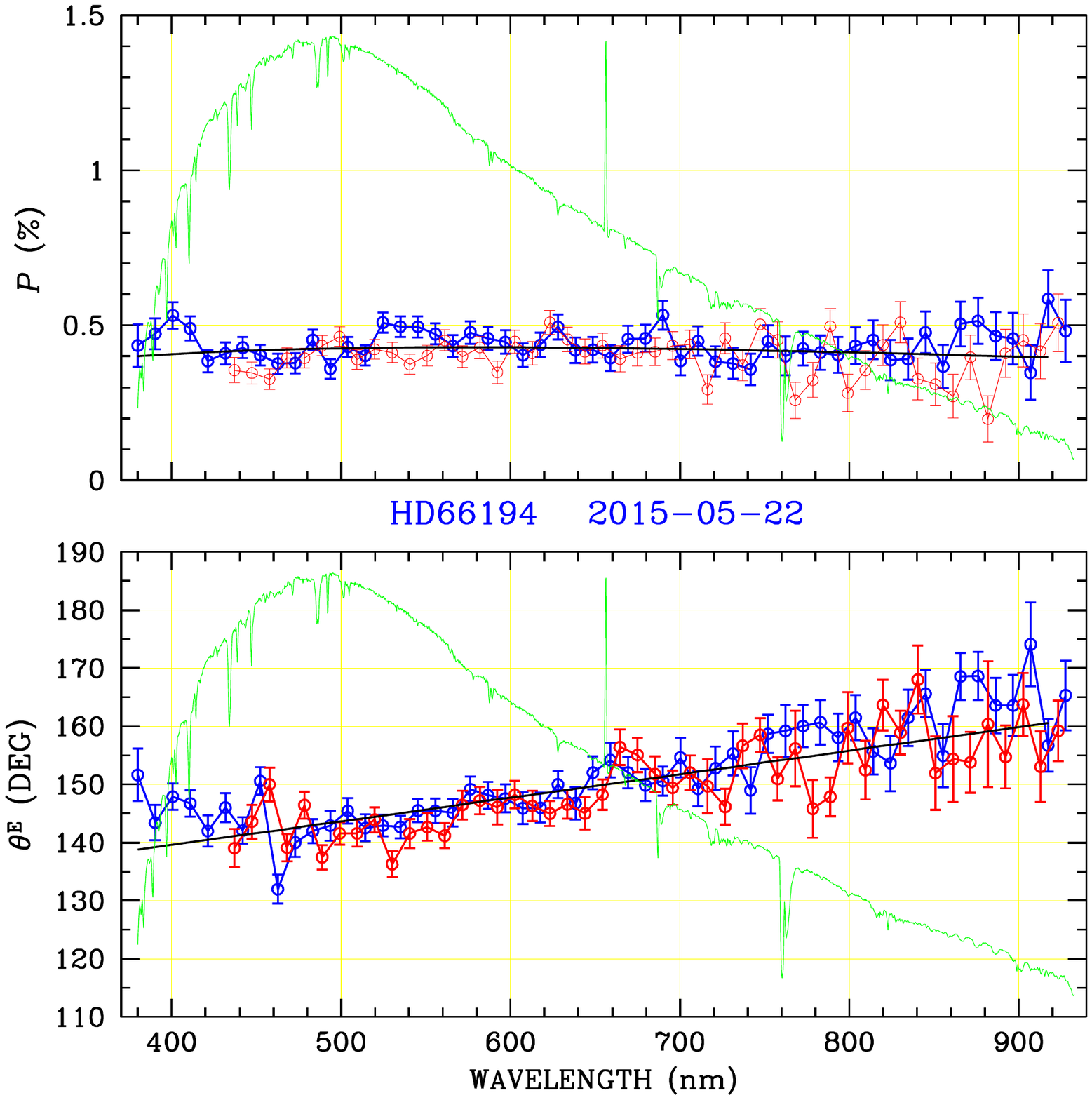}\\
  \includegraphics*[scale=0.42,trim={1.1cm 6.0cm 0.1cm 2.8cm},clip]{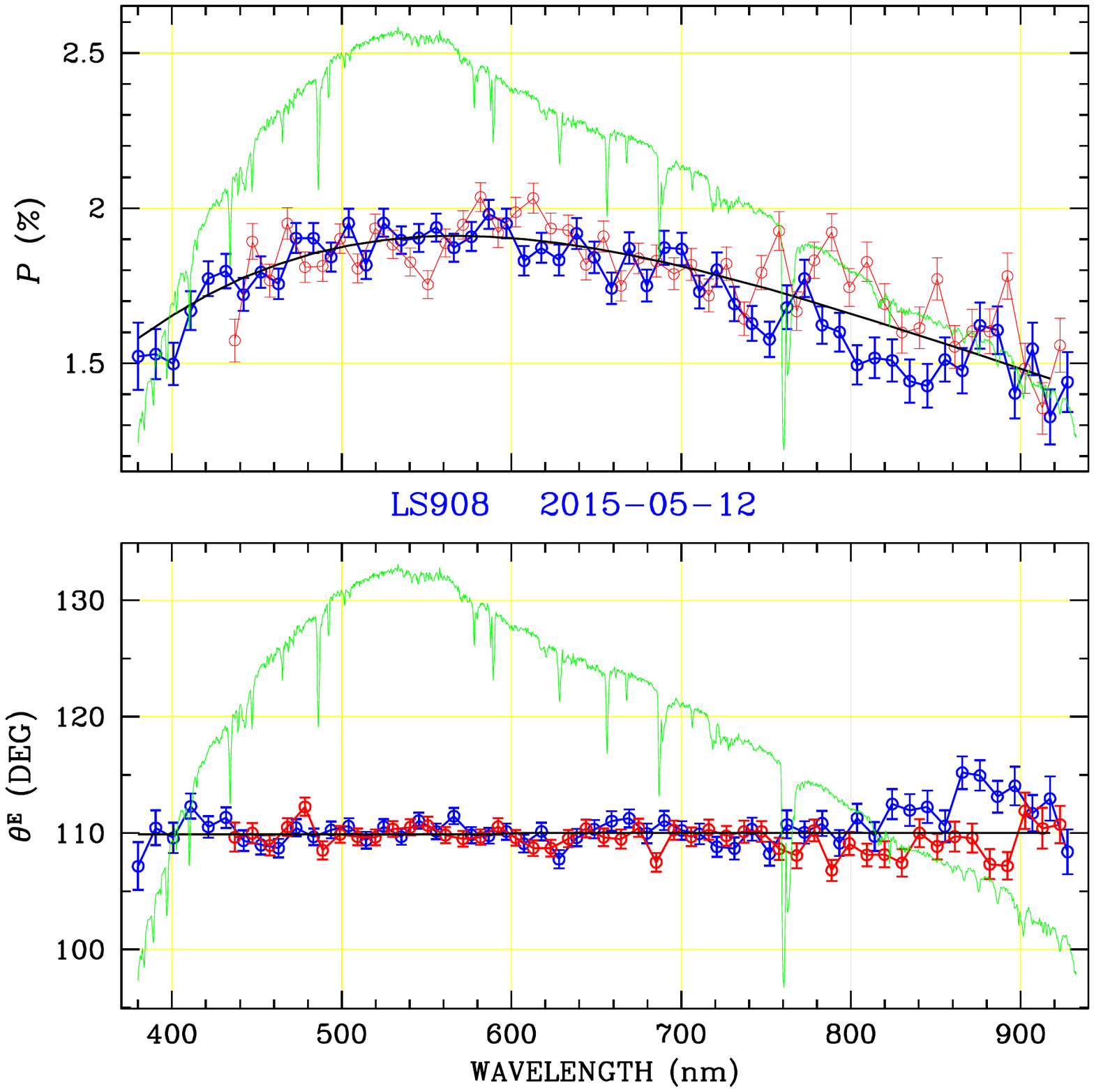}
  \includegraphics*[scale=0.42,trim={1.1cm 6.0cm 0.1cm 2.8cm},clip]{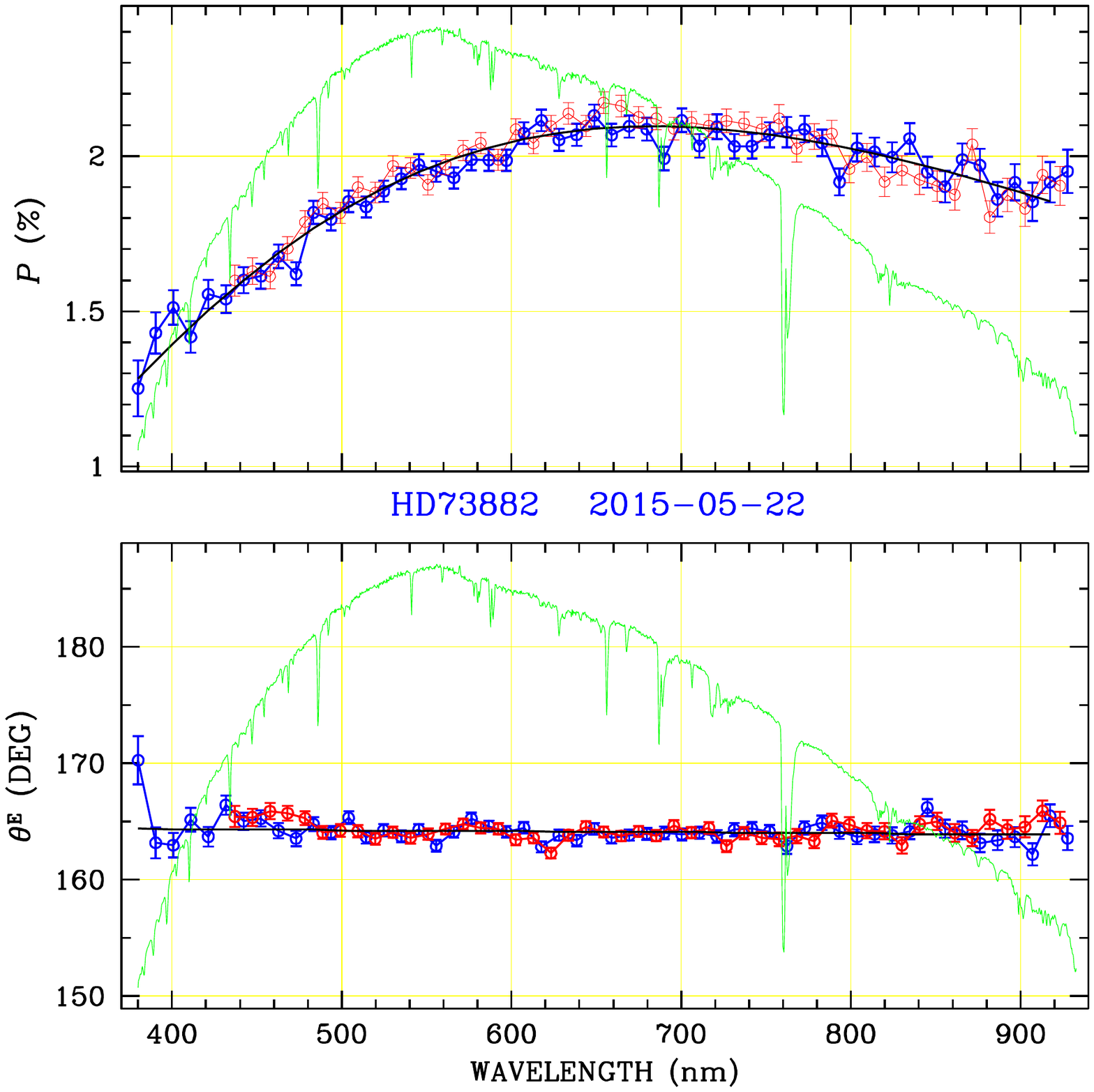}
\newpage

\noindent
  \includegraphics*[scale=0.42,trim={1.1cm 6.0cm 0.1cm 2.8cm},clip]{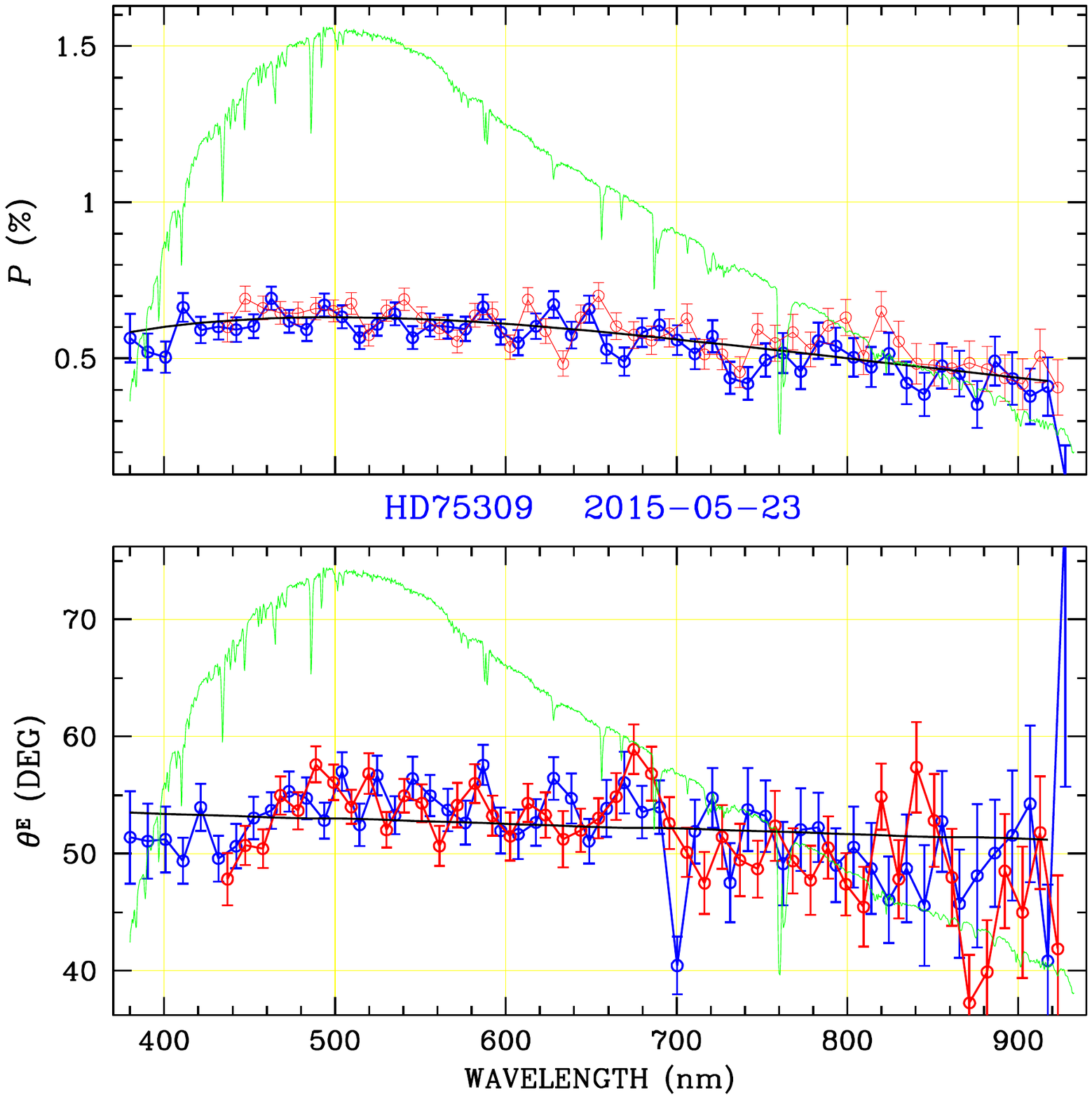}
  \includegraphics*[scale=0.42,trim={1.1cm 6.0cm 0.1cm 2.8cm},clip]{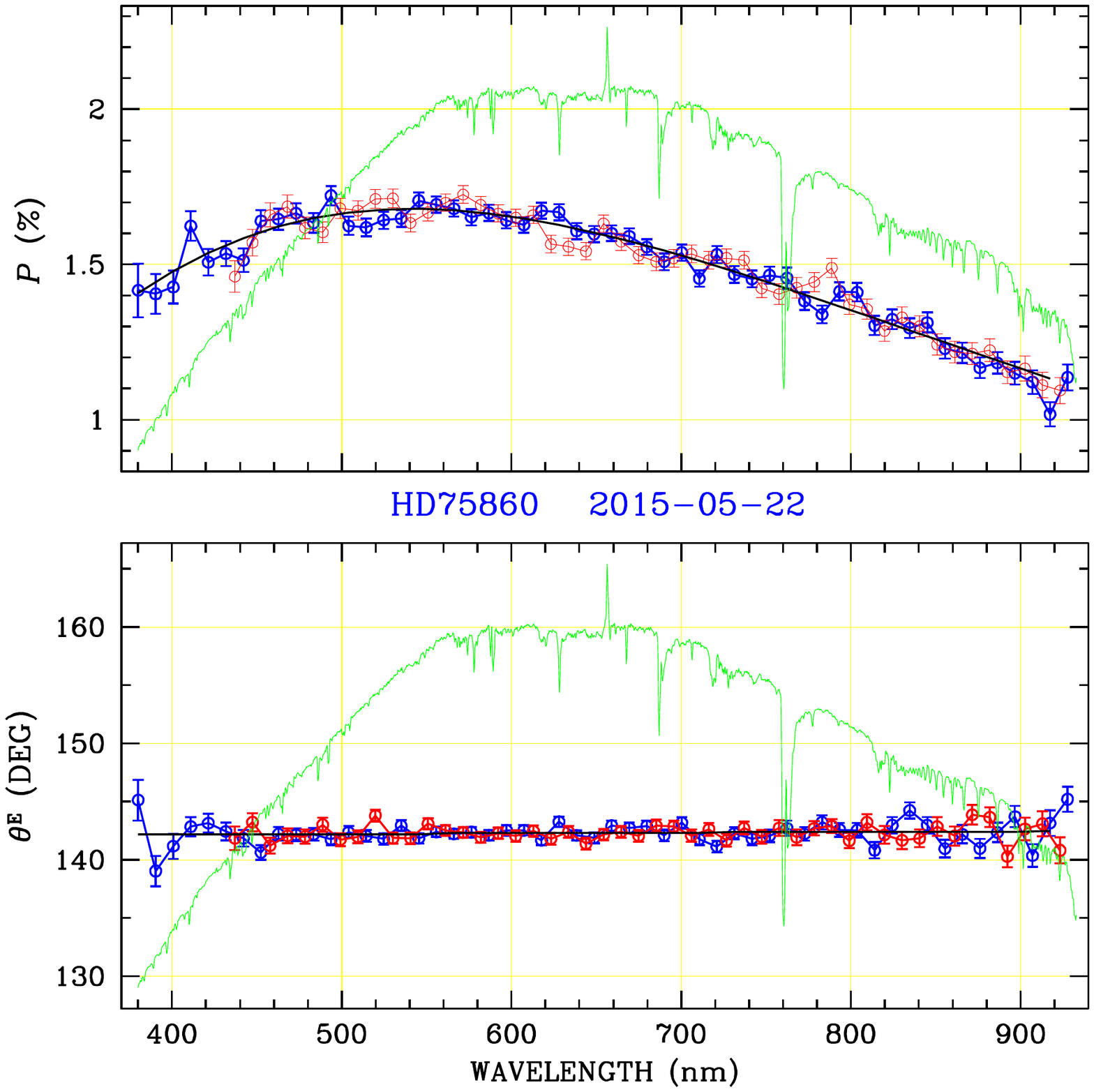}\\
  \includegraphics*[scale=0.42,trim={1.1cm 6.0cm 0.1cm 2.8cm},clip]{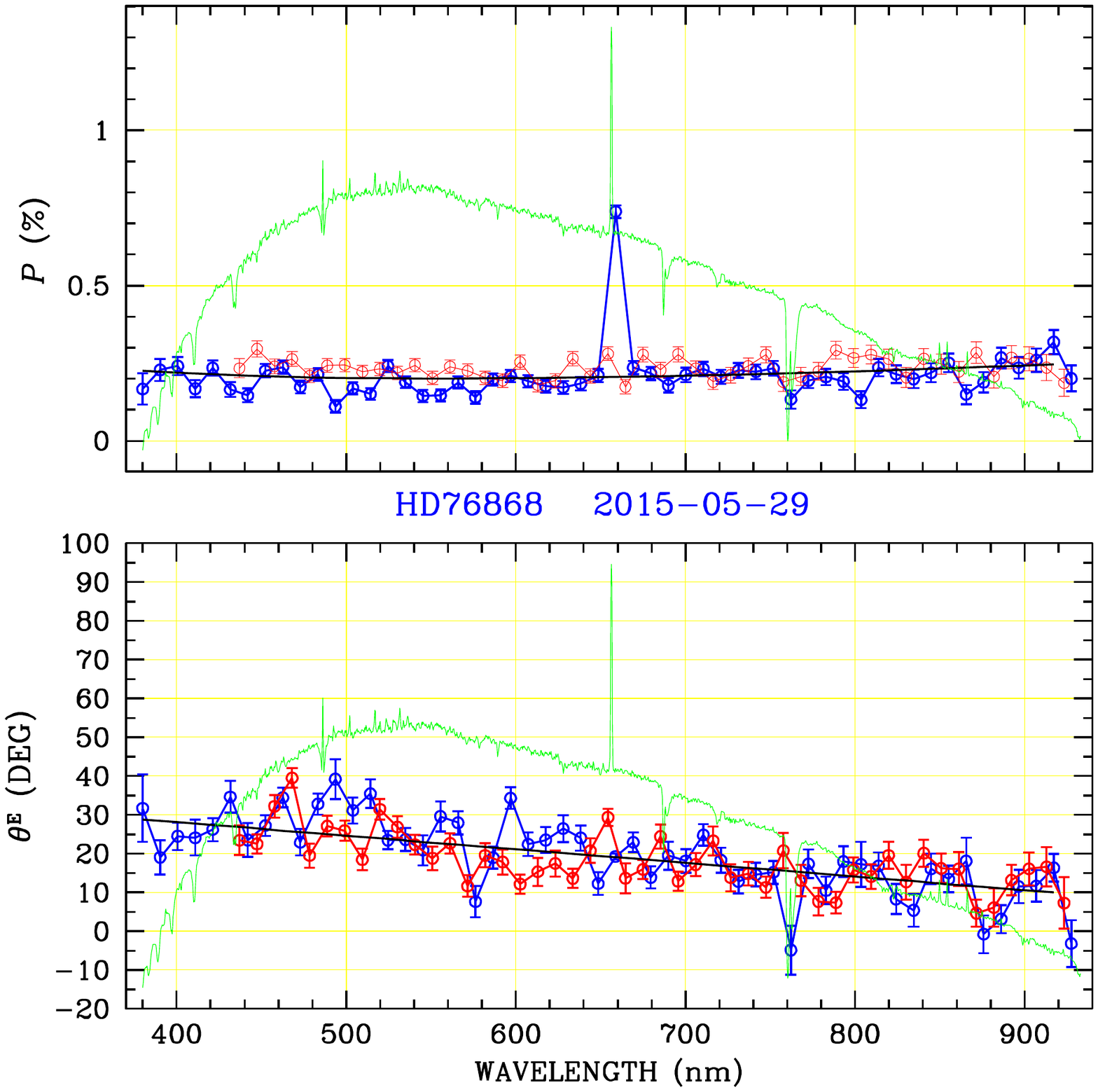}
  \includegraphics*[scale=0.42,trim={1.1cm 6.0cm 0.1cm 2.8cm},clip]{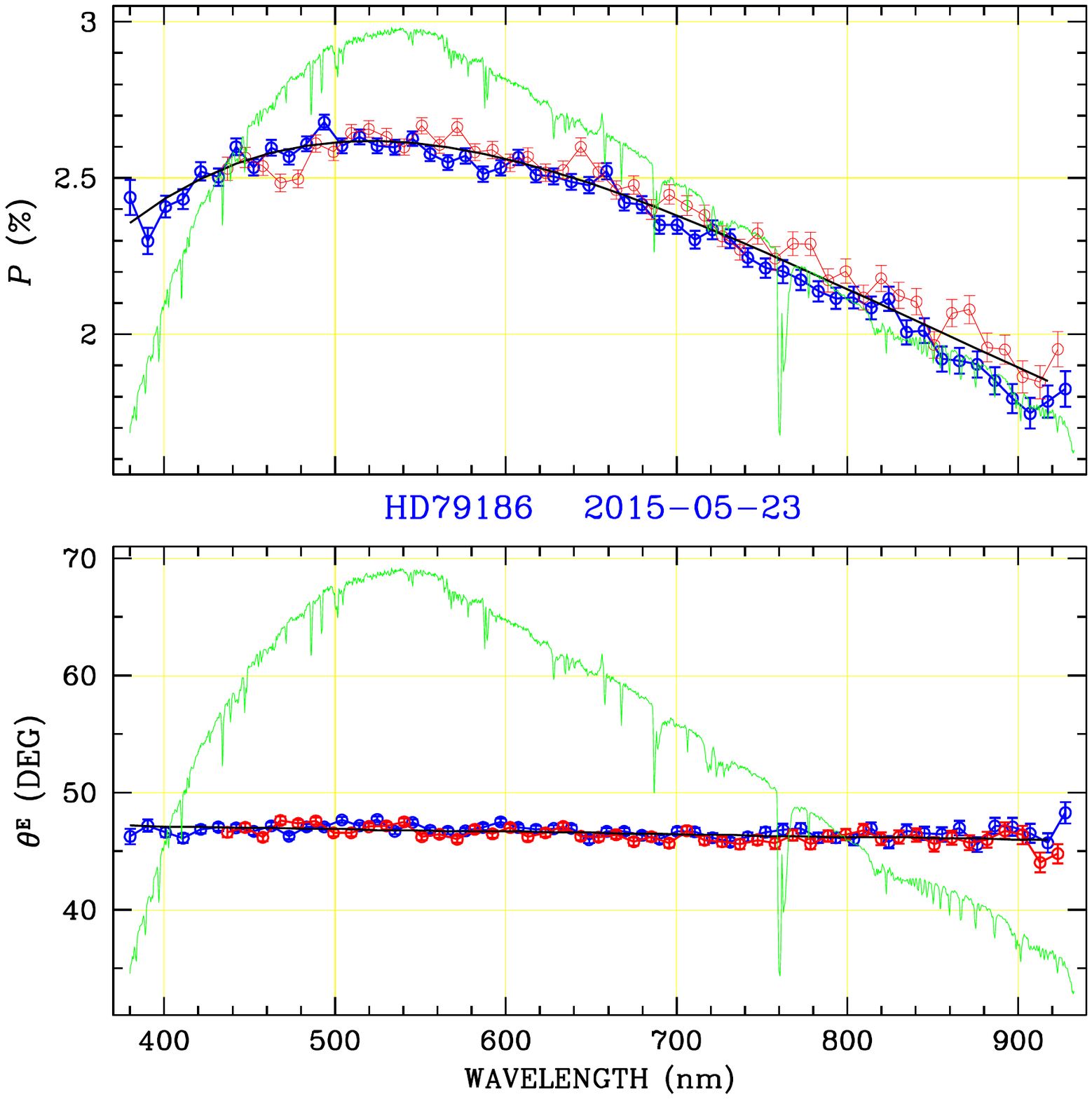}\\
  \includegraphics*[scale=0.42,trim={1.1cm 6.0cm 0.1cm 2.8cm},clip]{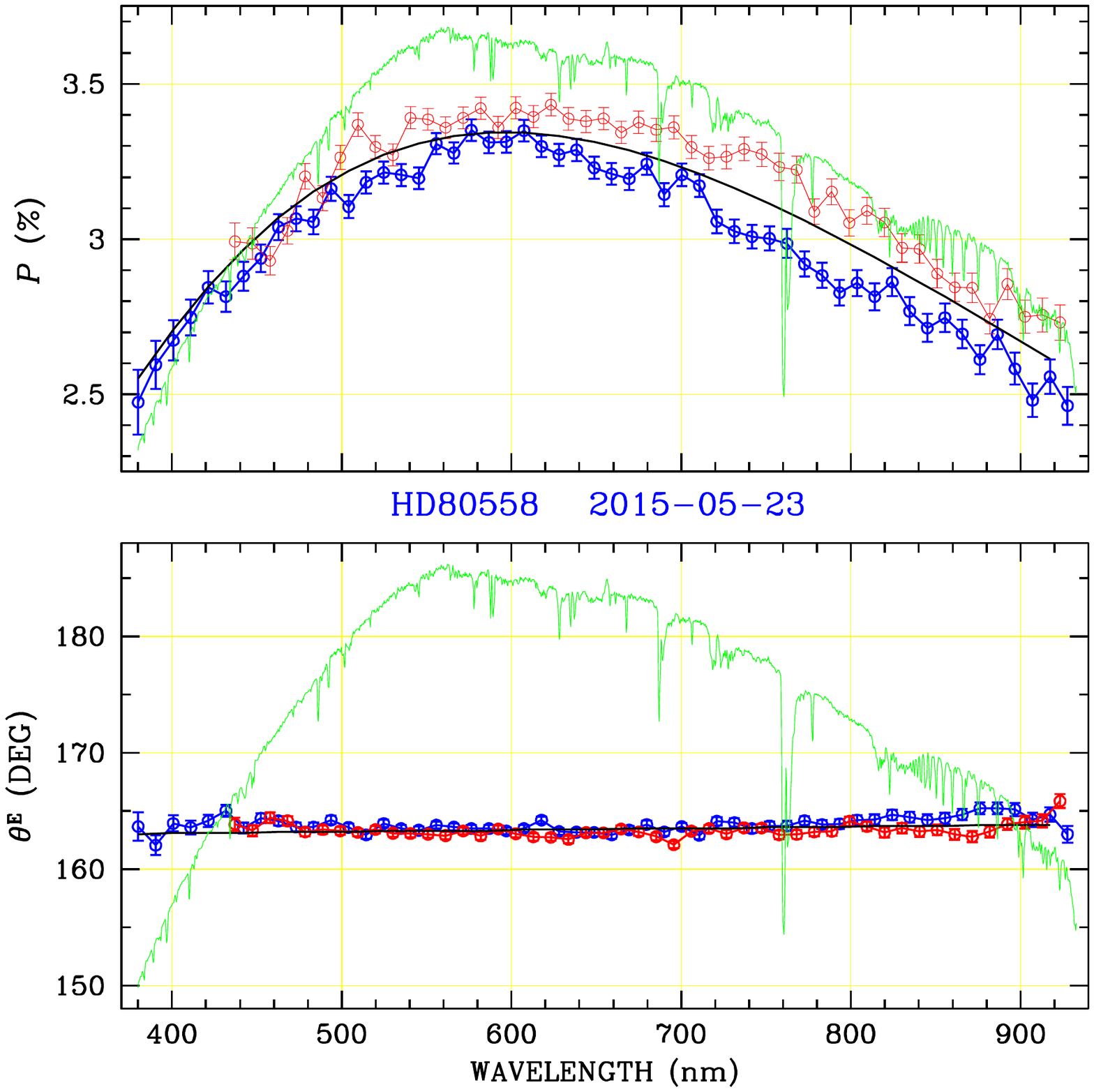}
  \includegraphics*[scale=0.42,trim={1.1cm 6.0cm 0.1cm 2.8cm},clip]{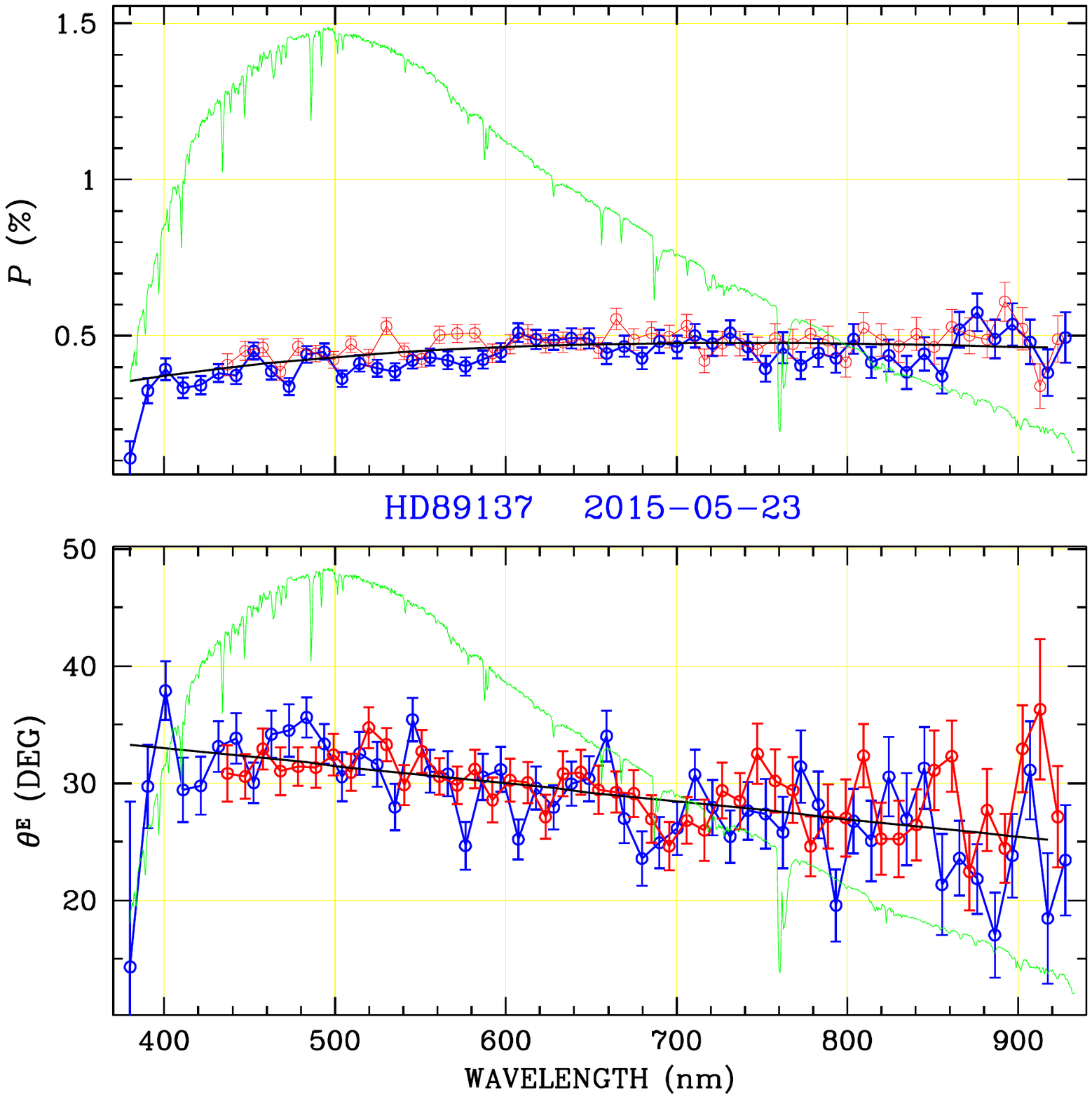}
\newpage

\noindent
  \includegraphics*[scale=0.42,trim={1.1cm 6.0cm 0.1cm 2.8cm},clip]{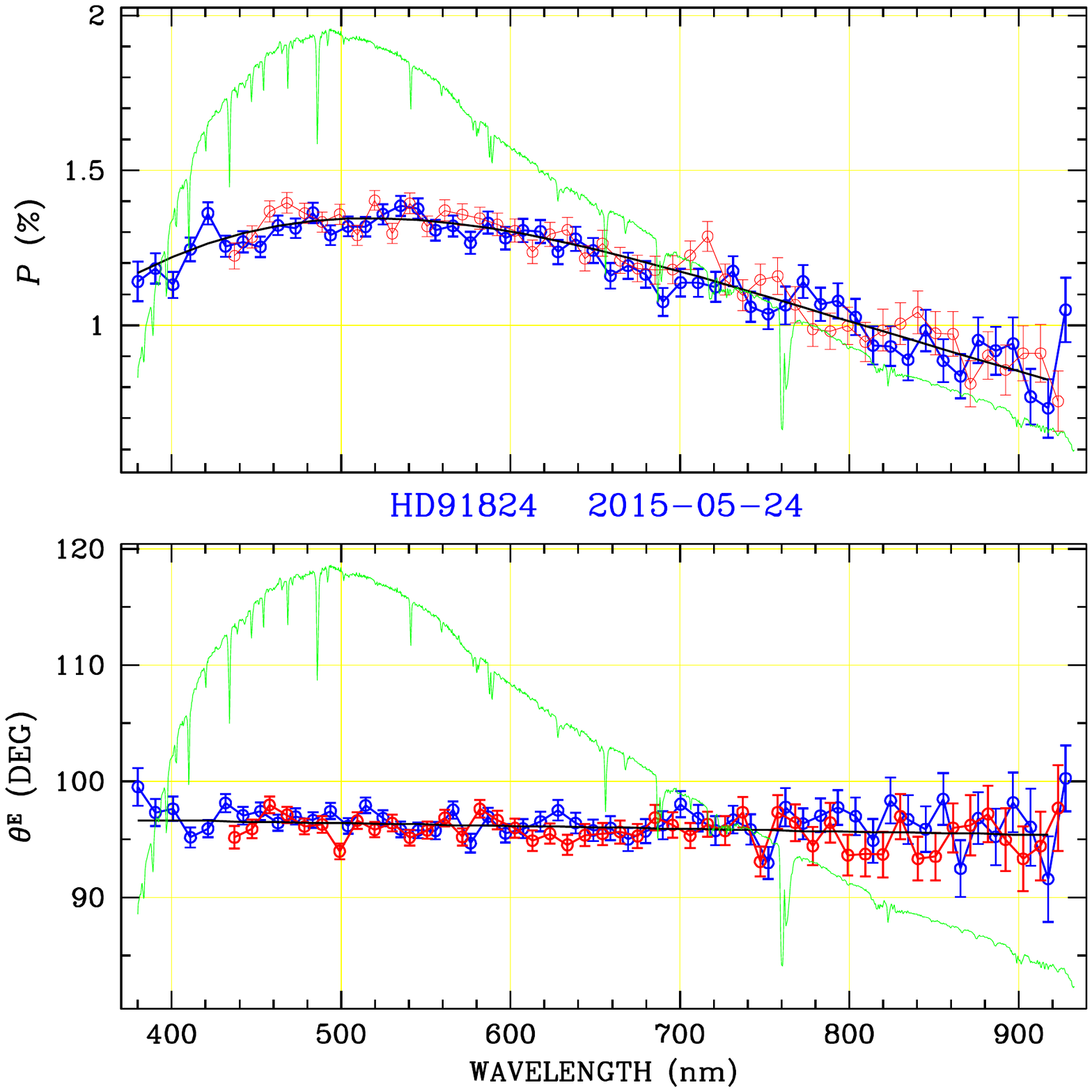}
  \includegraphics*[scale=0.42,trim={1.1cm 6.0cm 0.1cm 2.8cm},clip]{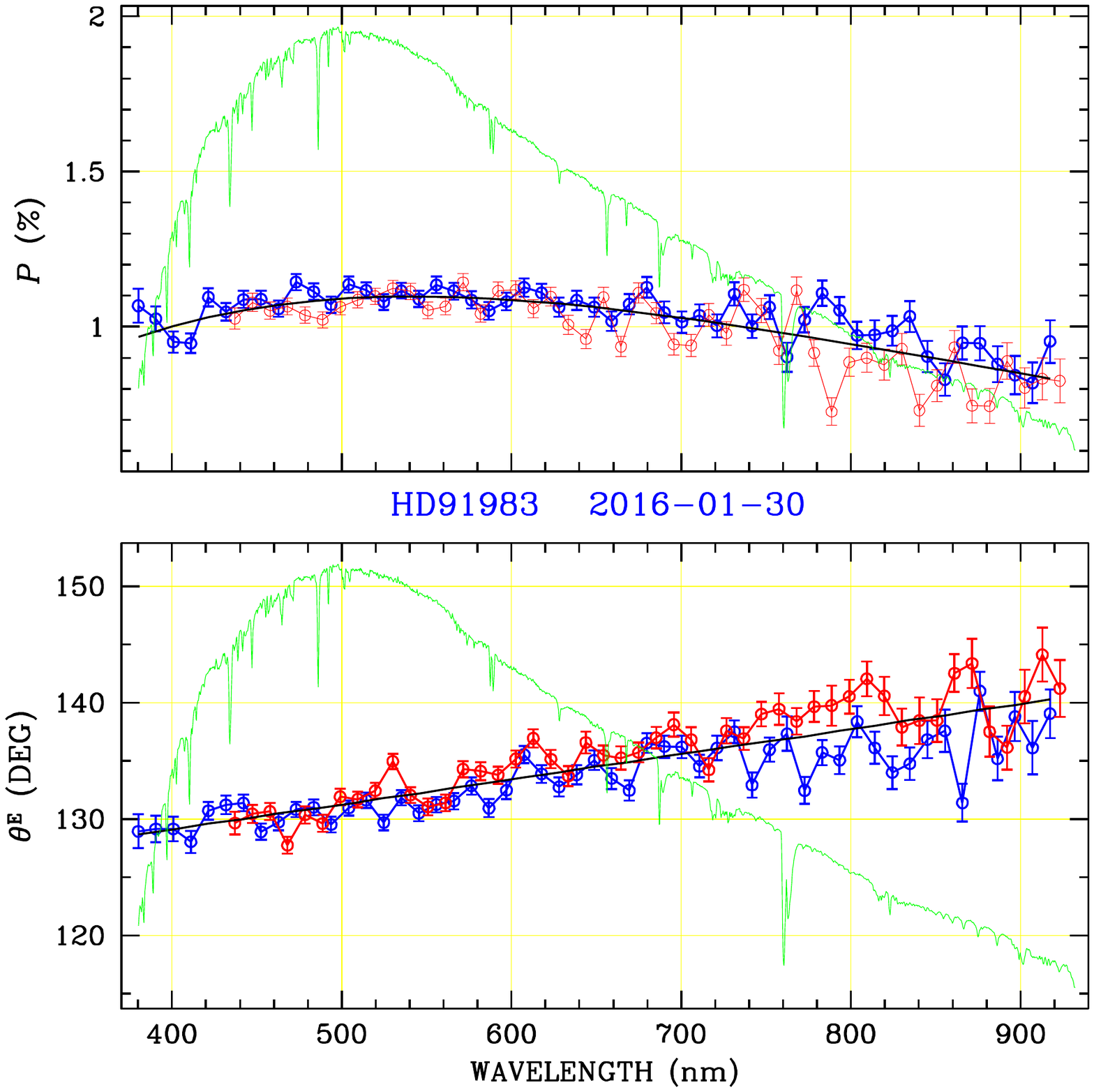}\\
  \includegraphics*[scale=0.42,trim={1.1cm 6.0cm 0.1cm 2.8cm},clip]{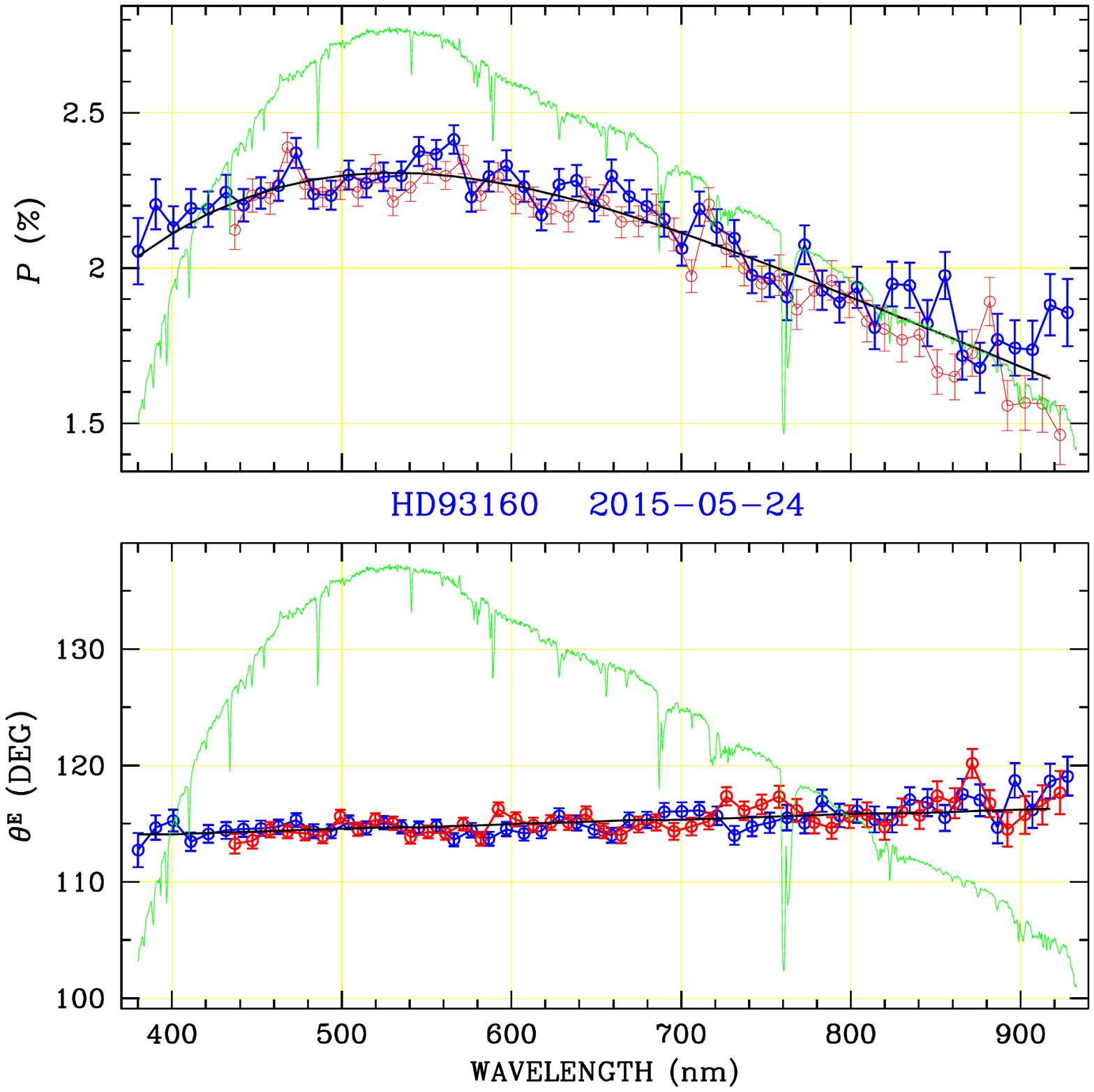}
  \includegraphics*[scale=0.42,trim={1.1cm 6.0cm 0.1cm 2.8cm},clip]{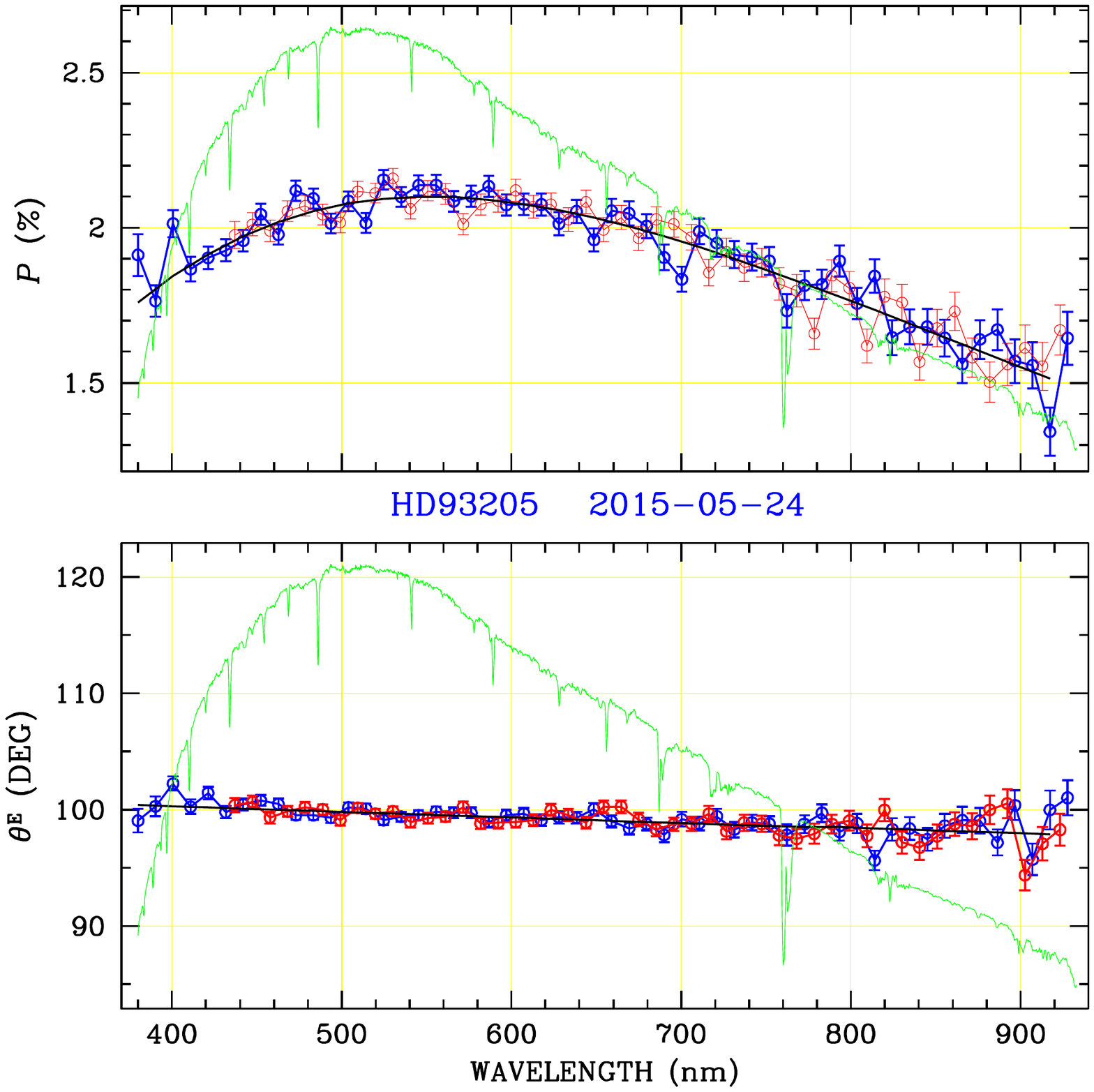}\\
  \includegraphics*[scale=0.42,trim={1.1cm 6.0cm 0.1cm 2.8cm},clip]{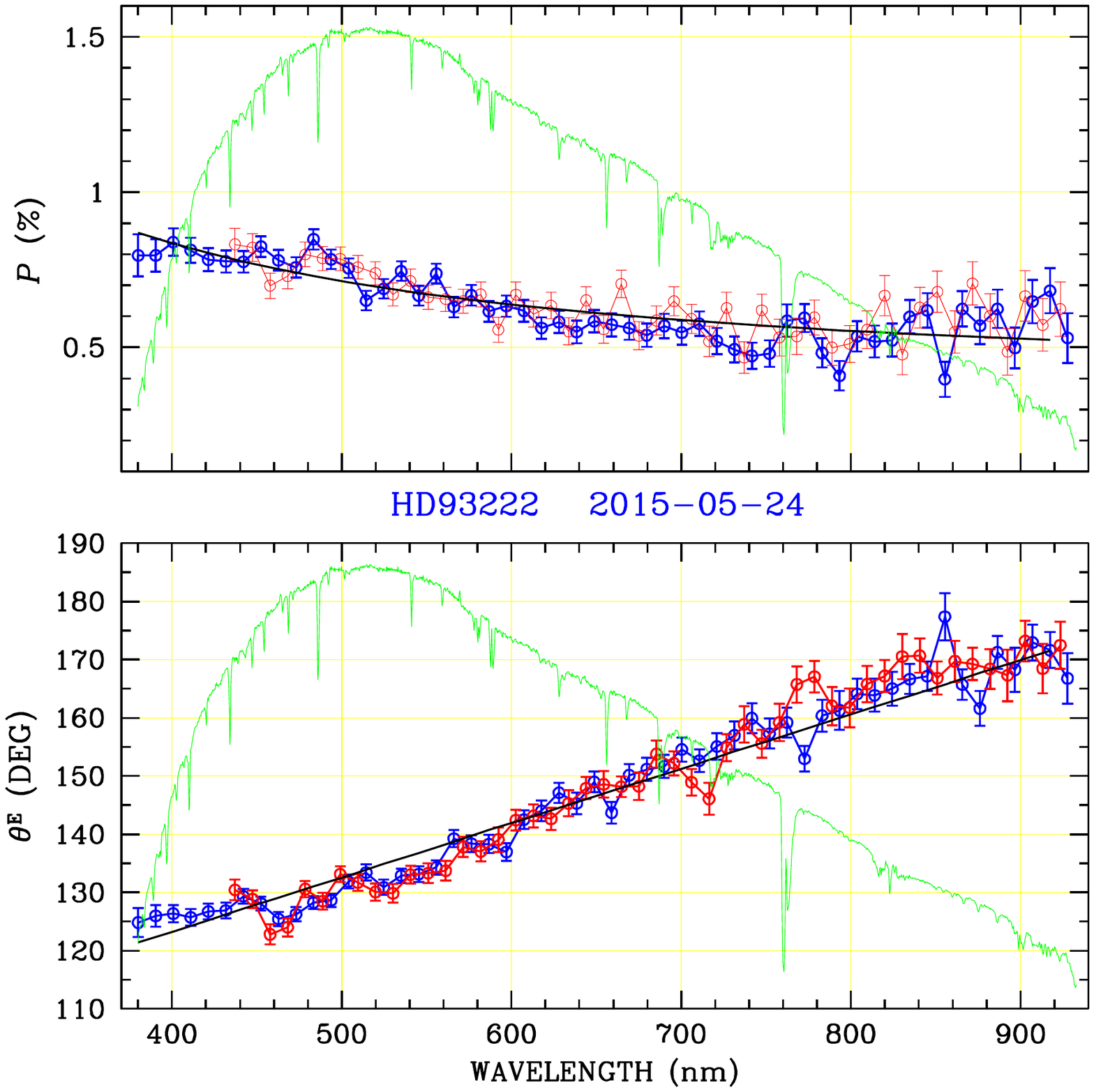}
  \includegraphics*[scale=0.42,trim={1.1cm 6.0cm 0.1cm 2.8cm},clip]{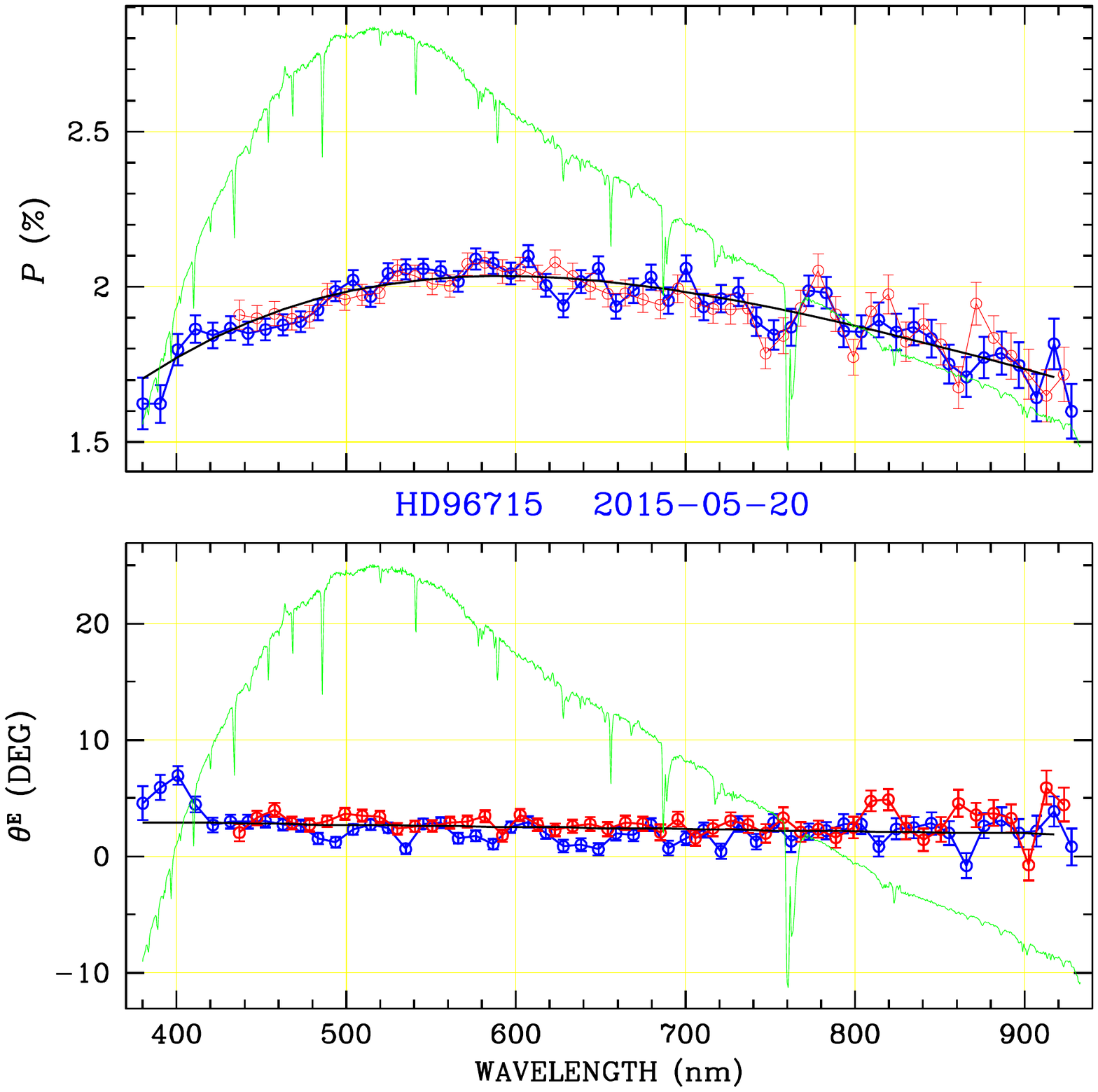}\\
\newpage

\noindent
  \includegraphics*[scale=0.42,trim={1.1cm 6.0cm 0.1cm 2.8cm},clip]{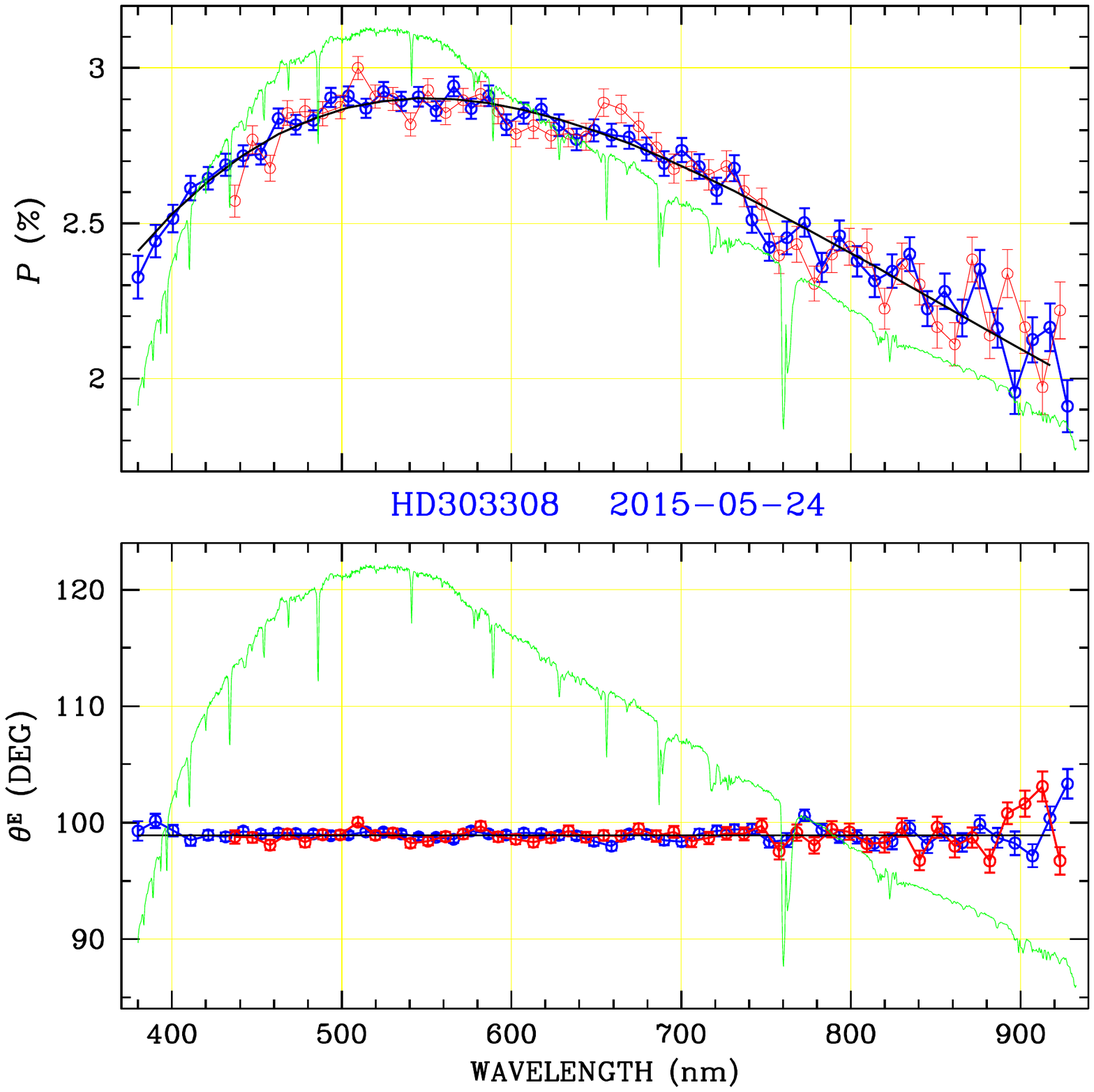}
  \includegraphics*[scale=0.42,trim={1.1cm 6.0cm 0.1cm 2.8cm},clip]{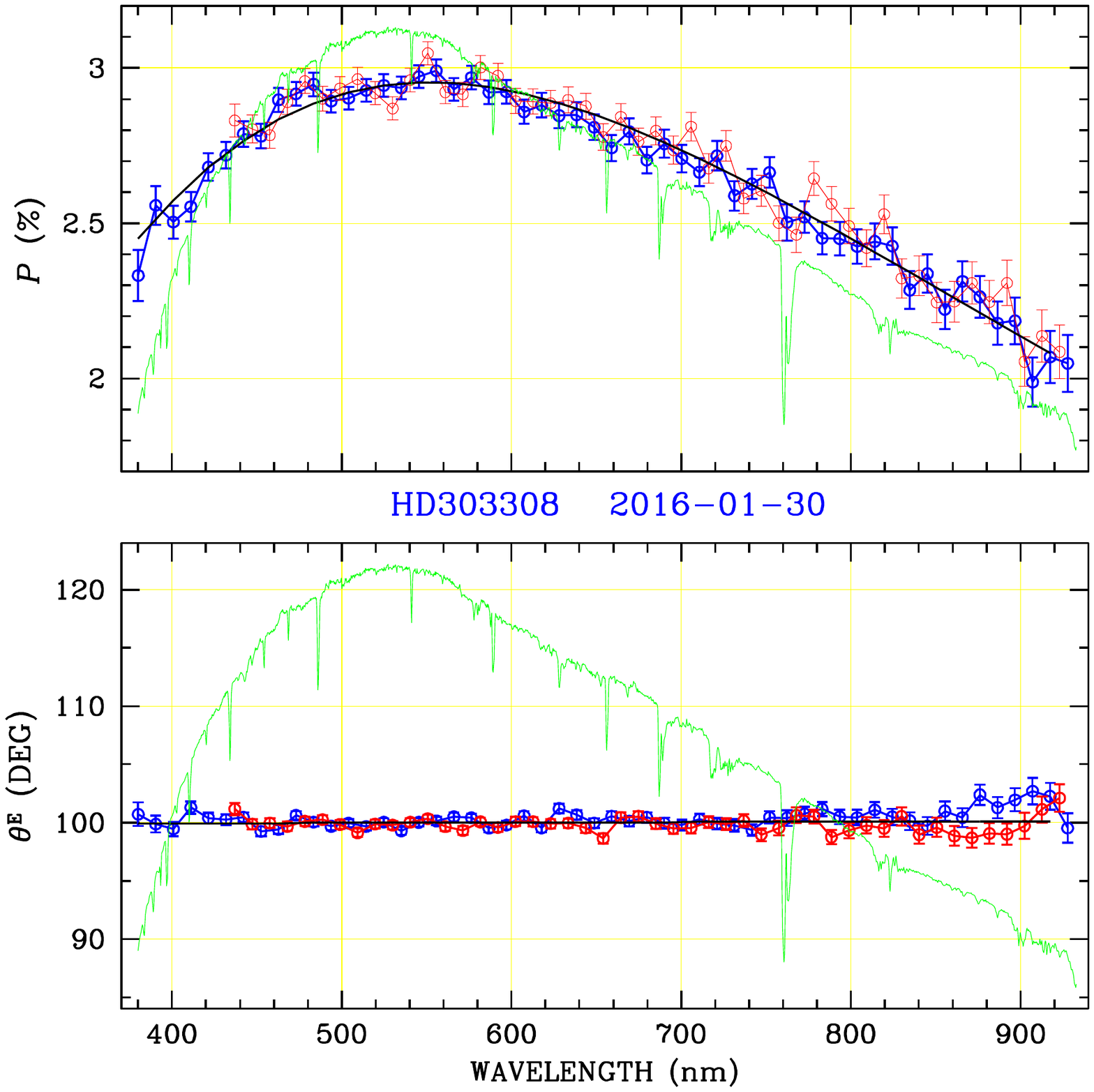}\\
  \includegraphics*[scale=0.42,trim={1.1cm 6.0cm 0.1cm 2.8cm},clip]{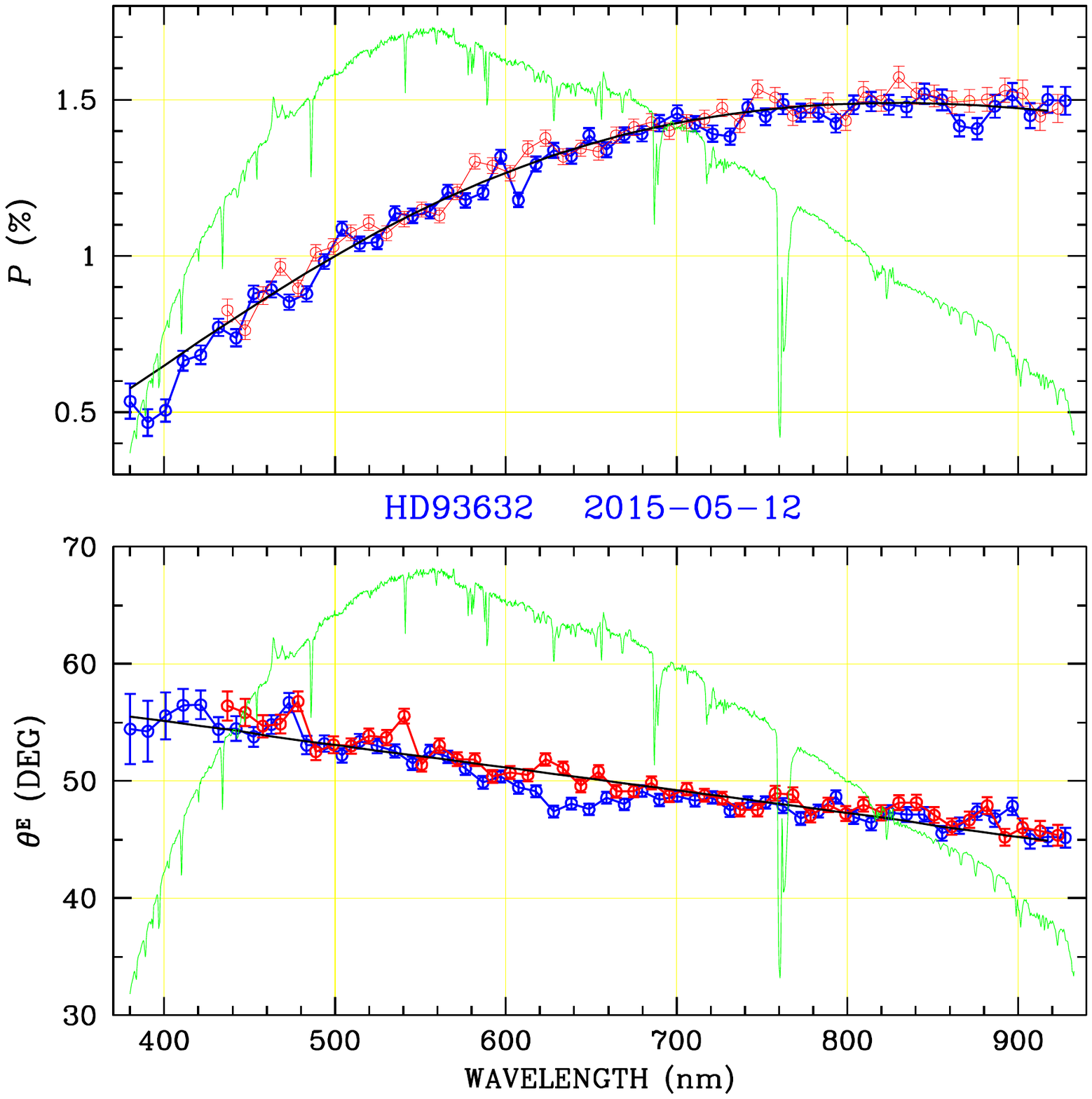}
  \includegraphics*[scale=0.42,trim={1.1cm 6.0cm 0.1cm 2.8cm},clip]{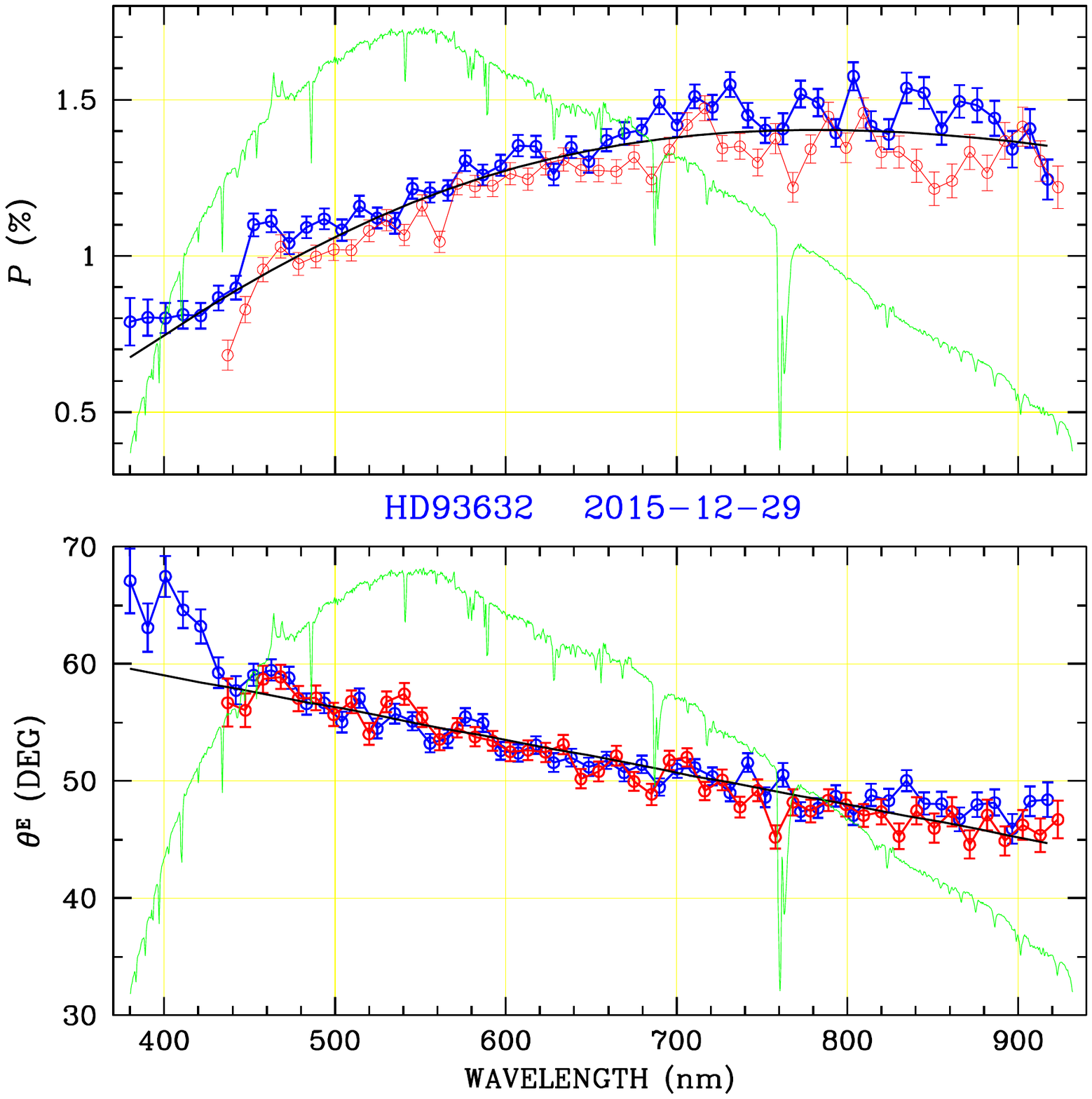}\\
  \includegraphics*[scale=0.42,trim={1.1cm 6.0cm 0.1cm 2.8cm},clip]{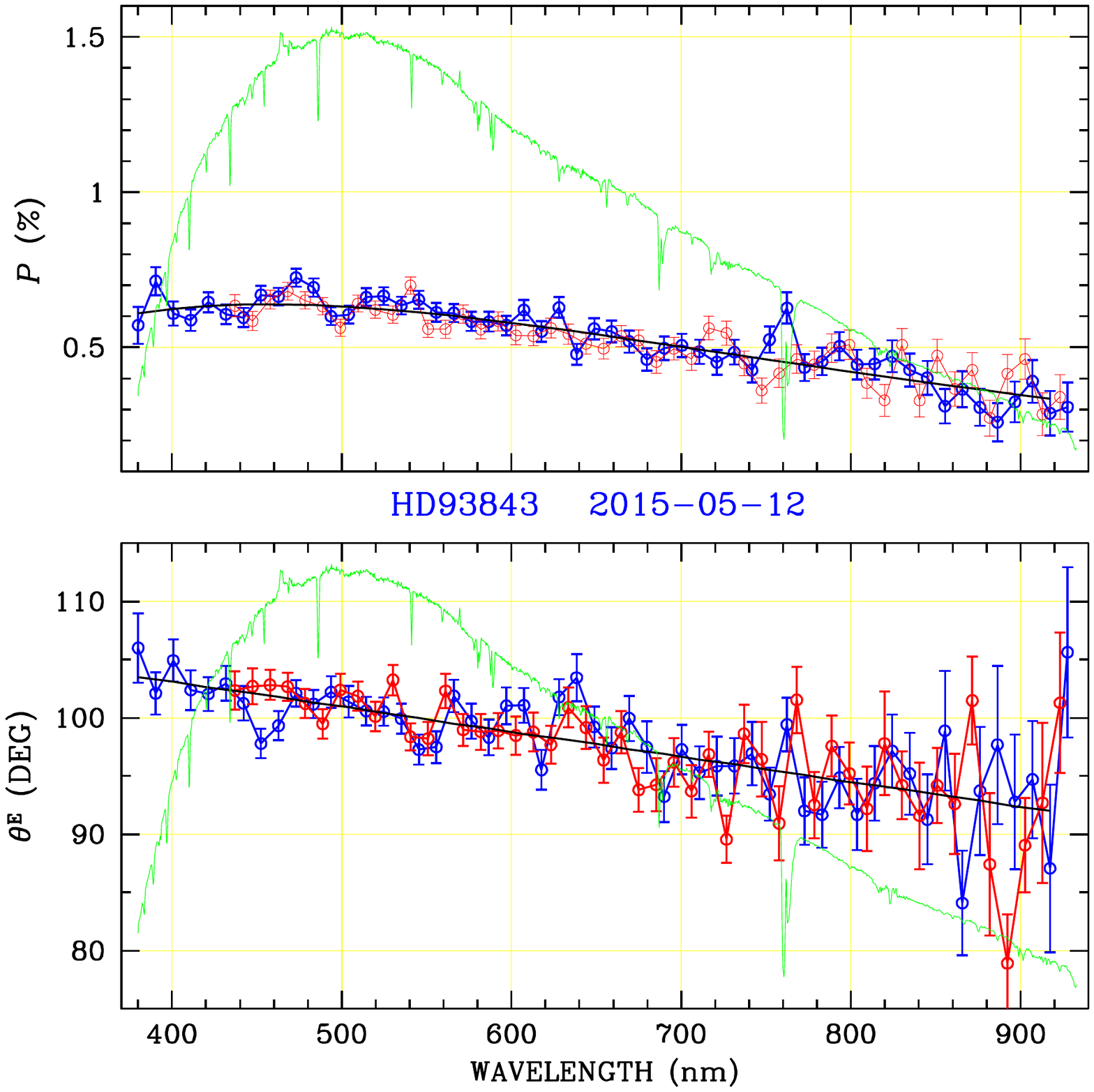}
  \includegraphics*[scale=0.42,trim={1.1cm 6.0cm 0.1cm 2.8cm},clip]{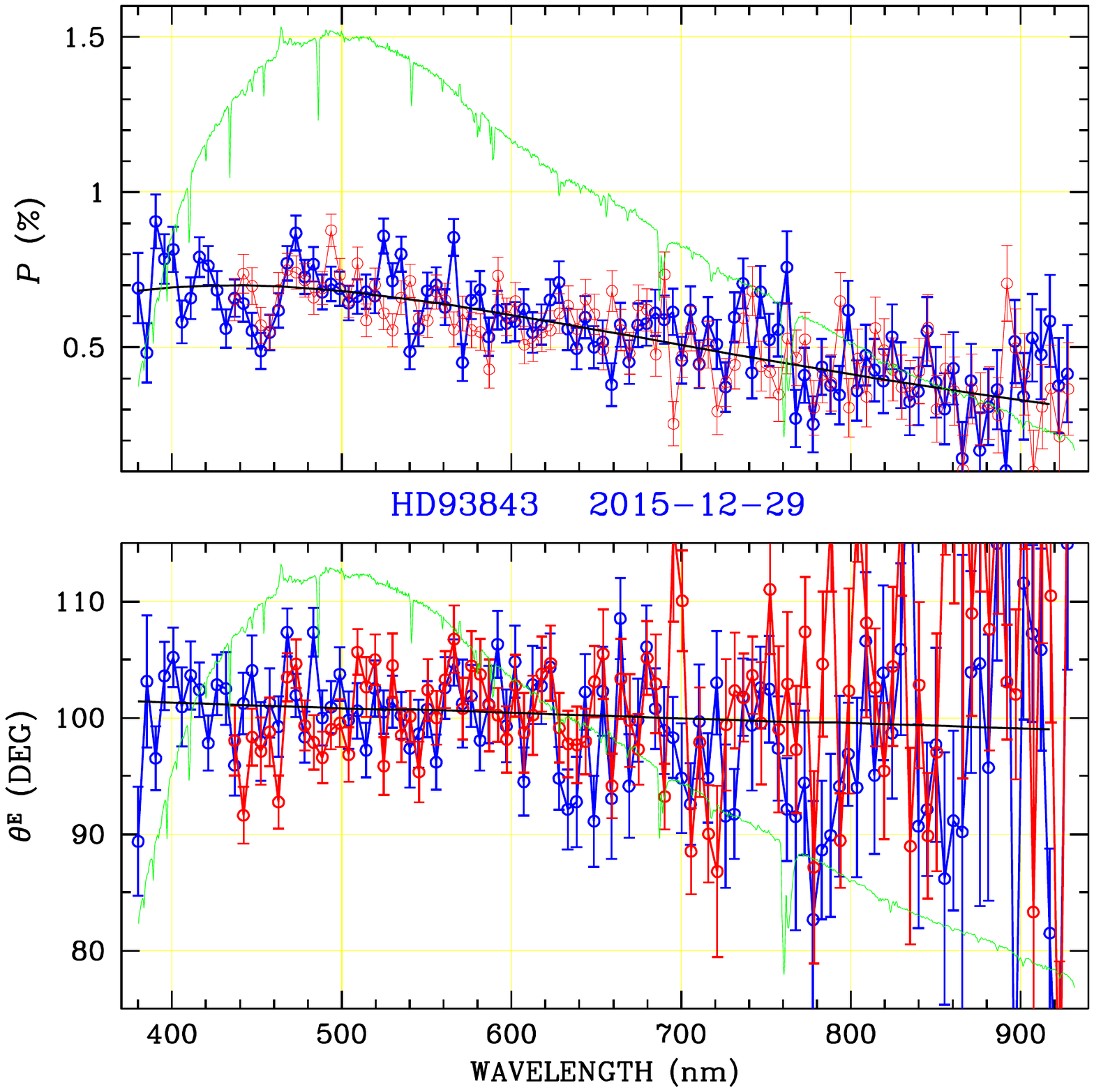}
\newpage

\noindent
  \includegraphics*[scale=0.42,trim={1.1cm 6.0cm 0.1cm 2.8cm},clip]{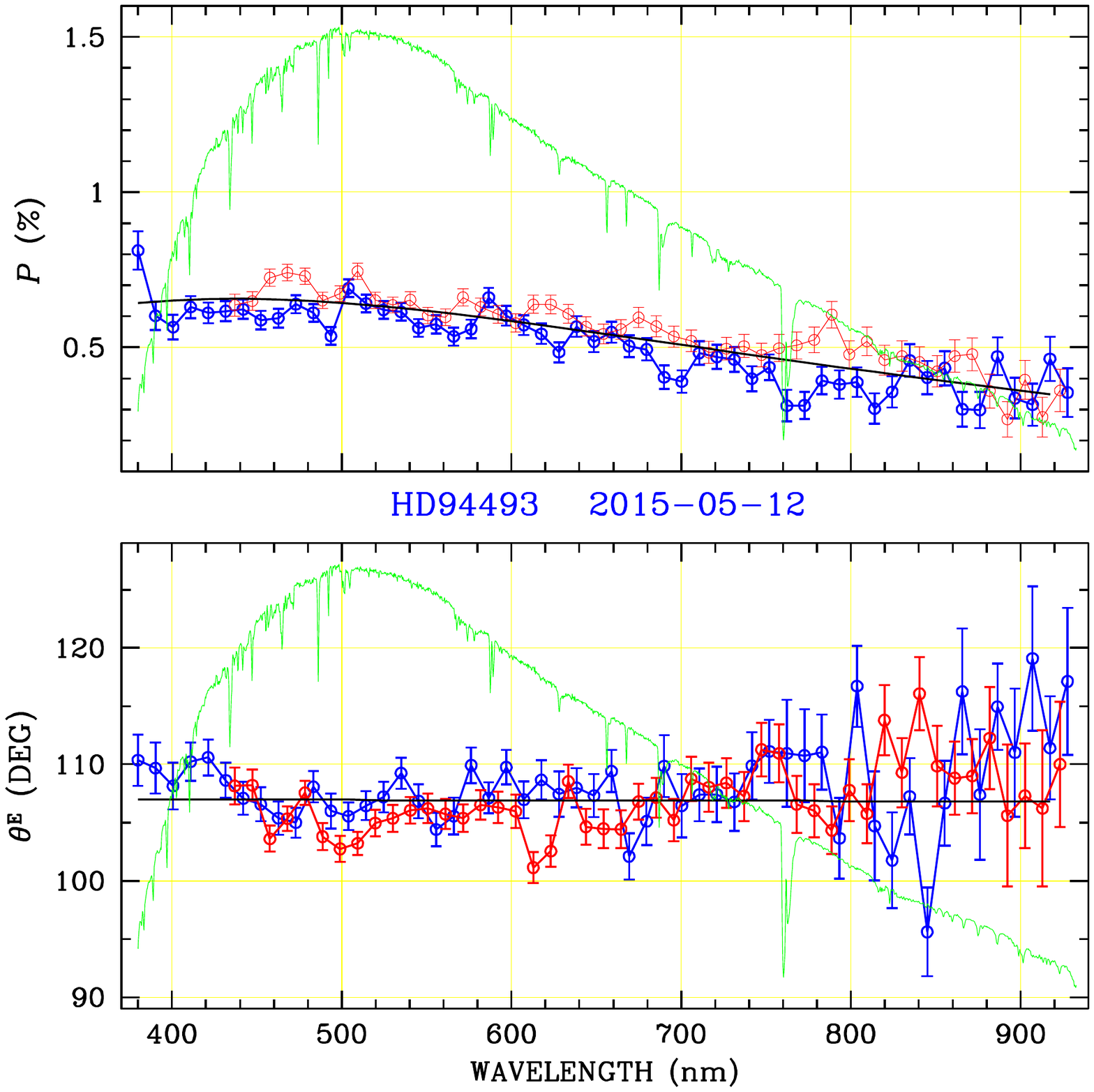}
  \includegraphics*[scale=0.42,trim={1.1cm 6.0cm 0.1cm 2.8cm},clip]{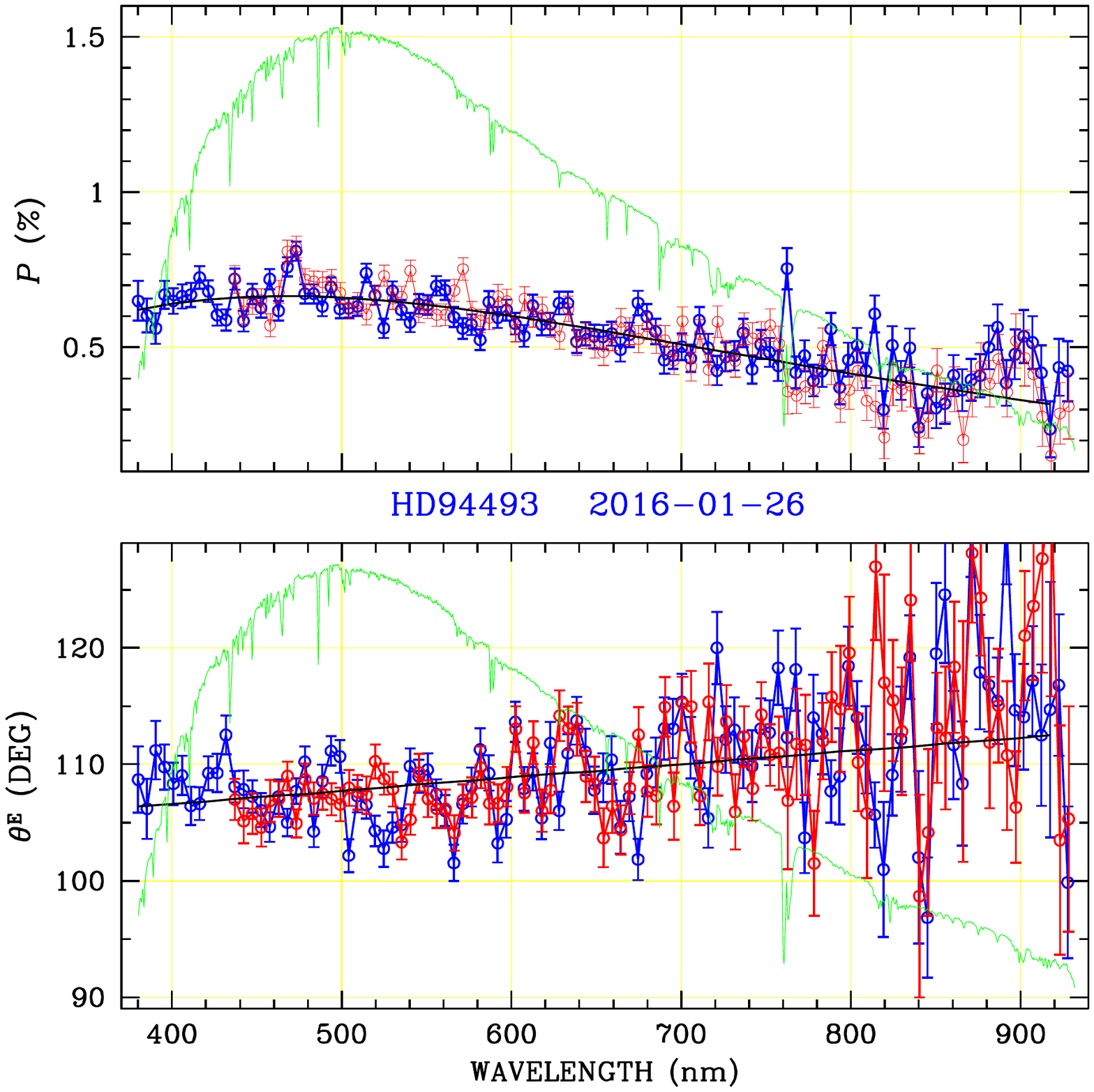}\\
  \includegraphics*[scale=0.42,trim={1.1cm 6.0cm 0.1cm 2.8cm},clip]{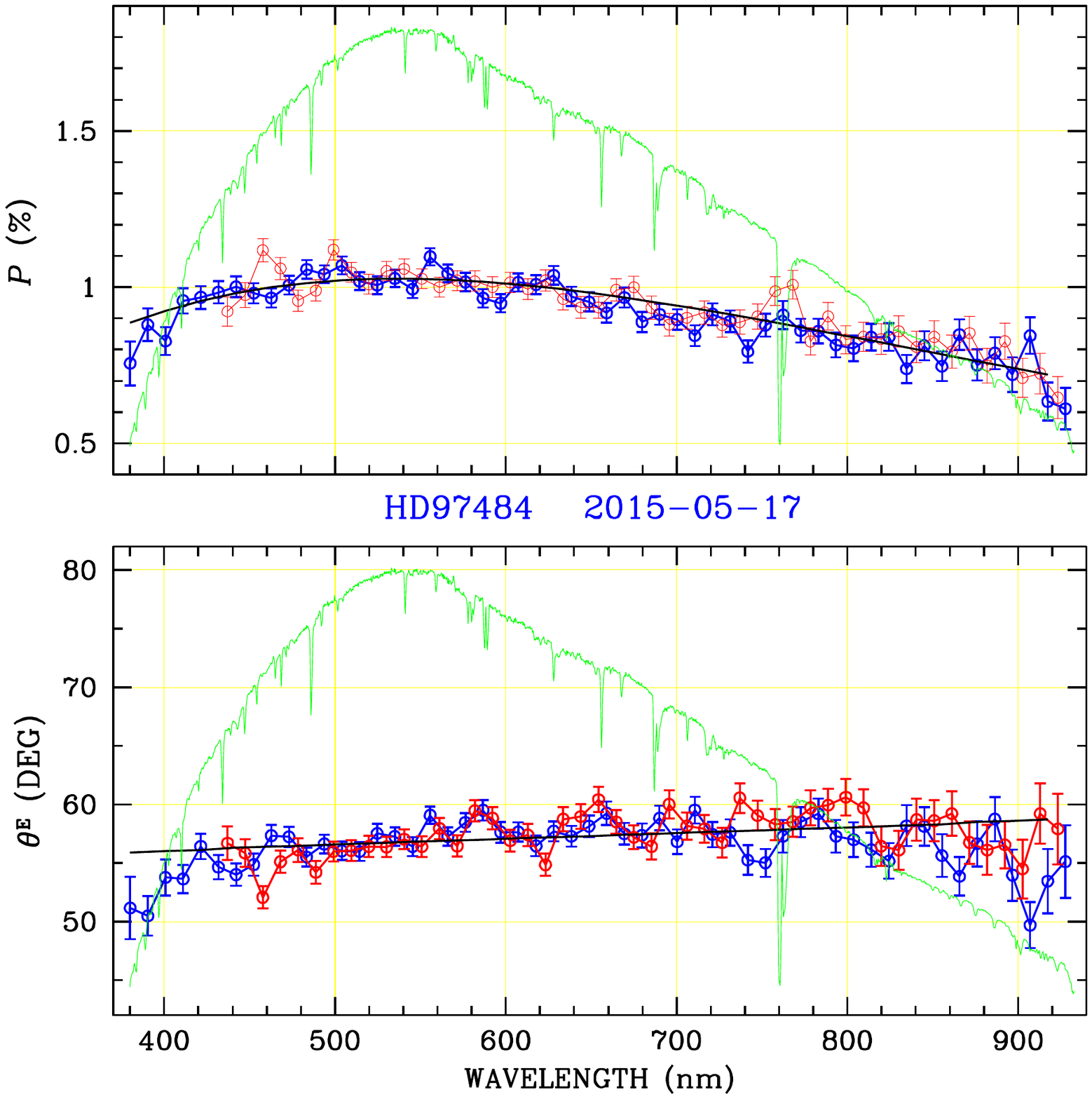}
  \includegraphics*[scale=0.42,trim={1.1cm 6.0cm 0.1cm 2.8cm},clip]{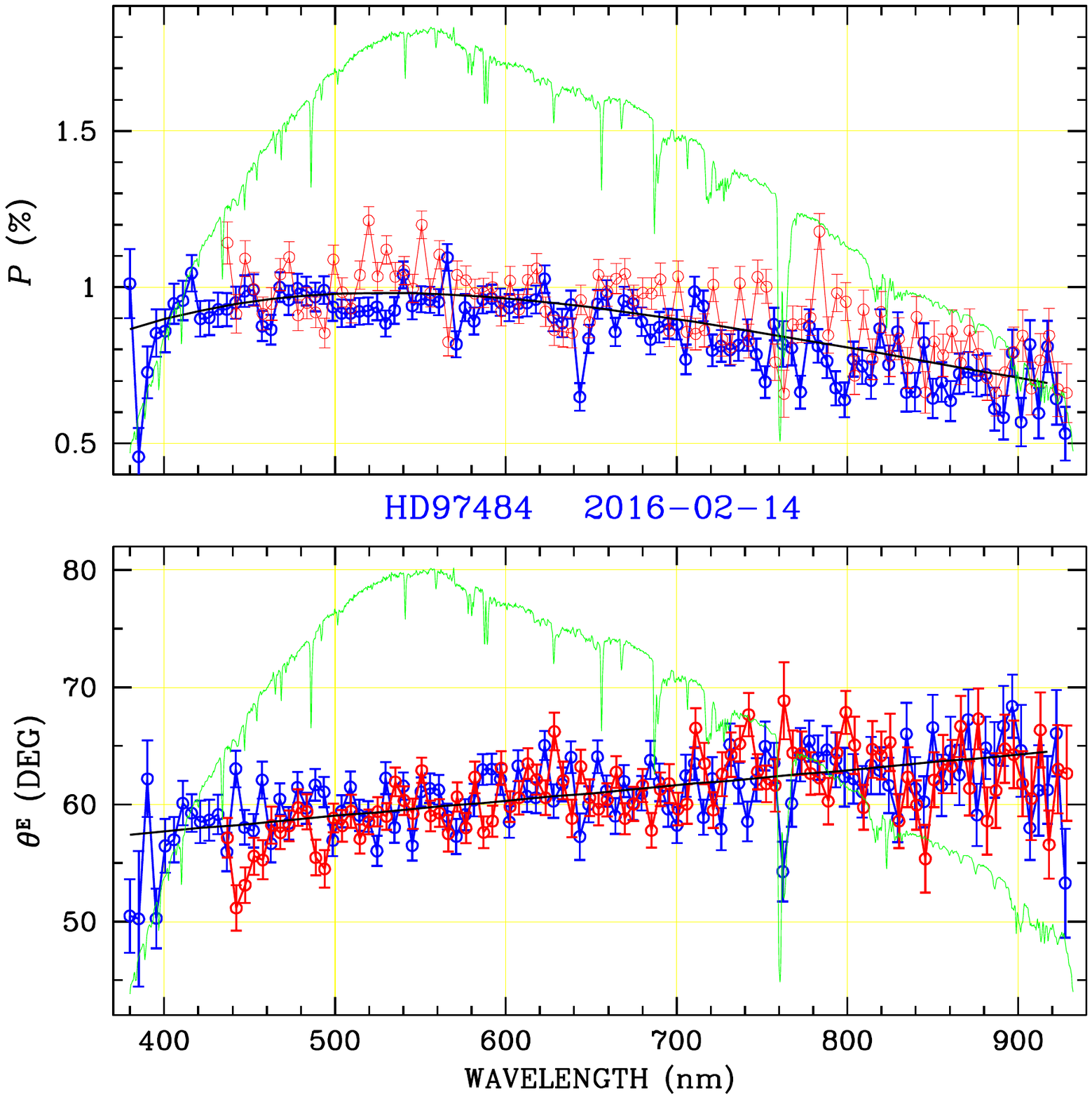}\\
  \includegraphics*[scale=0.42,trim={1.1cm 6.0cm 0.1cm 2.8cm},clip]{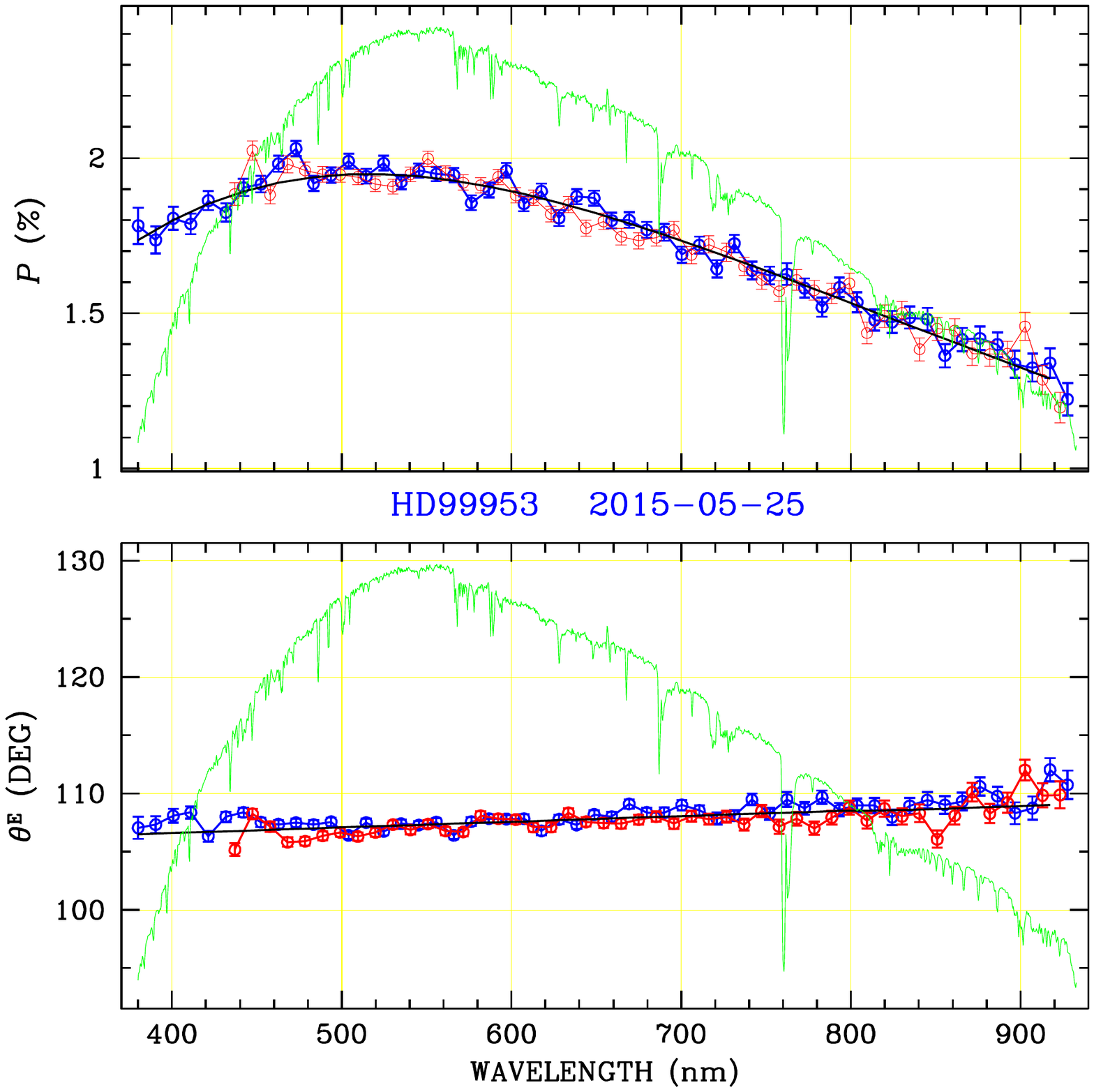}
  \includegraphics*[scale=0.42,trim={1.1cm 6.0cm 0.1cm 2.8cm},clip]{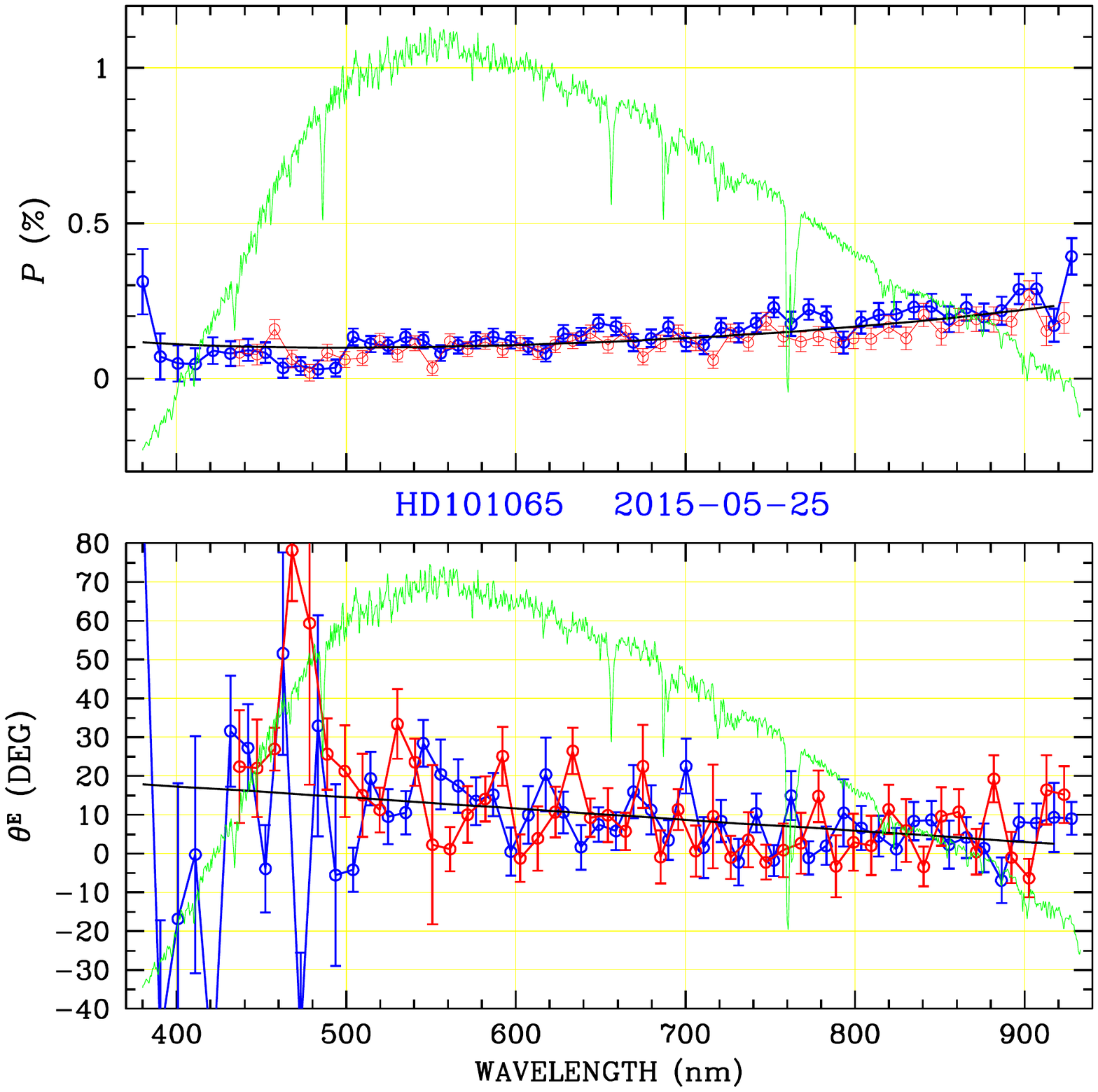}
\newpage

\noindent
  \includegraphics*[scale=0.42,trim={1.1cm 6.0cm 0.1cm 2.8cm},clip]{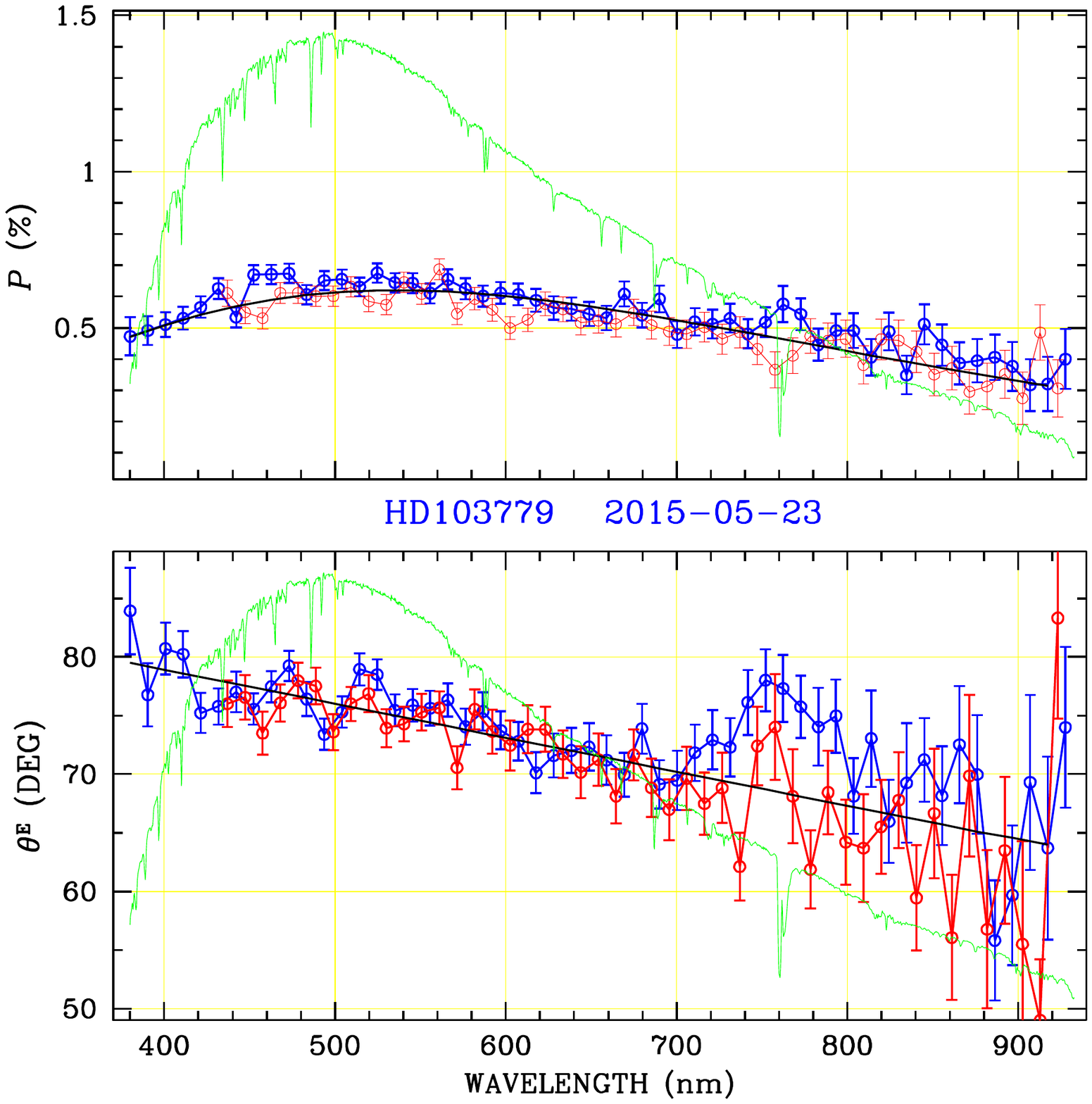}
  \includegraphics*[scale=0.42,trim={1.1cm 6.0cm 0.1cm 2.8cm},clip]{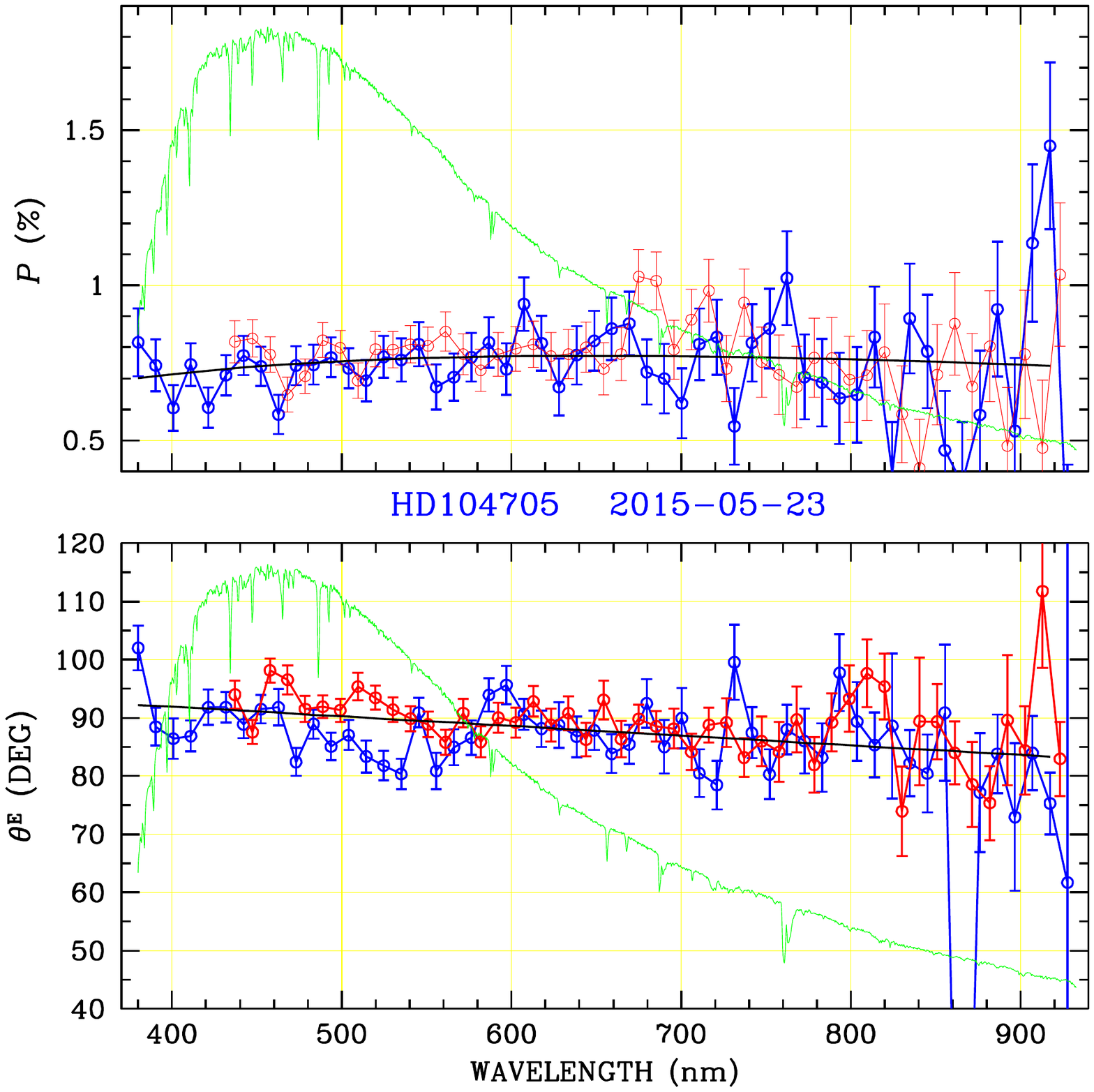}\\
  \includegraphics*[scale=0.42,trim={1.1cm 6.0cm 0.1cm 2.8cm},clip]{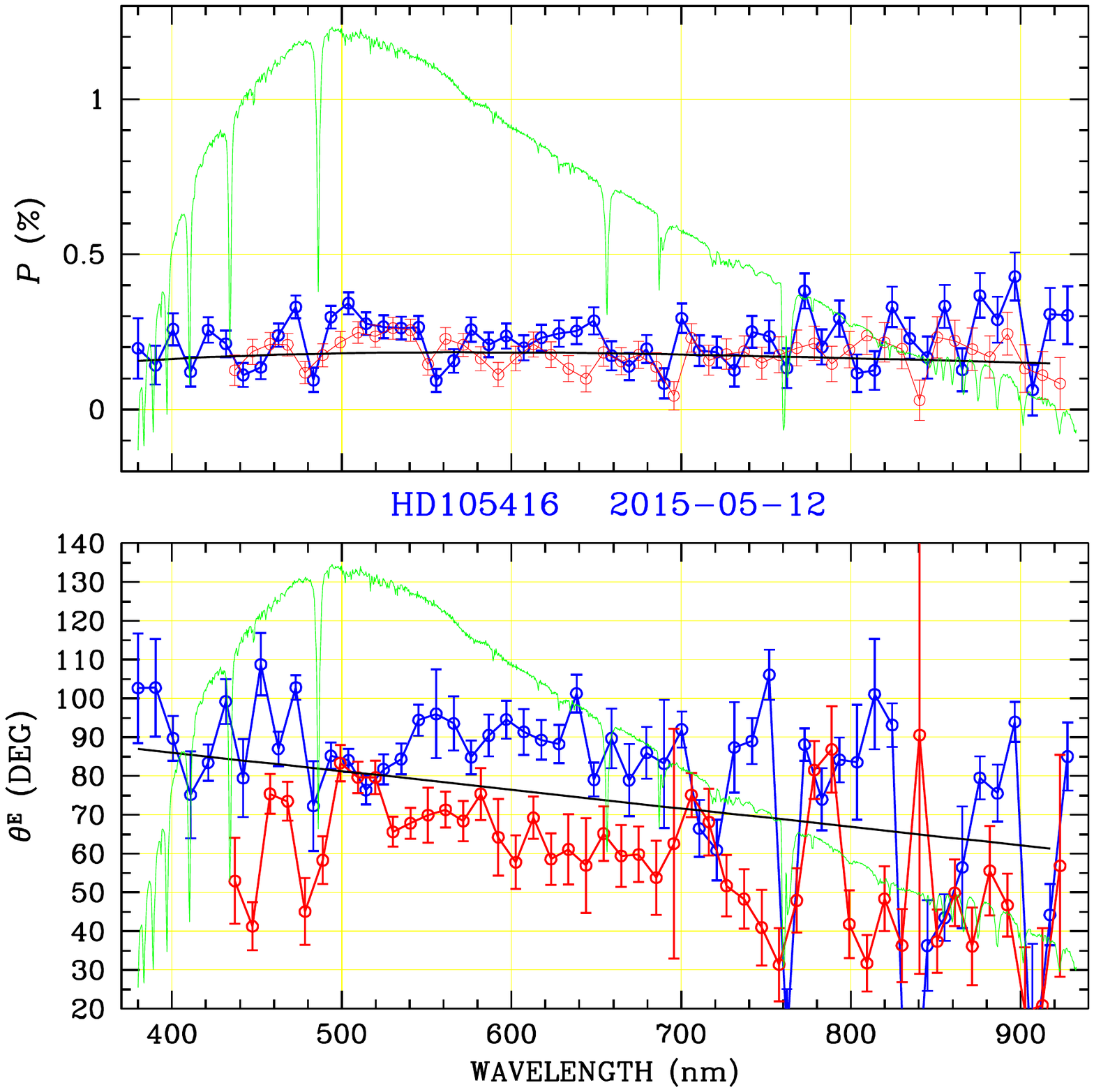}
  \includegraphics*[scale=0.42,trim={1.1cm 6.0cm 0.1cm 2.8cm},clip]{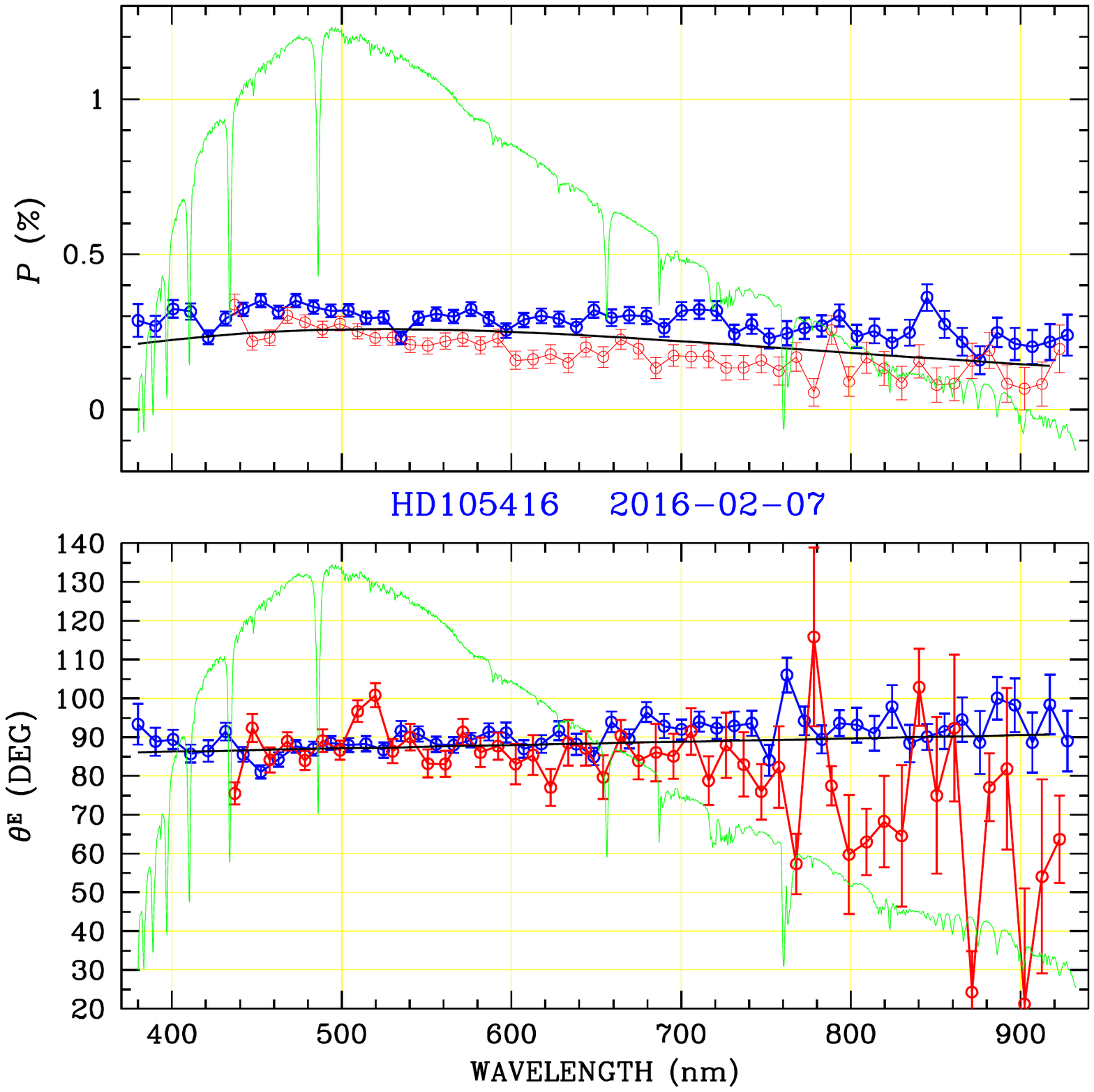}\\
  \includegraphics*[scale=0.42,trim={1.1cm 6.0cm 0.1cm 2.8cm},clip]{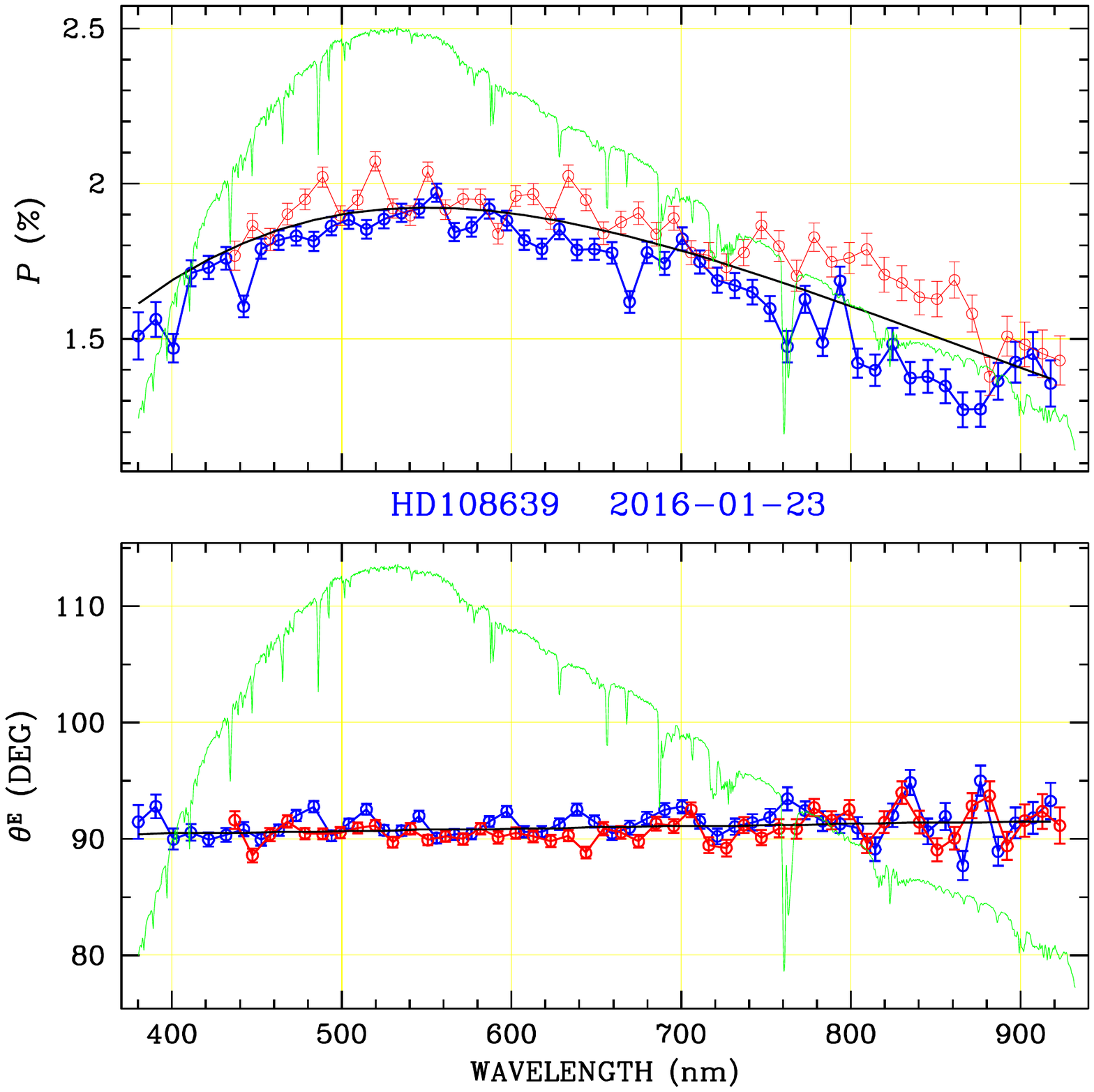}
  \includegraphics*[scale=0.42,trim={1.1cm 6.0cm 0.1cm 2.8cm},clip]{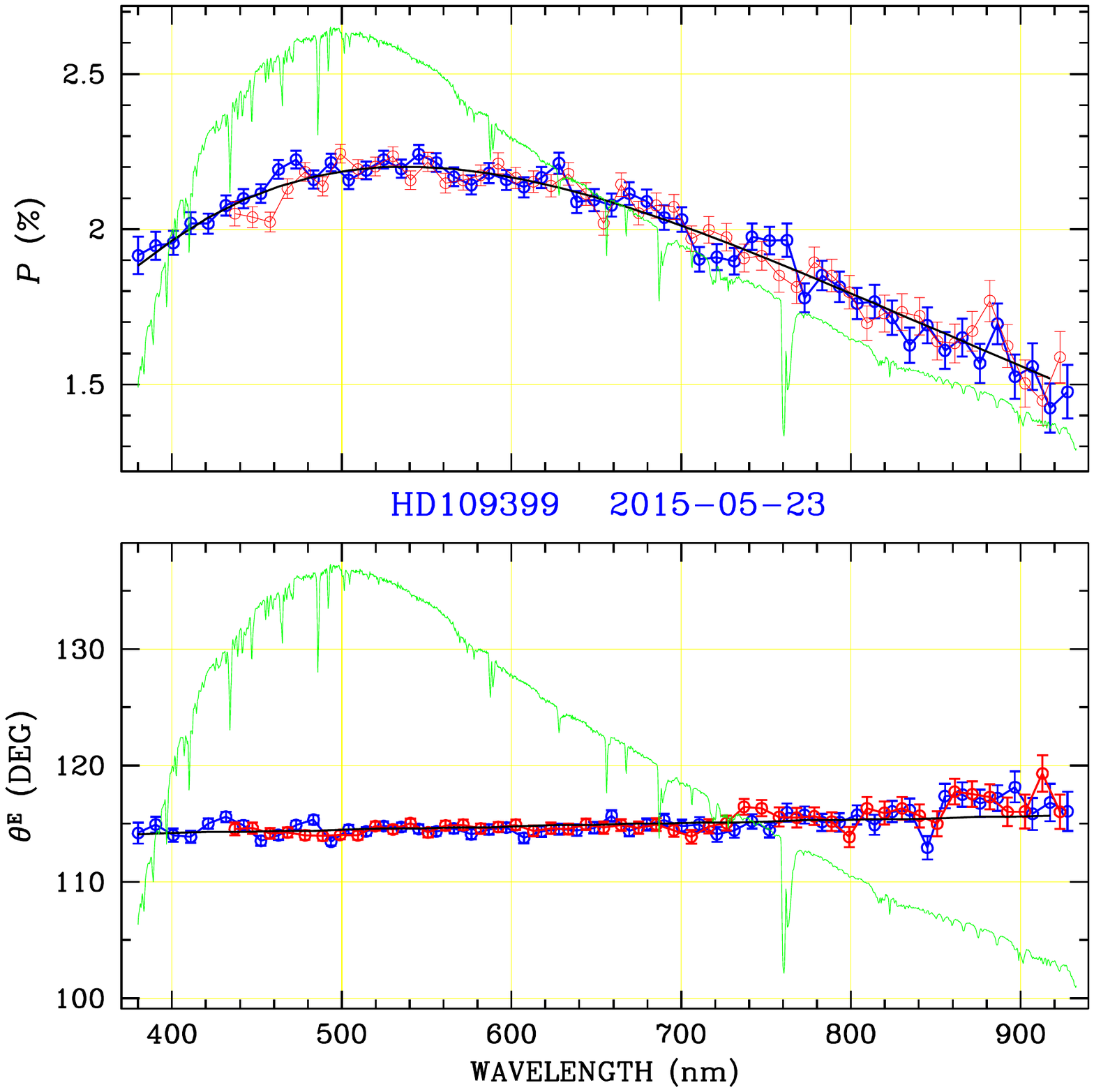}
\newpage

\noindent
  \includegraphics*[scale=0.42,trim={1.1cm 6.0cm 0.1cm 2.8cm},clip]{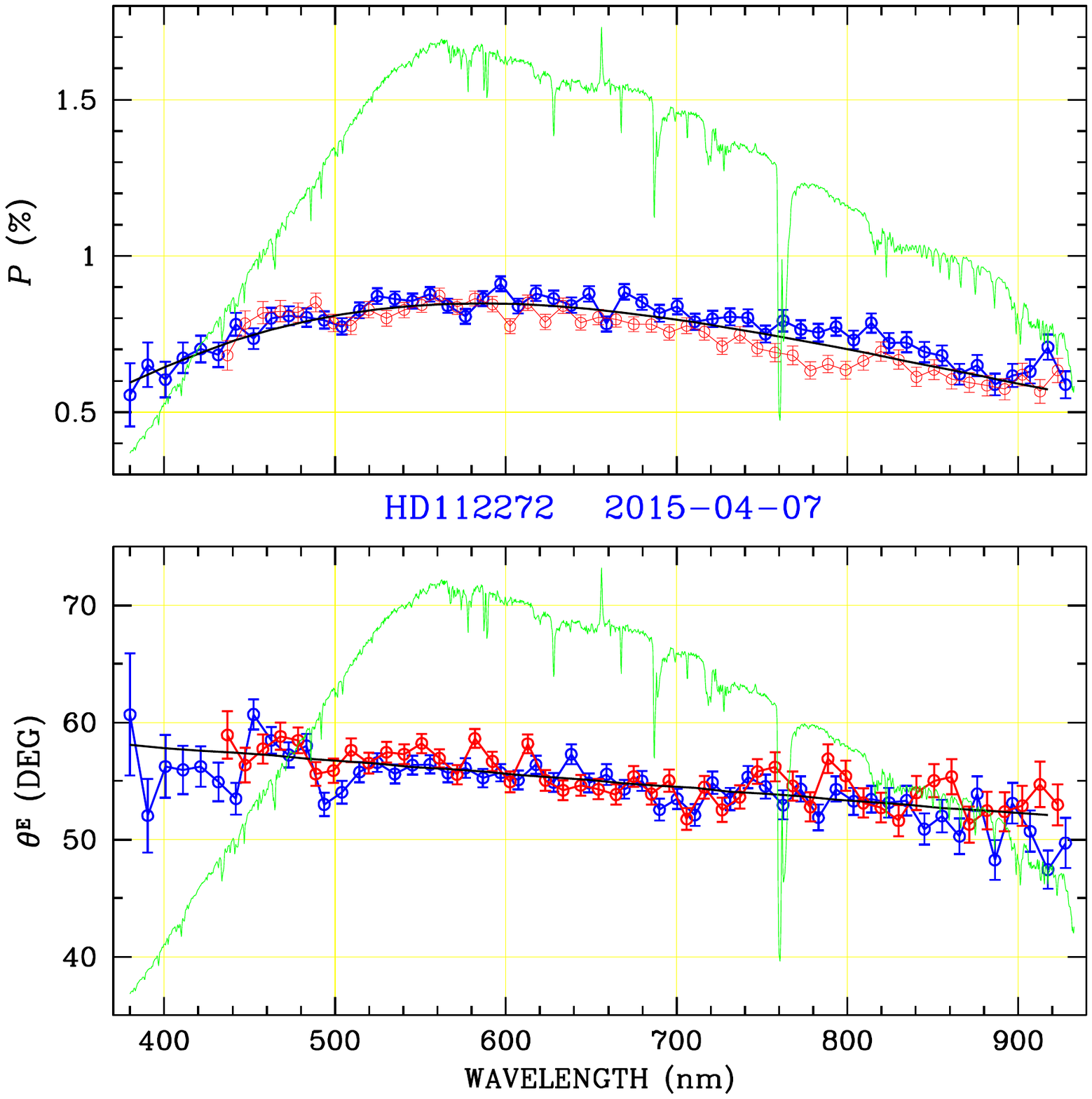}
  \includegraphics*[scale=0.42,trim={1.1cm 6.0cm 0.1cm 2.8cm},clip]{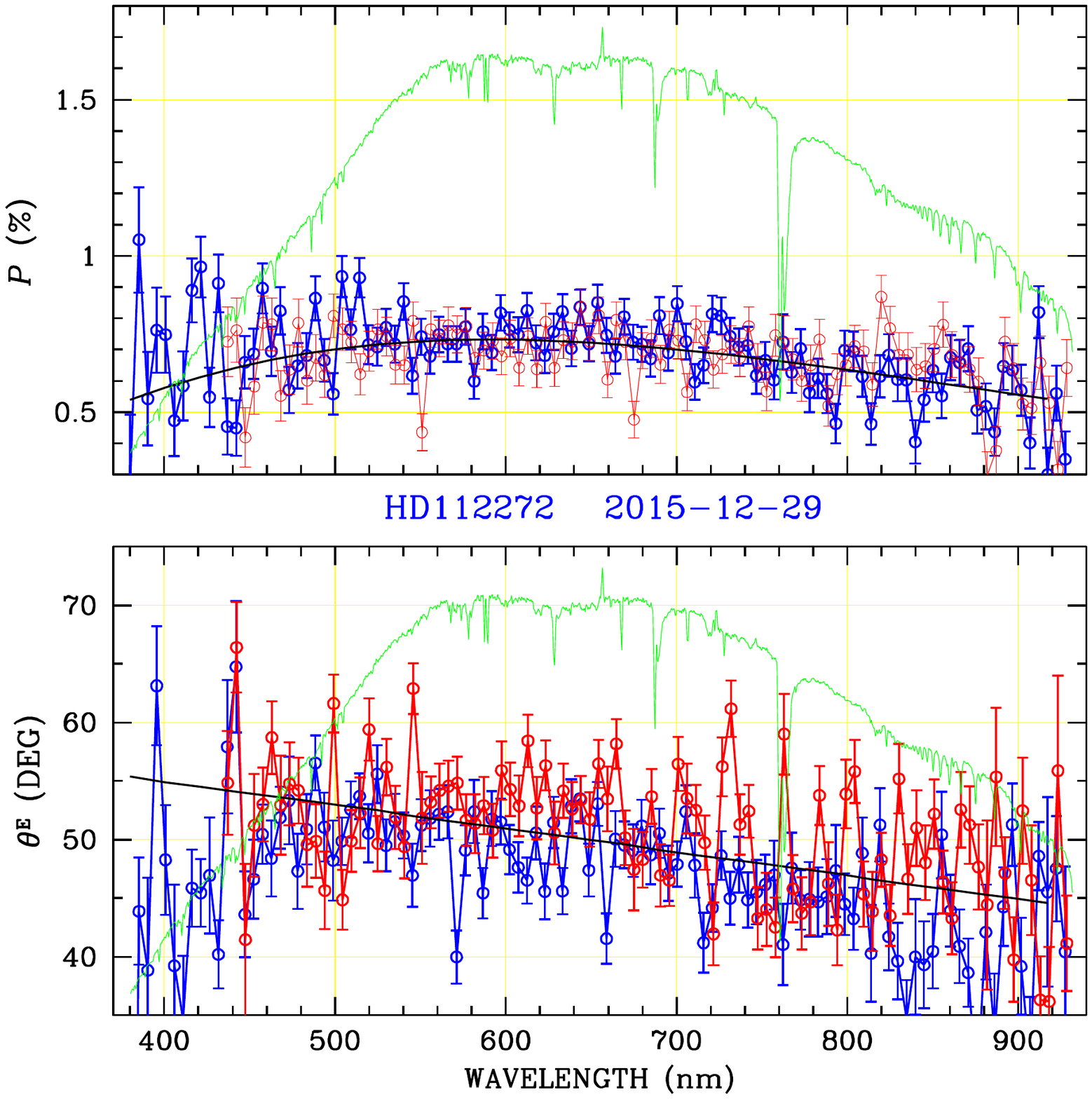}\\
  \includegraphics*[scale=0.42,trim={1.1cm 6.0cm 0.1cm 2.8cm},clip]{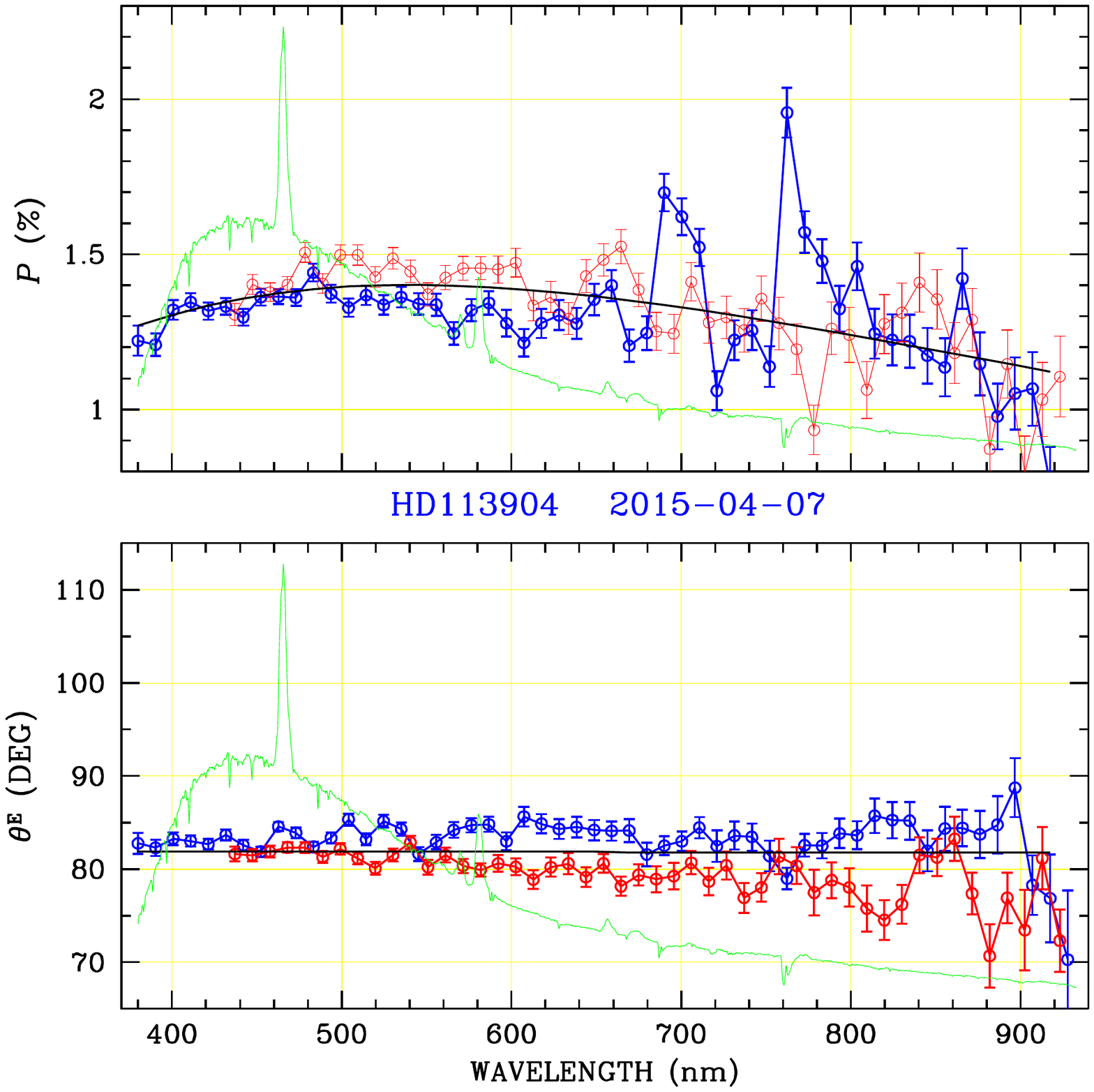}
  \includegraphics*[scale=0.42,trim={1.1cm 6.0cm 0.1cm 2.8cm},clip]{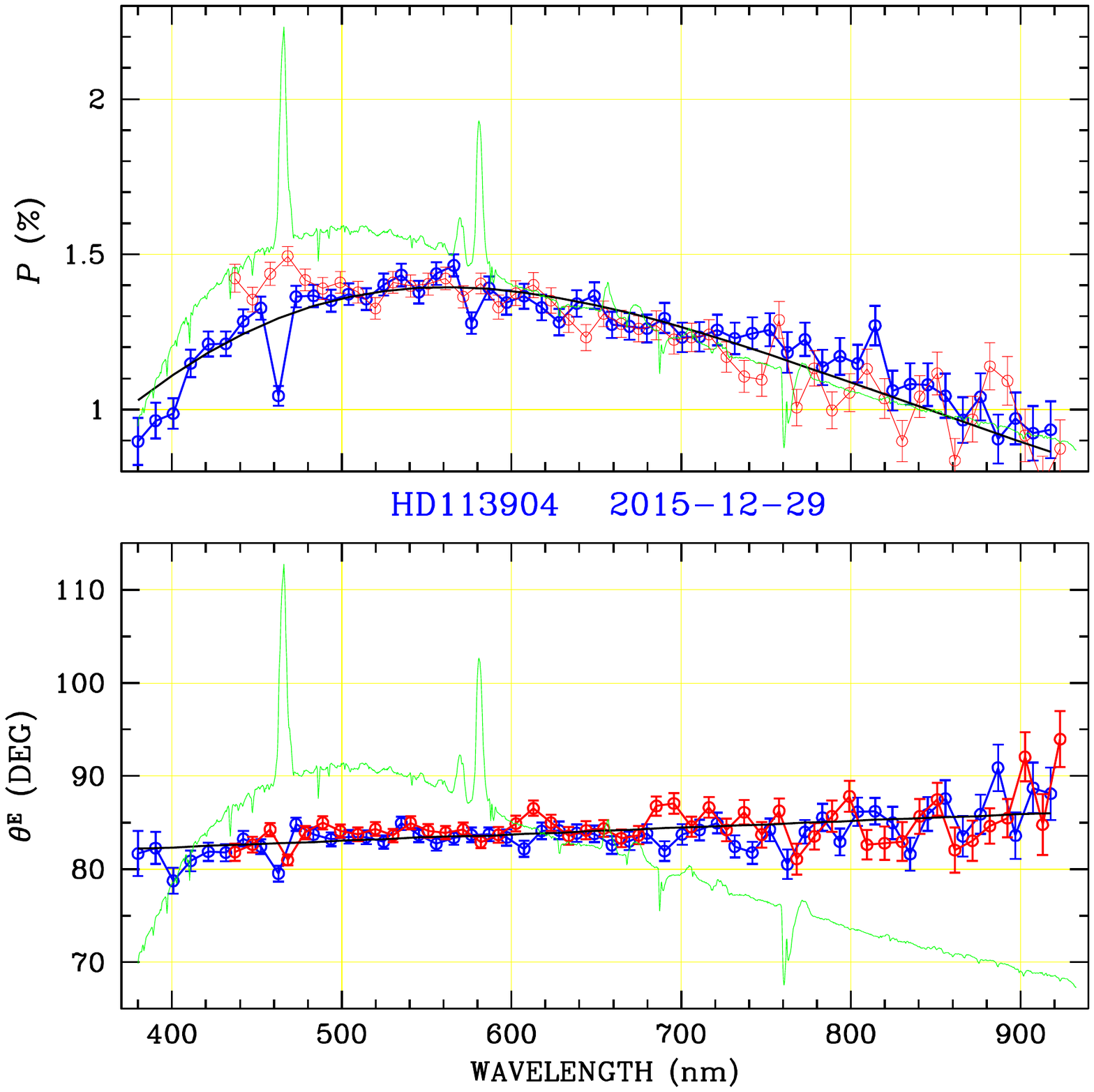}\\
  \includegraphics*[scale=0.42,trim={1.1cm 6.0cm 0.1cm 2.8cm},clip]{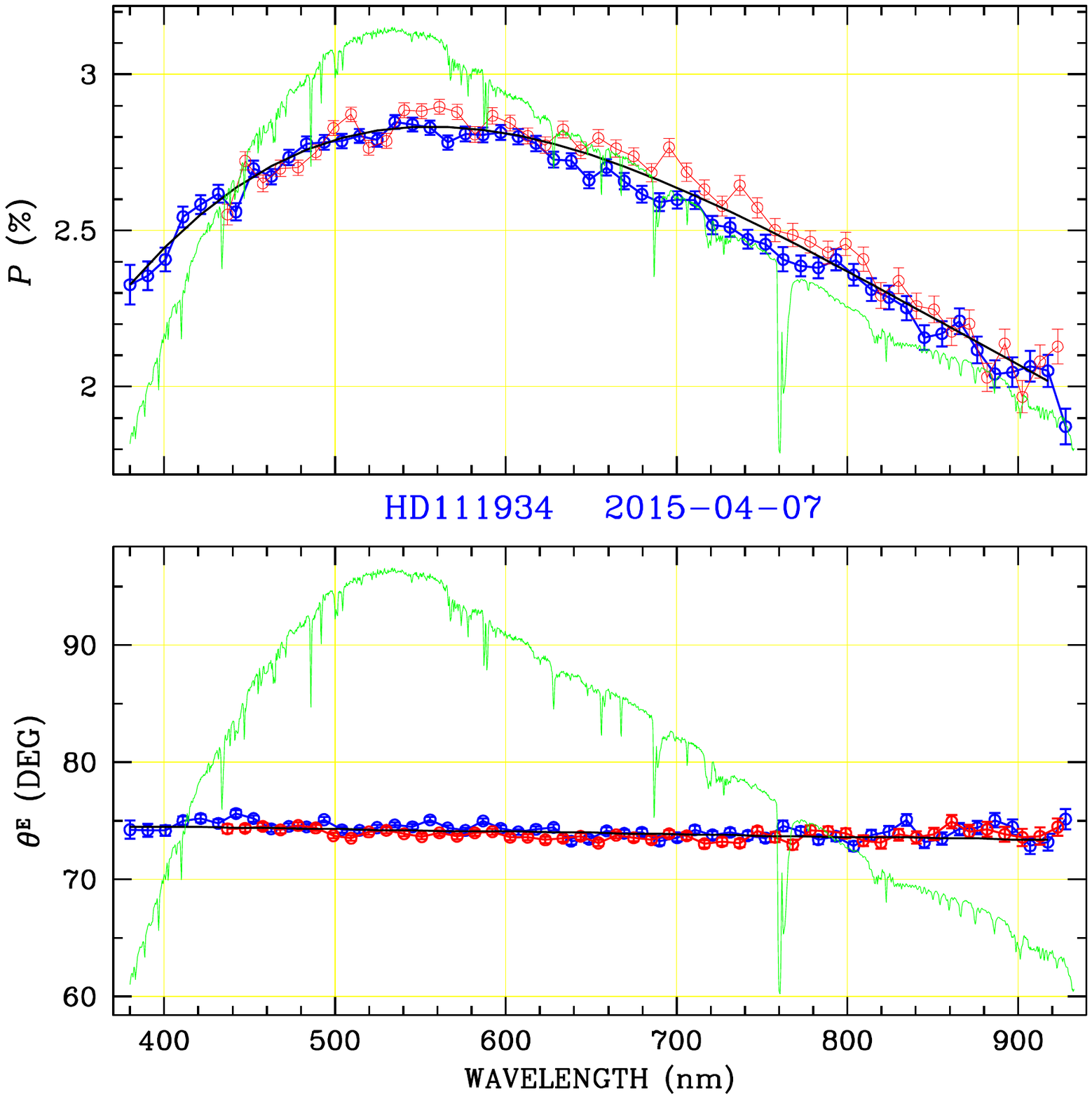}
  \includegraphics*[scale=0.42,trim={1.1cm 6.0cm 0.1cm 2.8cm},clip]{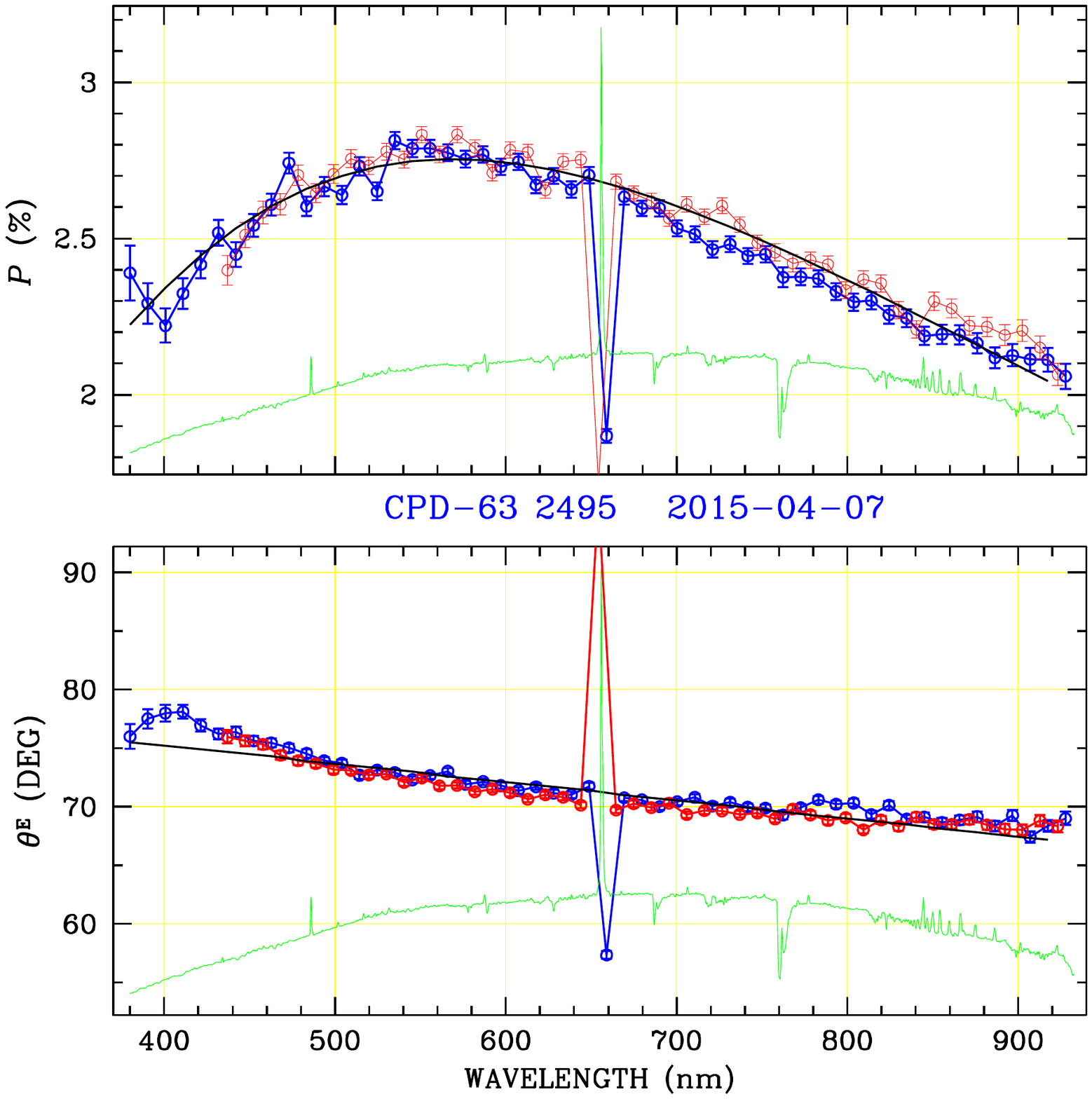}

\newpage
\noindent
  \includegraphics*[scale=0.42,trim={1.1cm 6.0cm 0.1cm 2.8cm},clip]{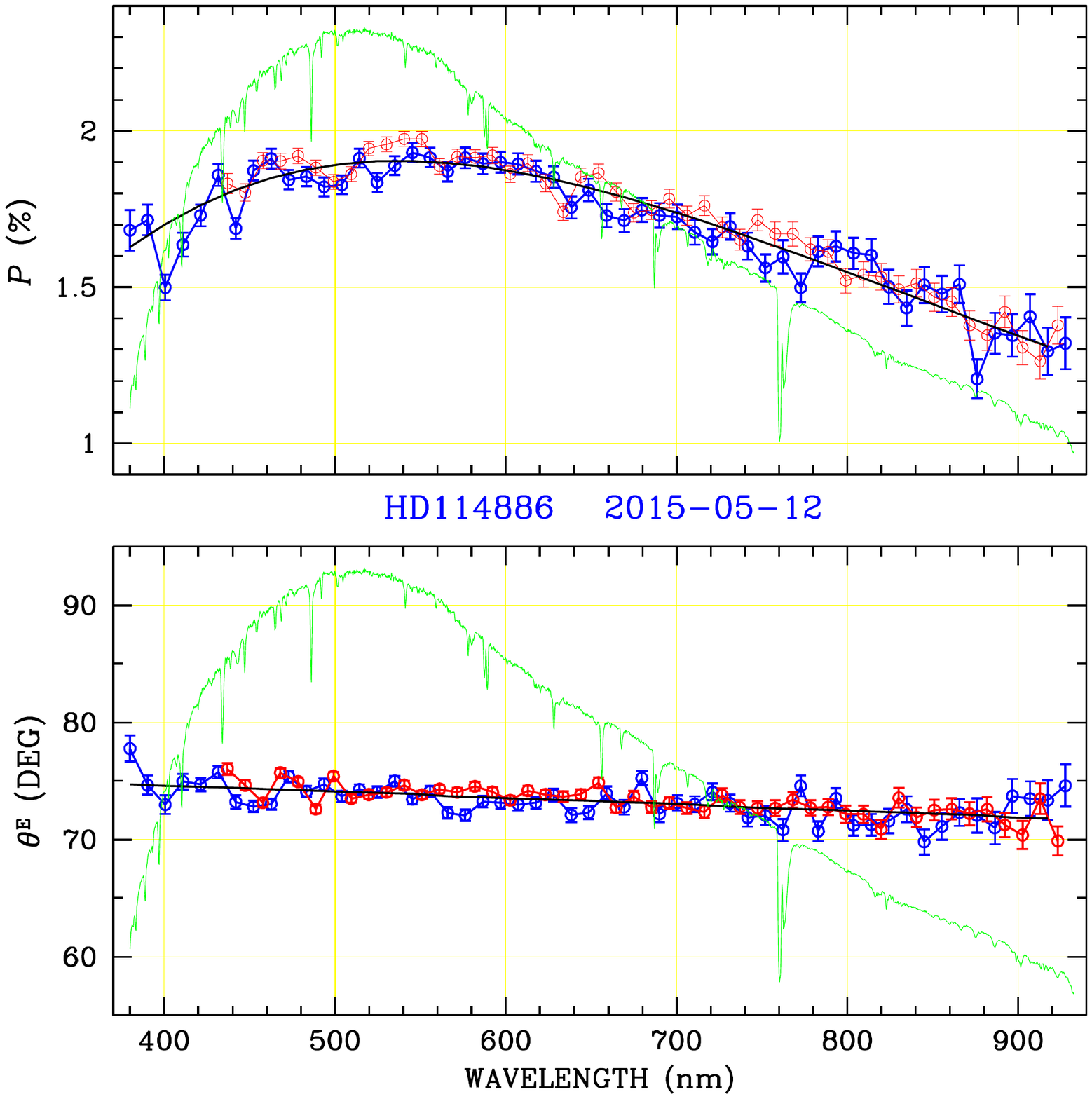}
  \includegraphics*[scale=0.42,trim={1.1cm 6.0cm 0.1cm 2.8cm},clip]{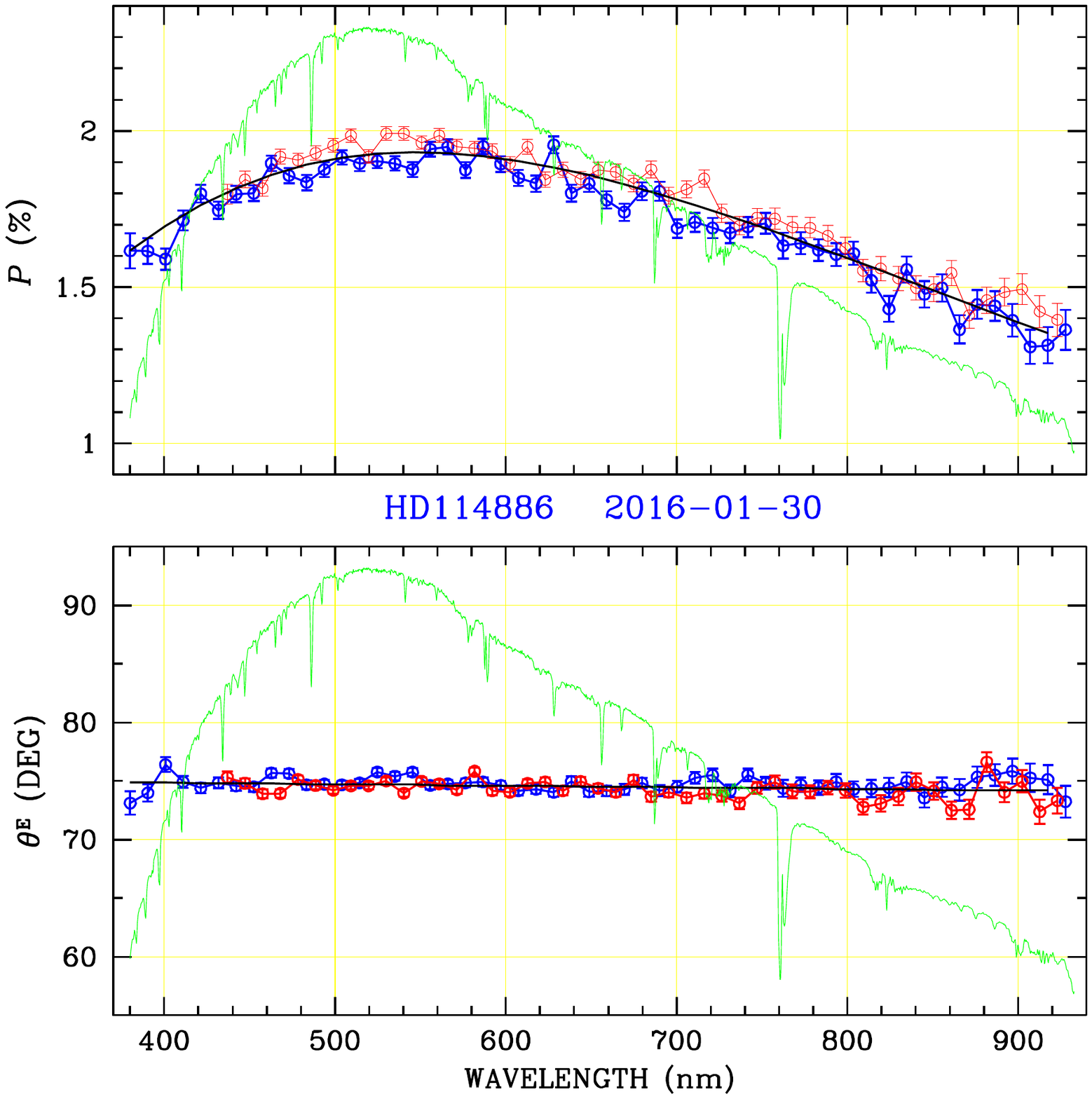}\\
  \includegraphics*[scale=0.42,trim={1.1cm 6.0cm 0.1cm 2.8cm},clip]{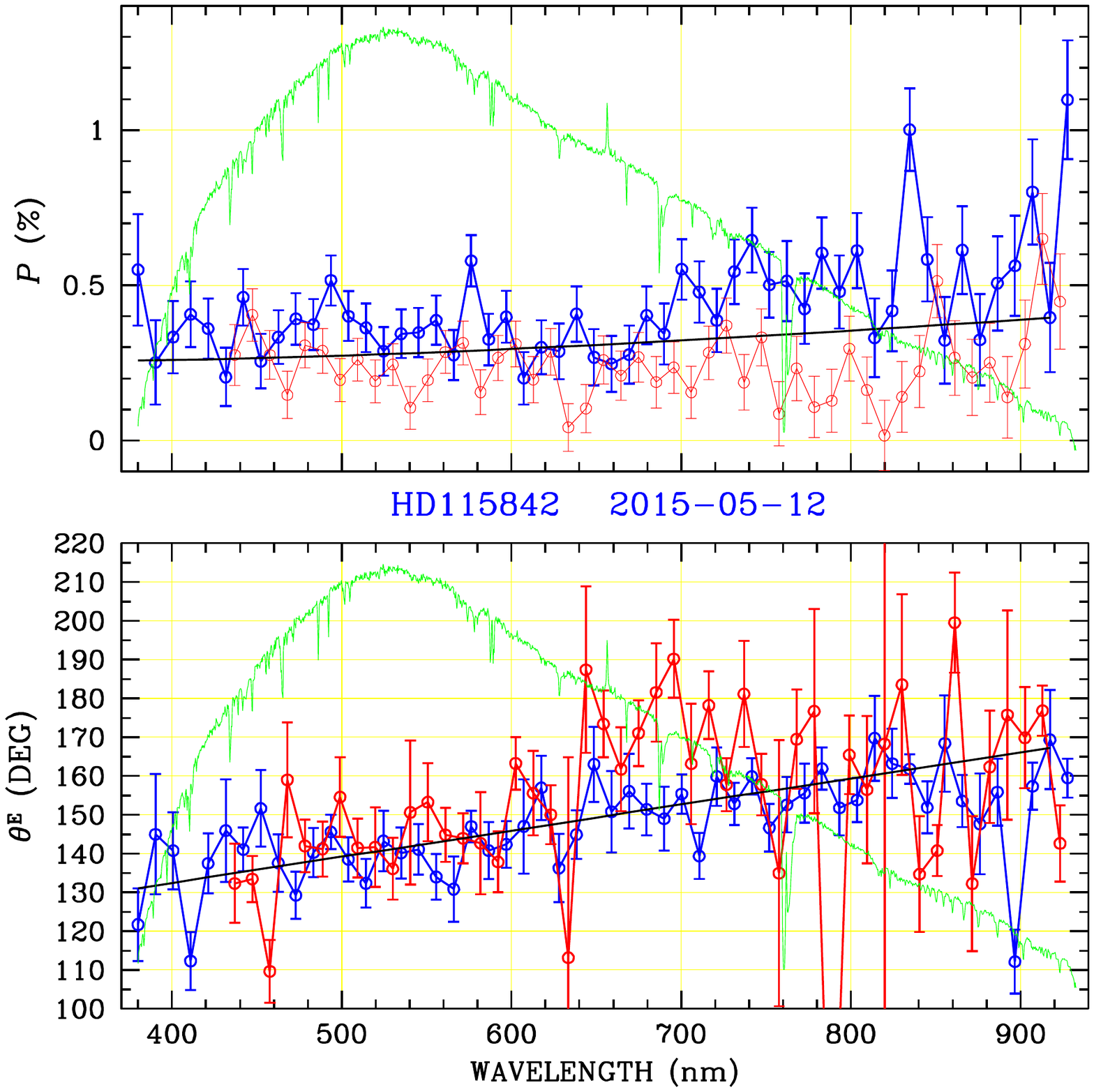}
  \includegraphics*[scale=0.42,trim={1.1cm 6.0cm 0.1cm 2.8cm},clip]{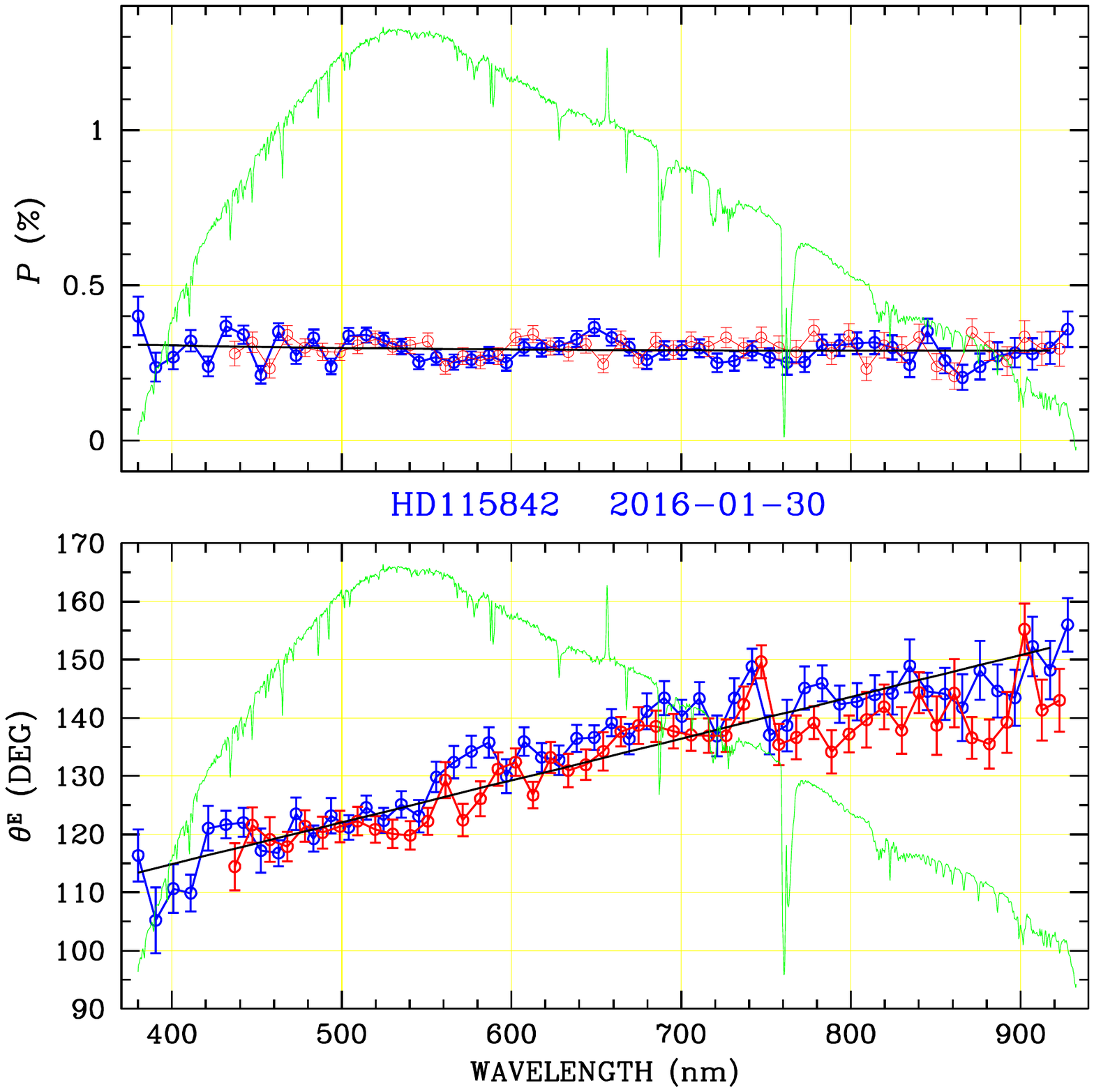}\\
  \includegraphics*[scale=0.42,trim={1.1cm 6.0cm 0.1cm 2.8cm},clip]{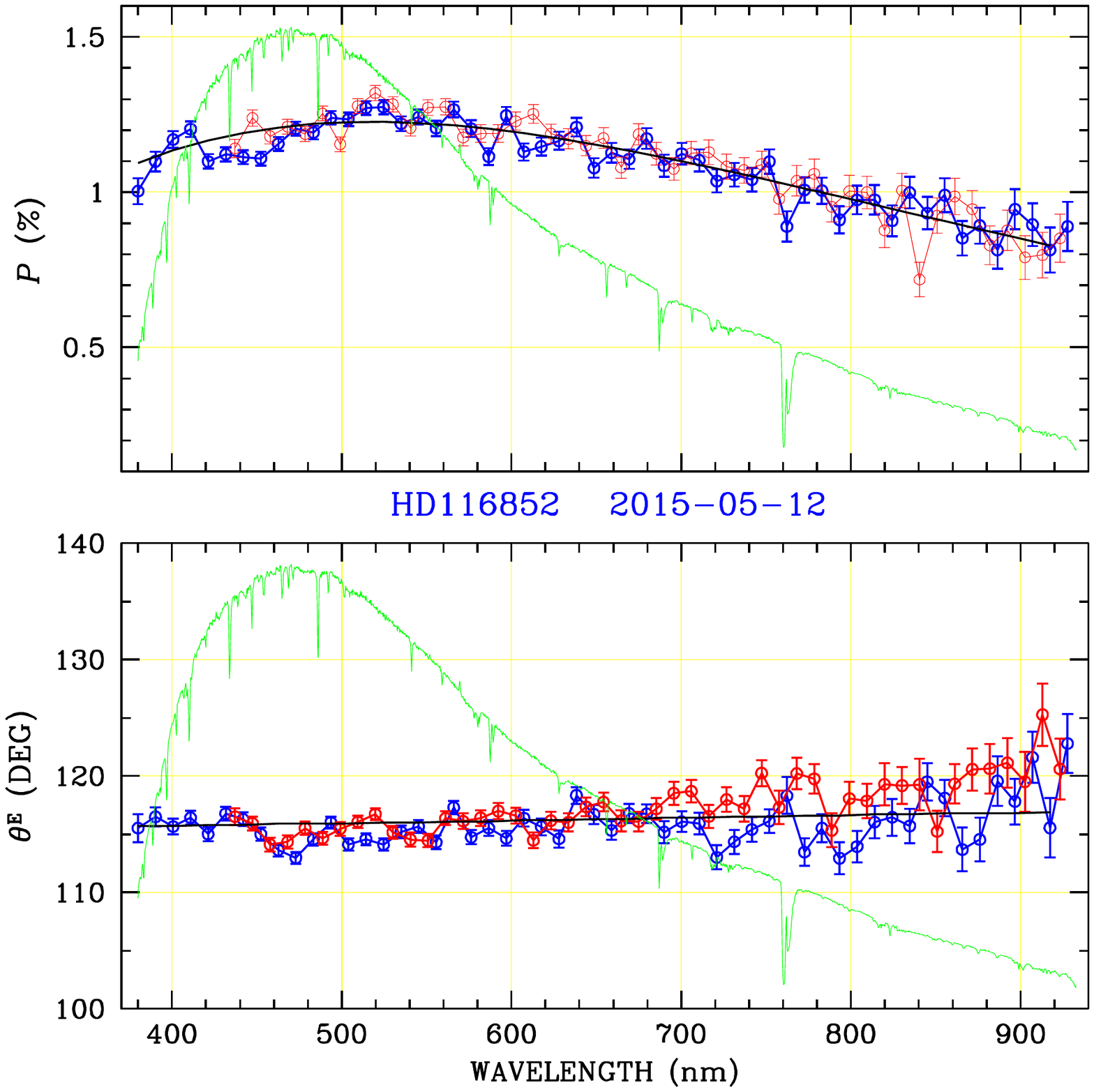}
  \includegraphics*[scale=0.42,trim={1.1cm 6.0cm 0.1cm 2.8cm},clip]{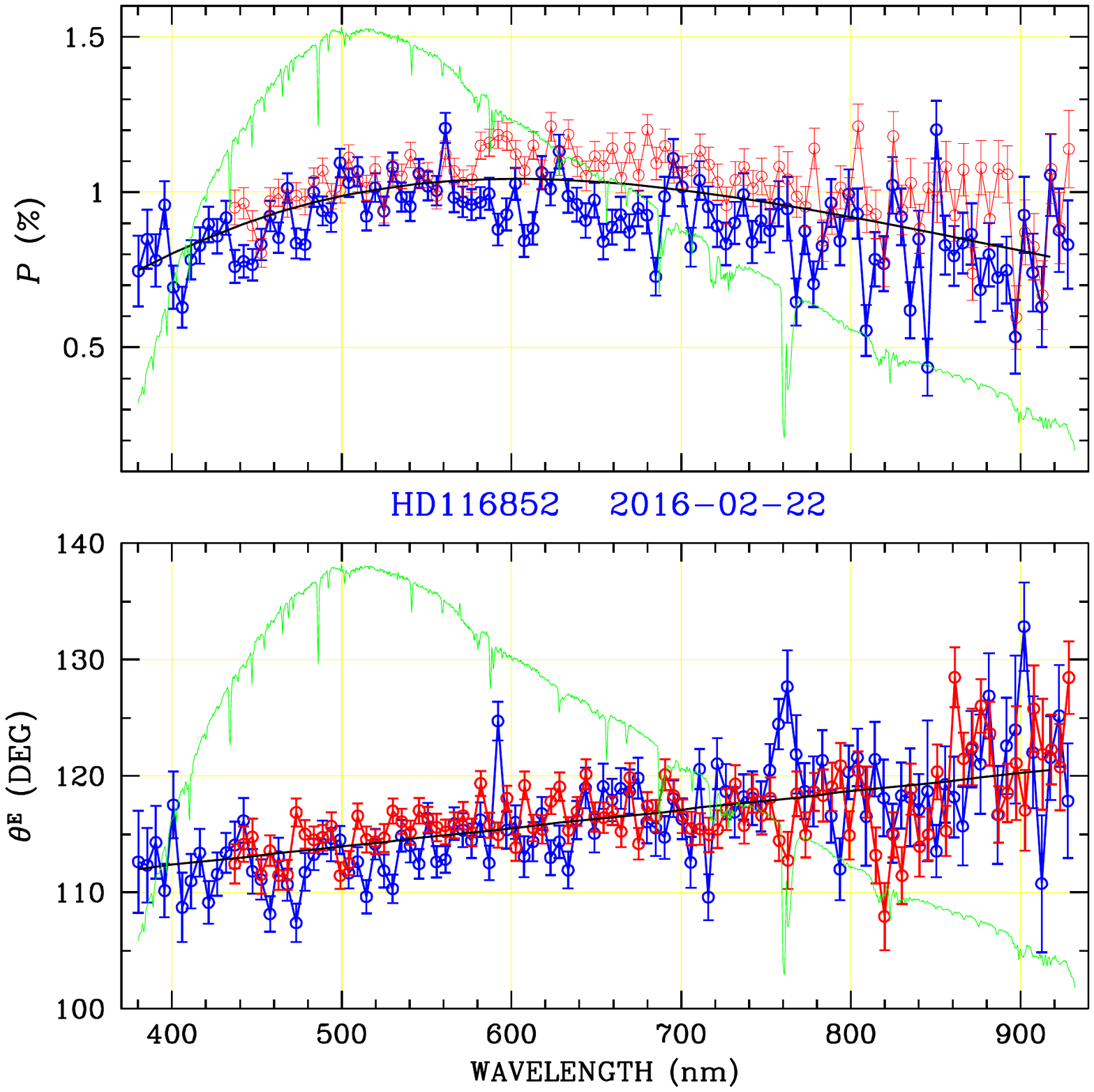}
\newpage

\noindent
  \includegraphics*[scale=0.42,trim={1.1cm 6.0cm 0.1cm 2.8cm},clip]{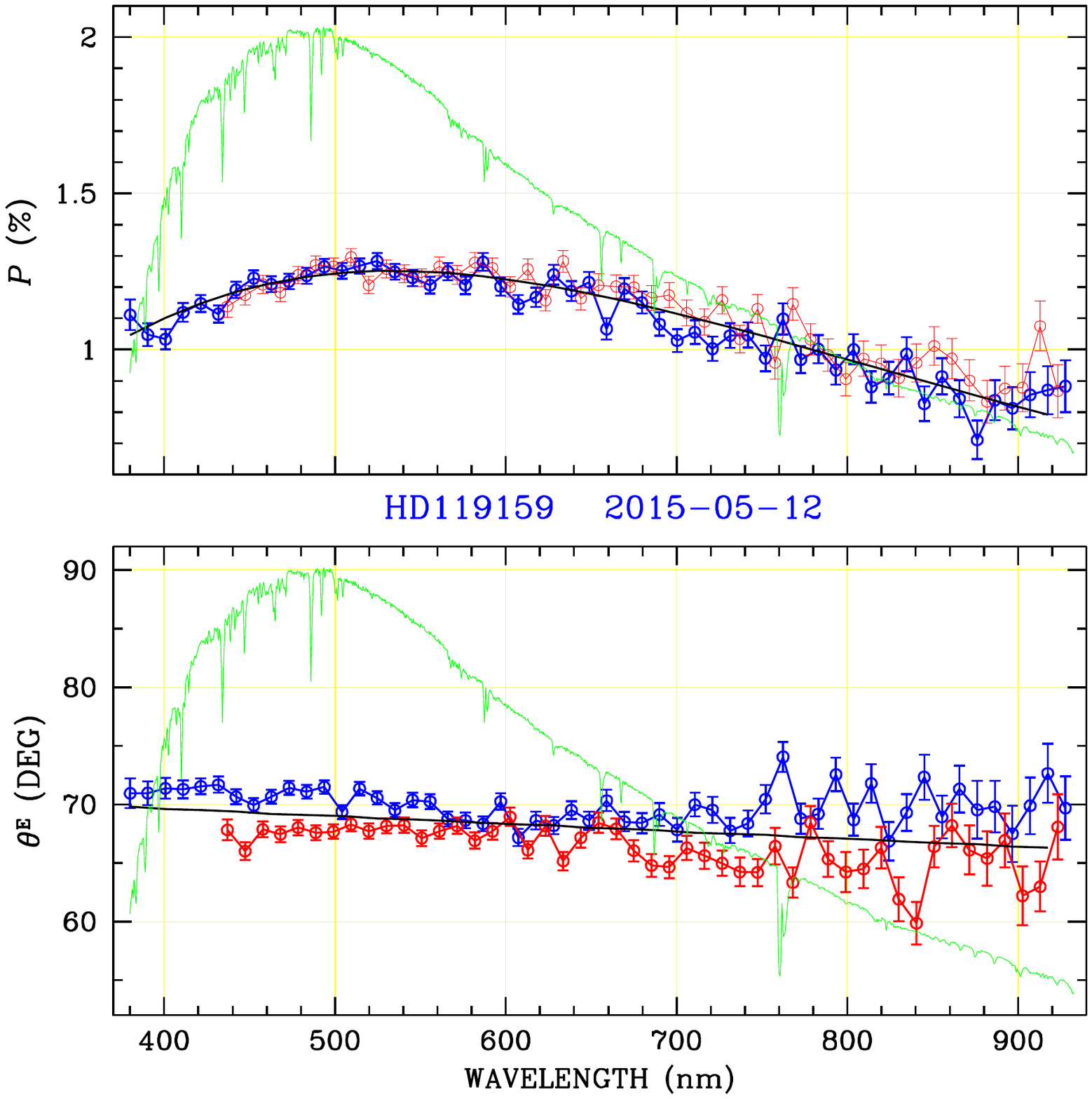}
  \includegraphics*[scale=0.42,trim={1.1cm 6.0cm 0.1cm 2.8cm},clip]{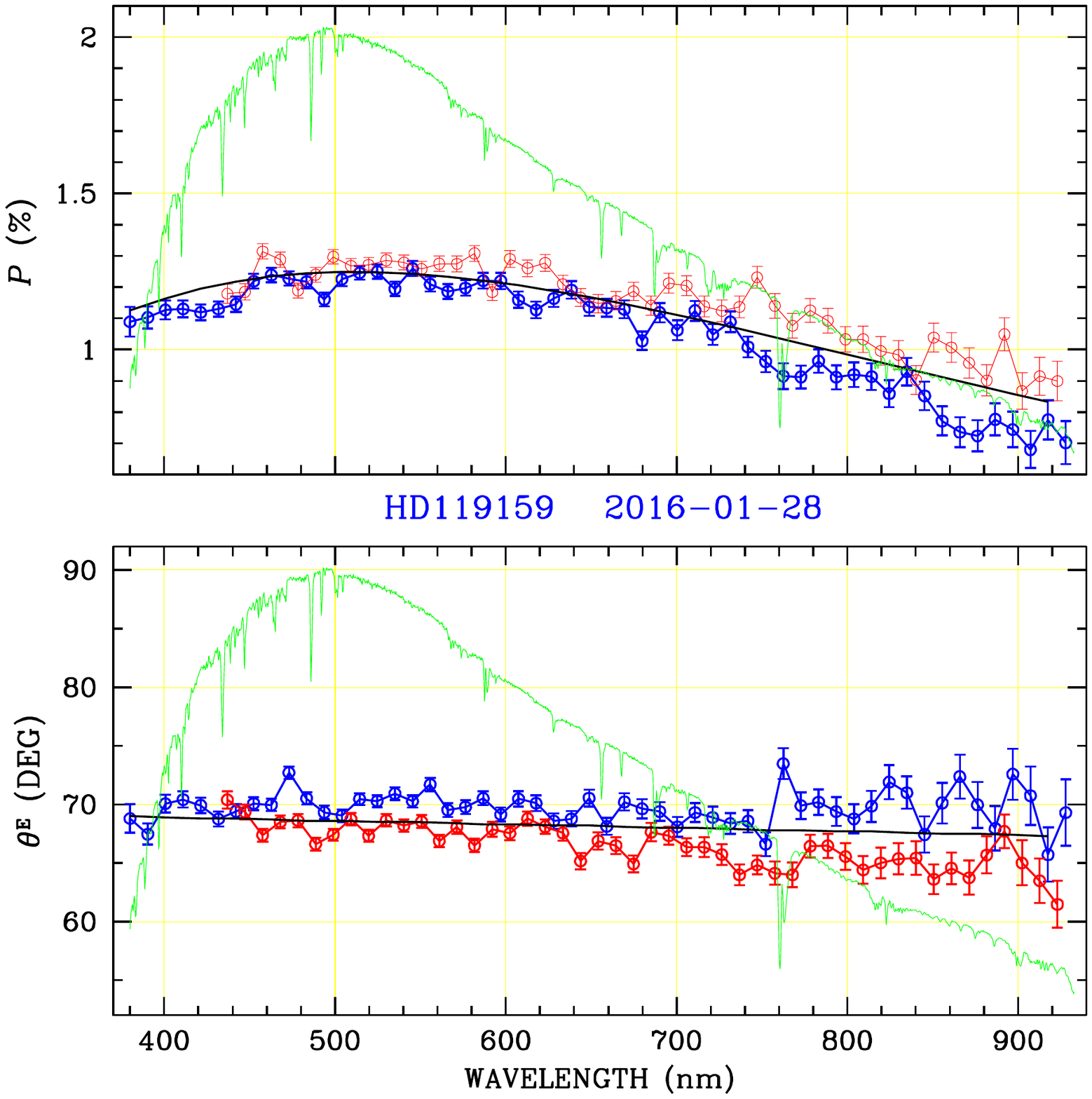}\\
  \includegraphics*[scale=0.42,trim={1.1cm 6.0cm 0.1cm 2.8cm},clip]{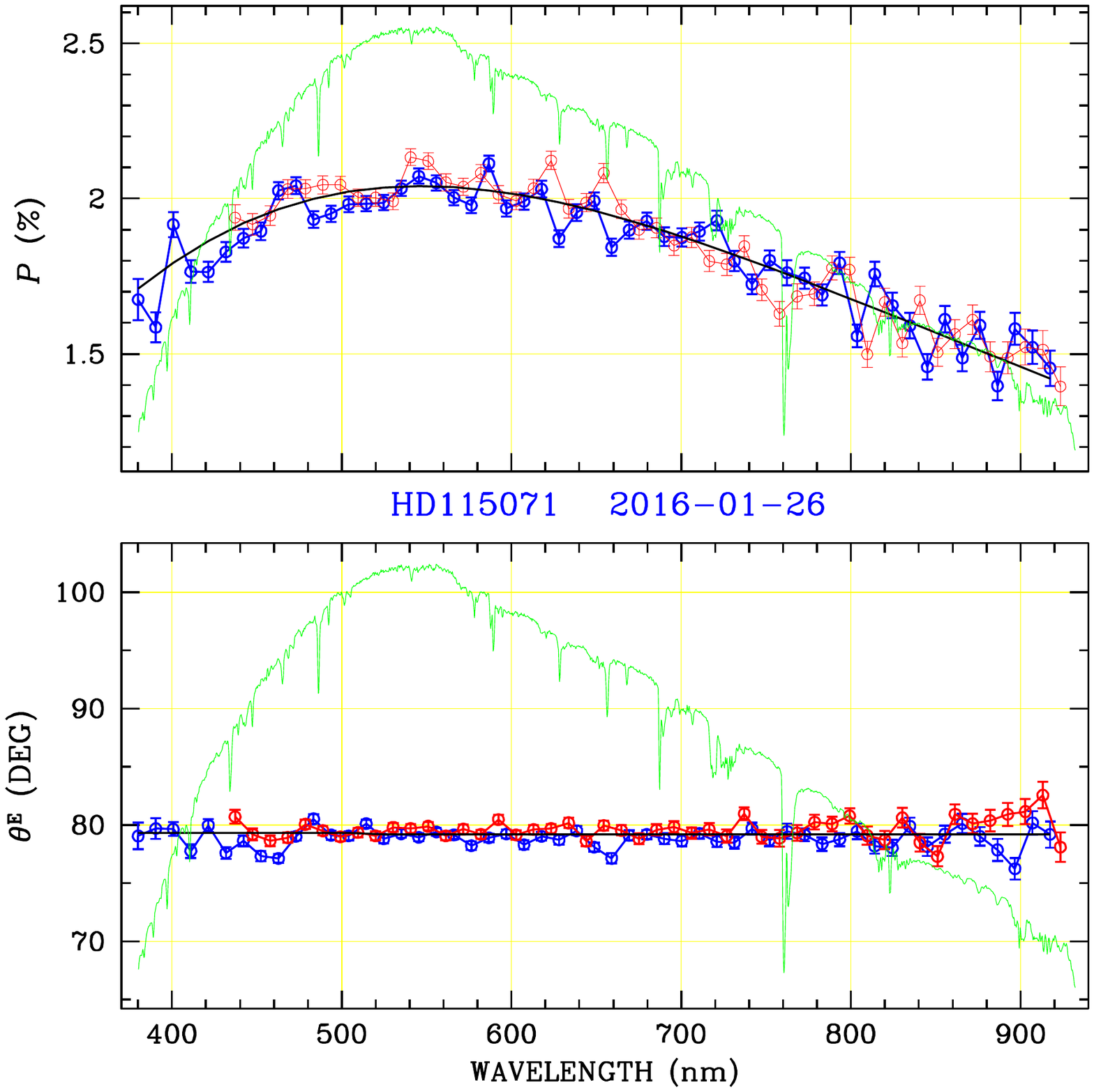}
  \includegraphics*[scale=0.42,trim={1.1cm 6.0cm 0.1cm 2.8cm},clip]{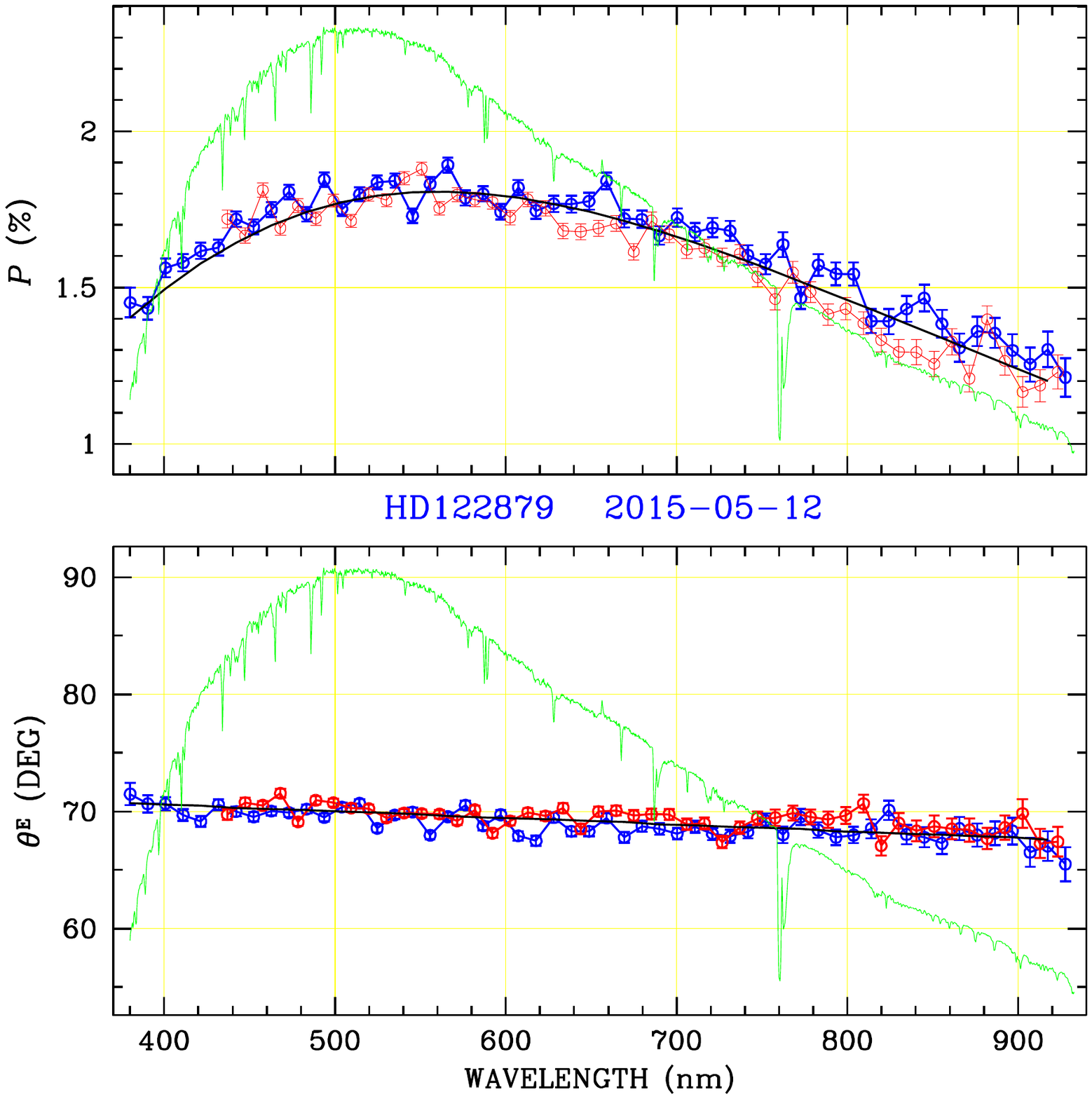}\\
  \includegraphics*[scale=0.42,trim={1.1cm 6.0cm 0.1cm 2.8cm},clip]{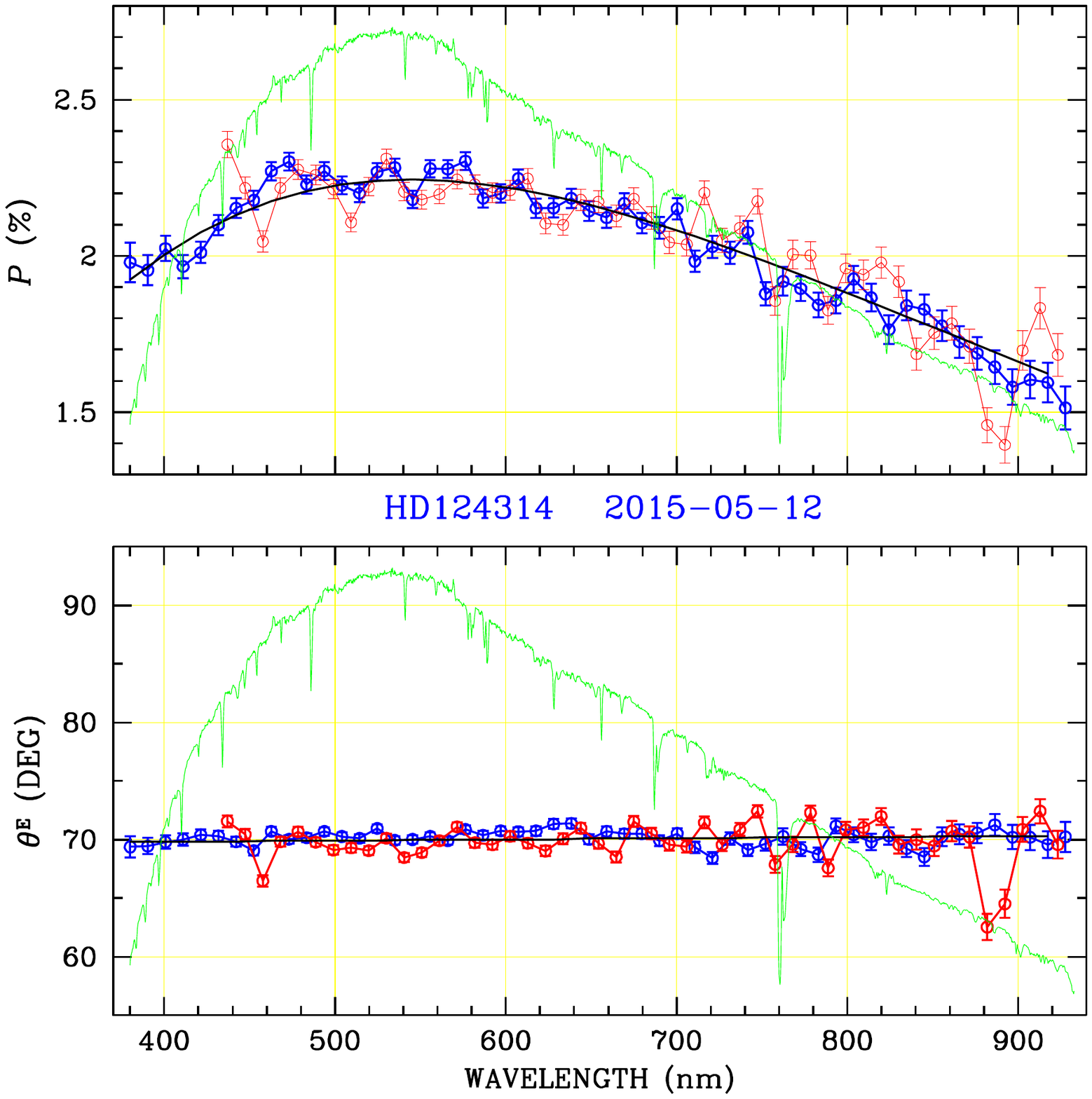}
  \includegraphics*[scale=0.42,trim={1.1cm 6.0cm 0.1cm 2.8cm},clip]{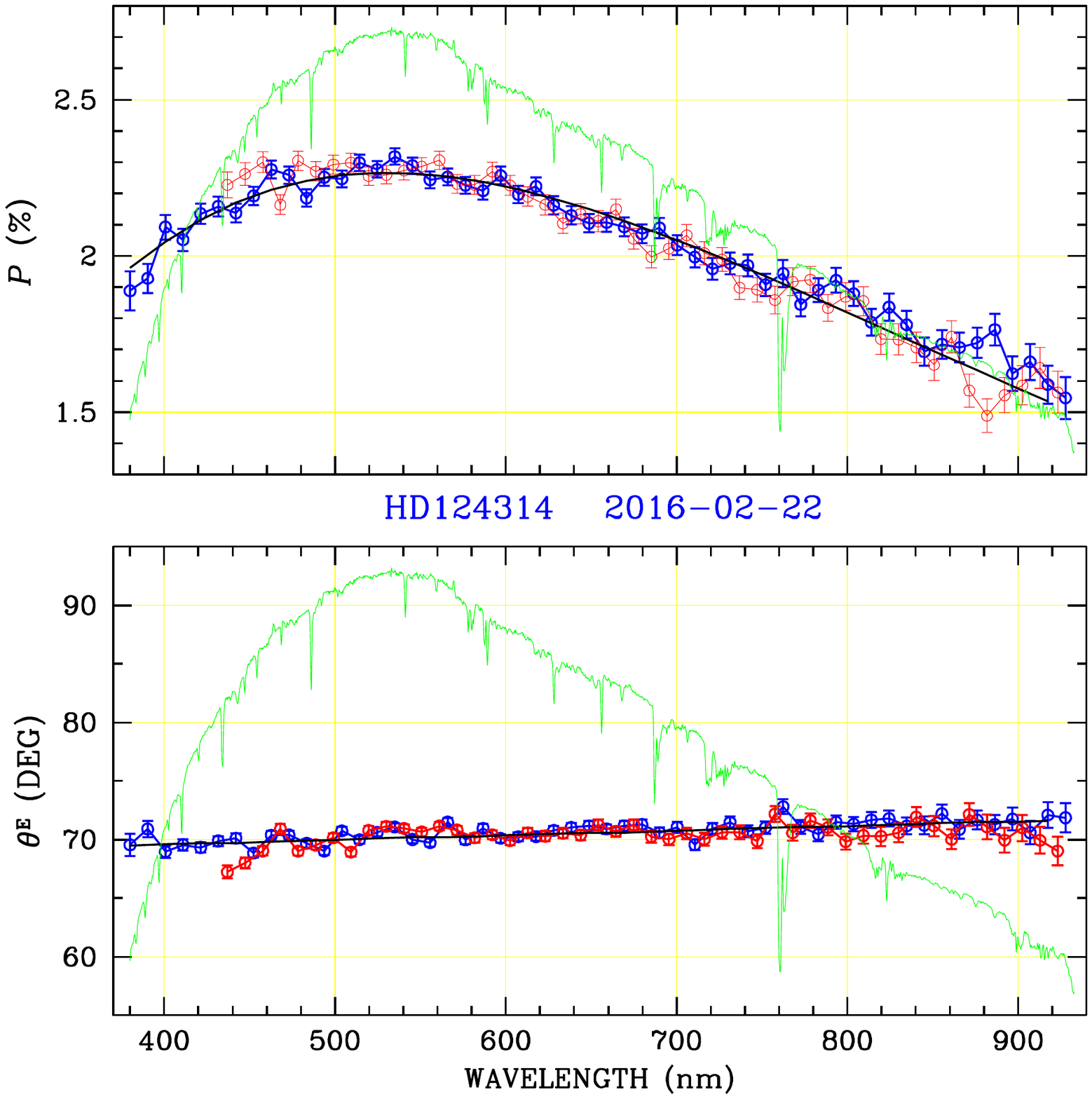}
\newpage

\noindent
  \includegraphics*[scale=0.42,trim={1.1cm 6.0cm 0.1cm 2.8cm},clip]{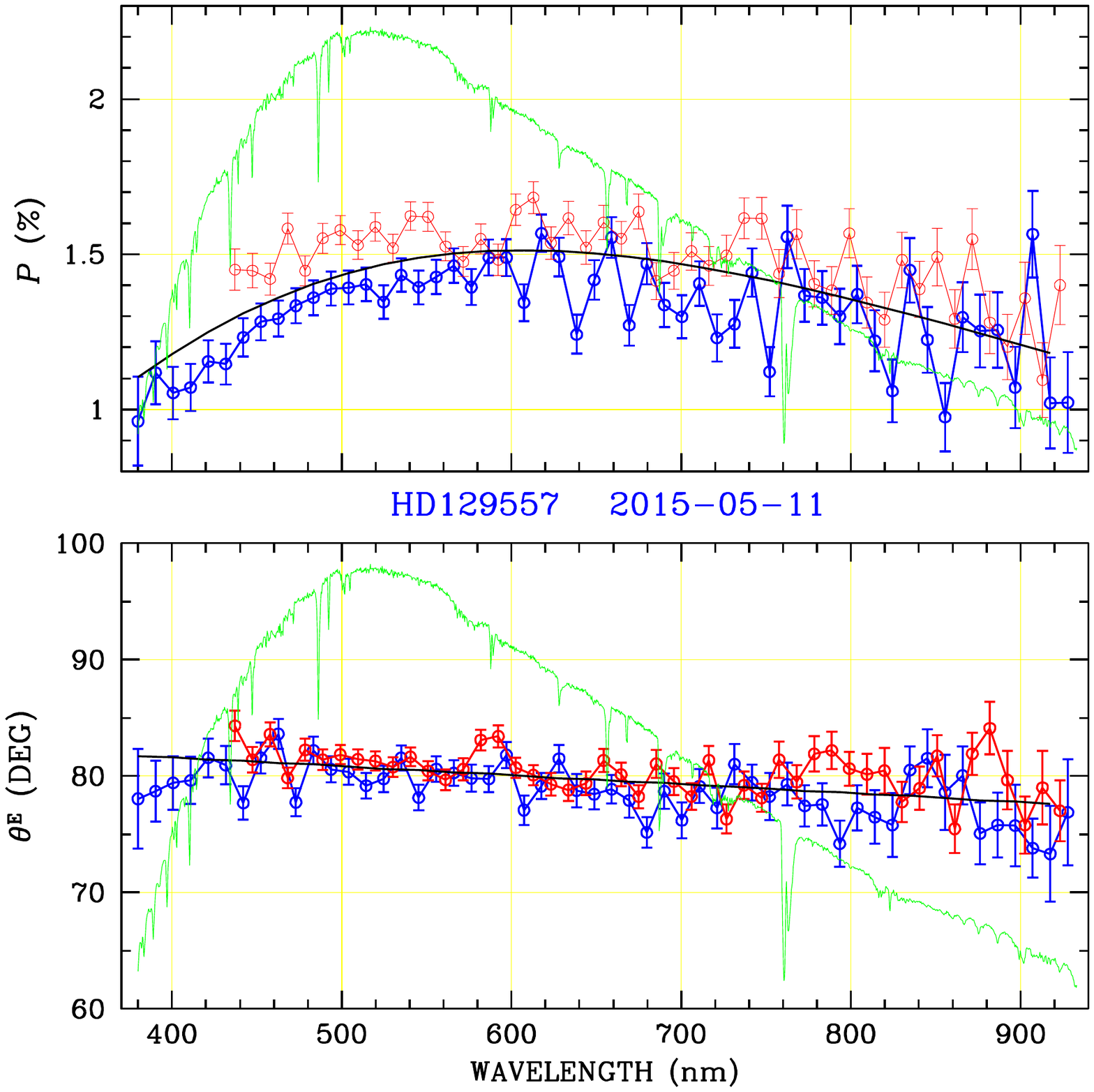}
  \includegraphics*[scale=0.42,trim={1.1cm 6.0cm 0.1cm 2.8cm},clip]{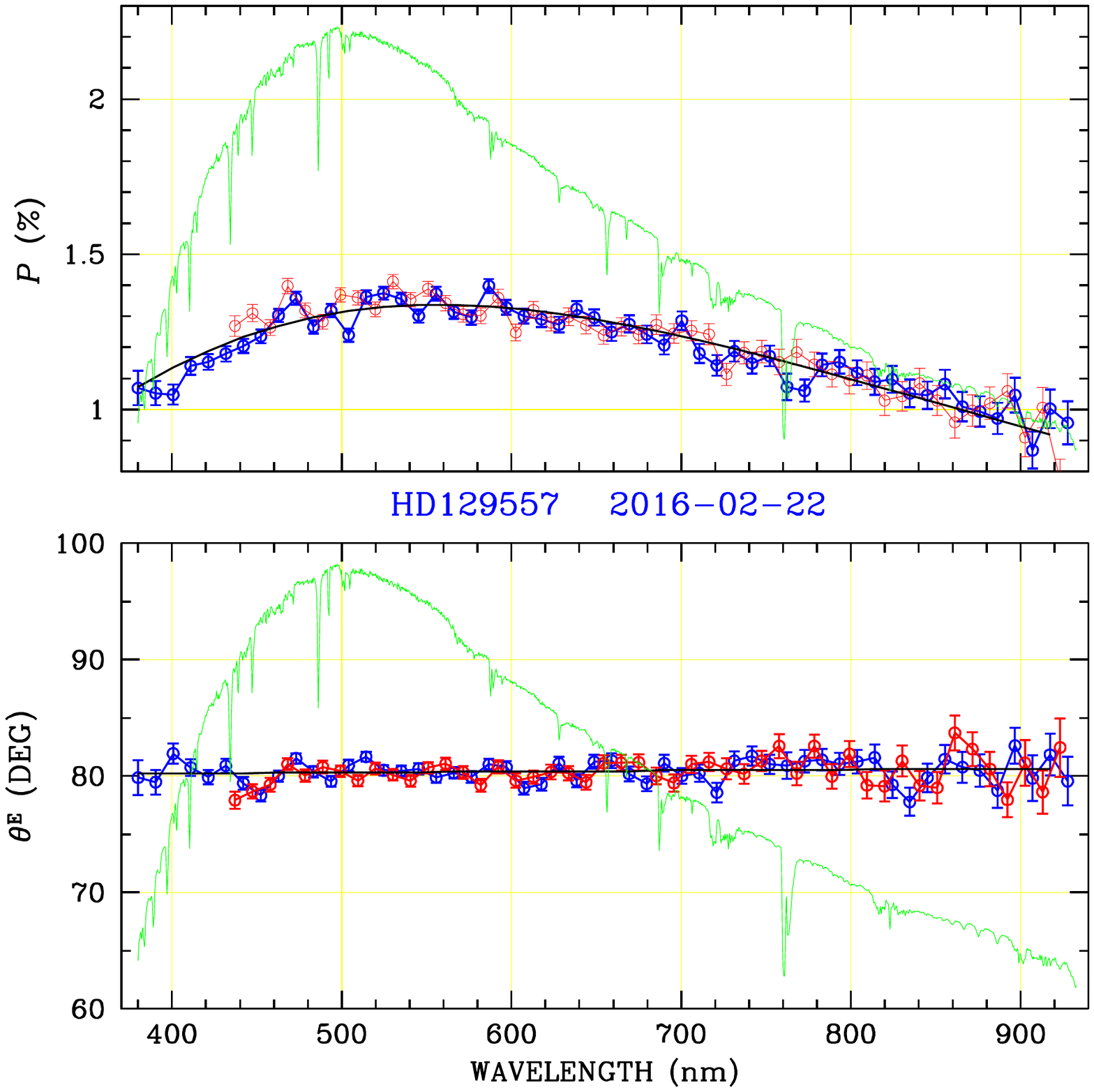}\\
  \includegraphics*[scale=0.42,trim={1.1cm 6.0cm 0.1cm 2.8cm},clip]{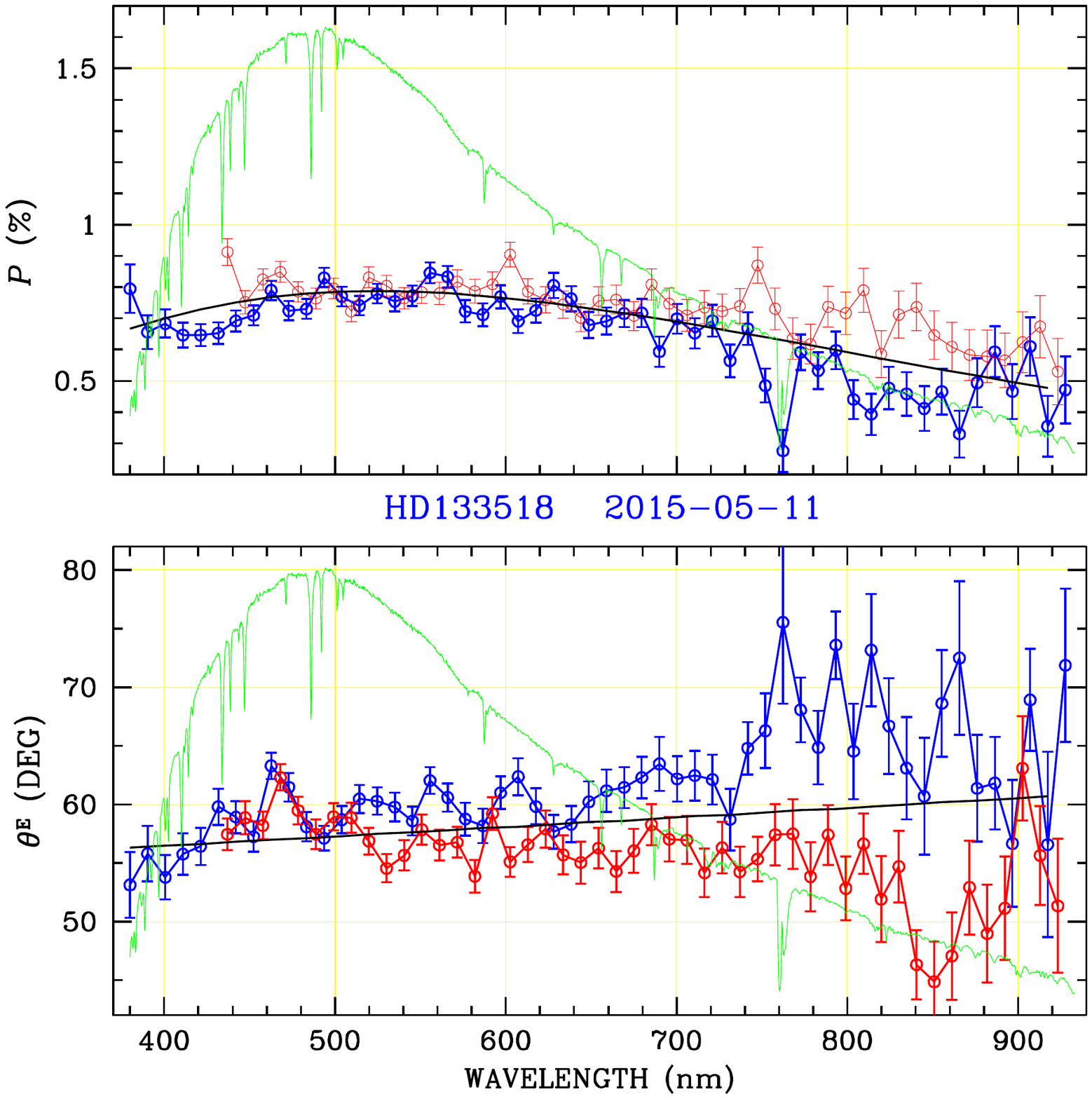}
  \includegraphics*[scale=0.42,trim={1.1cm 6.0cm 0.1cm 2.8cm},clip]{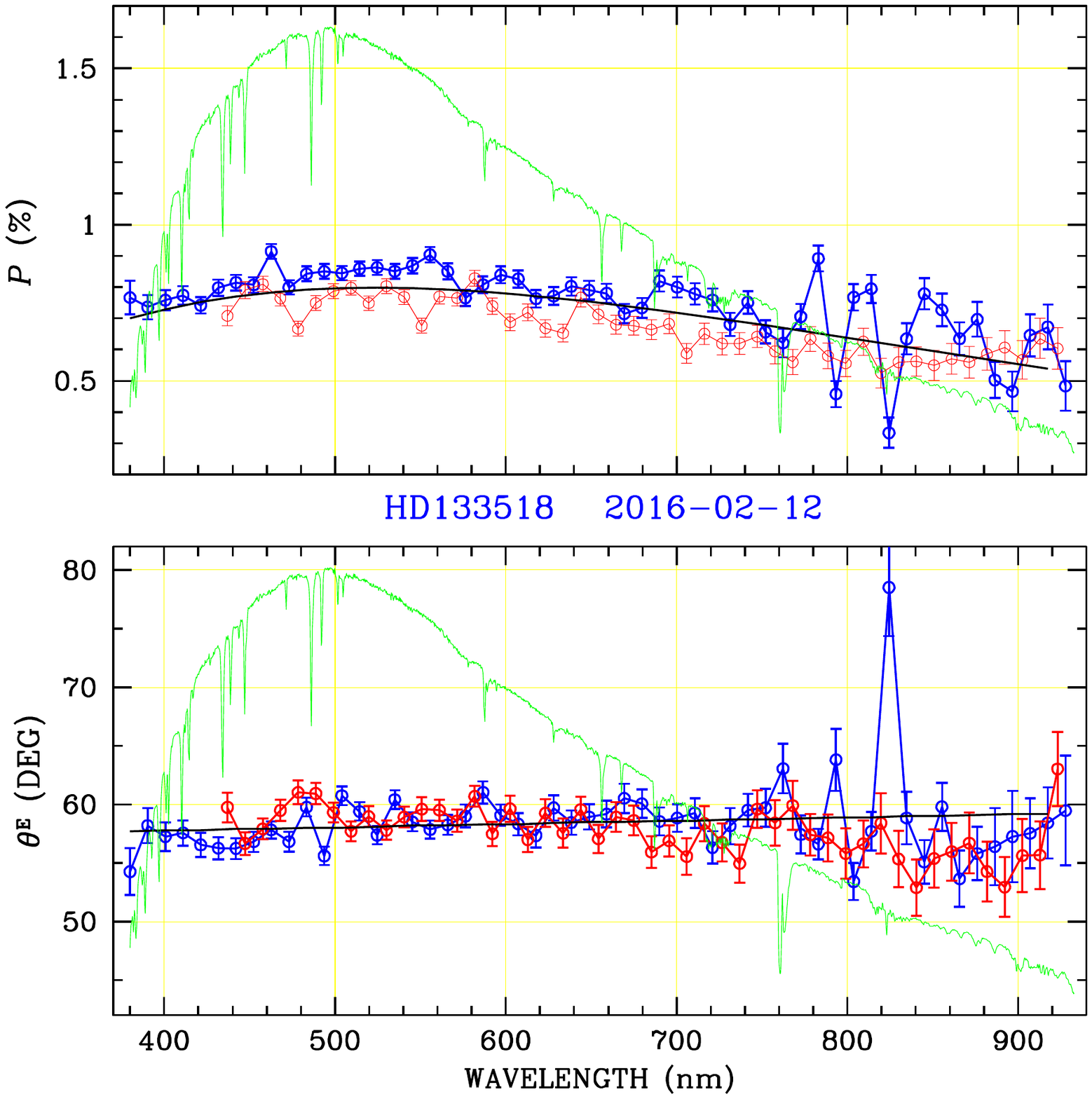}\\
  \includegraphics*[scale=0.42,trim={1.1cm 6.0cm 0.1cm 2.8cm},clip]{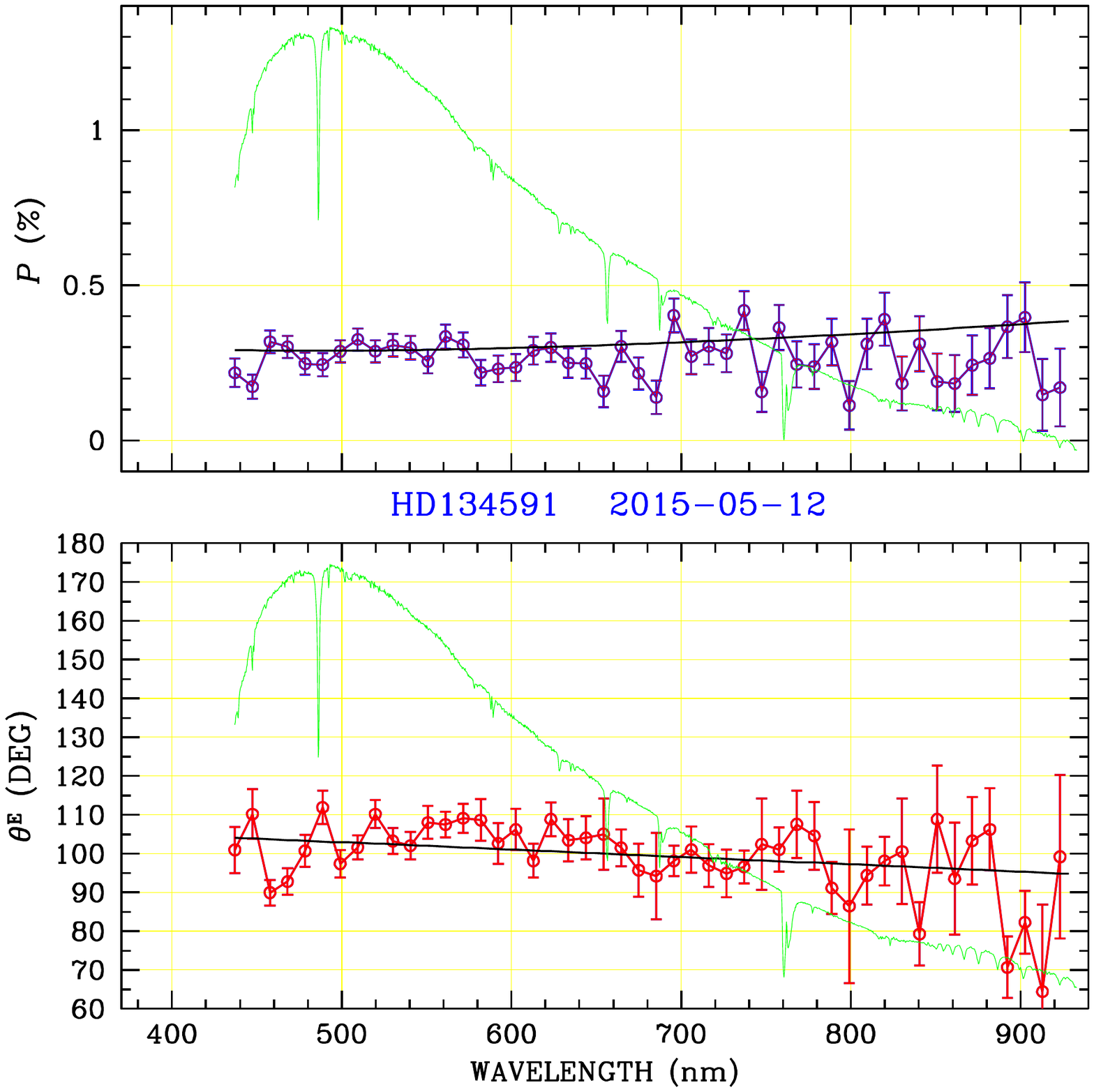}
  \includegraphics*[scale=0.42,trim={1.1cm 6.0cm 0.1cm 2.8cm},clip]{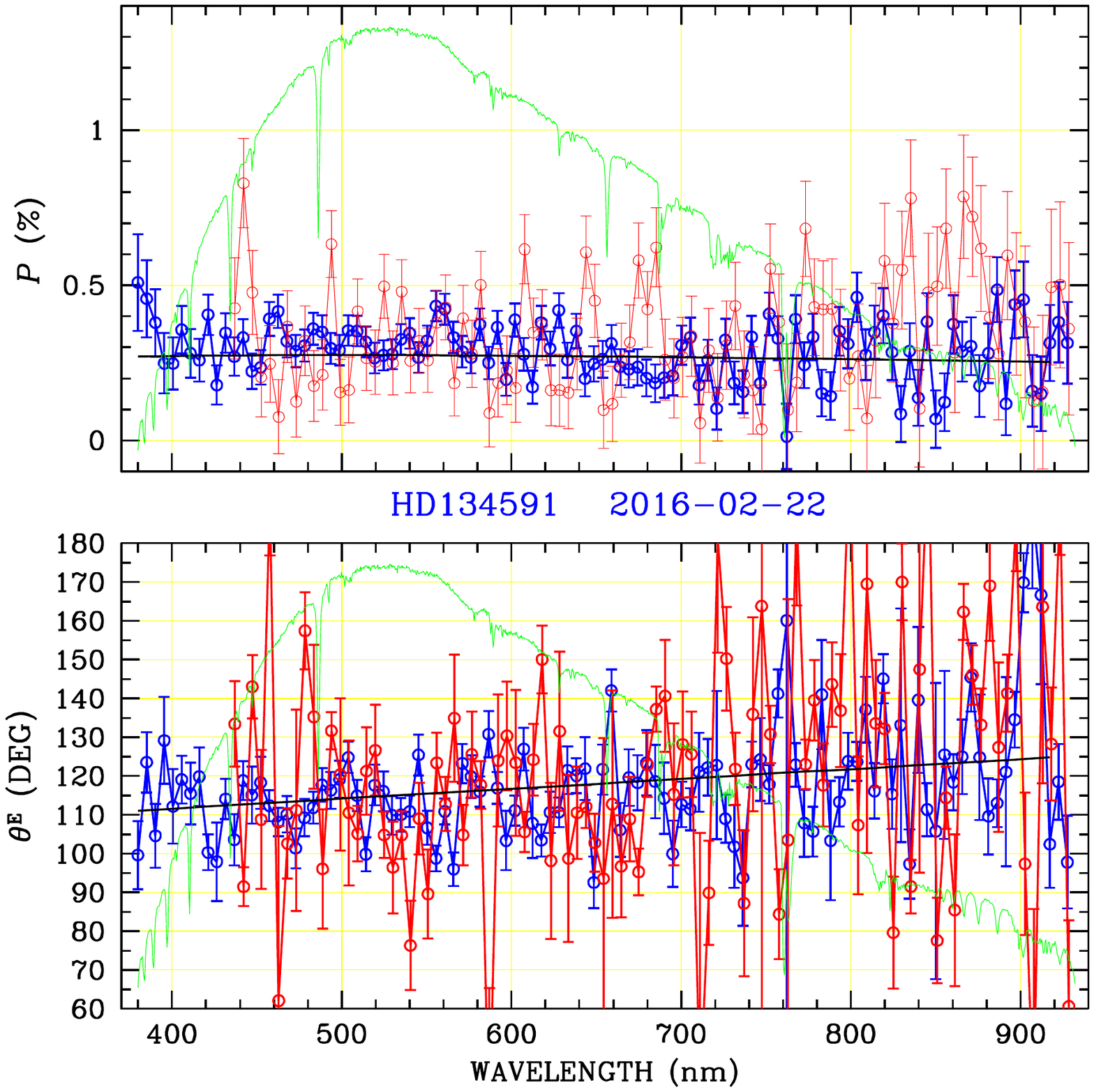}
\newpage

\noindent
  \includegraphics*[scale=0.42,trim={1.1cm 6.0cm 0.1cm 2.8cm},clip]{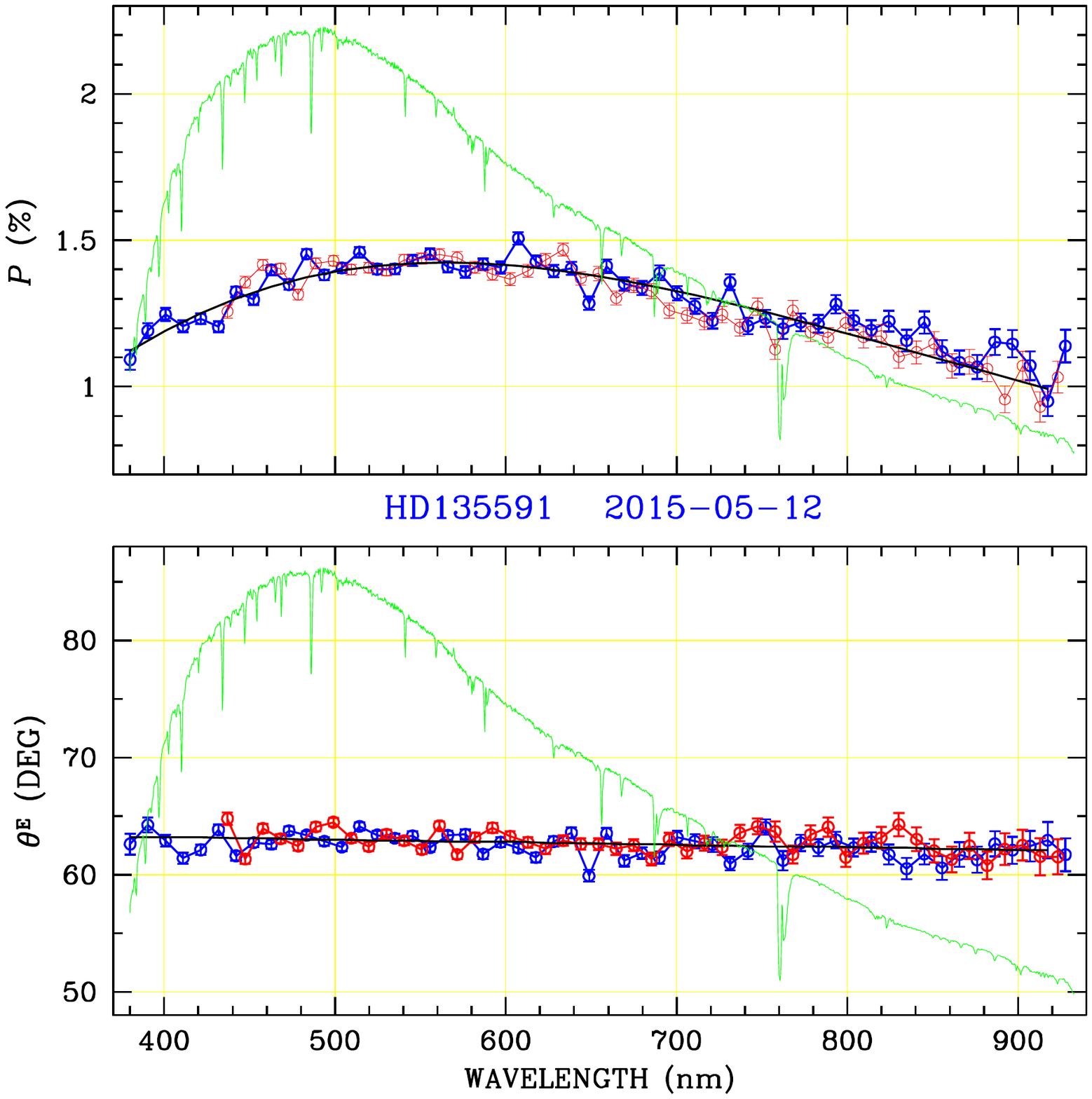}
  \includegraphics*[scale=0.42,trim={1.1cm 6.0cm 0.1cm 2.8cm},clip]{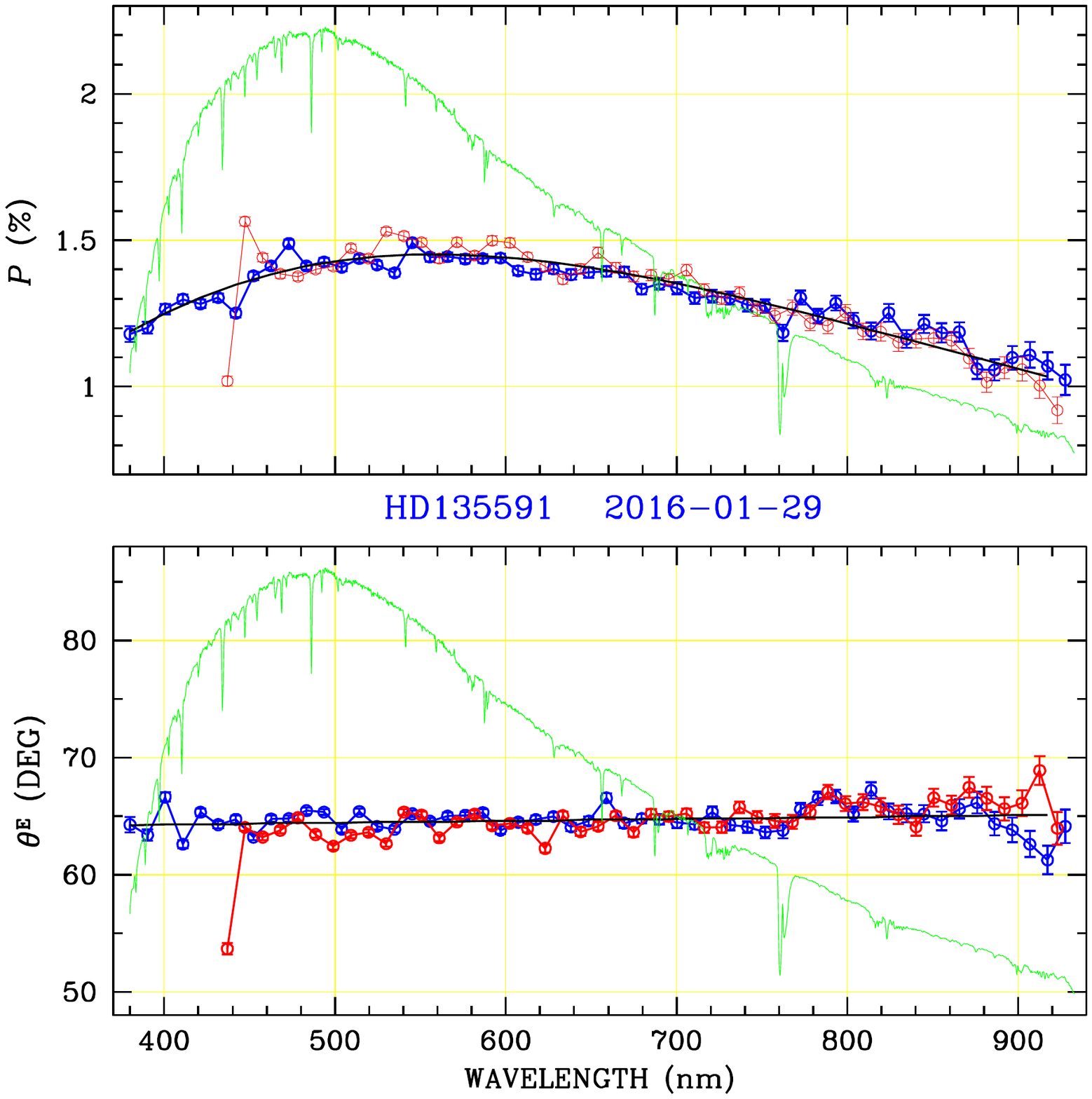}\\
  \includegraphics*[scale=0.42,trim={1.1cm 6.0cm 0.1cm 2.8cm},clip]{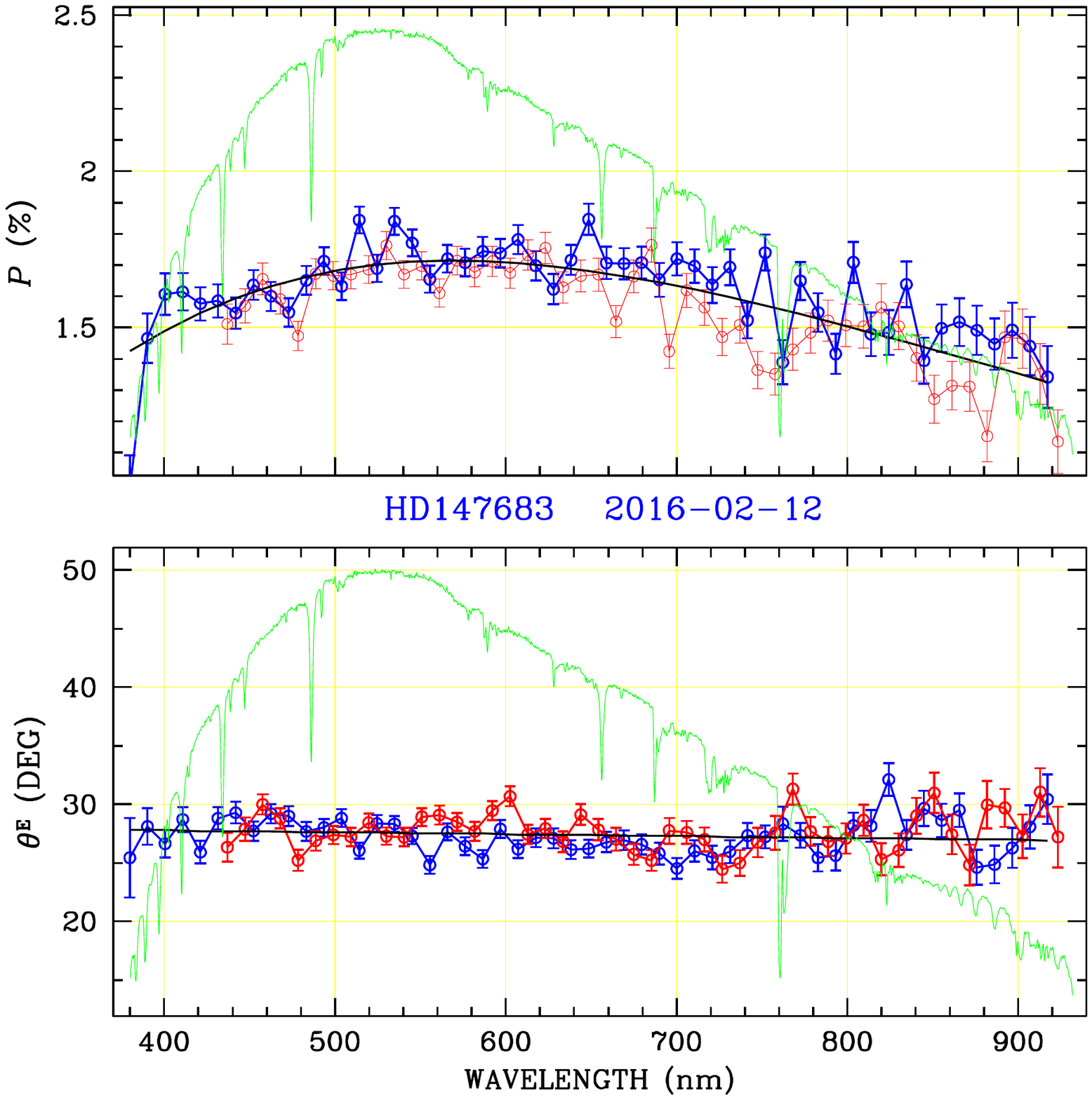}
  \includegraphics*[scale=0.42,trim={1.1cm 6.0cm 0.1cm 2.8cm},clip]{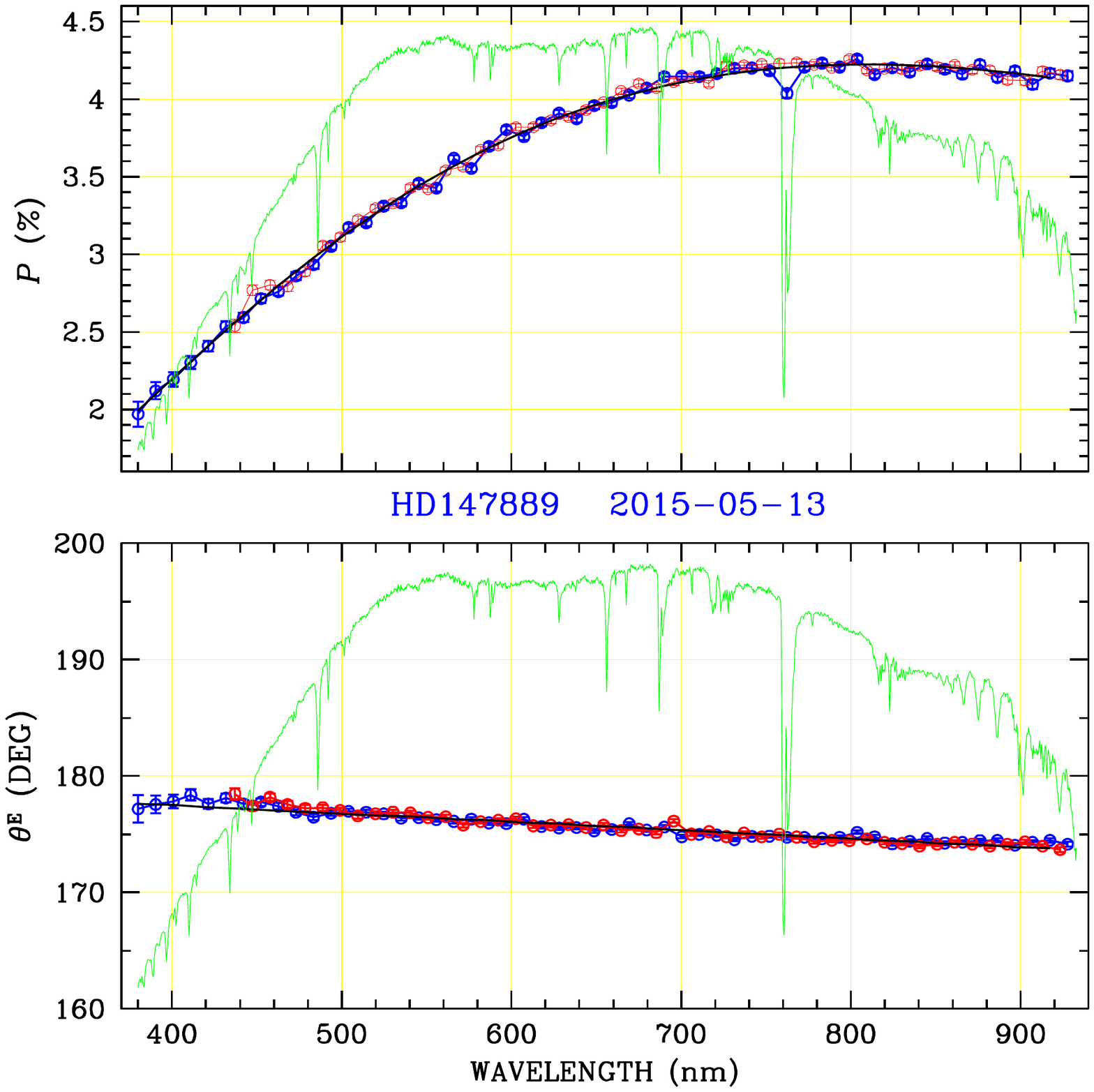}\\
  \includegraphics*[scale=0.42,trim={1.1cm 6.0cm 0.1cm 2.8cm},clip]{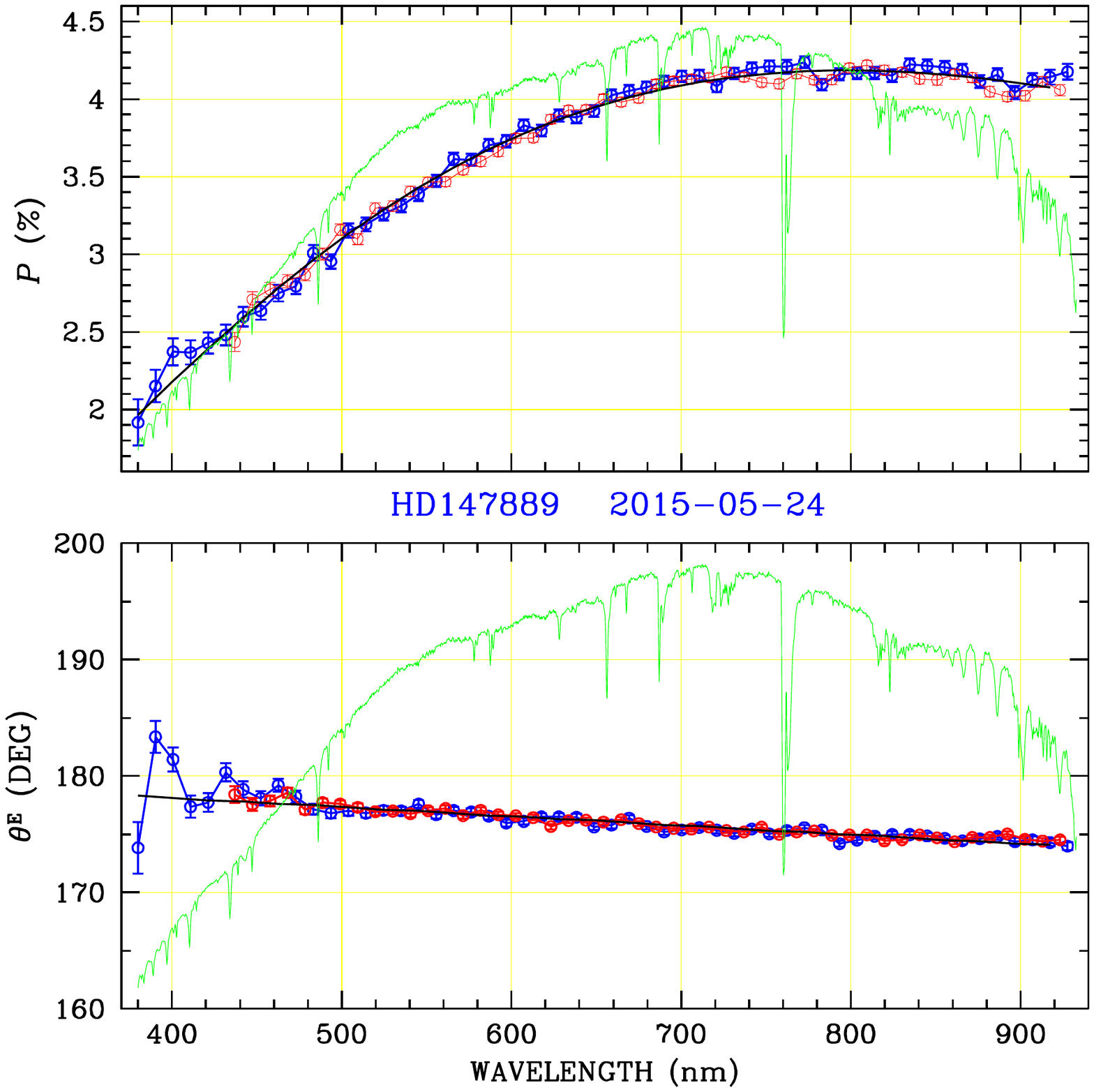}
  \includegraphics*[scale=0.42,trim={1.1cm 6.0cm 0.1cm 2.8cm},clip]{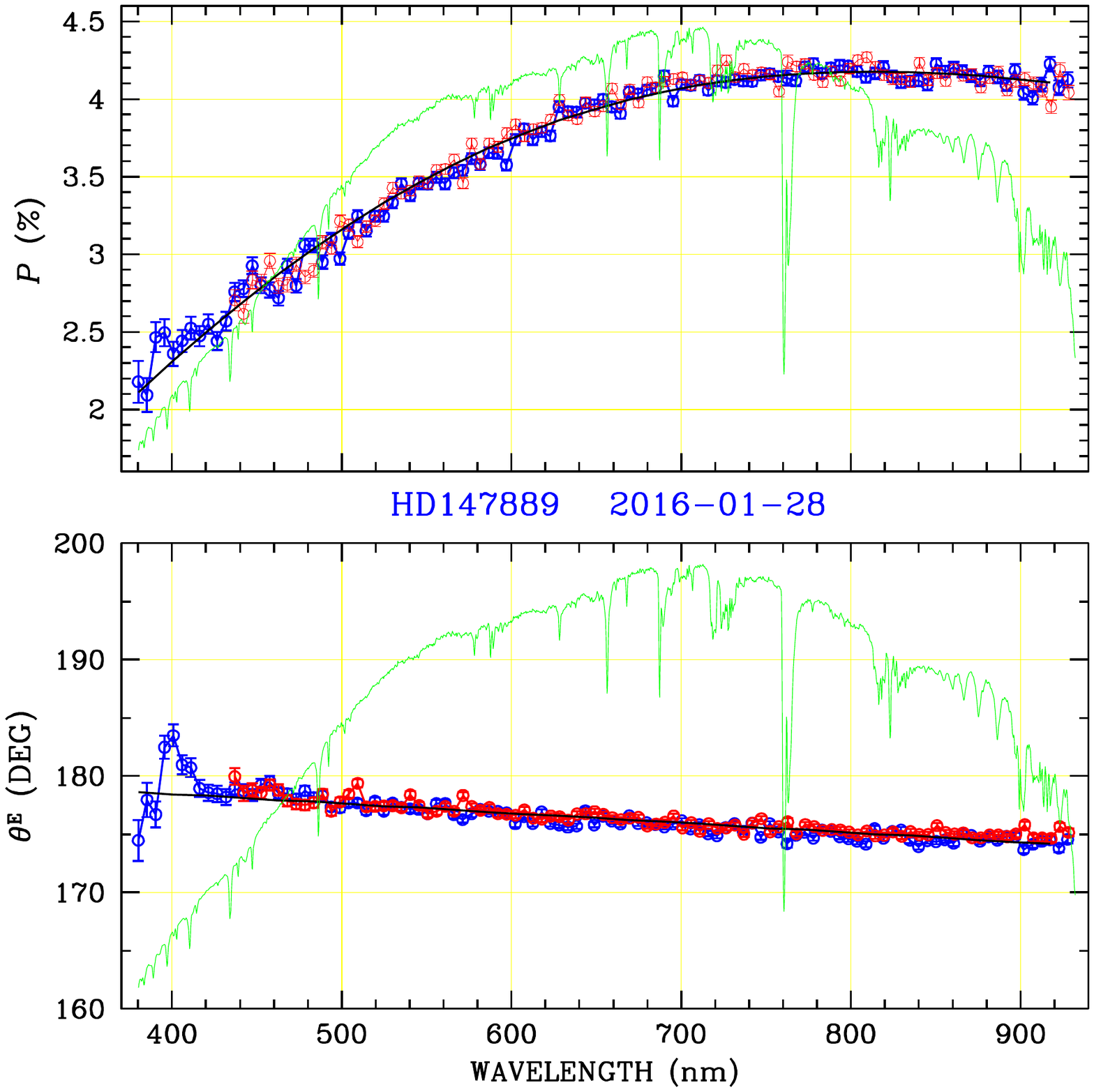}

\newpage
\noindent
  \includegraphics*[scale=0.42,trim={1.1cm 6.0cm 0.1cm 2.8cm},clip]{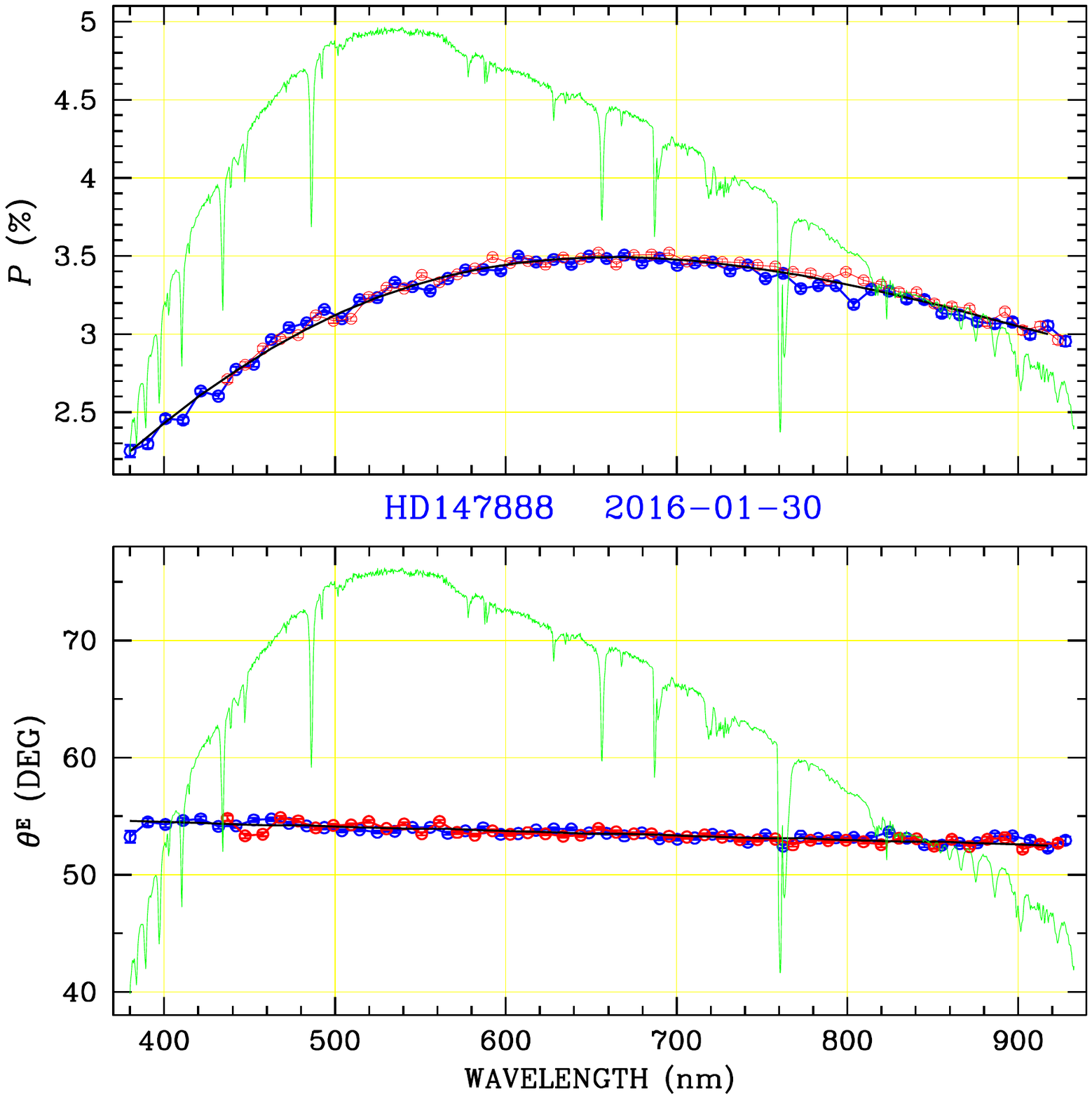}
  \includegraphics*[scale=0.42,trim={1.1cm 6.0cm 0.1cm 2.8cm},clip]{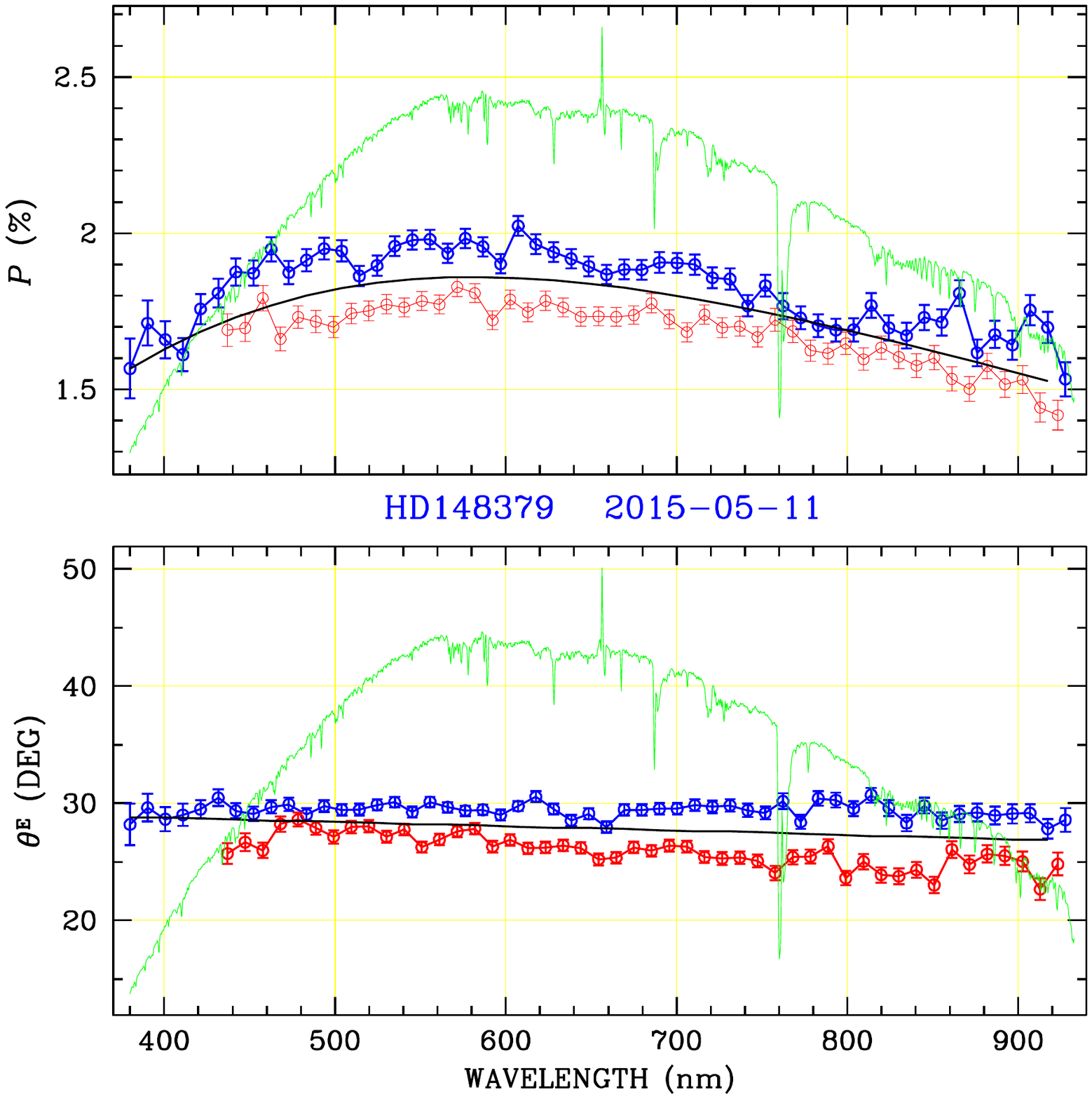}\\
  \includegraphics*[scale=0.42,trim={1.1cm 6.0cm 0.1cm 2.8cm},clip]{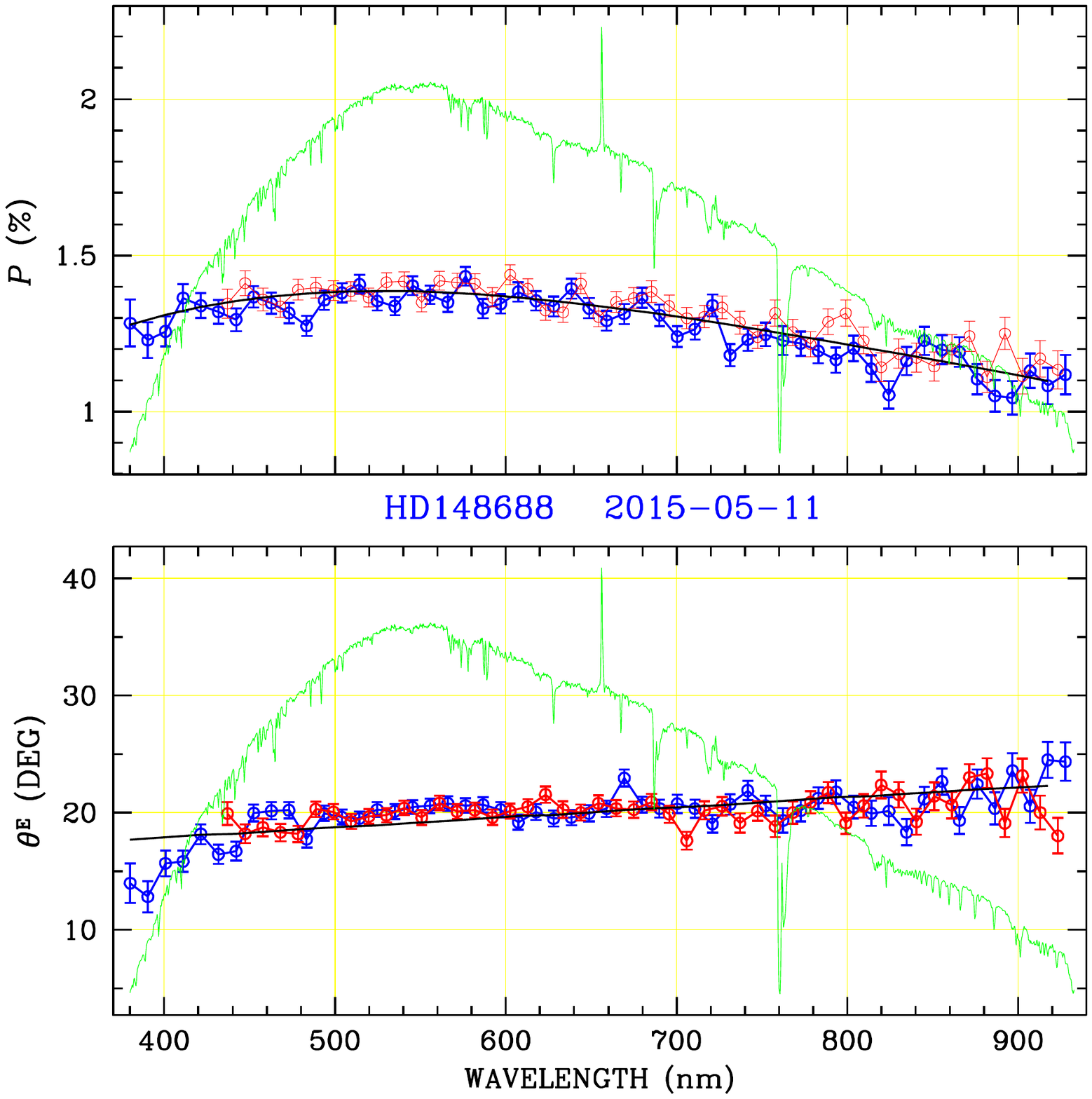}
  \includegraphics*[scale=0.42,trim={1.1cm 6.0cm 0.1cm 2.8cm},clip]{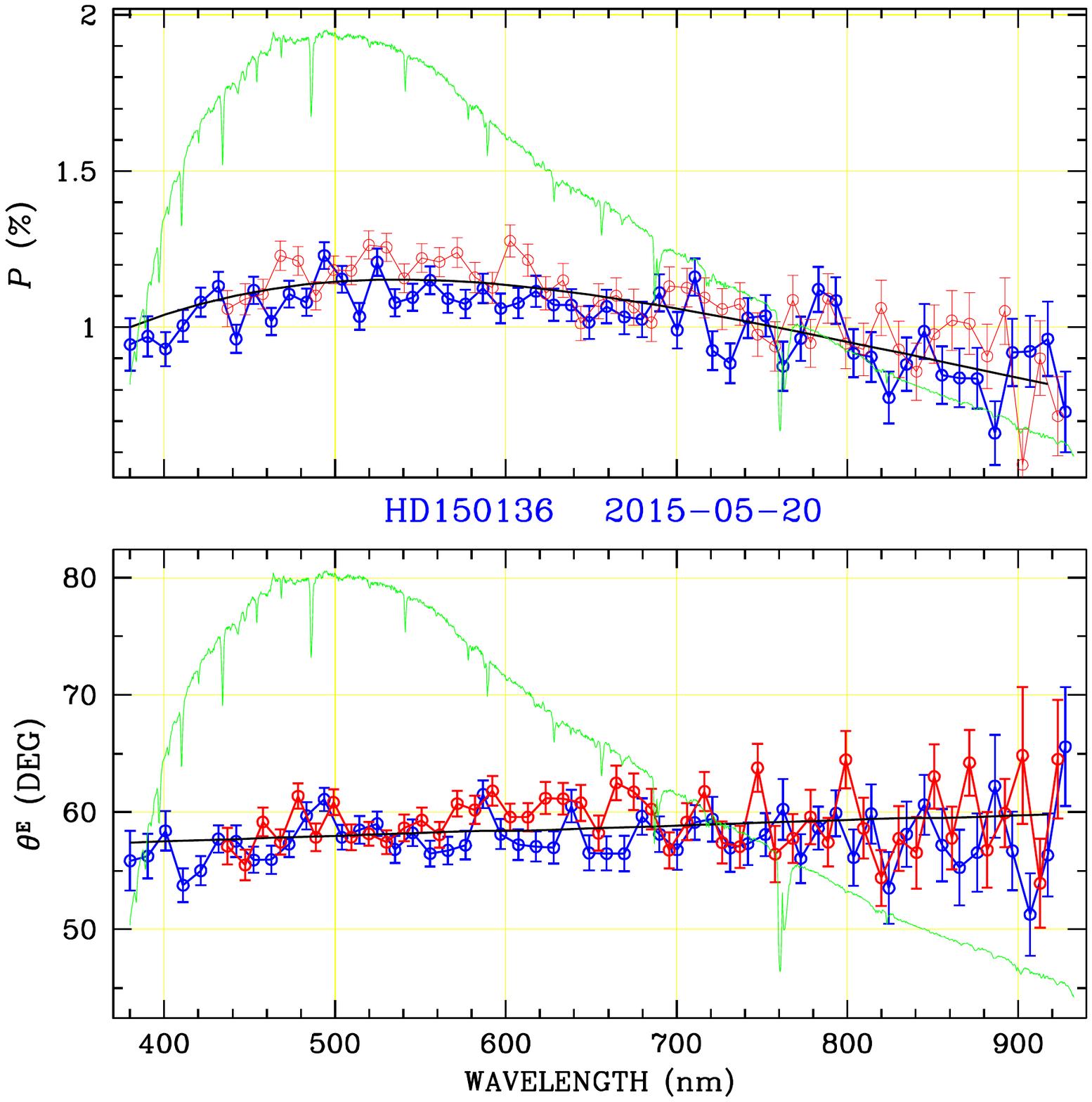}\\
  \includegraphics*[scale=0.42,trim={1.1cm 6.0cm 0.1cm 2.8cm},clip]{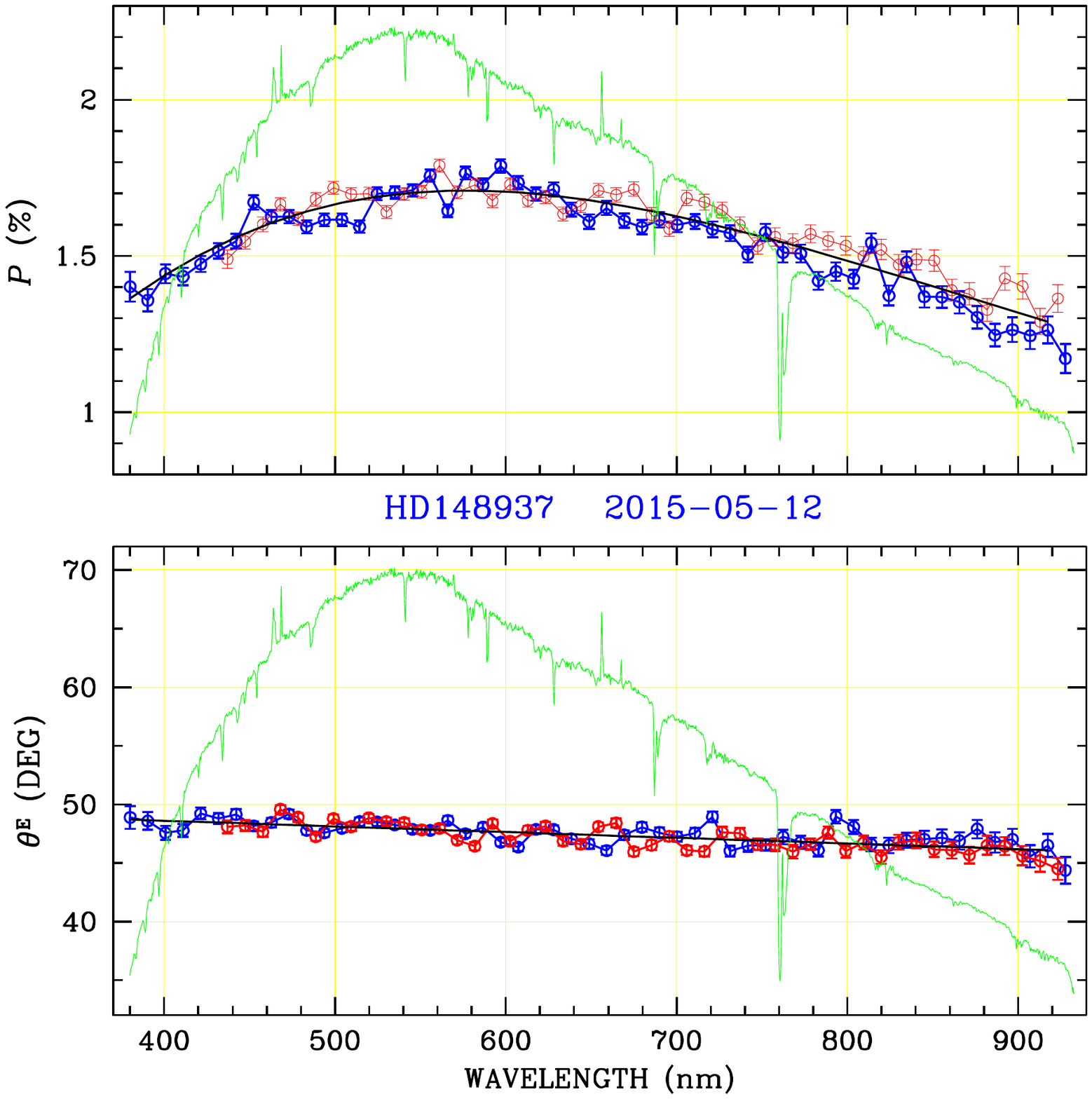}
  \includegraphics*[scale=0.42,trim={1.1cm 6.0cm 0.1cm 2.8cm},clip]{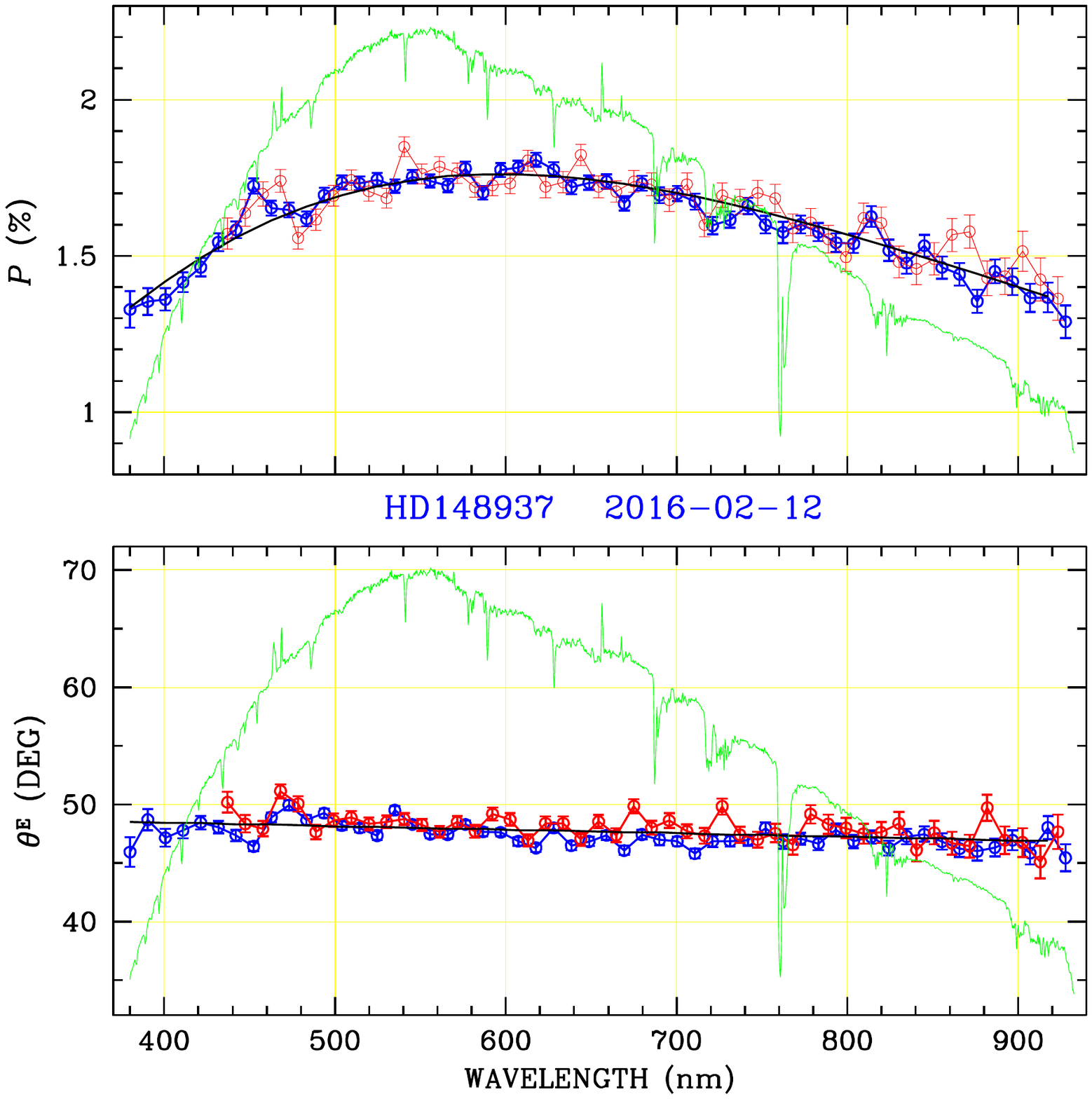}
\newpage

\noindent
  \includegraphics*[scale=0.42,trim={1.1cm 6.0cm 0.1cm 2.8cm},clip]{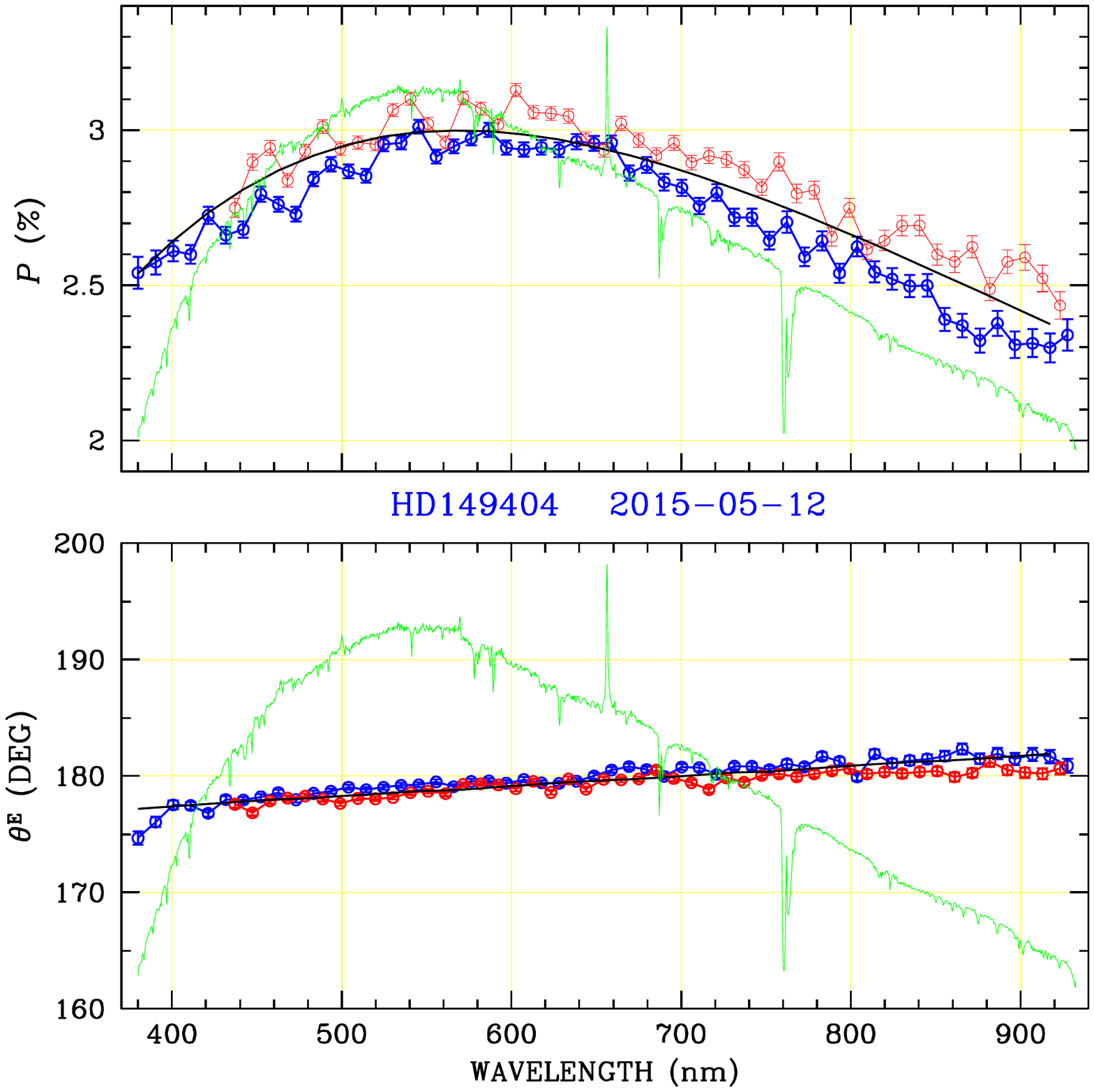}
  \includegraphics*[scale=0.42,trim={1.1cm 6.0cm 0.1cm 2.8cm},clip]{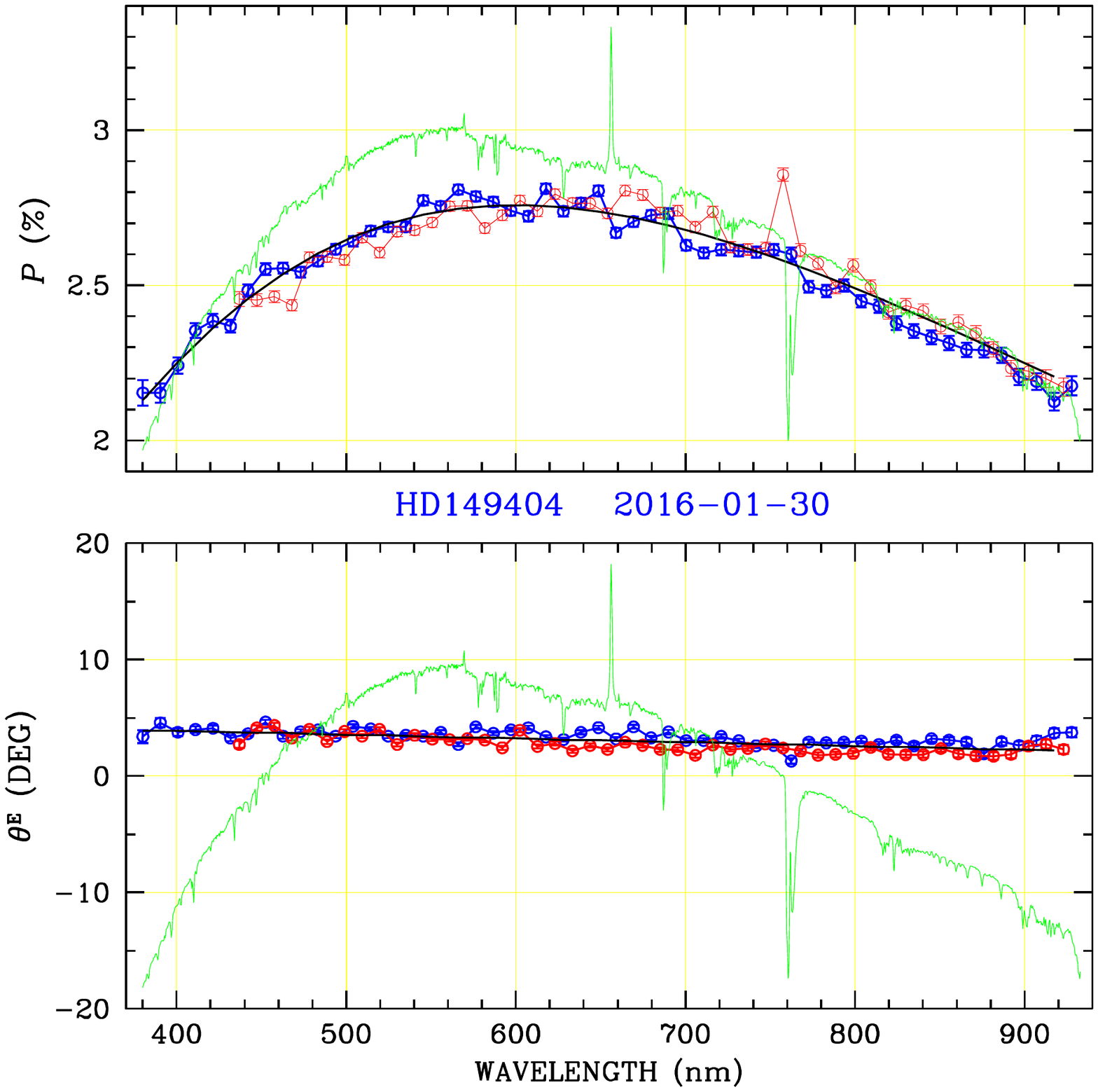}\\
  \includegraphics*[scale=0.42,trim={1.1cm 6.0cm 0.1cm 2.8cm},clip]{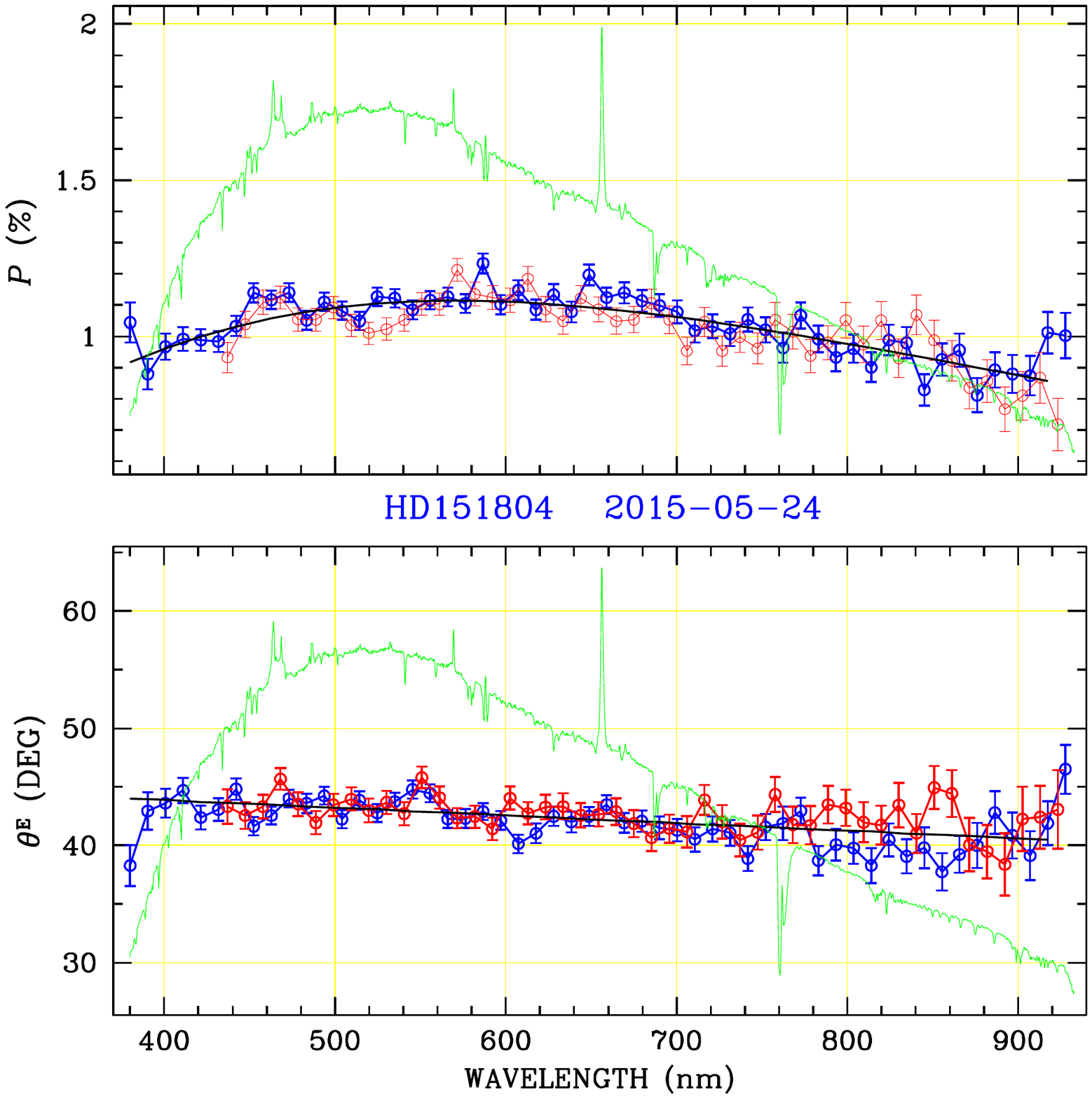}
  \includegraphics*[scale=0.42,trim={1.1cm 6.0cm 0.1cm 2.8cm},clip]{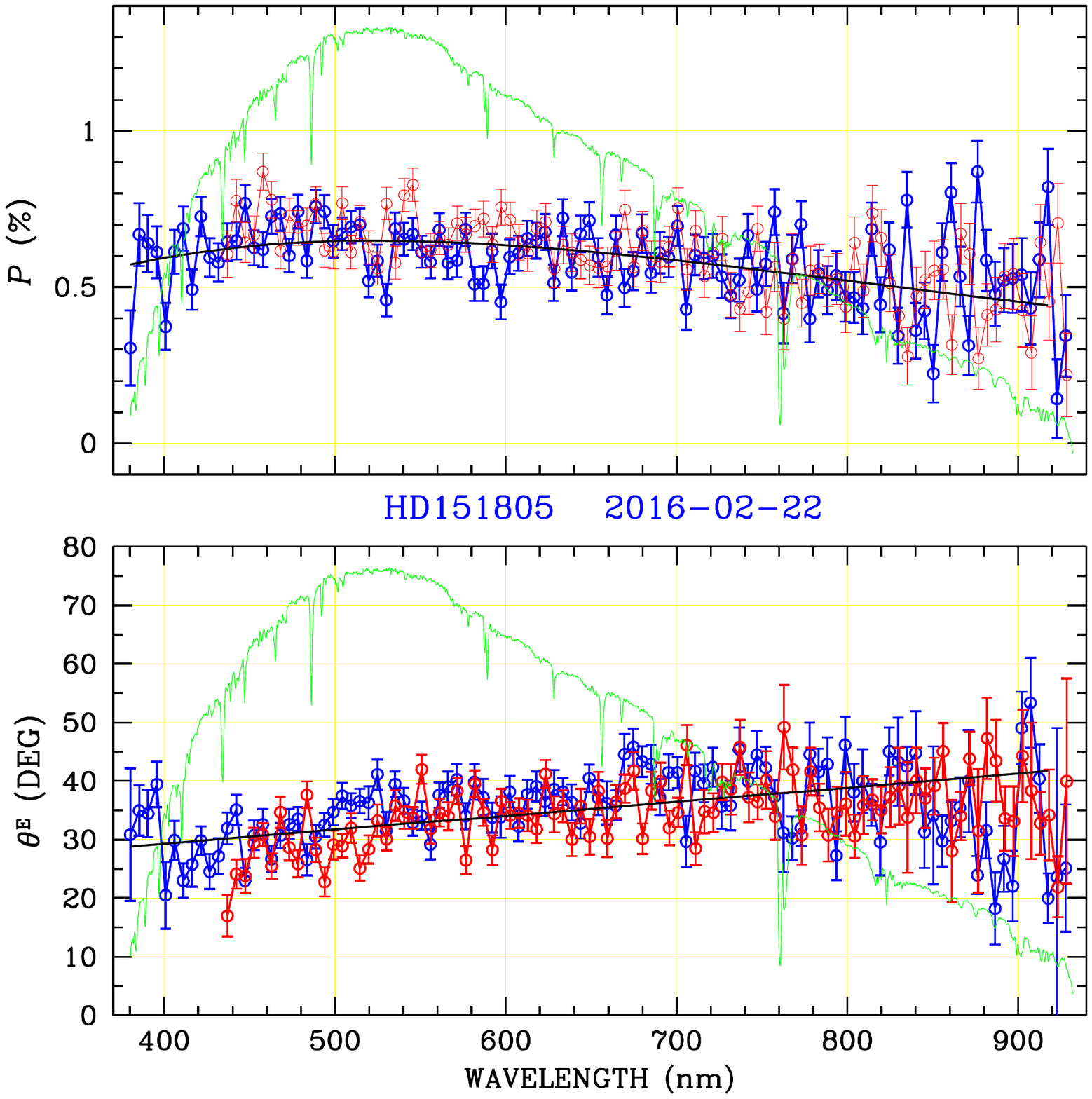}\\
  \includegraphics*[scale=0.42,trim={1.1cm 6.0cm 0.1cm 2.8cm},clip]{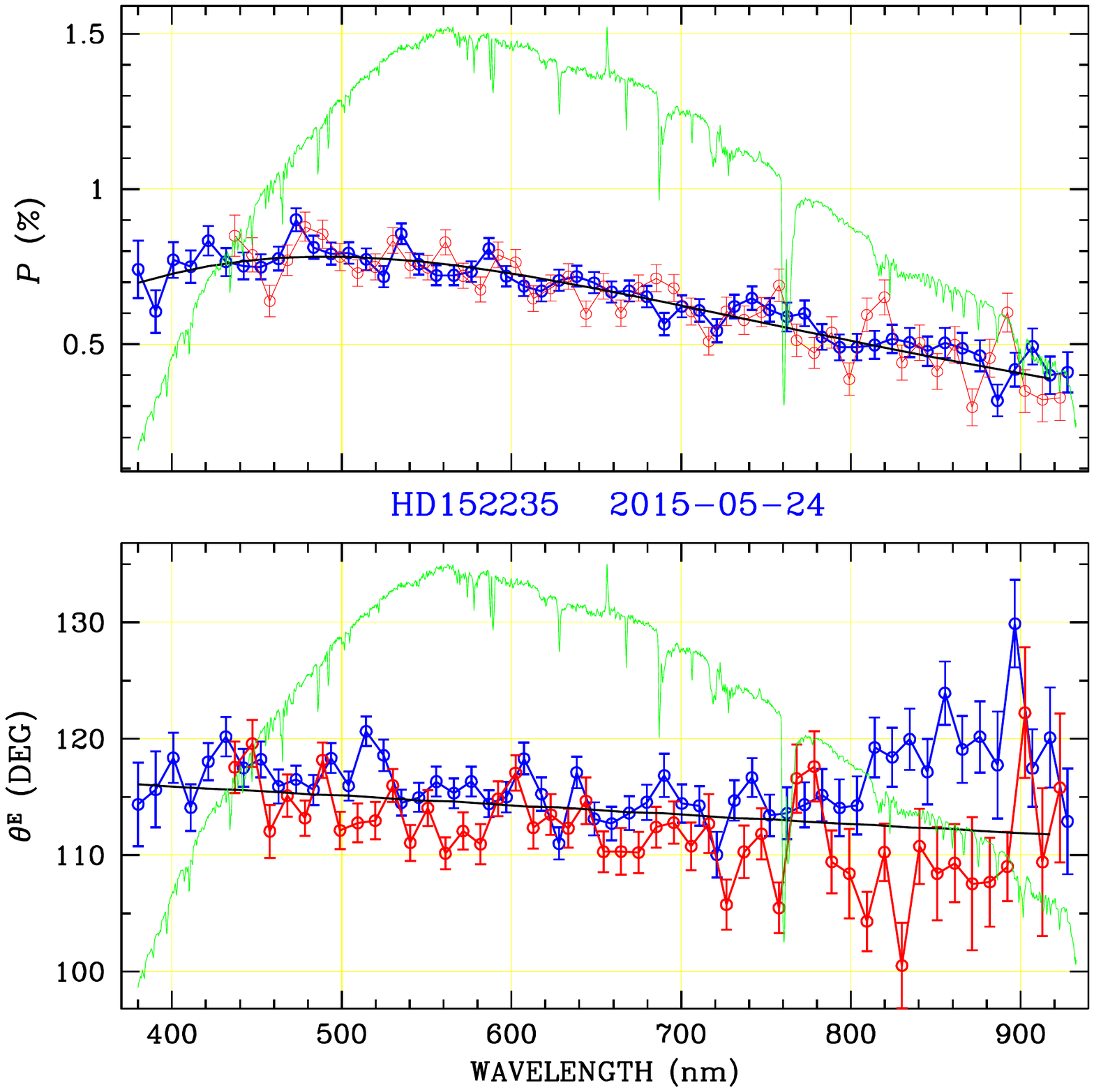}
  \includegraphics*[scale=0.42,trim={1.1cm 6.0cm 0.1cm 2.8cm},clip]{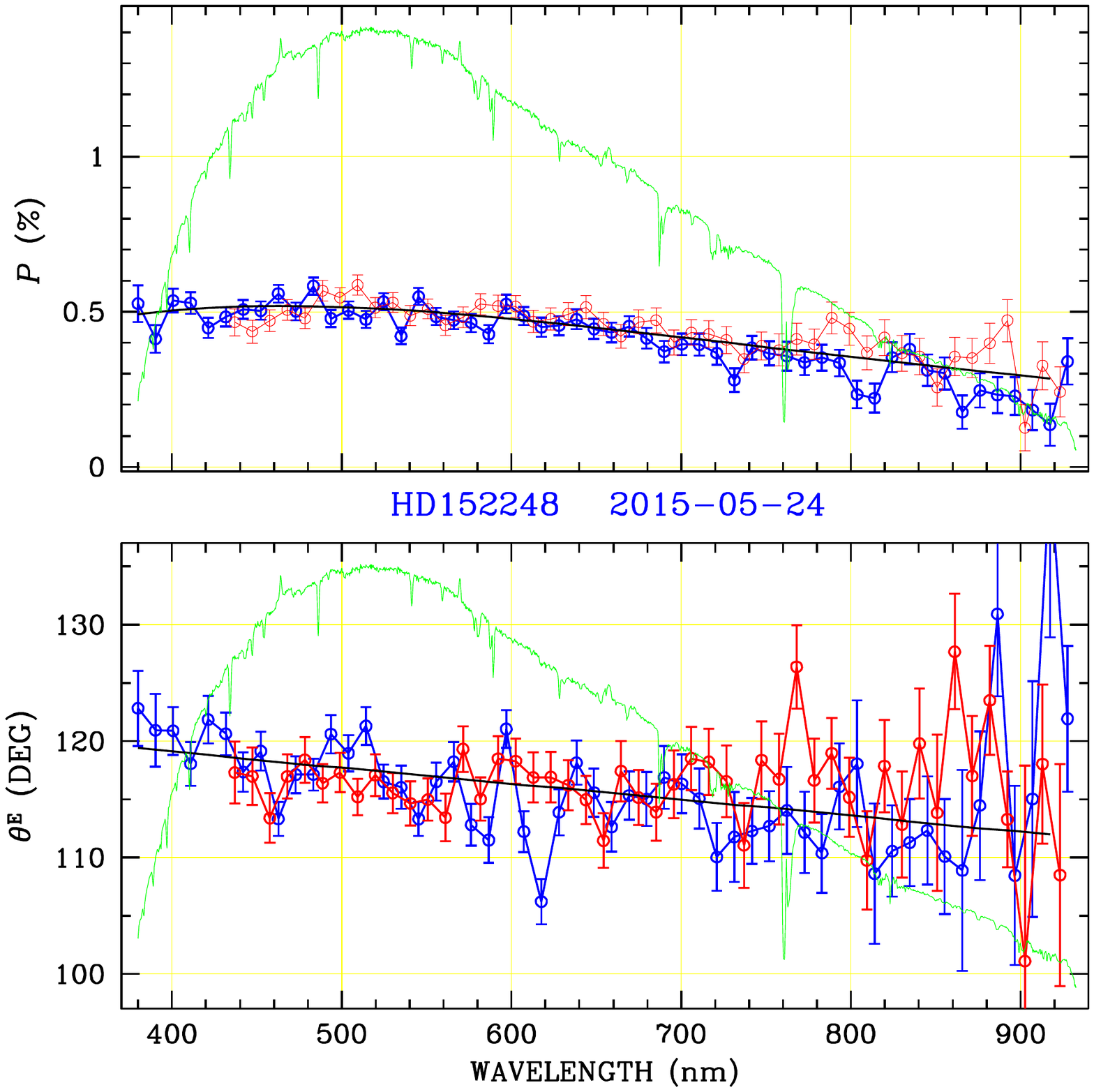}

\newpage
\noindent
  \includegraphics*[scale=0.42,trim={1.1cm 6.0cm 0.1cm 2.8cm},clip]{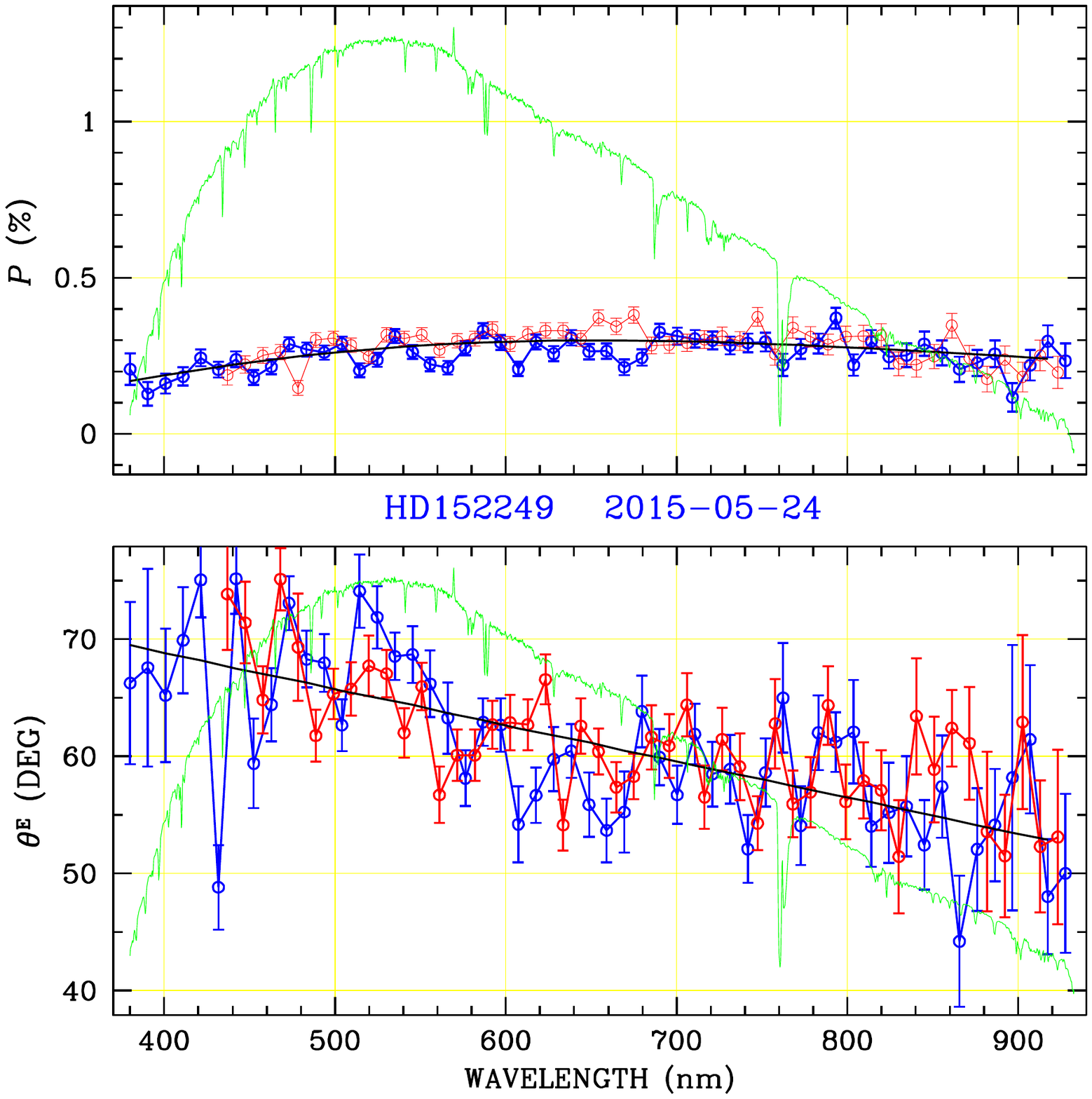}
  \includegraphics*[scale=0.42,trim={1.1cm 6.0cm 0.1cm 2.8cm},clip]{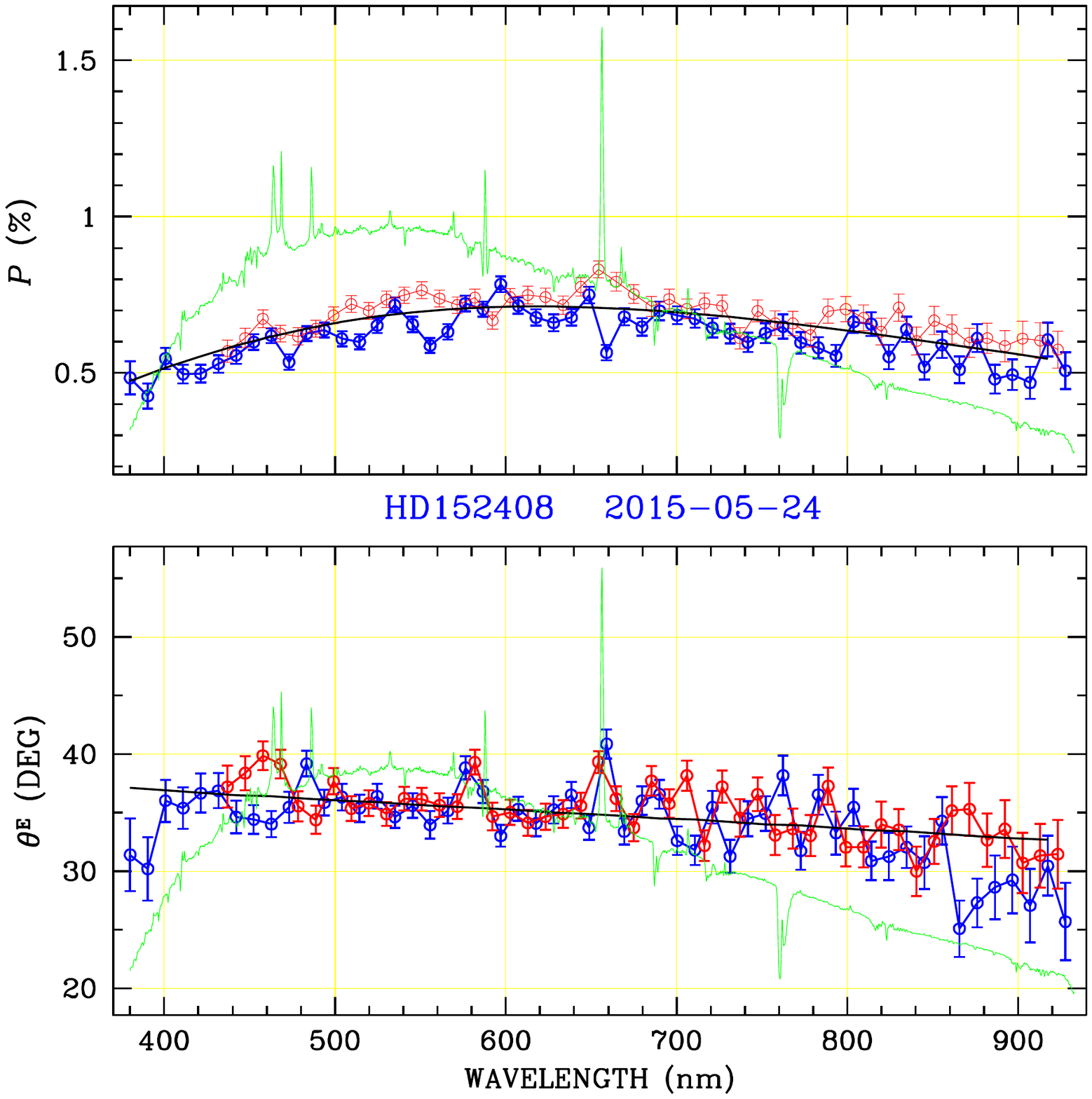}\\
  \includegraphics*[scale=0.42,trim={1.1cm 6.0cm 0.1cm 2.8cm},clip]{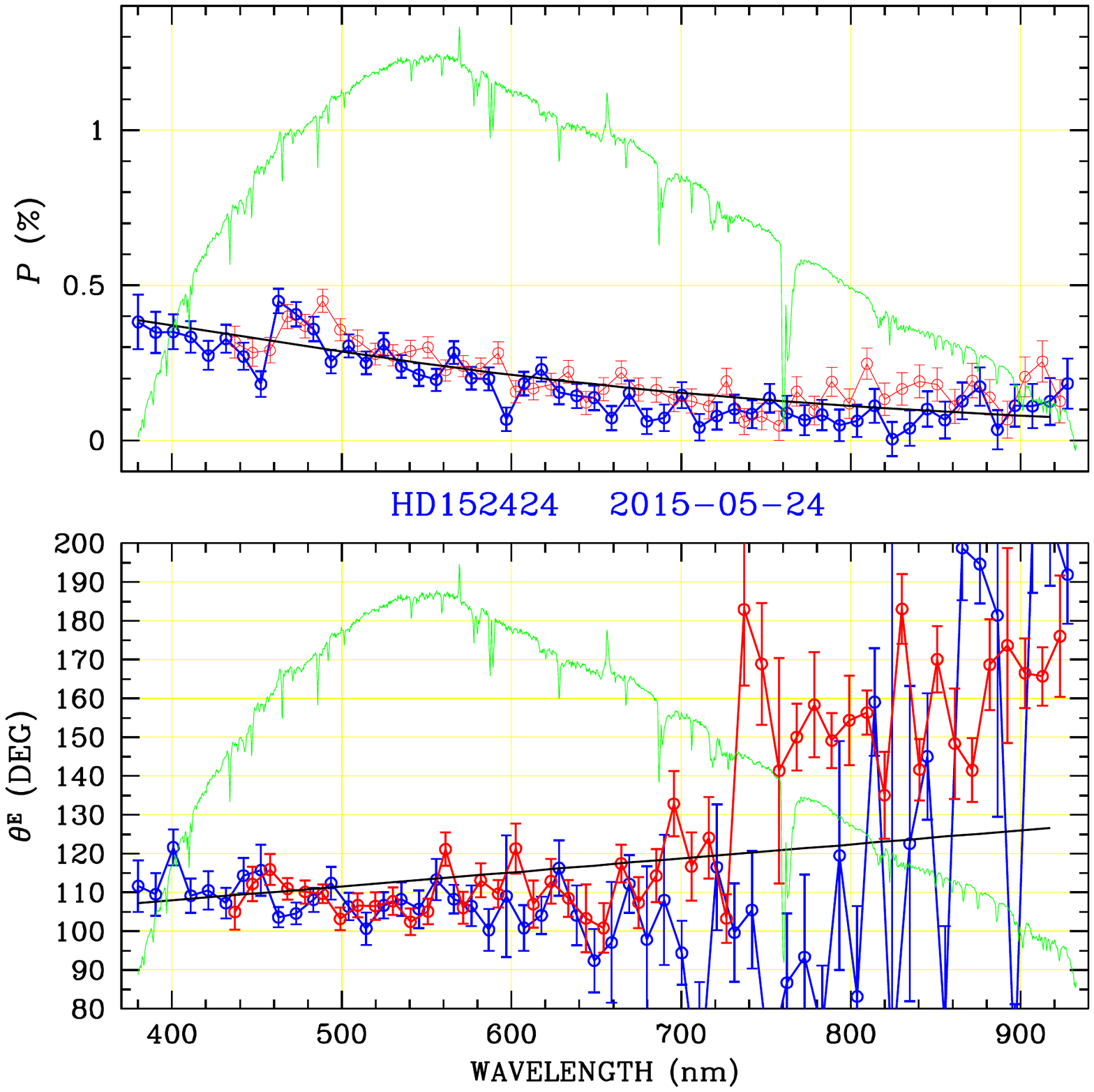}
  \includegraphics*[scale=0.42,trim={1.1cm 6.0cm 0.1cm 2.8cm},clip]{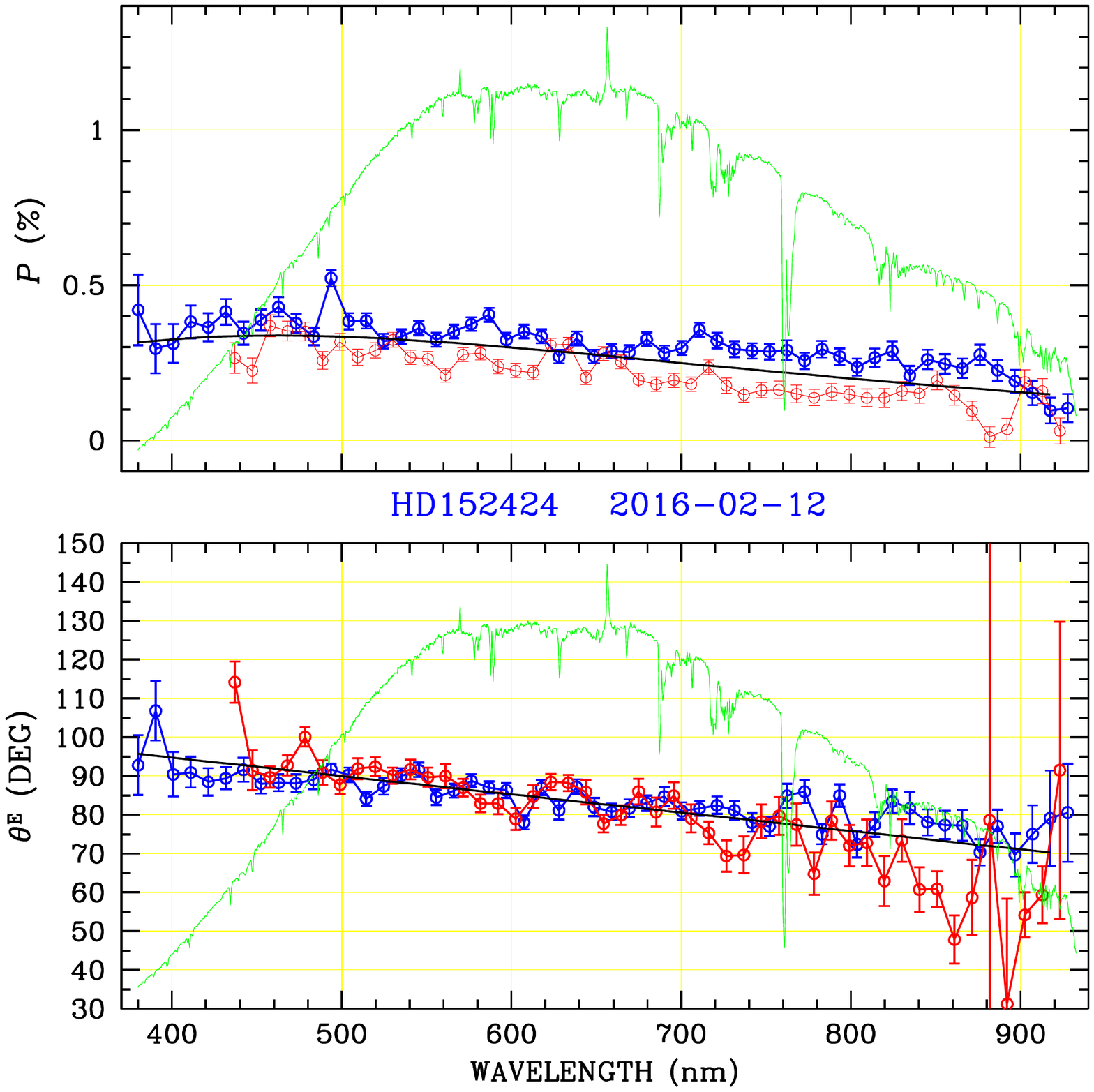}\\
  \includegraphics*[scale=0.42,trim={1.1cm 6.0cm 0.1cm 2.8cm},clip]{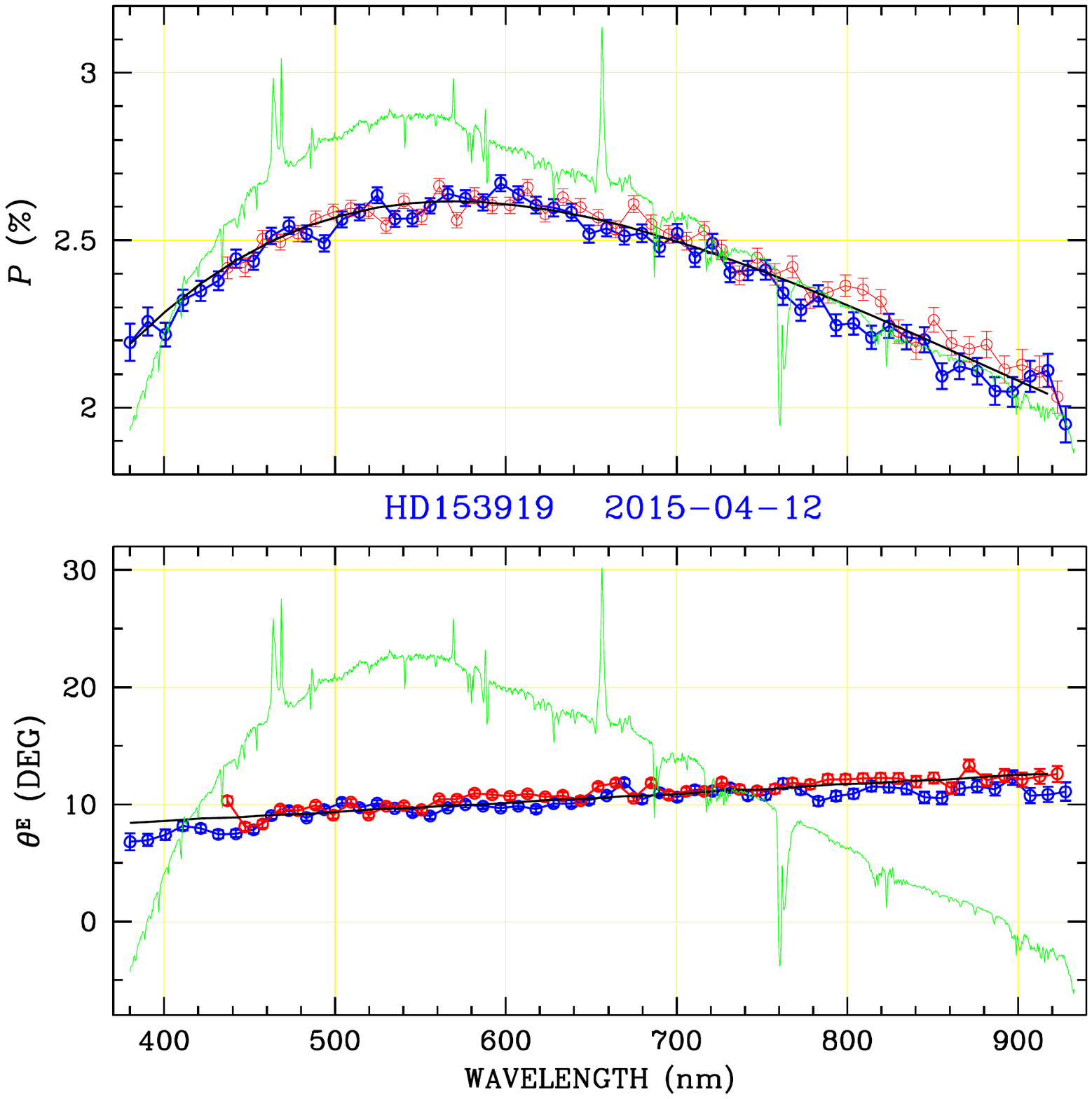}
  \includegraphics*[scale=0.42,trim={1.1cm 6.0cm 0.1cm 2.8cm},clip]{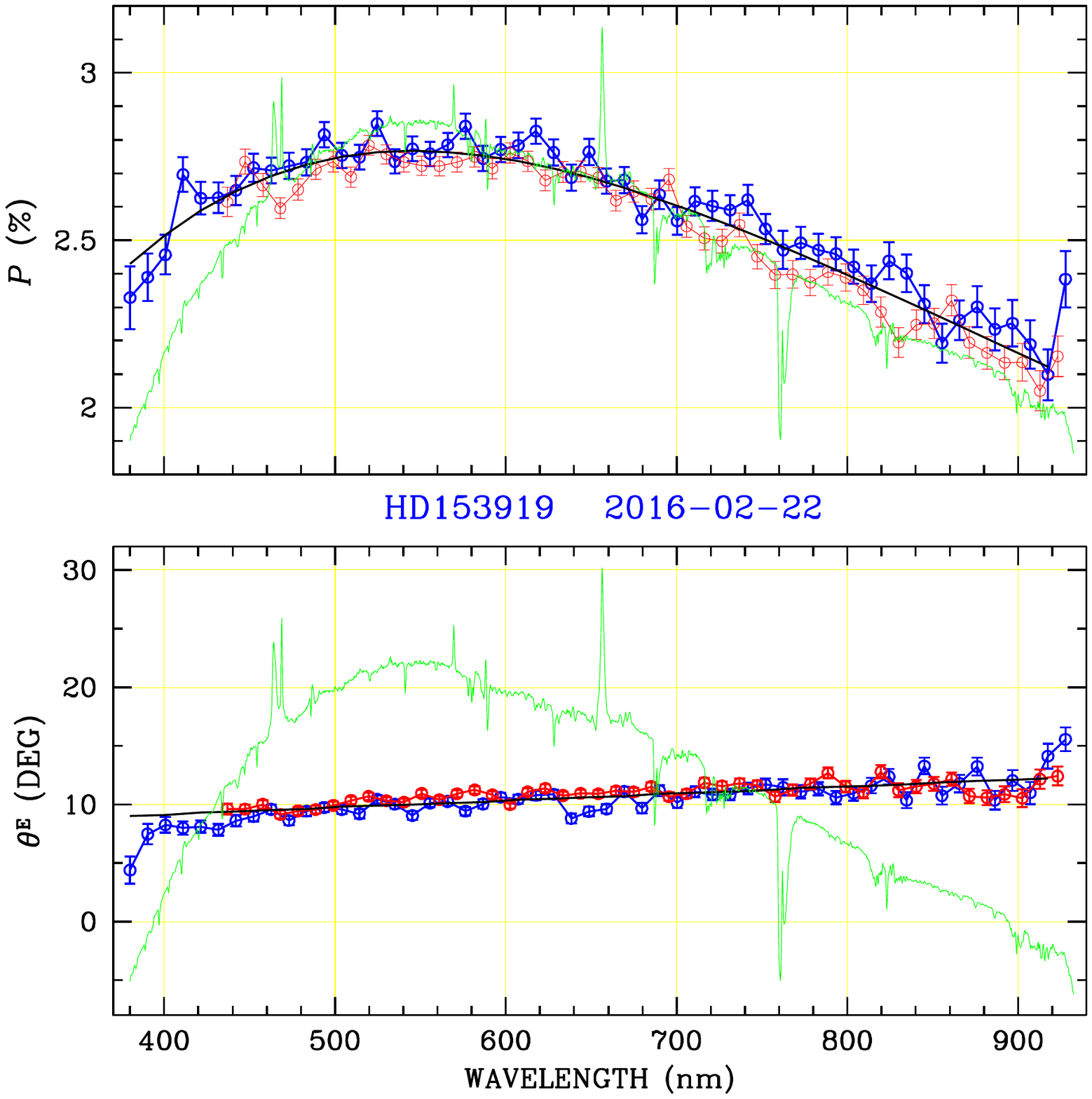}

\newpage
\noindent
  \includegraphics*[scale=0.42,trim={1.1cm 6.0cm 0.1cm 2.8cm},clip]{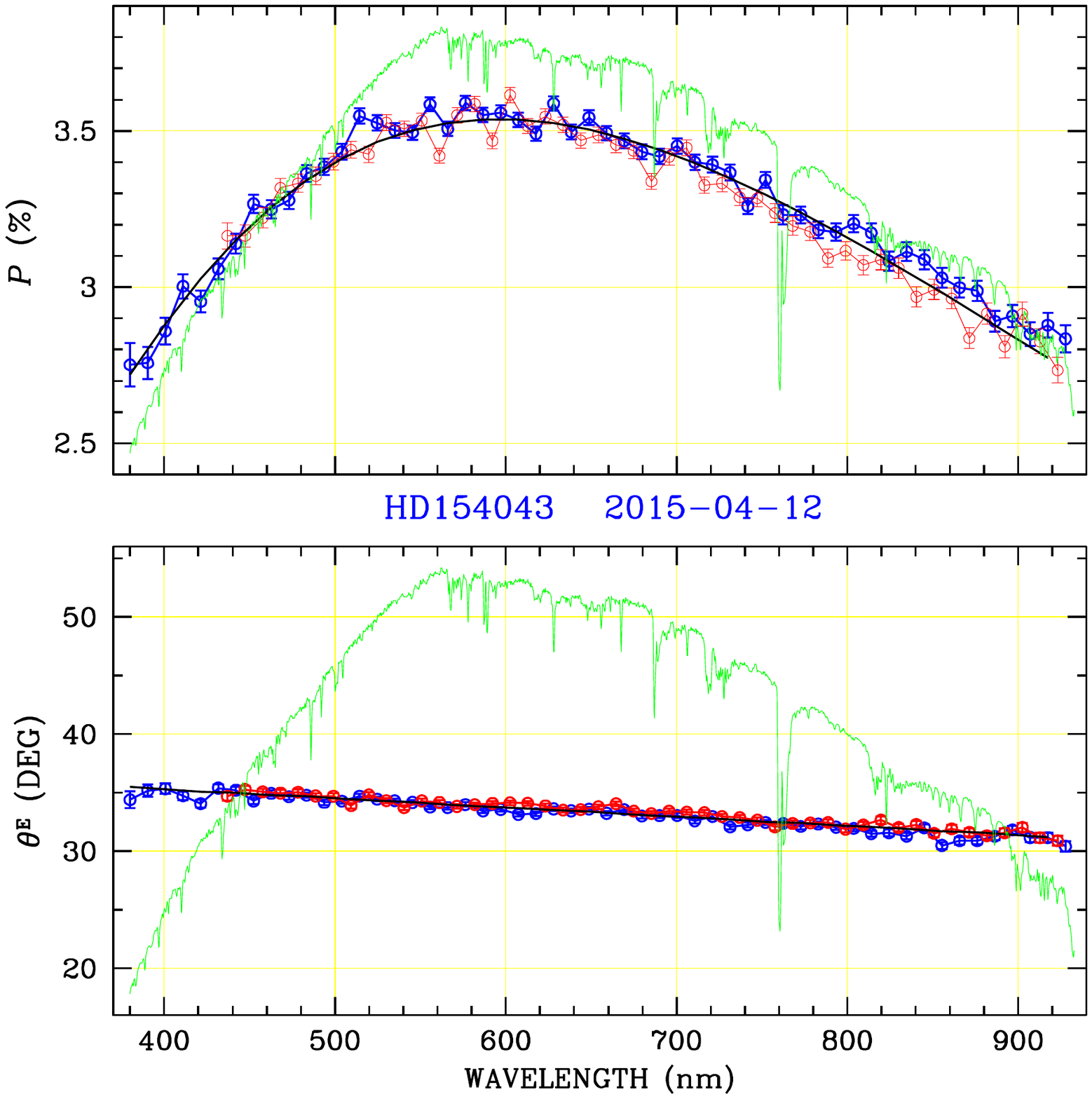}
  \includegraphics*[scale=0.42,trim={1.1cm 6.0cm 0.1cm 2.8cm},clip]{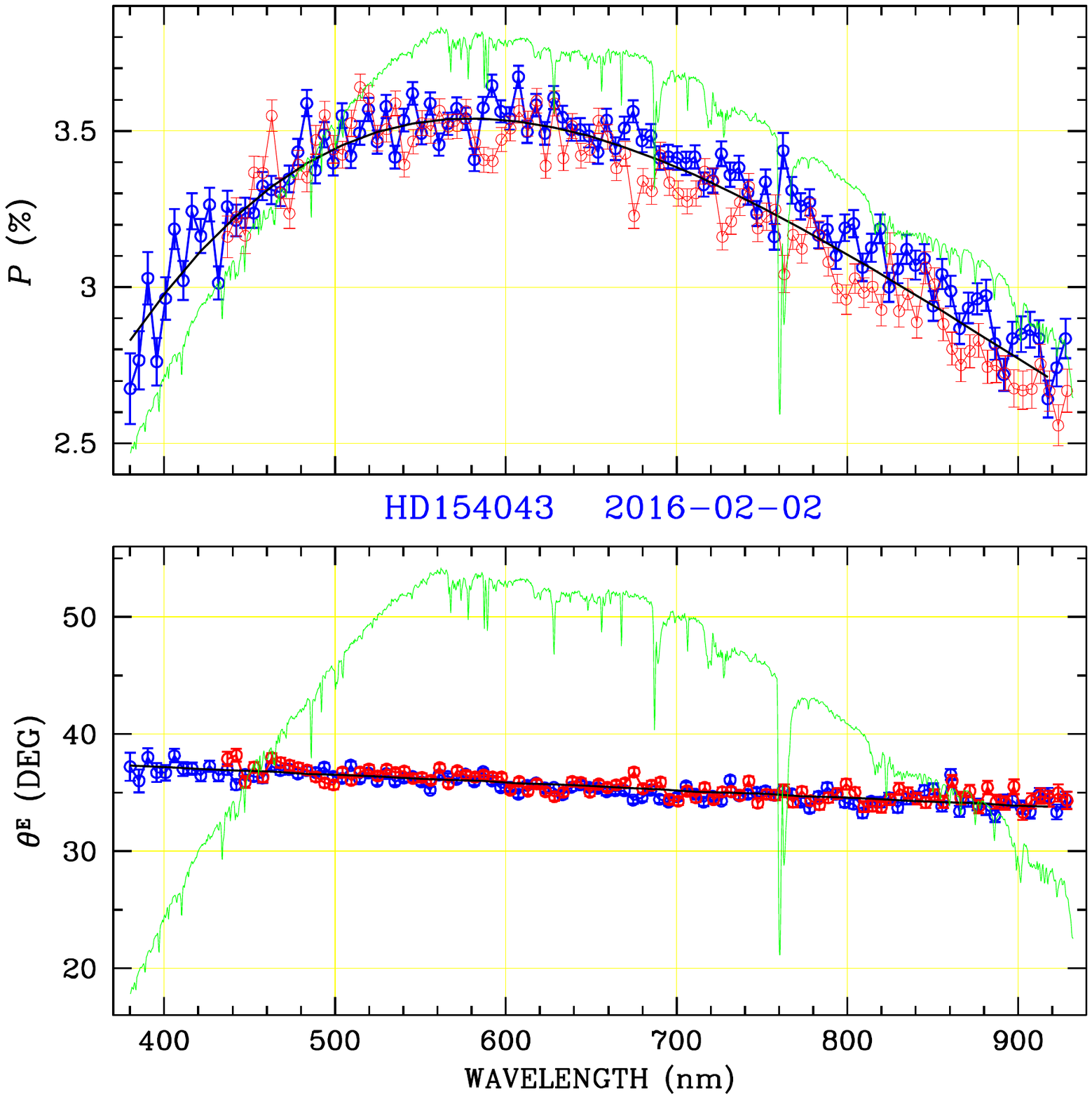}\\
  \includegraphics*[scale=0.42,trim={1.1cm 6.0cm 0.1cm 2.8cm},clip]{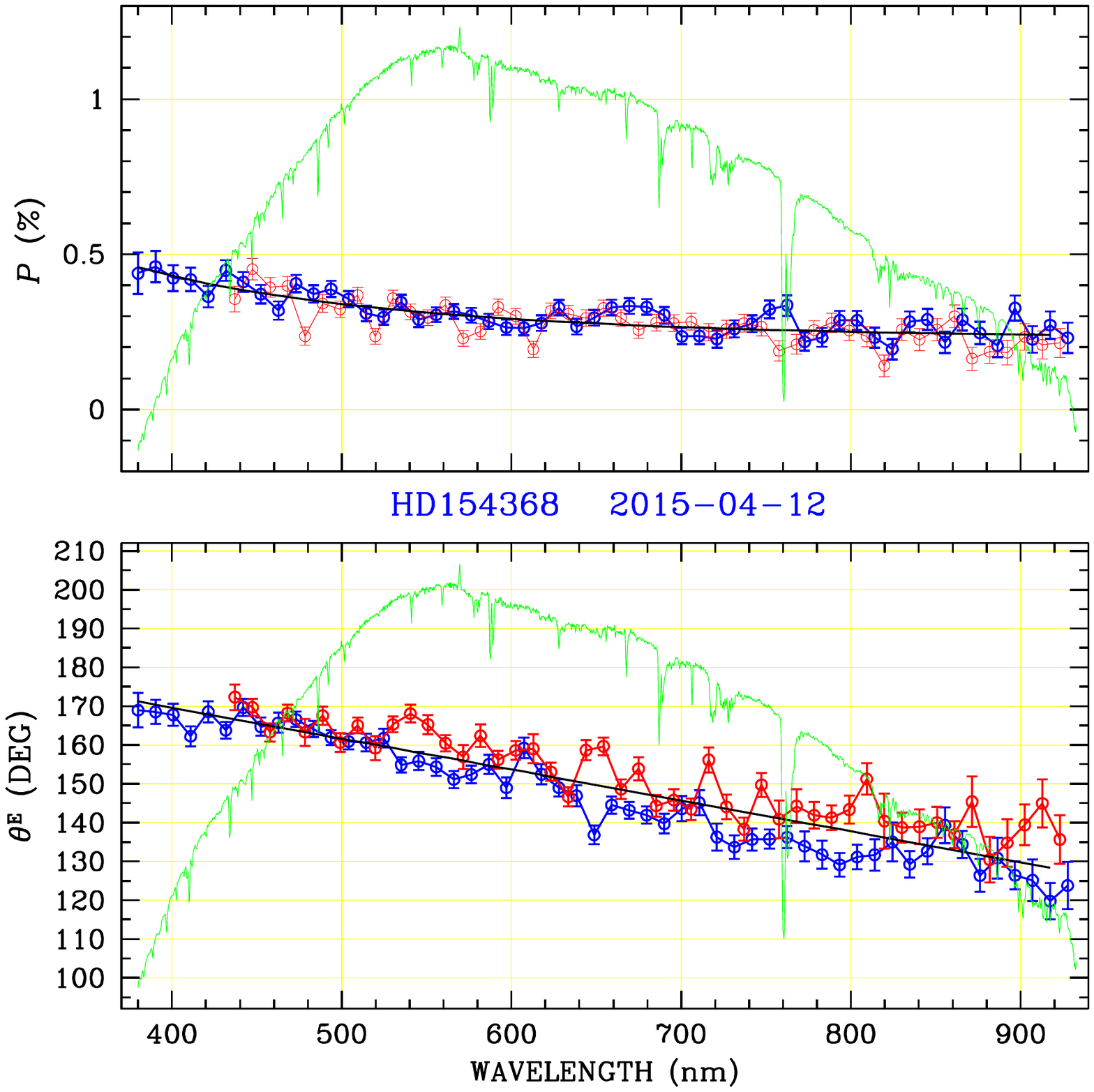}
  \includegraphics*[scale=0.42,trim={1.1cm 6.0cm 0.1cm 2.8cm},clip]{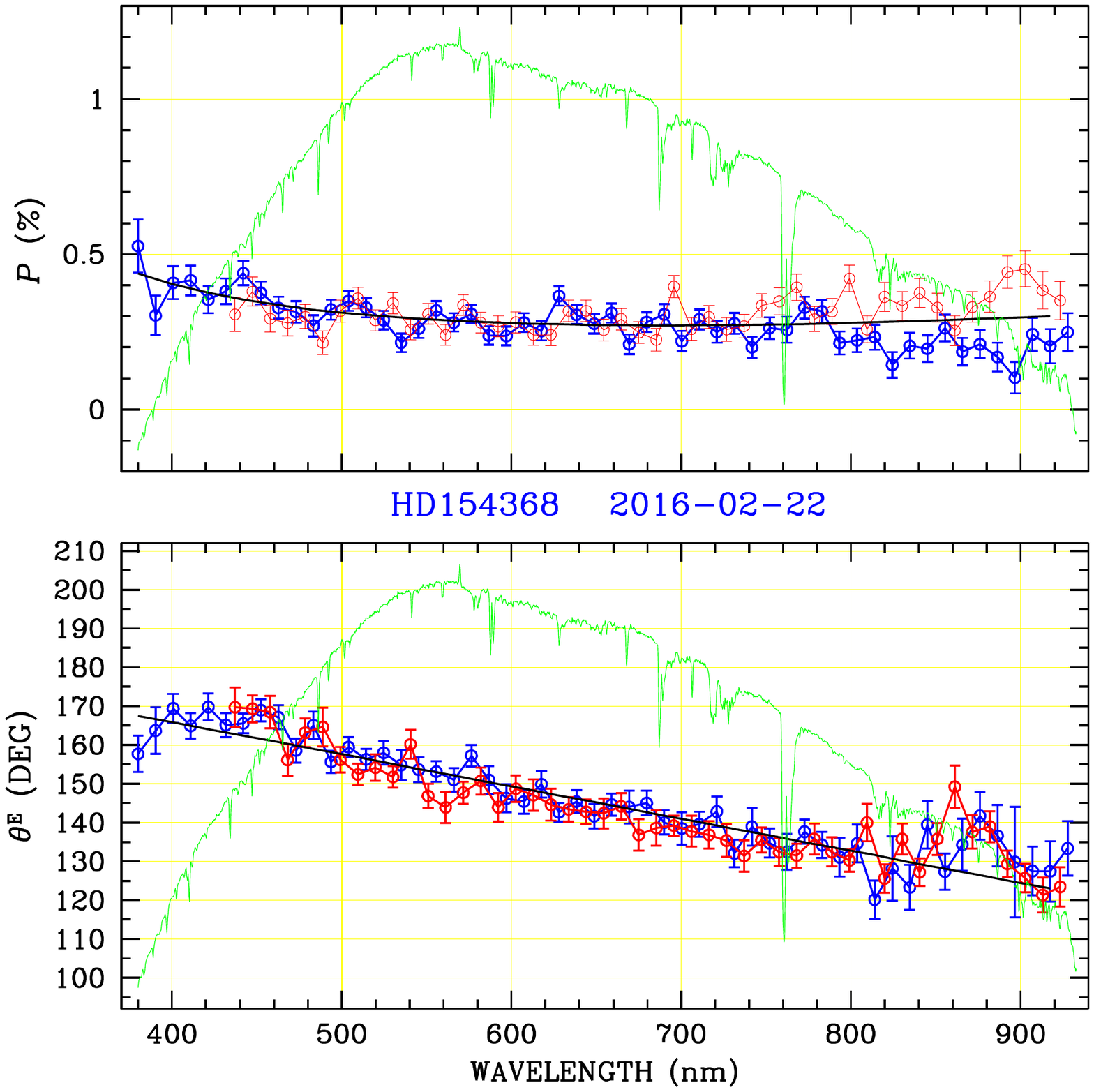}\\
  \includegraphics*[scale=0.42,trim={1.1cm 6.0cm 0.1cm 2.8cm},clip]{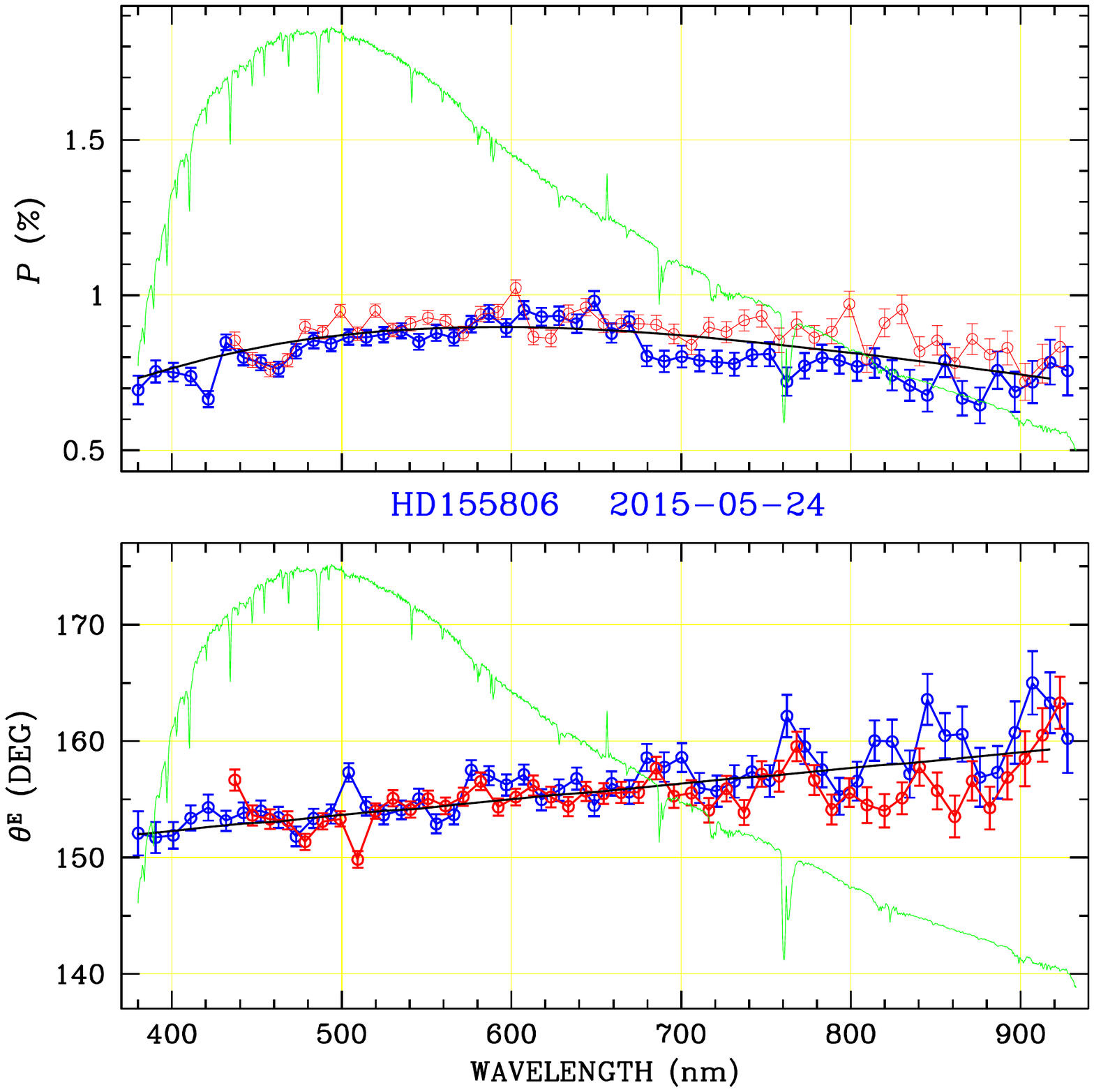}
  \includegraphics*[scale=0.42,trim={1.1cm 6.0cm 0.1cm 2.8cm},clip]{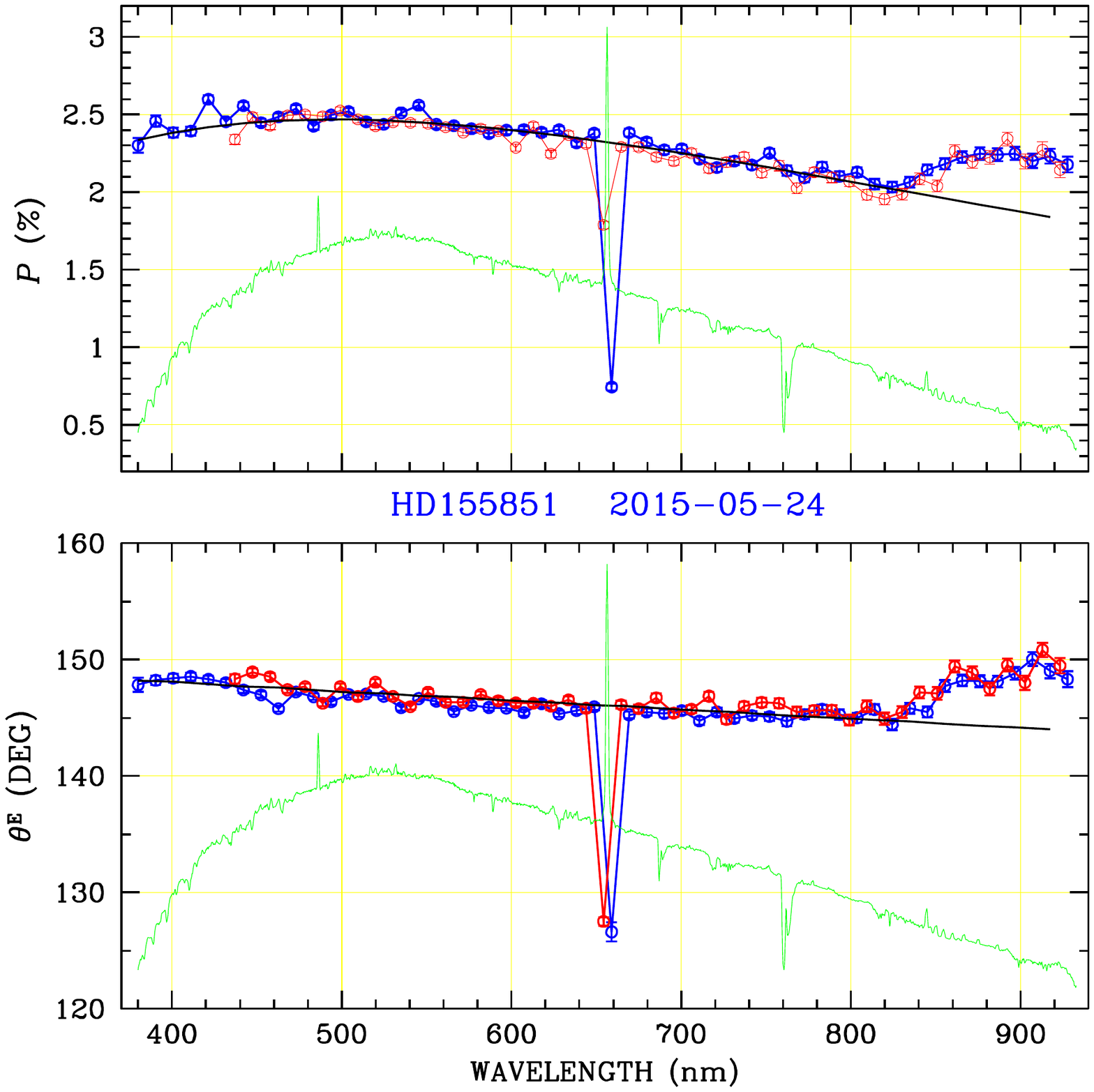}

\newpage
\noindent
  \includegraphics*[scale=0.42,trim={1.1cm 6.0cm 0.1cm 2.8cm},clip]{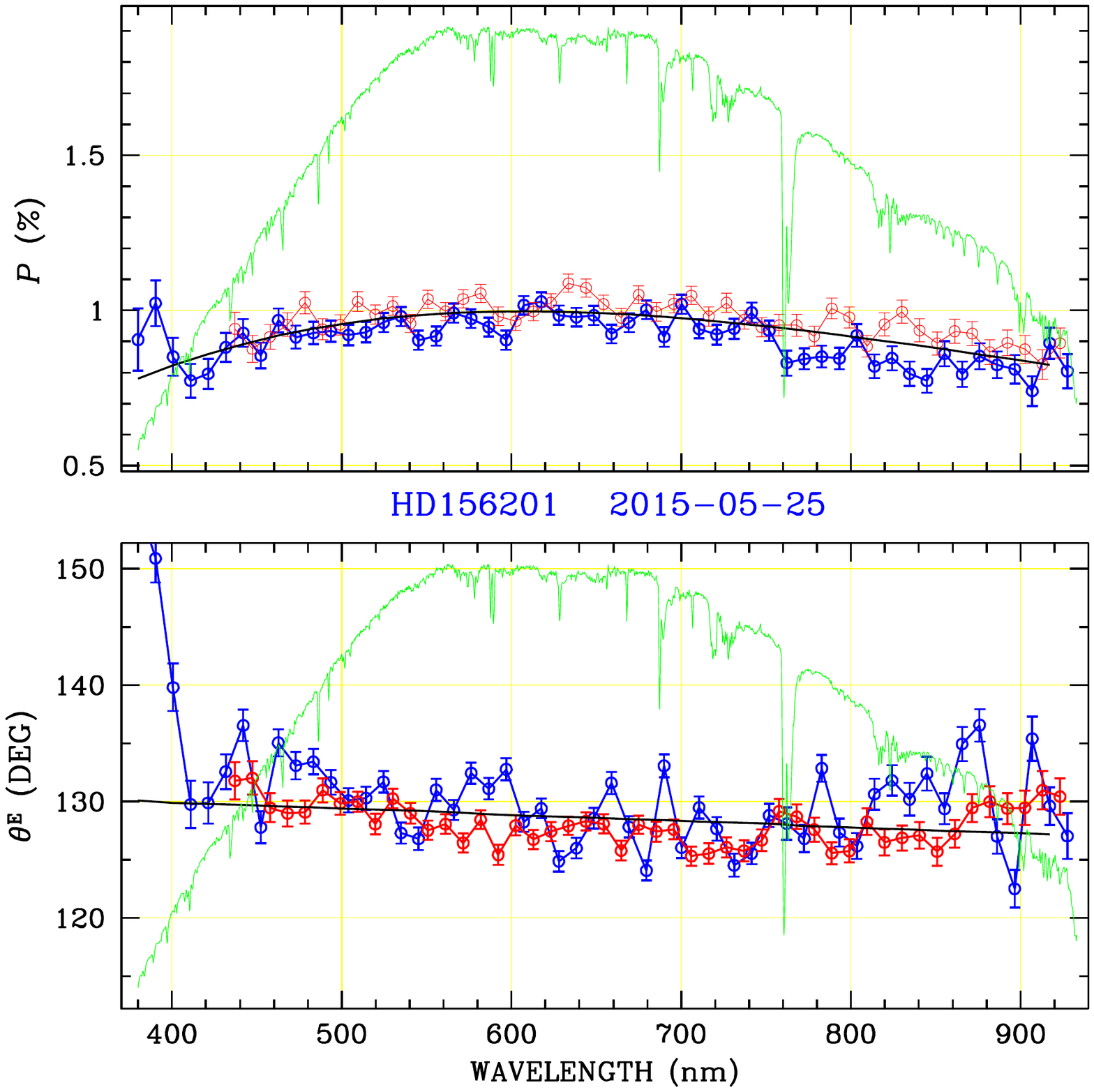}
  \includegraphics*[scale=0.42,trim={1.1cm 6.0cm 0.1cm 2.8cm},clip]{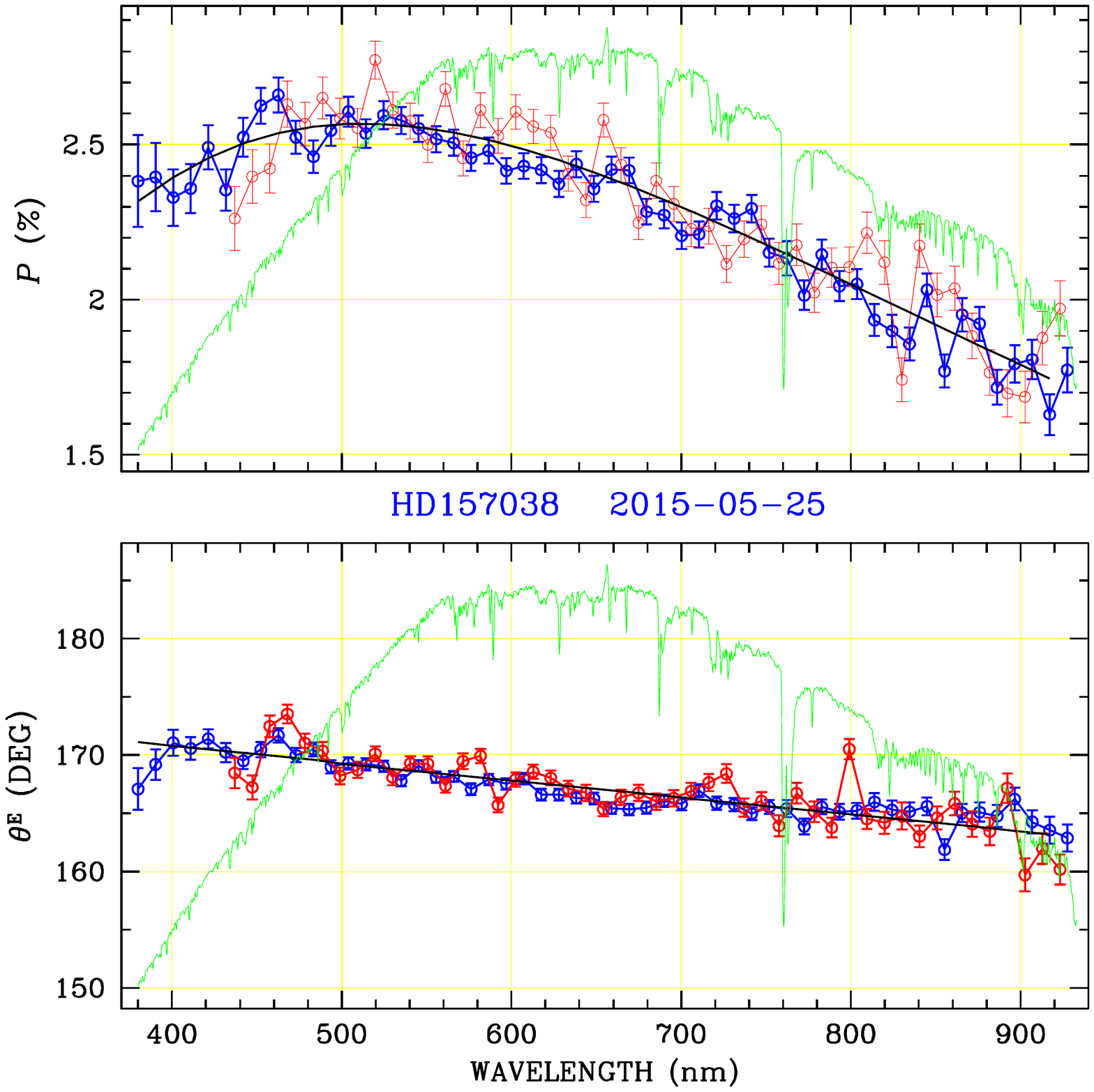}\\
  \includegraphics*[scale=0.42,trim={1.1cm 6.0cm 0.1cm 2.8cm},clip]{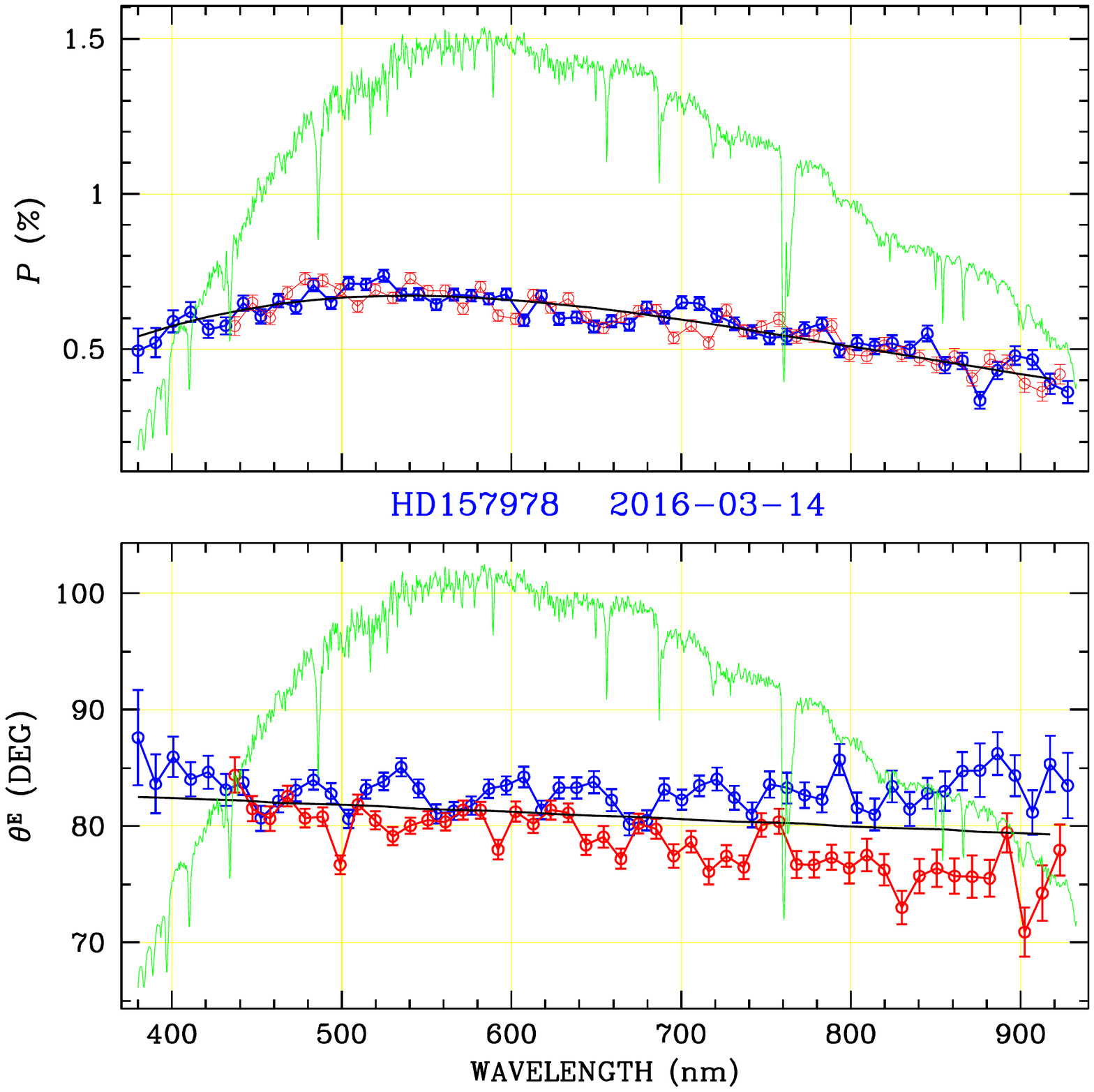}
  \includegraphics*[scale=0.42,trim={1.1cm 6.0cm 0.1cm 2.8cm},clip]{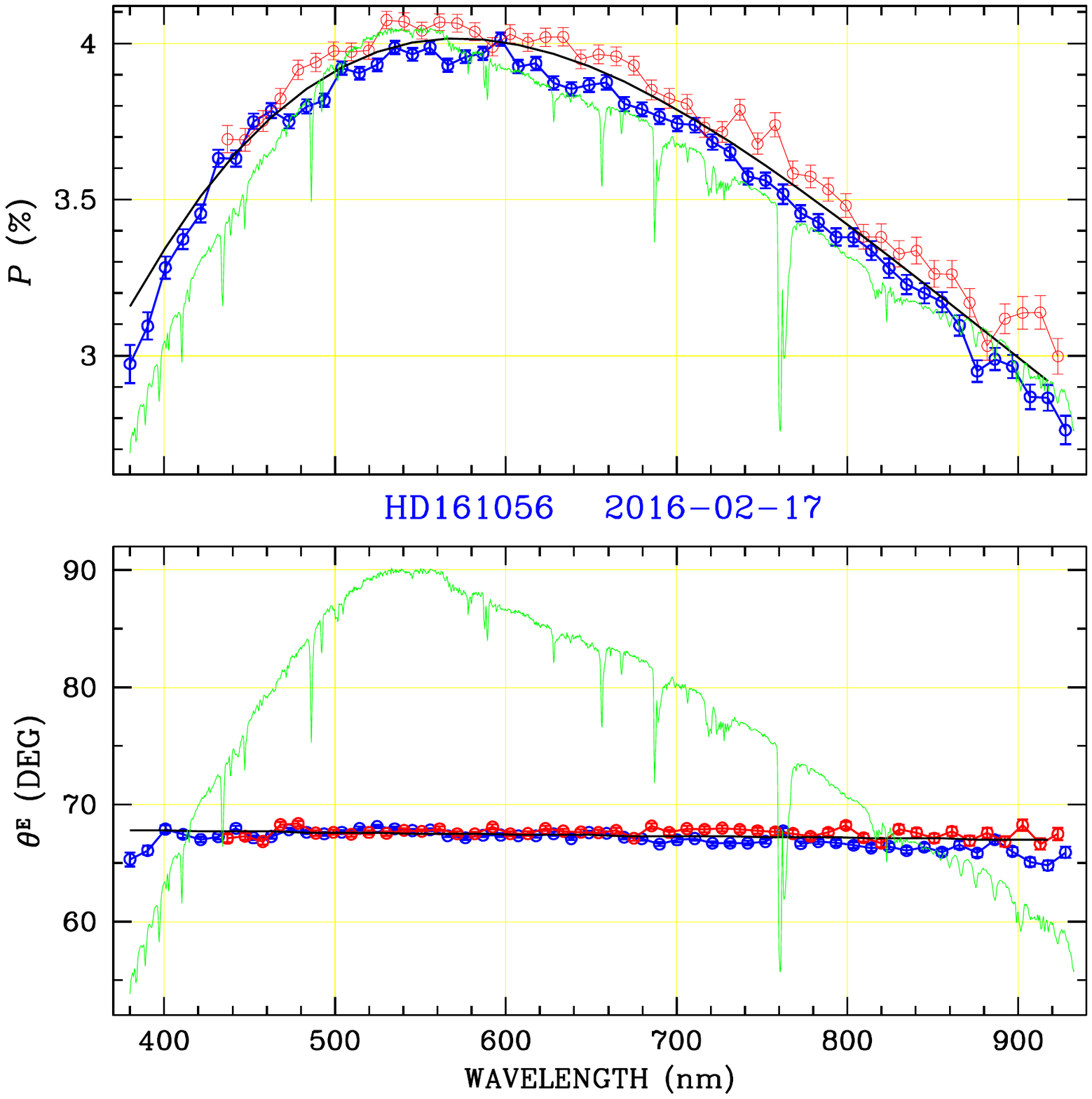}\\
  \includegraphics*[scale=0.42,trim={1.1cm 6.0cm 0.1cm 2.8cm},clip]{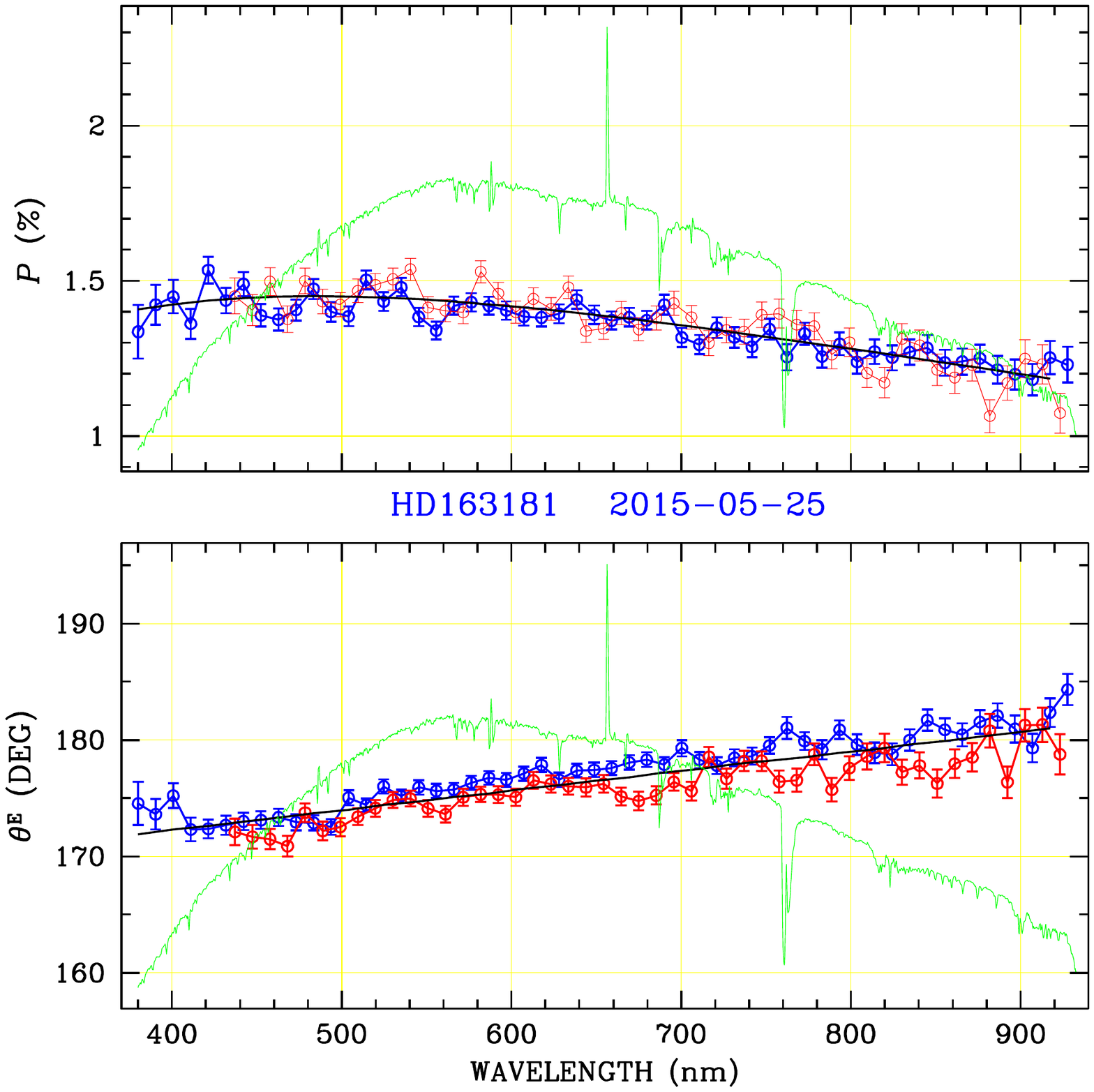}
  \includegraphics*[scale=0.42,trim={1.1cm 6.0cm 0.1cm 2.8cm},clip]{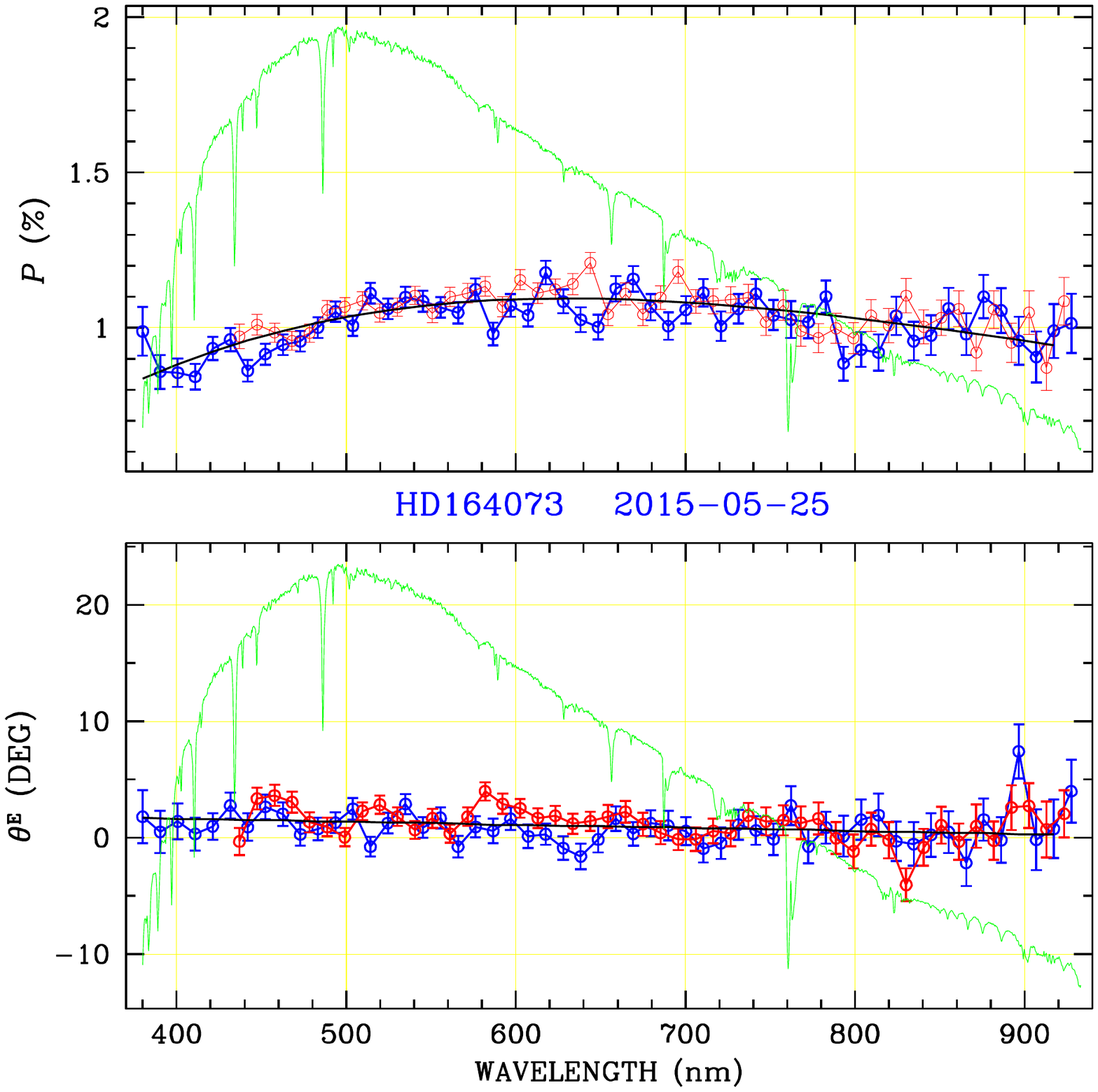}

\newpage
\noindent
  \includegraphics*[scale=0.42,trim={1.1cm 6.0cm 0.1cm 2.8cm},clip]{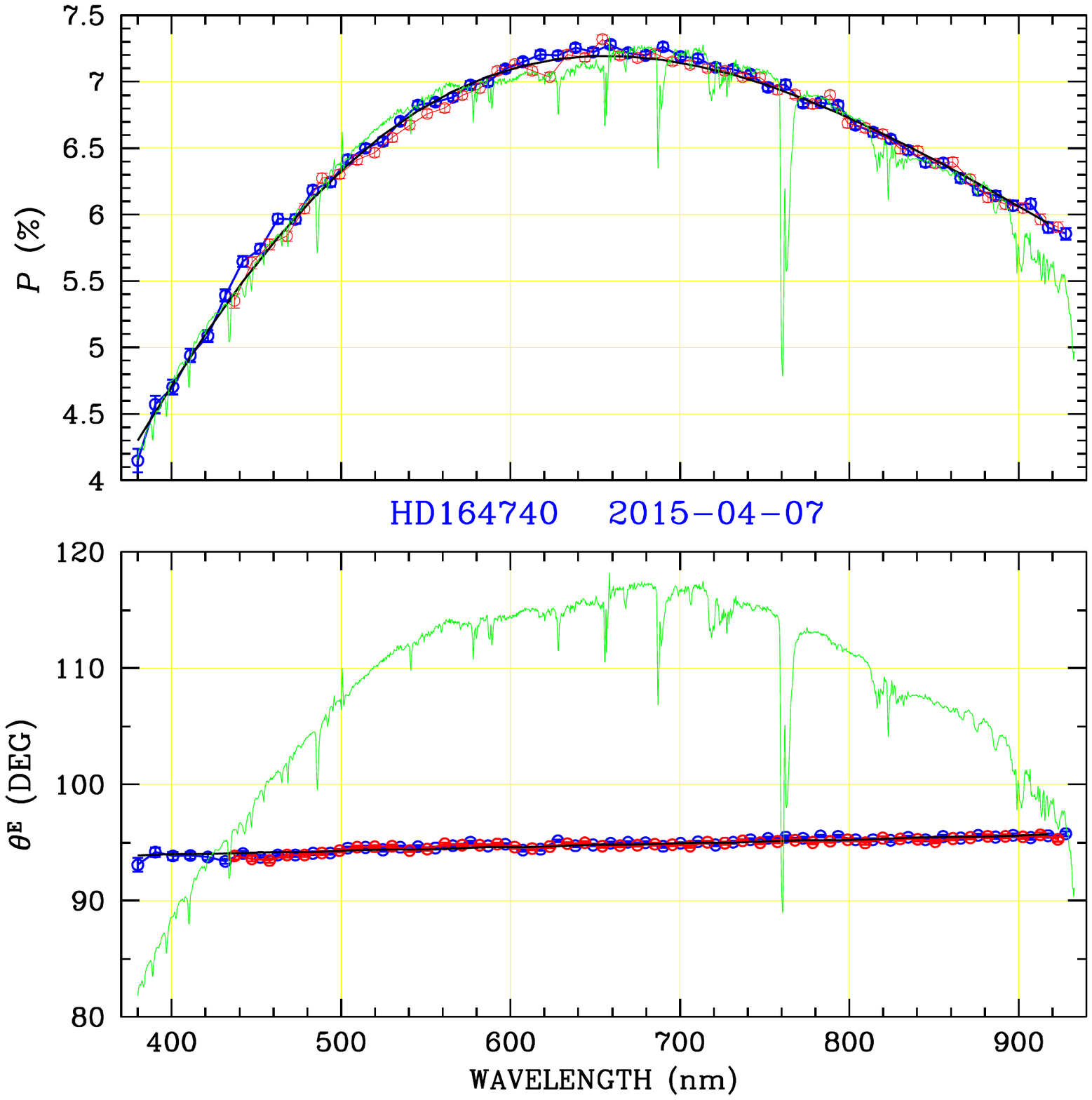}
  \includegraphics*[scale=0.42,trim={1.1cm 6.0cm 0.1cm 2.8cm},clip]{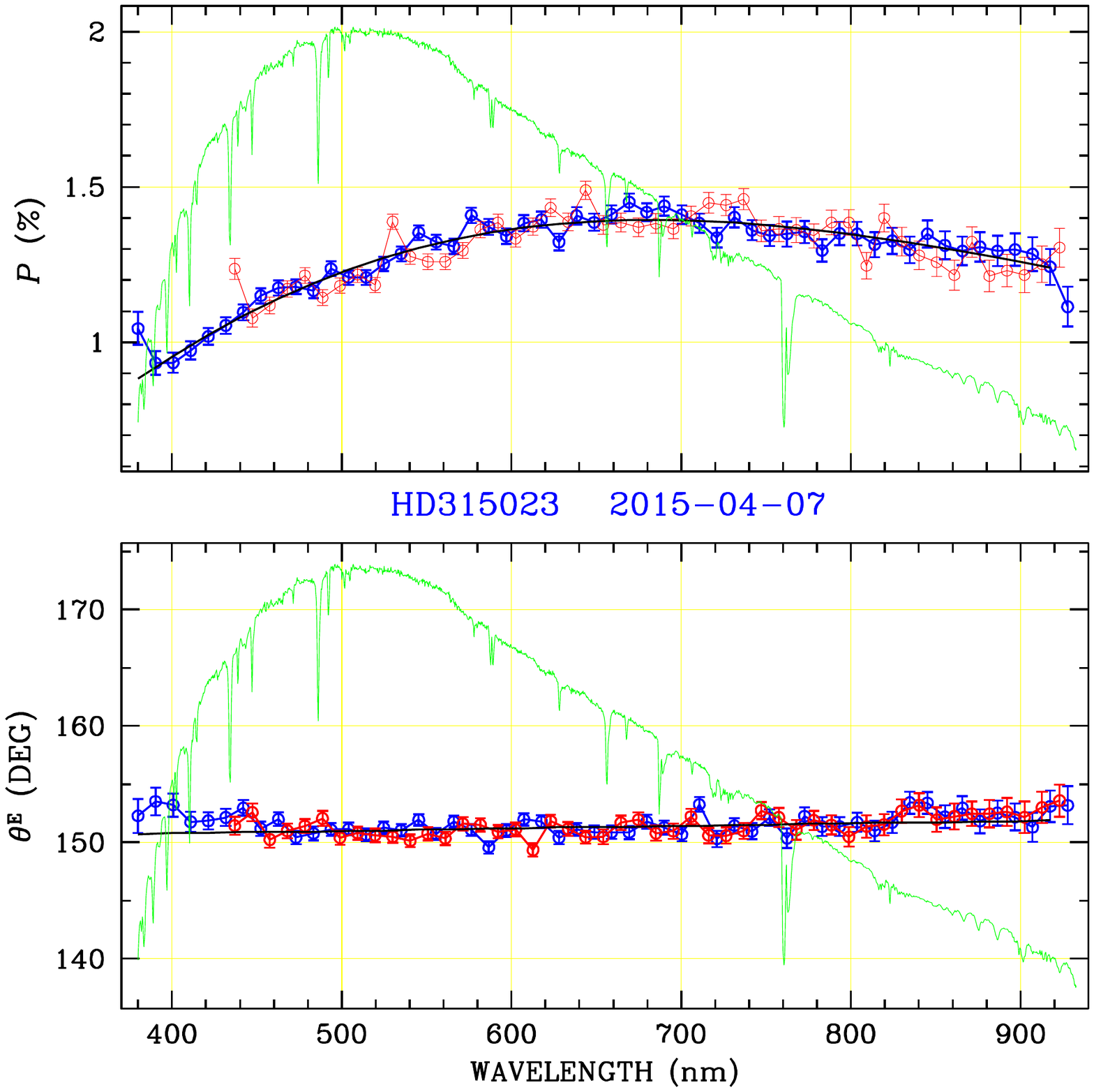}\\
  \includegraphics*[scale=0.42,trim={1.1cm 6.0cm 0.1cm 2.8cm},clip]{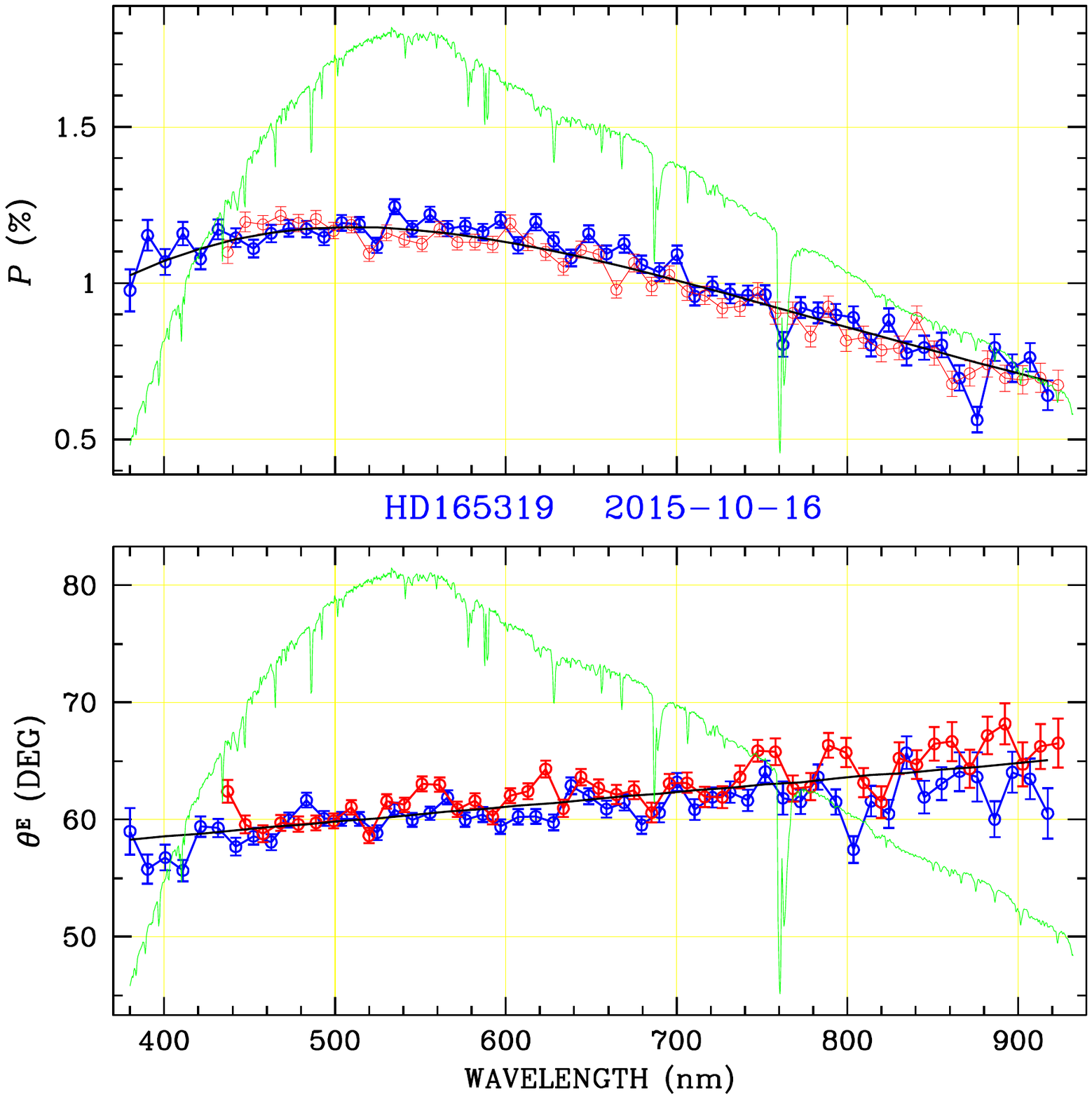}
  \includegraphics*[scale=0.42,trim={1.1cm 6.0cm 0.1cm 2.8cm},clip]{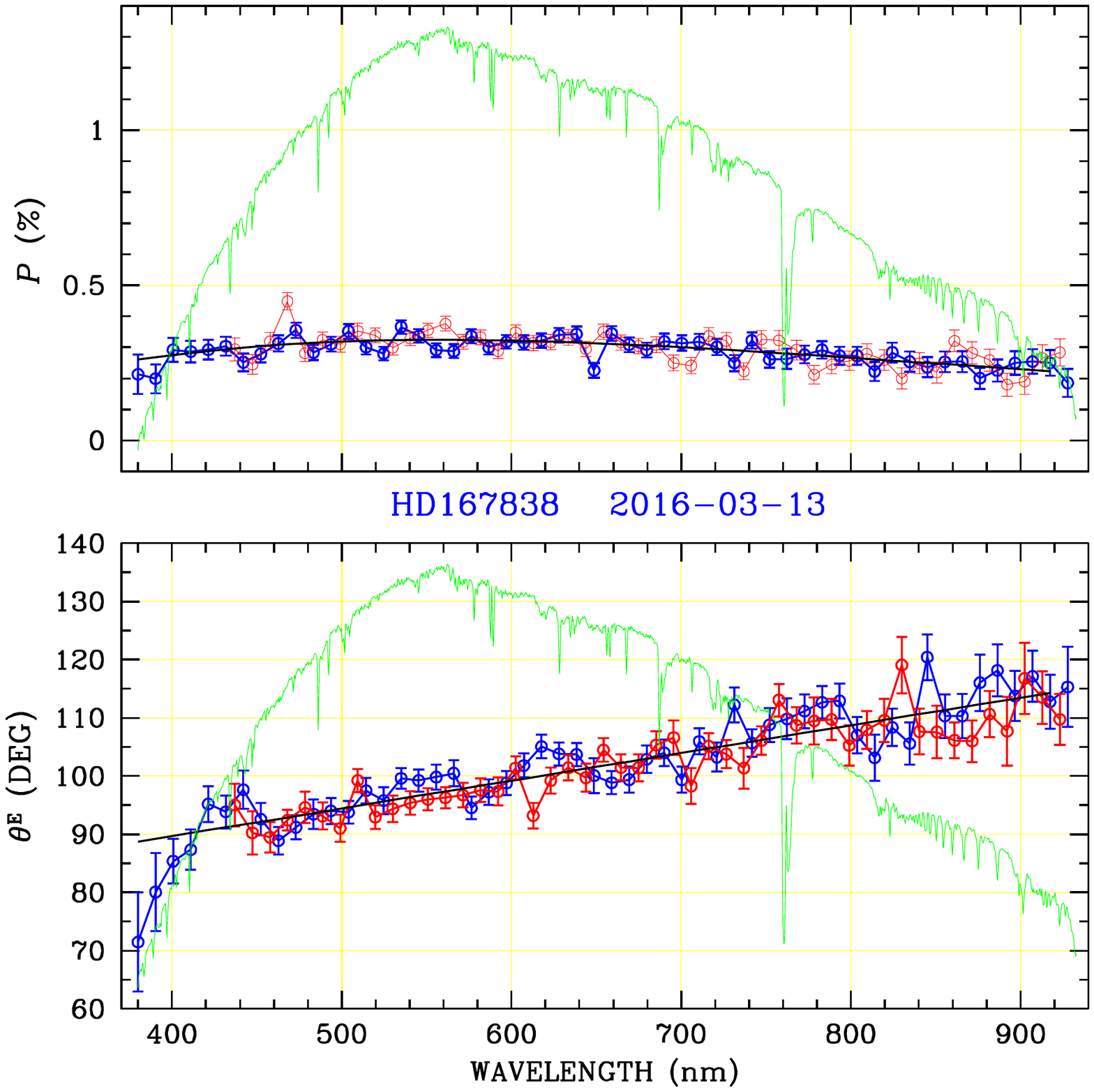}\\
  \includegraphics*[scale=0.42,trim={1.1cm 6.0cm 0.1cm 2.8cm},clip]{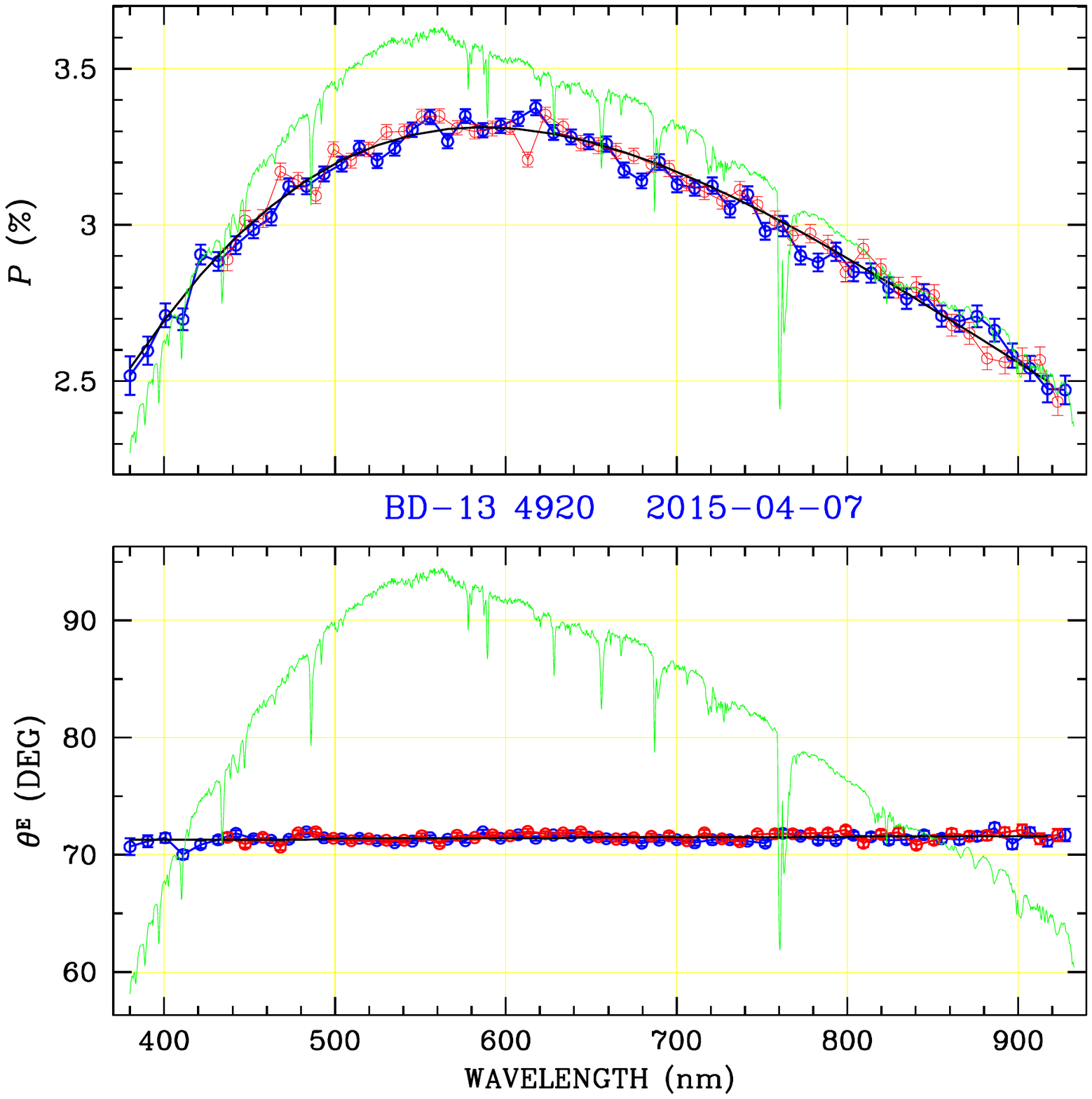}
  \includegraphics*[scale=0.42,trim={1.1cm 6.0cm 0.1cm 2.8cm},clip]{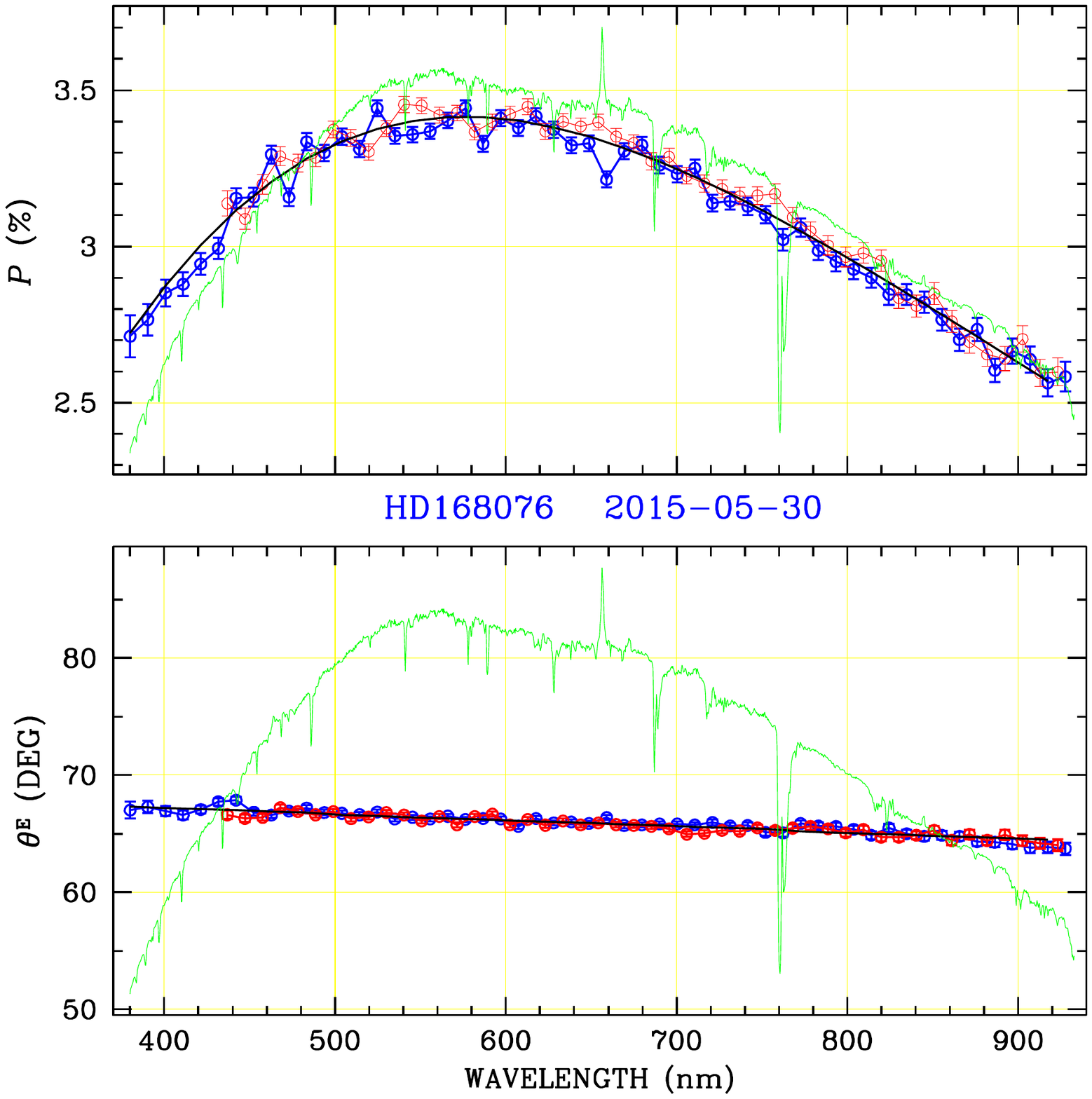}
\newpage

\noindent
  \includegraphics*[scale=0.42,trim={1.1cm 6.0cm 0.1cm 2.8cm},clip]{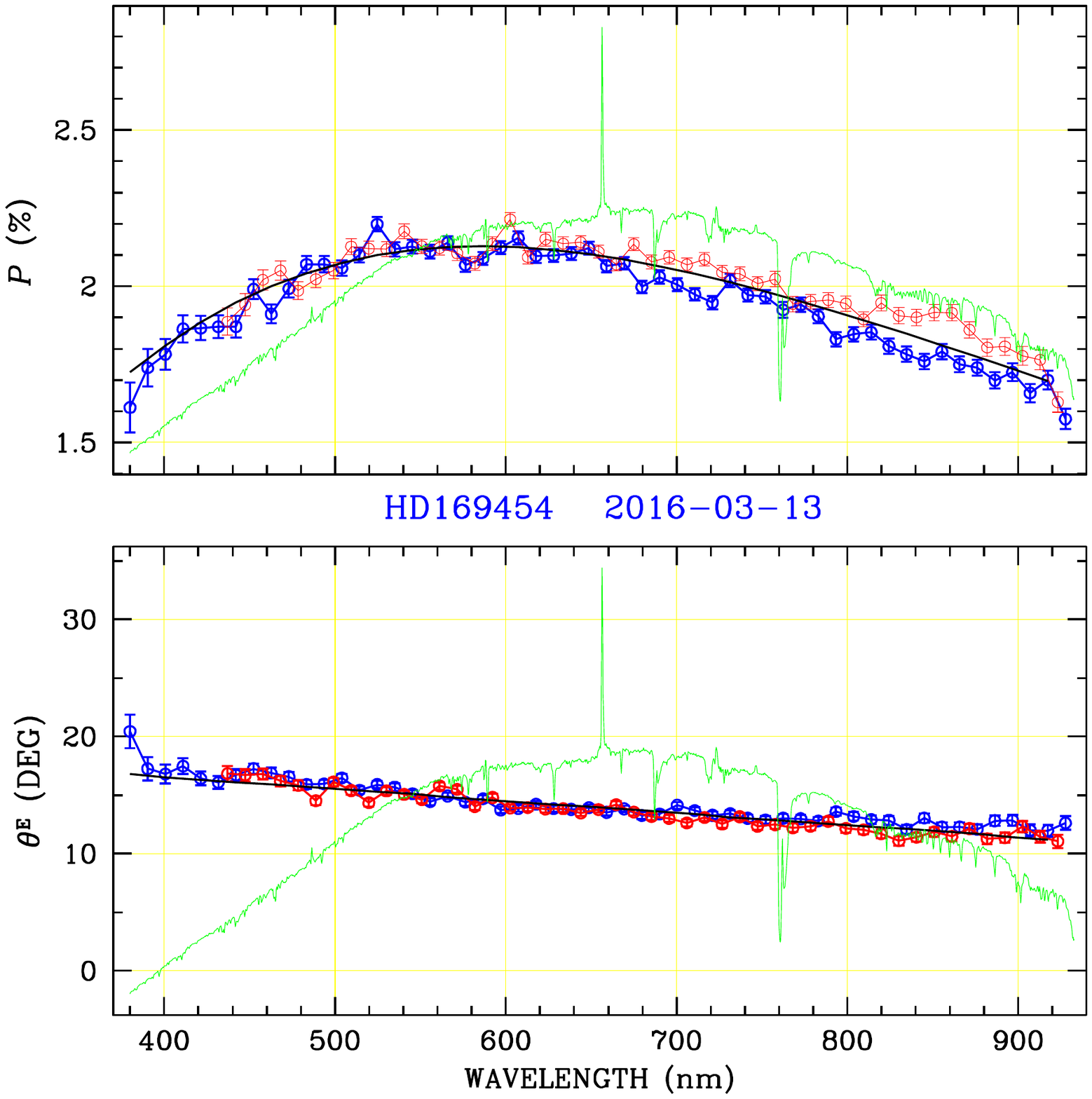}
  \includegraphics*[scale=0.42,trim={1.1cm 6.0cm 0.1cm 2.8cm},clip]{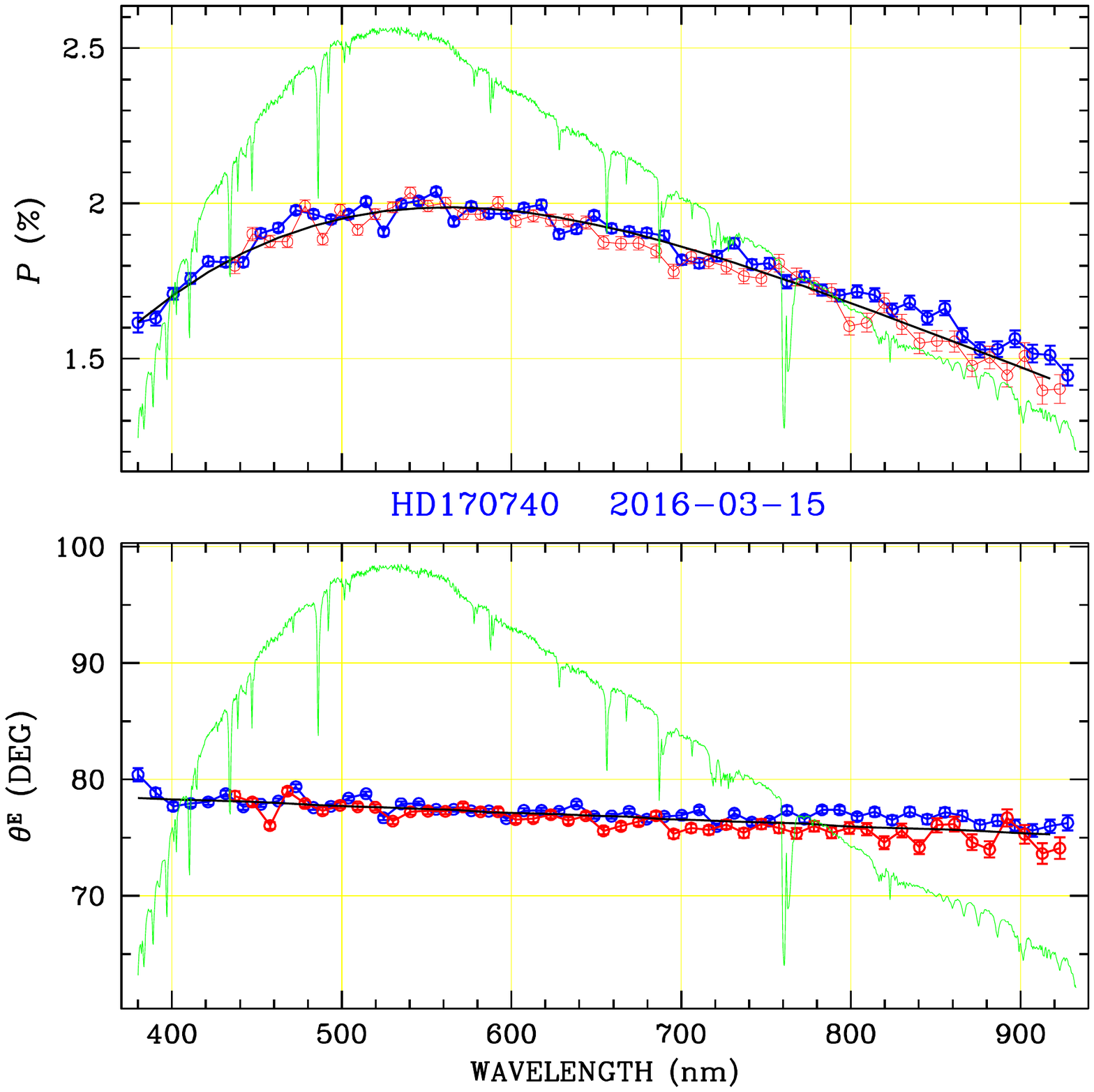}\\
  \includegraphics*[scale=0.42,trim={1.1cm 6.0cm 0.1cm 2.8cm},clip]{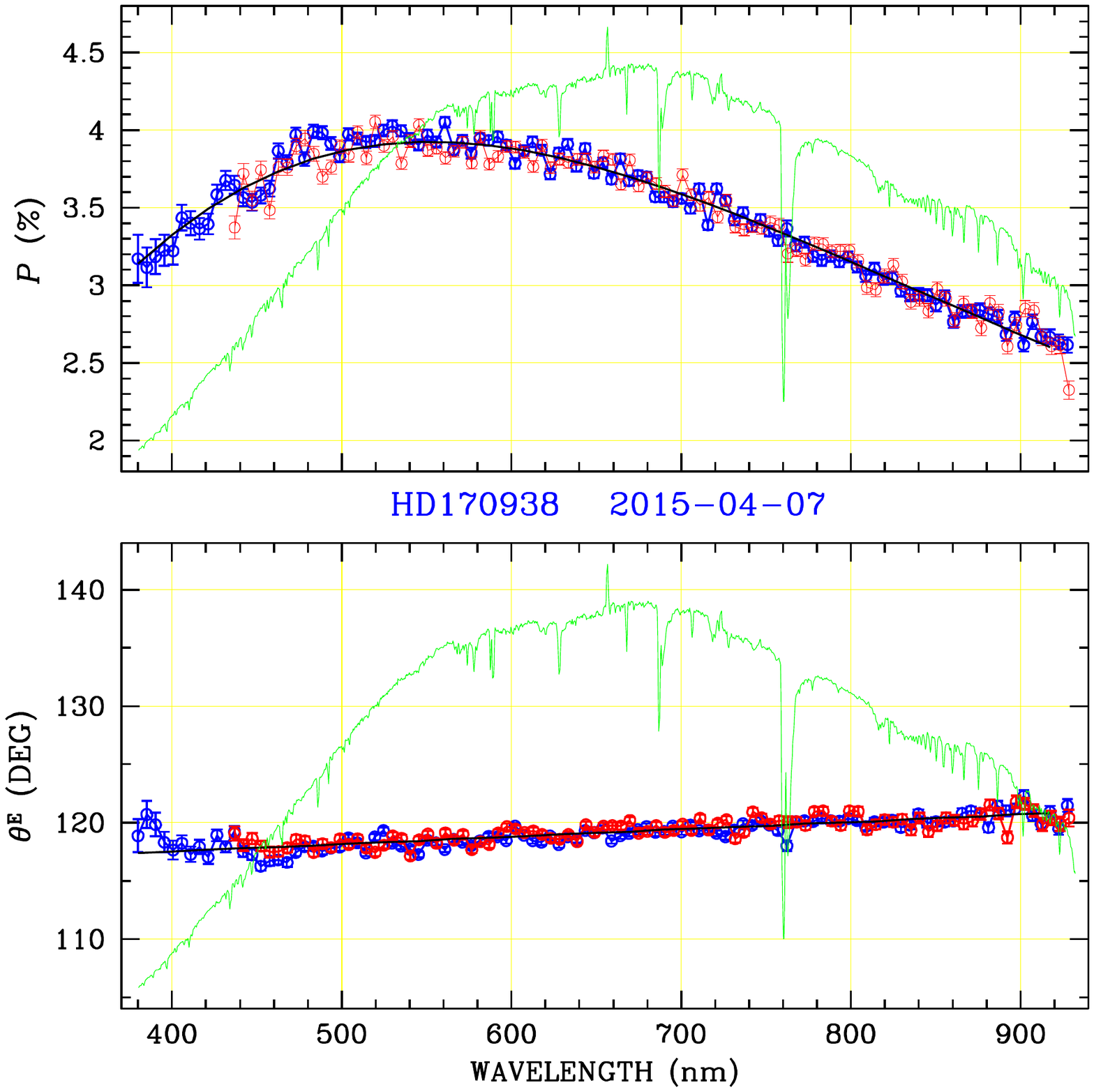}
  \includegraphics*[scale=0.42,trim={1.1cm 6.0cm 0.1cm 2.8cm},clip]{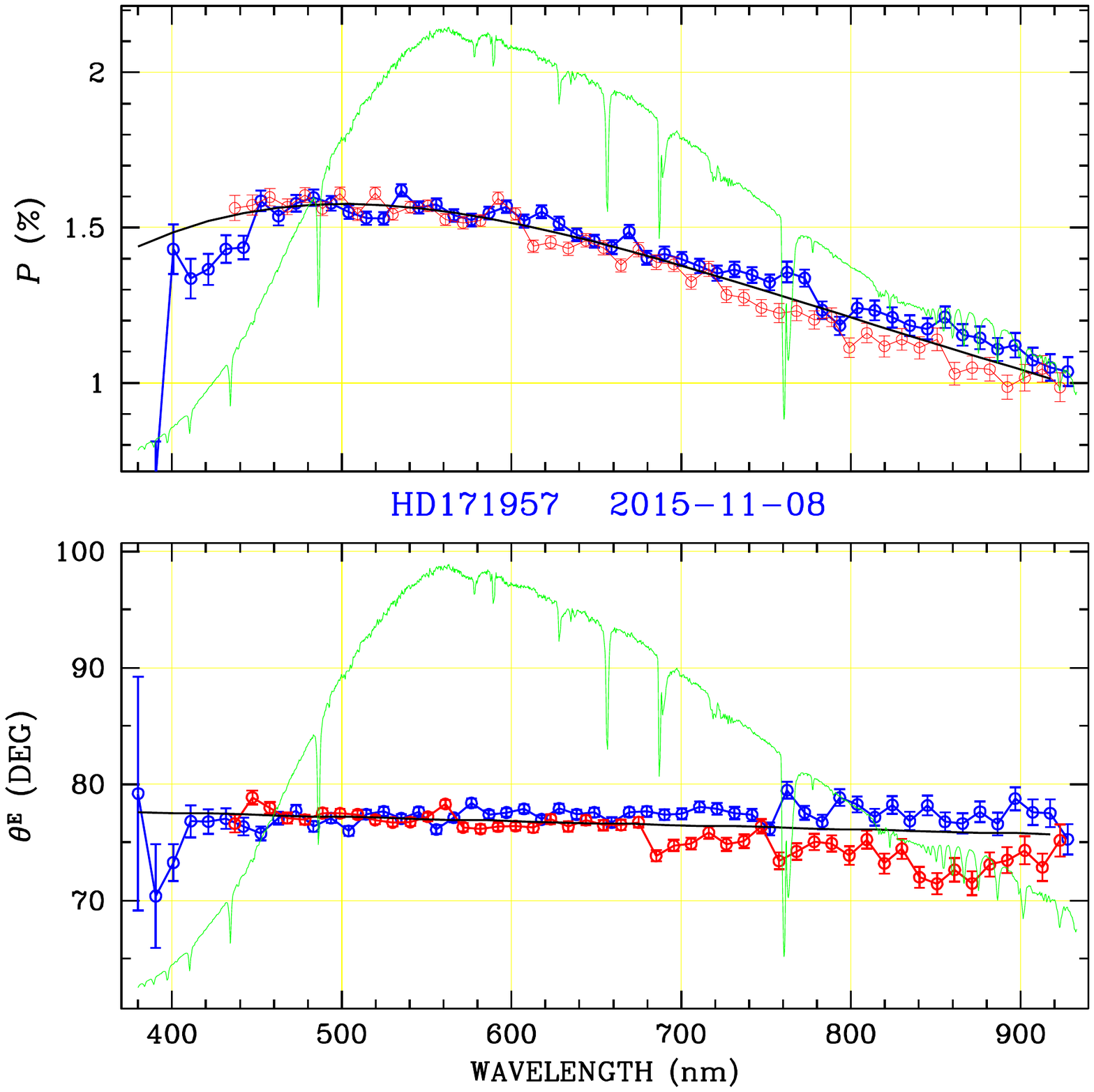}\\
  \includegraphics*[scale=0.42,trim={1.1cm 6.0cm 0.1cm 2.8cm},clip]{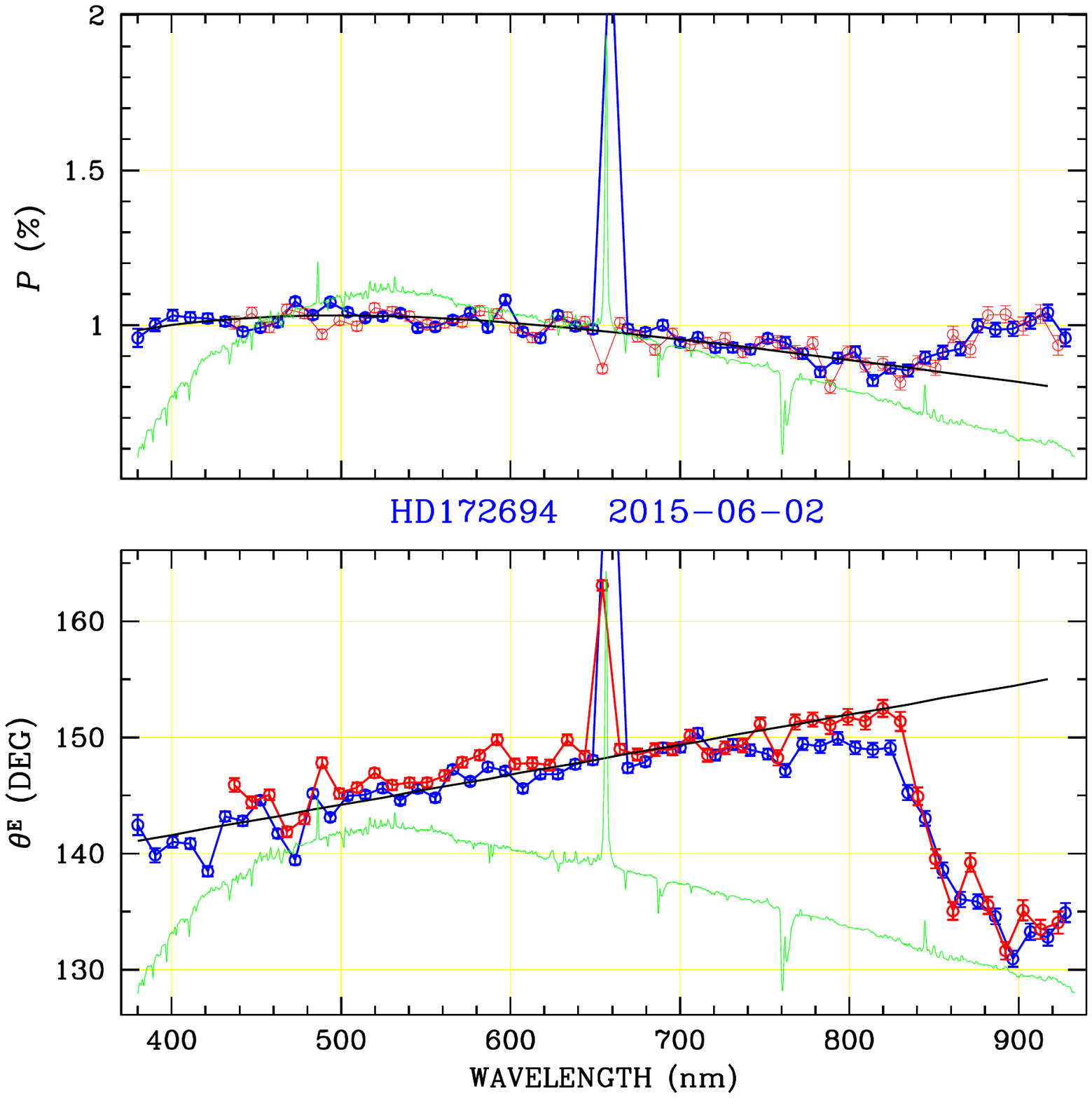}
  \includegraphics*[scale=0.42,trim={1.1cm 6.0cm 0.1cm 2.8cm},clip]{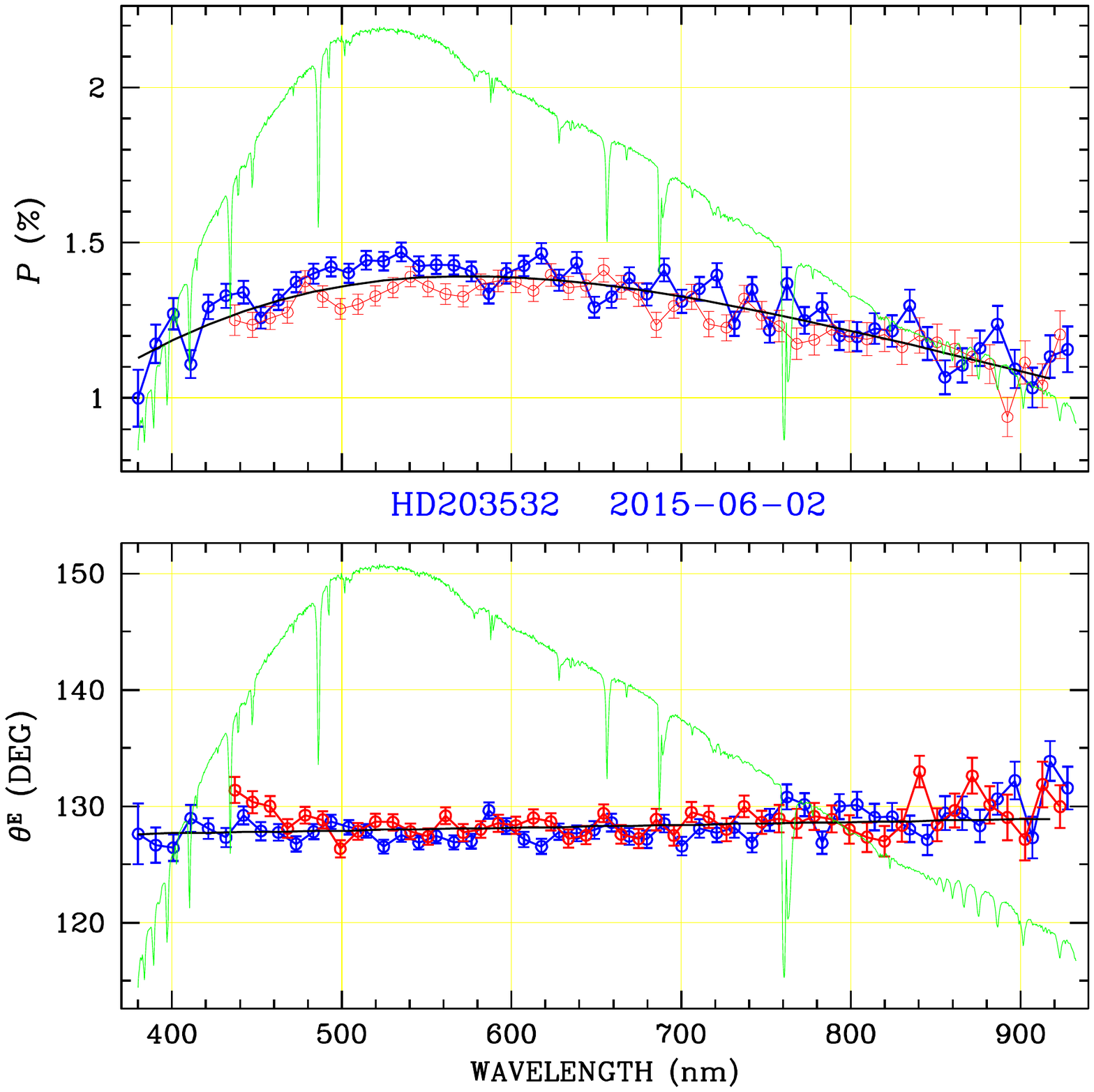}

  \newpage
\noindent  
  \includegraphics*[scale=0.42,trim={1.1cm 6.0cm 0.1cm 2.8cm},clip]{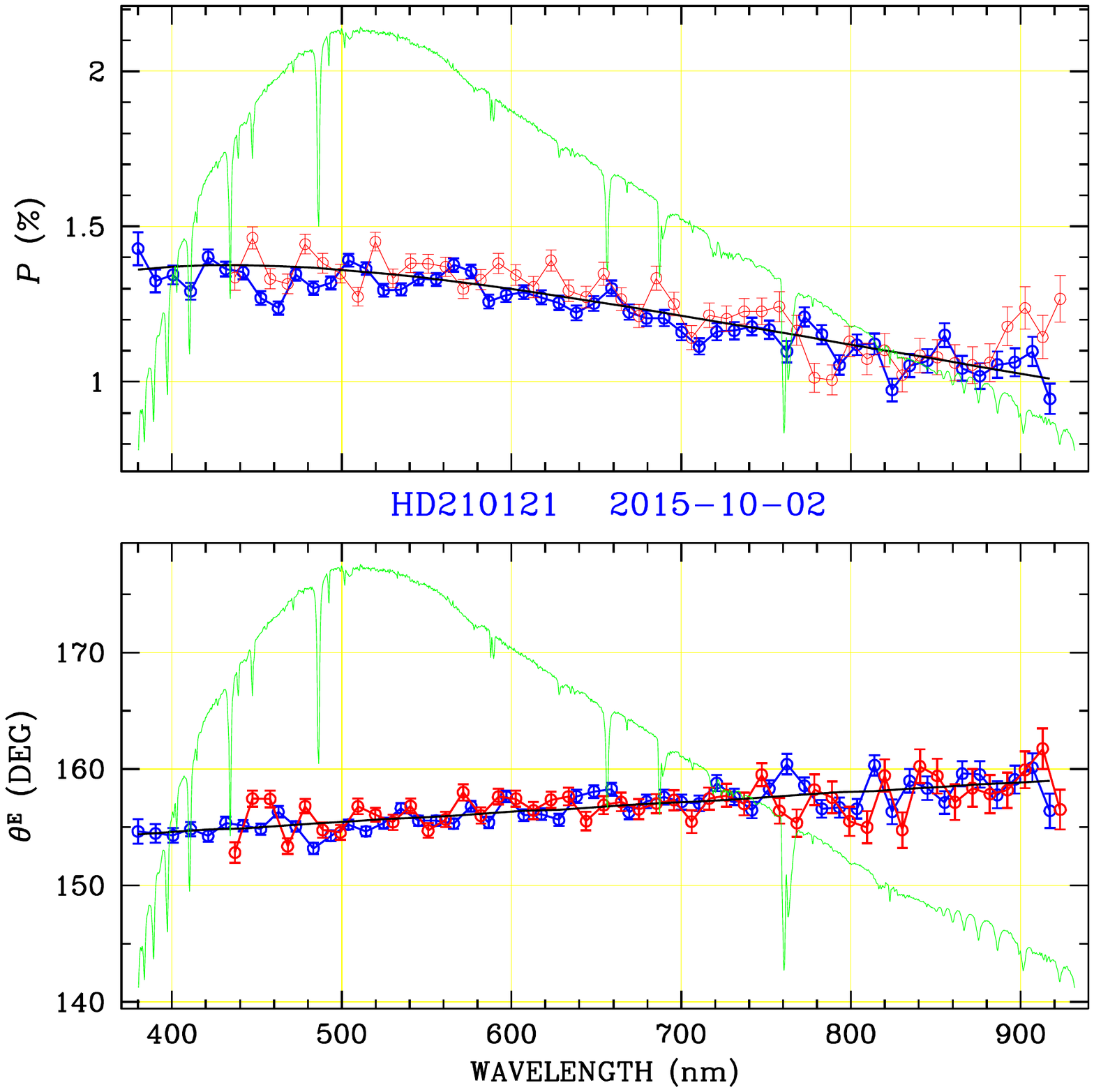}

\end{document}